\documentclass[prb,aps,nofootinbib,amssymb,twocolumn,superscriptaddress,10pt]{revtex4-2}
\usepackage[T1]{fontenc}
\usepackage{graphicx}
\usepackage{changepage}
\usepackage{xcolor}
\usepackage{dcolumn}
\usepackage{comment}
\usepackage{amsmath,amssymb,bbold,bm}
\usepackage{hyperref}
\usepackage{amsfonts}
\usepackage{listings}
\usepackage{braket}
\usepackage{float,latexsym}
\usepackage{multirow}
\usepackage{tikz}
\usepackage{color}
\usepackage{xr}
\usepackage[normalem]{ulem}
\usepackage{bibunits}
\usepackage{cleveref}
\usepackage{physics}
\usepackage{blindtext}
\usepackage{comment}
\usepackage{tabularx}
\usepackage{ltablex}
\usepackage{array}
\usepackage{pifont}
\usepackage[caption=false]{subfig}
\usepackage{siunitx}

\crefname{appsec}{Appendix}{Appendices}
\crefname{equation}{Eq.}{Eqs.}
\crefname{figure}{Fig.}{Figs.}
\crefname{table}{Table}{Tables}
\crefname{section}{Section}{Sections}

\def\ie{{\it i.e.}\ }

\let\vec\mathbf
\newcommand{\normord}[1]{:\mathrel{#1}:}
\newcommand{\cre}[2]{\hat{#1}^\dagger_{#2}}
\newcommand{\des}[2]{\hat{#1}_{#2}}
\newcommand{\expec}[1]{\left\langle {#1} \right\rangle}
\newcommand{\pauliVec}{\cdot \boldsymbol{\sigma}}
\newcommand{\pauliVecEta}{\cdot \boldsymbol{\sigma}^\eta}
\newcommand{\smallk}{\delta \vec{k}}
\newcommand{\bigOrd}[1]{\mathcal{O} \left( {#1} \right)}
\newcommand{\heaviside}[1]{\Theta \left( {#1} \right)}

\newcommand{\cmark}{\ding{51}}
\newcommand{\xmark}{\ding{55}}
\newcommand{\setLayer}{\left \lbrace 1, 3 \right\rbrace}
\newcommand{\UN}[1]{\mathrm{U} \left( #1 \right)}
\newcommand{\SUN}[1]{\mathrm{SU} \left( #1 \right)}

\newlength{\nsubht}

\newsavebox{\nsubbox}
\newcommand{\twoFigure}[4]{%
	\captionsetup[subfloat]{farskip=0pt}%
	\sbox\nsubbox{
		\resizebox{#3}{!}
		{%
			\includegraphics[height=6cm]{#1}%
			\includegraphics[height=6cm]{#2}%
		}%
	}%
	\setlength{\nsubht}{\ht\nsubbox}%
	\centering%
	\subfloat{\label{#4:a}\includegraphics[height=\nsubht]{#1}}%
	\subfloat{\label{#4:b}\includegraphics[height=\nsubht]{#2}}%
}

\newcommand{\twoFigureCBar}[5]{%
	\captionsetup[subfloat]{farskip=0pt}%
	\sbox\nsubbox{
		\resizebox{#4}{!}
		{%
			\includegraphics[height=6cm]{#1}%
			\includegraphics[height=6cm]{#2}%
			\includegraphics[height=6cm]{#3}%
		}%
	}%
	\setlength{\nsubht}{\ht\nsubbox}%
	\centering%
	\subfloat{\label{#5:a}\includegraphics[height=\nsubht]{#1}}%
	\subfloat{\label{#5:b}\includegraphics[height=\nsubht]{#2}}%
	\subfloat{\includegraphics[height=\nsubht]{#3}}%
	\addtocounter{subfigure}{-1}
}

\newcommand{\threeFigureCBar}[6]{%
	\captionsetup[subfloat]{farskip=0pt}%
	\sbox\nsubbox{
		\resizebox{#5}{!}
		{%
			\includegraphics[height=6cm]{#1}%
			\includegraphics[height=6cm]{#2}%
			\includegraphics[height=6cm]{#3}%
			\includegraphics[height=6cm]{#4}%
		}%
	}%
	\setlength{\nsubht}{\ht\nsubbox}%
	\centering%
	\subfloat{\label{#6:a}\includegraphics[height=\nsubht]{#1}}%
	\subfloat{\label{#6:b}\includegraphics[height=\nsubht]{#2}}%
	\subfloat{\label{#6:c}\includegraphics[height=\nsubht]{#3}}%
	\subfloat{\includegraphics[height=\nsubht]{#4}}%
	\addtocounter{subfigure}{-1}
}

\newcommand{\fiveFigureCBar}[8]{%
	\captionsetup[subfloat]{farskip=0pt}%
	\sbox\nsubbox{
		\resizebox{#7}{!}
		{%
			\includegraphics[height=6cm]{#1}%
			\includegraphics[height=6cm]{#2}%
			\includegraphics[height=6cm]{#3}%
			\includegraphics[height=6cm]{#4}%
			\includegraphics[height=6cm]{#5}%
			\includegraphics[height=6cm]{#6}%
		}%
	}%
	\setlength{\nsubht}{\ht\nsubbox}%
	\centering%
	\subfloat{\label{#8:a}\includegraphics[height=\nsubht]{#1}}%
	\subfloat{\label{#8:b}\includegraphics[height=\nsubht]{#2}}%
	\subfloat{\label{#8:c}\includegraphics[height=\nsubht]{#3}}%
	\subfloat{\label{#8:d}\includegraphics[height=\nsubht]{#4}}%
	\subfloat{\label{#8:e}\includegraphics[height=\nsubht]{#5}}%
	\subfloat{\includegraphics[height=\nsubht]{#6}}%
	\addtocounter{subfigure}{-1}
}
\allowdisplaybreaks

\begin{document}
\title{TSTG I: Single-Particle and Many-Body Hamiltonians and Hidden Non-local Symmetries of Trilayer Moir\'e Systems with and without Displacement Field}
\author{Dumitru C\u{a}lug\u{a}ru}%
\thanks{These authors contributed equally.}
\affiliation{Department of Physics, Princeton University, Princeton, New Jersey 08544, USA}
\author{Fang Xie}
\affiliation{Department of Physics, Princeton University, Princeton, New Jersey 08544, USA}
\author{Zhi-Da Song}
\affiliation{Department of Physics, Princeton University, Princeton, New Jersey 08544, USA}
\author{Biao Lian}
\affiliation{Department of Physics, Princeton University, Princeton, New Jersey 08544, USA}
\author{Nicolas Regnault}
\thanks{These authors contributed equally.}
\affiliation{Department of Physics, Princeton University, Princeton, New Jersey 08544, USA}
\affiliation{Laboratoire de Physique de l'Ecole normale superieure, ENS, Universit\'e PSL, CNRS, Sorbonne Universit\'e, Universit\'e Paris-Diderot, Sorbonne Paris Cit\'e, Paris, France}
\author{B. Andrei Bernevig}
\thanks{These authors contributed equally.}
\affiliation{Department of Physics, Princeton University, Princeton, New Jersey 08544, USA}
\date{\today}

\begin{abstract}

We derive the Hamiltonian for trilayer moir\'e systems with the Coulomb interaction projected onto the bands near the charge neutrality point. Motivated by the latest experimental results, we focus on the twisted symmetric trilayer graphene (TSTG) with a mirror-symmetry with respect to the middle layer. We provide a full symmetry analysis of the non-interacting Hamiltonian with a perpendicular displacement field coupling the band structure made otherwise of the twisted bilayer graphene (TBG) and the high velocity Dirac fermions, and we identify a hidden non-local symmetry of the problem. In the presence of this displacement field, we construct an approximate single-particle model, akin to the tripod model for TBG, capturing the essence of non-interacting TSTG. We also derive more quantitative perturbation schemes for the low-energy physics of TSTG with displacement field, obtaining the corresponding eigenstates. This allows us to obtain the Coulomb interaction Hamiltonian projected in the active band TSTG wavefunctions and derive the full many-body Hamiltonian of the system. We also provide an efficient parameterization of the interacting Hamiltonian. Finally, we show that the discrete symmetries at the single-particle level promote the $\UN{2} \times \UN{2}$ spin-valley symmetry to enlarged symmetry groups of the interacting problem under different limits. The interacting part of the Hamiltonian exhibits a large $\UN{4}\times \UN{4} \times \UN{4} \times \UN{4}$ symmetry in the chiral limit. Moreover, by identifying a new symmetry which we dub spatial many-body charge conjugation, we show that the physics of TSTG is symmetric around charge neutrality.
 
\end{abstract}

\maketitle

\section{Introduction}\label{sec:introduction}

As a result of its chemical versatility, an impressive number of stable carbon allotropes has been synthesized and investigated. One of the newest addition to the family, twisted bilayer graphene (TBG), has generated a lot of excitement in the condensed matter community. The resulting van der Waals heterostructure obtained by stacking two graphene layers with a small relative twist has been theoretically shown to host flat bands at certain so-called magic angles~\cite{LOP07,SUA10,BIS11}. Subsequent experimental studies have revealed various correlated insulating and superconducting phases in TBG near the first magic angle $\theta_{\mathrm{TBG}} \approx 1.05 ^\circ$, using both transport~\cite{CAO18,CAO20,CHE20a,LIU20a,LU19,LU20,PAR20a,POL19,SAI20,SAI21,SER20,STE20,WU20a,YAN19} and spectroscopy~\cite{CHO19,CHO20,KER19,NUC20,WON20,XIE19,JIA19,CHO21} experiments. In turn, these findings have inspired a wealth of theoretical investigations into the rich physics of TBG~\cite{KAN19,SEO19,BUL20,HEJ20,FER21,FER20,VEN18,POT21,ABO20,AHN19,BER20,BER20a,BER20b,BUL20a,CAO20b,CEA20,CHR20,CLA19,DA19,DA20,DAI16,DOD18,EFI18,EUG20,GON19,GUI18,GUO18,HEJ19,HEJ19a,HUA19,HUA20a,ISO18,JAI16,JUL20,KAN18,KAN20a,KEN18,KHA20,KON20,KOS18,LED20,LEW20,LIA19,LIA20,LIA20a,LIU12,LIU18,LIU19,LIU21,LIU21a,OCH18,PAD20,PEL18,PO18,PO19,REP20,REP20a,ROY19,SOE20,SON19,SON20b,TAR19,THO18,UCH14,VAF20,VEN18,WAN20,WIJ15,WIL20,WU18,WU19,WU19a,WU20b,XIE20,XIE20a,XIE20c,XIE20d,XU18,XU18b,YOU19,YUA18,ZHA20,ZOU18}.

Such progress on both the experimental and theoretical fronts has triggered a large effort into extending the family of moir\'e superlattices, promoting them as some of the most promising platforms to engineer strongly correlated quantum phases~\cite{KEN20}. The main driving force in investigating moir\'e materials beyond TBG is often the different band tunability properties of the former. Consequently, the extension to twisted multilayer graphene has already been widely studied theoretically~\cite{BUR19,CAO20b,CAR20,CAR20a,CEA19,GAR20,HAD20,KHA19,LEI20,LI19,LIU19a,LOP20,MOR19,PAR20b,SUA13,WU20c,ZHA20a,ZHU20,ZHU20a}. Later experiments have also revealed equally intriguing superconducting and insulating phases in moir\'e systems with three~\cite{CHE19,CHE19a,CHE20b,HAO21,PAR21,POL20,SHI20,TSA20} or four~\cite{BUR19,BUR20,CAO20a,LIU20b,SHE20} graphene layers.

Among the simplest moir\'e graphene systems beyond TBG, twisted symmetric trilayer graphene (TSTG)~\cite{MOR19,KHA19} has been recently experimentally realized in Refs.~\cite{PAR21,HAO21}. TSTG is comprised of three AAA-stacked graphene layers in which the middle layer is twisted slightly relative to the top and bottom ones. For this type of stacking, which was shown to be energetically favorable~\cite{CAR20}, the system is mirror-symmetric with respect to reflections in the plane of the middle graphene layer. As such, TSTG  decouples into mirror-symmetry sectors in the absence of interactions~\cite{KHA19} and can be thought of as being comprised of a ``TBG-like'' contribution with an interlayer coupling effectively enhanced by a factor of $\sqrt{2}$~\cite{KHA19}, and a high-velocity Dirac fermion~\cite{CAR20}. The renormalized interlayer coupling of the TBG fermions leads to a rescaling of the first magic angle by the same amount, yielding $\theta_{\mathrm{TSTG}}\approx 1.56^\circ$ in agreement with the recent experimental observations~\cite{PAR21,HAO21}. However, despite being independent at the single-particle level, the two mirror-symmetry sectors of TSTG are coupled by the electron-electron interactions, pointing to a potentially richer correlated physics compared to TBG. Moreover, the TBG and Dirac cone contributions can be hybridized by the application of a perpendicular displacement field~\cite{CAR20,HAO21,LEI20,PAR21}. This provides another knob to experimentally tune the TSTG band structure.

To unveil the above-mentioned richness, we here investigate both the single-particle Bistritzer-MacDonald model and the interaction Coulomb Hamiltonian for TSTG at the first magic angle, with or without displacement field. The main result of this paper is to derive expressions and effective models, as well as the symmetries of the interacting TSTG Hamiltonian under different limits. For this purpose, we discuss the discrete symmetries of the single-particle problem and show how they promote the $\UN{2} \times \UN{2}$ valley-spin rotation symmetry to enhanced rotation symmetries of the interacting problem. We uncover new non-local hidden symmetries of the system at both the single-particle and many-body level. At the same time, we also provide a series of approximations for the single-particle energy spectrum of TSTG in the presence of displacement field and show how it can be obtained in terms of the TBG flat band wave functions, whose properties have been extensively studied in Refs.~\cite{SON19,BER20,BER20a,SON20b}. Despite the addition of a Dirac degree of freedom, we find the symmetries of the many-body TSTG Hamiltonian to be enhanced from those of TBG.

The article is organized as follows. In \cref{sec:singlehamiltonian}, we review the single-particle TSTG Hamiltonian and derive a low-energy approximation. We then investigate its symmetries (including the hidden non-local symmetries) in \cref{sec:symmetries} under various limits with or without displacement field. \Cref{sec:singleparticlespectrum} focuses on the single-particle energy spectrum. We show that an approximate tripod model correctly captures the salient features of TSTG and we derive the single-particle projected Hamiltonian. \Cref{sec:interaction} is devoted to the interacting Hamiltonian, deriving the expression of the projected Coulomb interaction for the TSTG model. Finally, we discuss in \cref{sec:mbSym} the symmetries of the fully interacting projected TSTG Hamiltonian in several limits.

\section{Single-particle Hamiltonian}\label{sec:singlehamiltonian}

First, we outline the derivation of a Bistritzer-MacDonald model for TSTG~\cite{BIS11}. A more detailed exposition is provided in Appendix \ref{app:single_part_ham:a}. The main result of this section is to show that the TSTG Hamiltonian can be thought as a sum between a TBG Hamiltonian (with renormalized interlayer hopping amplitudes) and an independent Dirac cone Hamiltonian. Furthermore, we show that the hybridization between the TBG and Dirac cone fermions can be tuned by the addition of a perpendicular displacement field~\cite{HAO21,PAR21}.  

\subsection{Notations}\label{sec:singlehamiltonian:notation}
\begin{figure}[!t]
	\twoFigure{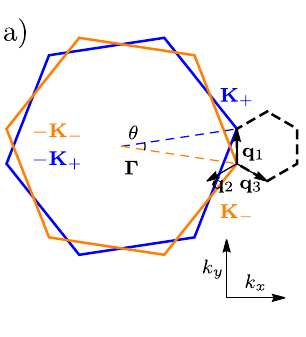}{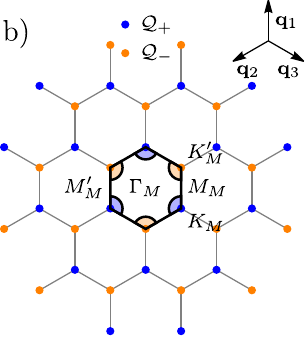}{\columnwidth}{fig:mbz_qlattice}
	\caption{The Moir\'e Lattice of TSTG. Panel (a) illustrates the BZs of the graphene layers, which are plotted in blue for the top ($l=3$) and bottom ($l=1$) layers and in orange for the middle layer ($l=2$). The $K$ ($K'$) point are located at $\vec{K}_+$ ($-\vec{K}_+$) for the top and bottom layer and at $\vec{K}_-$ ($-\vec{K}_-$). When the twist angle $\theta$ is small, an approximate translation symmetry arises, allowing us to define the MBZ (dashed black hexagon). The $\mathcal{Q}_{\pm}$ lattices are shown in panel (b). Inside the first MBZ (defined as the hexagonal region around $\Gamma_M$) we have plotted the regions $A_{\eta}^i$ defined in \cref{eqn:Dirac_zones} as filled blue ($\eta = +$) and orange ($\eta = -$) circular sectors.}
	\label{fig:mbz_qlattice}
\end{figure}
In the case of graphene, the twisted trilayer geometry was considered theoretically in Refs.~\cite{MOR19,KHA19}. Throughout this paper, however, we will follow the notation of Refs.~\cite{SON19,BER20,SON20b,BER20a,LIA20,BER20b,XIE20a}. We take $\cre{a}{\vec{p},\alpha,s,l}$ to represent the fermion operator in the plane wave basis for graphene layer $l=1,2,3$ (corresponding to the bottom, middle, and top layers, respectively). The momentum $\vec{p}$ is measured from the $\Gamma$ point of the monolayer graphene Brillouin Zone (BZ), as shown in \cref{fig:mbz_qlattice:a}, $\alpha=A,B$ is the sublattice index, and $s=\uparrow,\downarrow$ denotes the projection of the electron spin along the $\hat{z}$ direction. Within each graphene layer, the low-energy physics is concentrated around the two valleys, $K$ and $K'$, labeled by $\eta=\pm 1$ and located at momenta $\eta \vec{K}_{l}$. Owing to the mirror-symmetric arrangement of the graphene layers, we can introduce $\vec{K}_{+} \equiv \vec{K}_{1} =  \vec{K}_{3}$ to be the $K$ point in the bottom and top layer graphene BZ ($l=1,3$), and $\vec{K}_{-} \equiv \vec{K}_{2}$, to be the $K$ point of the middle layer graphene BZ ($l=2$). 

For convenience, we define the momenta $\vec{q}_{j} = C_{3z}^{j-1} \left(\vec{K}_{+}-\vec{K}_{-}\right)$, where $j=1,2,3$ and $C_{3z}$ represents the three-fold rotation transformation around the $\hat{z}$ axis. We can then define a moir\'e BZ (MBZ) for the TSTG moir\'e lattice $\mathcal{Q}_{0}=\mathbb{Z}\mathbf{b}_{M1}+\mathbb{Z}\mathbf{b}_{M2}$, which is generated by the reciprocal vectors $\vec{b}_{Mi}=\vec{q}_3-\vec{q}_i $ ($i=1,2$). We also define two shifted momentum lattices $\mathcal{Q}_{\pm}=\pm\mathbf{q}_{1}+\mathcal{Q}_{0}$, which together form a honeycomb lattice, as seen in \cref{fig:mbz_qlattice:b}. We can then introduce the low-energy fermion operators defined on the moir\'e lattice as $\cre{a}{\vec{k},\vec{Q},\eta,\alpha,s,l} \equiv \cre{a}{\eta \vec{K}_l + \vec{k} - \vec{Q}, \alpha, s, l}$ for $ \vec{Q} \in \mathcal{Q}_{\eta,l}$ with $\vec{k}$ measured from the $\Gamma_{M}$ point, and $\mathcal{Q}_{\eta,l} = \mathcal{Q}_{\eta}$ for $l=1,3$ and $\mathcal{Q}_{\eta,l} = \mathcal{Q}_{-\eta}$ for $l=2$. 

The expression of the TSTG single-particle Hamiltonian in terms of the $\cre{a}{}$ operators given in \cref{si:eqn:singlePart1} of Appendix \ref{app:single_part_ham:a} can be simplified by introducing a basis transformation: in the absence of a perpendicular displacement field, a TSTG sample is symmetric under mirror $m_z$ reflections with respect to the middle graphene layer plane. This allows us to define a set of mirror-symmetric and mirror-antisymmetric operators, which are respectively given by
\begin{equation}
	\label{eqn:ckQ}
	\cre{c}{\vec{k},\vec{Q},\eta,\alpha,s}=\begin{cases}
		\frac{1}{\sqrt{2}} \left( \cre{a}{\vec{k},\vec{Q},\eta,\alpha,s,3} + \cre{a}{\vec{k},\vec{Q},\eta,\alpha,s,1}  \right) & \vec{Q} \in \mathcal{Q}_{\eta} \\
		\cre{a}{\vec{k},\vec{Q},\eta,\alpha,s,2} & \vec{Q} \in \mathcal{Q}_{-\eta}
	\end{cases},
\end{equation}
and
\begin{equation}
	\label{eqn:bkQ+}
	\cre{b}{\vec{k},\vec{Q},\eta,\alpha,s} = \frac{1}{\sqrt{2}} \left( \cre{a}{\vec{k},\vec{Q},\eta,\alpha,s,3} - \cre{a}{\vec{k},\vec{Q},\eta,\alpha,s,1}  \right)  \vec{Q} \in \mathcal{Q}_{\eta}.
\end{equation}

\subsection{Hamiltonian}\label{sec:singlehamiltonian:hamiltonian}

When written in with the aid of the $\cre{b}{}$ and $\cre{c}{}$ operators, the single-particle Hamiltonian can be separated into three terms
\begin{equation}
	\label{eqn:singlePartTSTG}
	\hat{H}_{0} = \hat{H}_{\mathrm{TBG}} + \hat{H}_{D} + \hat{H}_{U}.
\end{equation}
In \cref{eqn:singlePartTSTG}, the mirror-symmetric low-energy operators give rise to the term
\begin{equation}
	\label{eqn:singlePartTBG}
	\hat{H}_{\mathrm{TBG}} = \sum_{\substack{\vec{k} \in \text{MBZ} \\ \eta, \alpha, \beta, s \\ \vec{Q},\vec{Q}' \in \mathcal{Q}_{\pm}}} \left[h^{\left(\eta\right)}_{\vec{Q},\vec{Q}'} \left( \vec{k} \right) \right]_{\alpha \beta} \cre{c}{\vec{k},\vec{Q},\eta,\alpha,s} \des{c}{\vec{k},\vec{Q}',\eta,\beta,s},
\end{equation}
which is similar to the ordinary TBG Hamiltonian~\cite{SON19,KHA19}, but with a tunneling amplitude which is rescaled by a factor of $\sqrt{2}$, corresponding to
\begin{equation}
	\label{eqn:firstQuantSinglePartTBG}
	h^{\left(\eta\right)}_{\vec{Q},\vec{Q}'} \left( \vec{k} \right) = h^{D,\eta}_{\vec{Q}} \left( \vec{k} \right) \delta_{\vec{Q},\vec{Q}'} + \sqrt{2} h^{I,\eta}_{\vec{Q},\vec{Q}'}.
\end{equation}
The first-quantized Hamiltonians $h^{D,\eta}_{\vec{Q}} \left( \vec{k} \right)$ and $h^{I,\eta}_{\vec{Q},\vec{Q}'}$ from \cref{eqn:firstQuantSinglePartTBG}, whose exact forms are given in Appendix \ref{app:single_part_ham:a}, denote a Dirac cone contribution with Fermi velocity $v_F$ folded inside the first MBZ and an interlayer hopping term, respectively. In particular, there are two parameters $w_0$ and $w_1$ in $h^{I,\eta}_{\vec{Q},\vec{Q}'}$, which correspond to the interlayer hoppings at the $AA$ and $AB / BA$ stacking centers, respectively. Generically, one has $0 \leq w_0 < w_1$ due to lattice relaxation and corrugation effects~\cite{UCH14,WIJ15,DAI16,JAI16,SON20b}. At the same time, the mirror-symmetric operators, which are only defined for $\vec{Q} \in \mathcal{Q}_{\eta}$, correspond to a solitary Dirac cone contribution
\begin{equation}
	\label{eqn:singlePartDirac}
		\hat{H}_{D} = \sum_{\substack{\vec{k} \in \text{MBZ} \\ \eta, \alpha, \beta, s}} \sum_{\vec{Q} \in \mathcal{Q}_{\eta}} \left[h^{D,\eta}_{\vec{Q}} \left( \vec{k} \right) \right]_{\alpha \beta} \cre{b}{\vec{k},\vec{Q},\eta,\alpha,s} \des{b}{\vec{k},\vec{Q},\eta,\beta,s}.
\end{equation}
Additionally, in \cref{eqn:singlePartTSTG}, we have introduced a perpendicular displacement field, which is equivalent to an onsite potential of $U/2$, $0$, $-U/2$ in the top, middle, and bottom layers, respectively. The displacement field contribution couples the TBG-like and the Dirac cone fermions giving rise to 
\begin{equation}
	\label{eqn:singlePartDisplacement}
		\hat{H}_{U} = \sum_{\substack{\vec{k} \in \text{MBZ} \\ \eta, \alpha, s}} \sum_{\vec{Q} \in \mathcal{Q}_{\eta}} \frac{U}{2} \left( \cre{b}{\vec{k},\vec{Q},\eta,\alpha,s} \des{c}{\vec{k},\vec{Q},\eta,\alpha,s} + \mathrm{h.c.} \right),
\end{equation}
which explicitly breaks the mirror $m_z$ symmetry. In what follows, we will find it convenient to employ dimensionless units in which momentum ($\vec{k}$) and energy ($E$) are rescaled according to 
\begin{equation}
	\label{eqn:nonDimRescale}
	\vec{k} \to \frac{\vec{k}}{k_{\theta}}, \qquad
	E \to \frac{E}{v_F k_{\theta}},
\end{equation}
where $k_{\theta} = \abs{\vec{K}_+ - \vec{K}_-}$. This essentially amounts to setting $v_F = 1$, as well as $\abs{\vec{q}_i} = 1$ ($i=1,2,3$).

\subsection{Low-energy approximation}\label{sec:singlehamiltonian:approximation}
The low-energy physics of TSTG with displacement field near the magic angle arises from the interplay between the almost flat (\ie with a bandwidth much smaller than one, in non-dimensional units) bands of $\hat{H}_{\mathrm{TBG}}$ and the MBZ-folded high-velocity Dirac cone bands of $\hat{H}_{D}$. The only states of $\hat{H}_D$ which can efficiently perturb and hybridize the flat-band modes of the TBG-like sector are the ones which have an energy significantly smaller than one. As a low-energy approximation, we can thus restrict ourselves to the momentum points where $\abs{h_{\vec{Q}}^{D,\eta}\left( \vec{k} \right)} \ll 1$ in \cref{eqn:singlePartDirac}, which is equivalent to $\vec{Q} \in \left\lbrace \eta \vec{q}_i \right\rbrace$ and $\vec{k}$ belonging to one of the three zones $A^{i}_{\eta}$ (where $i=1,2,3$) defined for each valley $\eta$ as
\begin{equation}
	\label{eqn:Dirac_zones}
	A^{i}_{\eta} = \left\lbrace \vec{k} \in \mathrm{MBZ} \mid \abs{\vec{k} - \vec{\eta q_i}} \leq \Lambda  \right\rbrace. 
\end{equation}
Effectively, we consider the Dirac cone contribution in the MBZ only within a small distance $\Lambda$ from the Dirac points of $\hat{H}_D$, as shown in \cref{fig:mbz_qlattice:b}. Typically, the cutoff $\Lambda$ is smaller than the gap between the TBG active and passive bands, but bigger than the bandwidth of the flat bands of $\hat{H}_{\mathrm{TBG}}$. For $0 \leq U \leq 0.3$, we find that $\Lambda \leq 0.2$ (see Appendix \ref{app:approx_single_part:spec}). With these approximations, we can write the Dirac cone Hamiltonian projected into the low-energy degrees of freedom as 
\begin{equation}
	\label{eqn:Dirac_approx}
		H_D = \sum_{\eta, \alpha, \beta, s} \sum_{i=1}^3 \sum_{\vec{k} \in A^{i}_{\eta}} \left[h^{D,\eta}_{\eta \vec{q}_i} \left( \vec{k} \right) \right]_{\alpha \beta} \cre{b}{\vec{k},\eta \vec{q}_i,\eta,\alpha,s} \des{b}{\vec{k},\eta \vec{q}_i,\eta,\beta,s},
\end{equation} 
which is denoted without the ``hat'' to distinguish it from the unprojected $\hat{H}_D$.

\section{Symmetries of the single-particle Hamiltonian}\label{sec:symmetries}

This section outlines the symmetries of the TSTG single-particle Hamiltonian from \cref{eqn:singlePartTSTG}. The reader is referred to Appendix \ref{app:symmetries} for a more in-depth discussion. In the case of zero displacement field, TSTG is symmetric under mirror reflections with the mirror plane parallel to the graphene layers, enabling us to discuss the symmetries of the system for each independent mirror-symmetry sector. Finally, we identify which symmetries of TSTG survive the hybridization between the Dirac cone and TBG fermions in presence of the applied displacement field. 

\subsection{Symmetry transformations}\label{sec:symmetries:a}

Due to its negligible spin-orbit coupling, single-layer graphene admits a series of spinless symmetry transformations, some of which are inherited by the single-particle TSTG Hamiltonian from \cref{eqn:singlePartTSTG}. To keep the discussion general, we can consider the action of these transformations on a generic fermion operator $\cre{f}{\vec{k},\vec{Q},\eta,\alpha,s}$ defined on the moir\'e lattice (where $\cre{f}{} = \cre{b}{},\cre{c}{}$). The unitary discrete symmetry transformations $C_{2z}$, $C_{3z}$, and $C_{2x}$  respectively denote a two-fold rotation around the $\hat{z}$ axis, a three-fold rotation around the $\hat{z}$ axis, and a two-fold rotation around the $\hat{x}$ axis. Their action on the moir\'e lattice fermion operators is given by 
\begin{equation}
	\begin{split}
		C_{2z} \cre{f}{\vec{k},\vec{Q},\eta,\alpha,s} C_{2z}^{-1} &= \sum_{\beta} \left( \sigma_x \right)_{\beta \alpha} \cre{f}{-\vec{k},-\vec{Q},-\eta,\beta,s}, \\
		C_{3z} \cre{f}{\vec{k},\vec{Q},\eta,\alpha,s} C_{3z}^{-1} &= \sum_{\beta}\left(e^{i\eta  \frac{2\pi}{3}\sigma_{z}} \right)_{\beta\alpha} \cre{f}{ C_{3z}\vec{k}, C_{3z}\vec{Q}, \eta,\beta,s}, \\
		C_{2x} \cre{f}{\vec{k},\vec{Q},\eta,\alpha,s} C_{2x}^{-1} &= \sum_{\beta} \left( \sigma_x \right)_{\beta \alpha} \cre{f}{C_{2x}\vec{k},C_{2x}\vec{Q},\eta,\beta,s}.
	\end{split}
\end{equation}
 
We also introduce the spinless mirror symmetry $m_z$ acting on the two fermion flavors as 
\begin{equation}
	\begin{split}
		m_z \cre{c}{\vec{k},\vec{Q},\eta,\alpha,s} m_z^{-1} = \cre{c}{\vec{k},\vec{Q},\eta,\alpha,s}, \\
		m_z \cre{b}{\vec{k},\vec{Q},\eta,\alpha,s} m_z^{-1} =-\cre{b}{\vec{k},\vec{Q},\eta,\alpha,s}.
	\end{split}
\end{equation}
Finally, we define the action of the spinless anti-unitary time-reversal operator 
\begin{equation}
 	T \cre{f}{\vec{k},\vec{Q},\eta,\alpha,s} T^{-1} = \cre{f}{-\vec{k},-\vec{Q},-\eta,\alpha,s}.
\end{equation}

The above operators represent commuting symmetries of the single-layer graphene Hamiltonian. In addition, there are three useful transformations which give rise to anticommuting symmetries, reflecting a relation between the positive and negative energy spectra of the Hamiltonians: a unitary particle-hole symmetry $P$ and two chiral transformations $C$ and $C'$, the latter two being only valid for different limits of the values of $w_0 /w_1$ (respectively $w_0 = 0$ and $w_1 = 0$). Their action on the moir\'e lattice fermions is given by 
\begin{equation}
	\begin{split}
		P \cre{f}{\vec{k},\vec{Q},\eta,\alpha,s} P^{-1} &= \zeta_{\vec{Q}} \cre{f}{-\vec{k},-\vec{Q},\eta,\alpha,s}, \\
		C \cre{f}{\vec{k},\vec{Q},\eta,\alpha,s} C^{-1} &= \left( \sigma_z \right)_{\beta \alpha} \cre{f}{\vec{k},\vec{Q},\eta,\beta,s}, \\
		C' \cre{f}{\vec{k},\vec{Q},\eta,\alpha,s} C^{\prime -1} &= \zeta_{\vec{Q}} \left( \sigma_z \right)_{\beta \alpha} \cre{f}{\vec{k},\vec{Q},\eta,\beta,s}, \\
	\end{split}
\end{equation}
where $\zeta_{\vec{Q}} = \pm 1$ for $\vec{Q} \in \mathcal{Q}_\pm$.

\subsection{Symmetries in different limits}

\begin{table*}[!ht]
\begin{tabular}{c c c c c c | c c c | c c c c c} 
	 & $C_{2z}$ & $C_{3z}$ & $C_{2x}$ & $m_{z}$ & $T$ & $P$ & $C_{2x} P$ & $m_z C_{2x} P$ & $C$ & $C'$ & $m_z C$ & $m_z C'$\\ 
\hline \hline 
	$\hat{H}_{\mathrm{TBG}}$ & \cmark & \cmark & \cmark & \cmark & \cmark & \cmark & \cmark & \cmark & \cmark $(w_0 = 0)$ & \cmark $(w_1 = 0)$ & \cmark $(w_0 = 0)$ & \cmark $(w_1 = 0)$\\ 
	$\hat{H}_{D}$ & \cmark & \cmark & \xmark & \cmark & \cmark & \xmark & \cmark & \cmark & \cmark & \cmark & \cmark & \cmark\\ 
	$\hat{H}_{0}$ & \cmark & \cmark & \xmark & \xmark & \cmark & \xmark & \xmark & \cmark & \xmark & \xmark & \cmark $(w_0 = 0)$ & \cmark $(w_1 = 0)$\\ 
\end{tabular} 
\caption{Commuting ($C_{2z}$, $C_{3z}$, $C_{2z}$, $m_{z}$, $T$) and anticommuting ($P$, $C_{2x} P$, $m_z C_{2x} P$, $C$, $C'$, $m_z C$, $m_z C'$) symmetries of the TSTG Hamiltonain under different limits. The presence or absence of a given symmetry is respectively indicated by \cmark or \xmark. Some transformation denote symmetries only for some given parameter choices (which are specified in parentheses). For $\hat{H}_0$, we indicate the symmetries for $U \neq 0$. The symmetries for the case $U=0$ can be deduced from the symmetries of $\hat{H}_{\mathrm{TBG}}$ and $\hat{H}_D$.}
\label{tab:sym_all_ham}
\end{table*}

We now briefly outline the symmetries of the single-particle Hamiltonian from \cref{eqn:singlePartTSTG}. The reader can find a more in-depth discussion in Appendix \ref{app:symmetries}. We will first consider the case without displacement field and discuss the symmetries of the system for each mirror-symmetry sector individually. Finally, we will explore how the introduction of a non-zero $U$ breaks or preserves the various symmetries from the $U = 0$ case.
\subsubsection{Symmetries in the $U=0$ case}
In the absence of displacement field, the Hamiltonian $\hat{H}_{\mathrm{TBG}}$ is symmetric under $C_{2z}$, $C_{3z}$, $C_{2x}$, $m_z$, and $T$~\cite{SON19}. In comparison, the mirror-antisymmetric sector $\hat{H}_{D}$ has only the $C_{2z}$, $C_{3z}$, $m_z$, and $T$ symmetries (\ie it is not symmetric under $C_{2x}$). Each graphene layer has an $\SUN{2}$ spin-rotational symmetry, owing to the negligible spin-oribt coupling. In conjunction with the charge $\UN{1}$ symmetry of each graphene valley, this leads to a $\UN{2} \times \UN{2}$ continuous symmetry for each of the two Hamiltonians $\hat{H}_{D}$ and $\hat{H}_{\mathrm{TBG}}$. As the two mirror-symmetry sectors are decoupled in the absence of displacement field, this results in a flavor-valley-spin $\left[ \UN{2} \times \UN{2}\right]_{\hat{c}} \times \left[ \UN{2} \times \UN{2}\right]_{\hat{b}}$ symmetry for $\hat{H}_{0}$ when $U = 0$. Here and in what follows, we will always employ $\left[ \dots \right]_{\hat{f}}$ to denote the continuous symmetry groups that act only within a certain fermion flavor $\hat{f} = \hat{b},\hat{c}$. 

Besides the above commuting symmetries, the mirror-symmetric sector Hamiltonian is particle-hole symmetric~\cite{SON19,SON20b}
\begin{equation}
	\left\lbrace \hat{H}_{\mathrm{TBG}} , P \right\rbrace = 0.
\end{equation}
For some parameter choices, it also has a chiral symmetry: $\left\lbrace \hat{H}_{\mathrm{TBG}} , C \right\rbrace = 0$, for $w_0 = 0$ (the \emph{first} chiral limit) or $\left\lbrace \hat{H}_{\mathrm{TBG}} , C' \right\rbrace = 0$, for $w_1 = 0$ (the \emph{second} chiral limit)~\cite{BER20a,TAR19}.

In contrast, the mirror-antisymmetric sector Hamiltonian is not particle-hole symmetric, but anticommutes with the combined $C_{2x} P$ transformation
\begin{equation}
	\left\lbrace \hat{H}_{D} , C_{2x} P \right\rbrace = 0.
\end{equation} 
Moreover, as opposed to $\hat{H}_{\mathrm{TBG}}$, $\hat{H}_{D}$ always satisfies the chiral symmetry, anticommuting with both $C$ and $C'$ irrespective of $w_0$ and $w_1$. When acting on the $\cre{b}{}$ operators, the two chiral operators are however identical up to a valley-charge rotation, as shown in Appendix \ref{app:symmetries:a}, and hence they do not generate distinct symmetries. 

The projected Dirac cone Hamiltonian $H_D$ features another low-energy non-crystalline symmetry $L$, obeying $\left\lbrace H_{D} , L \right\rbrace = 0$. To define its action, we first note that due to the Bloch periodicity property $\cre{b}{\vec{k},\vec{Q},\eta,\alpha,s}=\cre{b}{\vec{k}-\vec{G},\vec{Q}+\vec{G},\eta,\alpha,s}$, the projected Dirac cone Hamiltonian from \cref{eqn:Dirac_approx} can be cast into a simpler, albeit less symmetric form
\begin{equation}
	H_D = \sum_{\eta,\alpha,\beta,s} \sum_{\substack{\vec{k} \\ \abs{\vec{k} - \eta \vec{q}_1} \leq \Lambda}} \left[h^{D,\eta}_{\eta \vec{q}_1} \left( \vec{k} \right) \right]_{\alpha \beta} \cre{b}{\vec{k},\eta \vec{q}_1,\eta,\alpha,s} \des{b}{\vec{k},\eta \vec{q}_1,\eta,\beta,s},
\end{equation}
with $\Lambda \leq 0.2$. The action of the $L$ operators can be defined as
\begin{equation}
	L \cre{b}{\delta \vec{k} + \eta \vec{q}_1, \vec{Q},\eta,\alpha,s} L^{-1} = \cre{b}{-\delta \vec{k} + \eta \vec{q}_1, \vec{Q},\eta,\alpha,s},
\end{equation}
for any $\abs{\delta \vec{k}} \leq \Lambda$. Since $L$ maps $\delta \vec{k} + \eta \vec{q}_1$ to $-\delta \vec{k} + \eta \vec{q}_1$, two momentum points which are not related by any crystalline symmetry, it represents an emerging effective low-energy symmetry of $\hat{H}_D$.

\subsubsection{Symmetries in the $U \neq 0$ case}
The introduction of a displacement field breaks the $C_{2x}$ and $m_z$ symmetries of TSTG and only $C_{2z}$, $C_{3z}$, and $T$ remain good symmetries of $\hat{H}_0$. The flavor-valley-spin $\left[\UN{2} \times \UN{2} \right]_{\hat{c}} \times \left[\UN{2} \times \UN{2} \right]_{\hat{b}}$ rotation symmetry is also broken to a valley-spin $\UN{2} \times \UN{2}$ symmetry in the $U \neq 0$ case (see Appendix \ref{app:symmetries:b}). The combined particle-hole transformation $m_z C_{2x} P$ remains a good anticommuting symmetry of $\hat{H}_0$, obeying
\begin{equation}
	\left\lbrace \hat{H}_0, m_z C_{2x} P \right\rbrace=0 .
\end{equation}
Finally, the TSTG Hamiltonian in the presence of displacement breaks the chiral transformations $C$ and $C'$, but preserves the combined operations $m_z C$ and $m_z C'$, having a (modified) chiral symmetry for the same parameter choices as $\hat{H}_{\mathrm{TBG}}$.

\subsection{Summary of symmetries}

In the absence of displacement field the TSTG Hamiltonian splits into mirror-symmetry sectors for which both the commuting and the anticommuting symmetries can be individually discussed. The addition of displacement field breaks the $m_z$ symmetry and couples the $\cre{b}{}$ and $\cre{c}{}$ fermion flavors (see Appendix \ref{app:symmetries:b}). This effectively breaks some of the symmetry transformations of $\hat{H}_0$ in the $U=0$ case to combined operations for $U \neq 0$, as shown in \cref{tab:sym_all_ham}.

\section{Single-particle Spectrum}\label{sec:singleparticlespectrum}
This section focuses on understanding the low-energy single-particle spectrum of TSTG with or without a perpendicular displacement field. While the main results are presented here, the more detailed exposition can be found in Appendix \ref{app:approx_single_part}. After introducing the energy band basis for TSTG, we show how a non-zero $U$ hybridizes the TBG and Dirac cone fermions by building a simplified tripod model~\cite{BIS11}. For the experimentally relevant values of the displacement field~\cite{HAO21}, corresponding to $U<0.3$, we can develop a perturbation theory in $U$ for the hybridization between the two mirror-symmetry sectors of TSTG. The final result of this section is an expression for the low-energy projected TSTG Hamiltonian.
\subsection{Energy band basis}\label{sec:singleparticlespectrum:ebasis}
For the low-energy spectrum of TSTG, it is useful to introduce the energy band basis for the two mirror-symmetry sectors (see also Appendix \ref{app:single_part_ham:b}) of the system. For each band $n$ (where $n>0$ denotes the $n$-th conduction band, while $n<0$ labels $\abs{n}$-th valence band), we define the single-particle wave functions $u^{\hat{c}}_{\vec{Q} \alpha; n \eta} \left( \vec{k} \right)$ and corresponding band energies $\epsilon^{\hat{c}}_{n,\eta} \left( \vec{k} \right)$ for the first-quantized TBG Hamiltonian $h^{(\eta)}_{\vec{Q},\vec{Q}'}\left( \vec{k} \right)$ from \cref{eqn:singlePartTBG} according to 
\begin{equation}
	\label{eqn:singlePartWavfTBG}
		\sum_{\vec{Q}', \beta} \left[h^{\left(\eta\right)}_{\vec{Q},\vec{Q}'} \left( \vec{k} \right) \right]_{\alpha \beta} u^{\hat{c}}_{\vec{Q}' \beta; n \eta} \left( \vec{k} \right) = \epsilon^{\hat{c}}_{n,\eta} \left( \vec{k} \right) u^{\hat{c}}_{\vec{Q} \alpha; n \eta} \left( \vec{k} \right).
\end{equation}
Similarly, the single-particle wave functions $u^{\hat{b}}_{\vec{Q} \alpha; n \eta}$ and corresponding band energies $\epsilon^{\hat{b}}_{n,\eta} \left( \vec{k} \right)$ of the Dirac Hamiltonian $h^{D,\eta}_{\vec{Q}} \left( \vec{k} \right)$ from \cref{eqn:singlePartDirac} must obey
\begin{equation}
	\label{eqn:singlePartWavfDirac}
	\sum_{\beta} \left[h^{D,\eta}_{\vec{Q}} \left( \vec{k} \right) \right]_{\alpha \beta} u^{\hat{b}}_{\vec{Q} \beta; n \eta} \left( \vec{k} \right) = \epsilon^{\hat{b}}_{n,\eta} \left( \vec{k} \right) u^{\hat{b}}_{\vec{Q} \alpha; n \eta} \left( \vec{k} \right),
\end{equation}
allowing us to define the energy band basis for both mirror-symmetry sectors of TSTG
\begin{equation}
	\label{eqn:singlePartWavf}
	\begin{split}
		\cre{c}{\vec{k},n,\eta,s} = \sum_{\vec{Q}\in \mathcal{Q}_{\pm},\alpha} u^{\hat{c}}_{\vec{Q} \alpha; n \eta} \left( \vec{k} \right) \cre{c}{\vec{k},\vec{Q},\eta,\alpha,s},\\
		\cre{b}{\vec{k},n,\eta,s} = \sum_{\vec{Q}\in \mathcal{Q}_{\eta},\alpha} u^{\hat{b}}_{\vec{Q} \alpha; n \eta} \left( \vec{k} \right) \cre{b}{\vec{k},\vec{Q},\eta,\alpha,s}.
	\end{split}	
\end{equation}
The commuting and anticommuting symmetries presented in \cref{sec:symmetries} impose certain relations between the single-particle TSTG wave functions. Throughout this paper, we adopt the gauge-fixing convention presented in Appendix \ref{app:gauge} and in Ref.~\cite{BER20a} to fix the relative phase of the energy band operators and corresponding wave functions in \cref{eqn:singlePartWavf}.

\subsection{An approximate tripod model for TSTG}\label{sec:singleparticlespectrum:tripod}

\begin{figure}[!t]
	\twoFigure{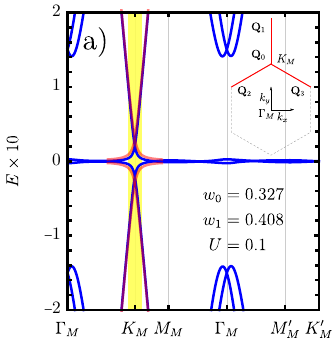}{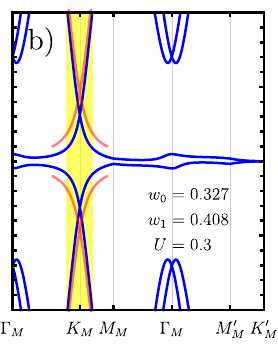}{\columnwidth}{fig:main_spectrum}
	\caption{The low-energy band structure of TSTG in the presence of a perpendicular displacement field. The blue lines show the TSTG energy spectrum near charge neutrality for valley $\eta = +$, computed by numerical diagonalization of the Hamiltonian in \cref{eqn:singlePartTSTG} along high symmetry momentum lines. The tunneling parameters $w_0$ and $w_1$, as well as the displacement field $U$ are specified inside each panel. Additionally, in panel (a) we illustrate schematically the four plane waves of the tripod model from \cref{eqn:tripod_matrix}. We also highlight in yellow the $A_{+}^1$ region as defined in \cref{eqn:Dirac_zones}, where the hybridization between the TBG active bands and the Dirac Hamiltonian is significant. The quantitative (qualitative) features of the TSTG band structure for $U=0.1$ ($U=0.3$) are accurately captured by the approximate dispersion of the tripod model in \cref{eqn:tripod_bands}, shown in red.}
	\label{fig:main_spectrum}
\end{figure}

We now consider a simplified model for the low-energy physics of TSTG near the $K_M$ point at $\vec{k}= \vec{q}_1$. We employ the TBG tripod model~\cite{BIS11} that we further modify by coupling with a Dirac cone Hamiltonian, as required by \cref{eqn:singlePartDirac,eqn:singlePartDisplacement}. Focusing on the $\eta=+$ valley and restricting to the four $\vec{Q}$-points (\ie four plane wave states) mandated by the tripod model (see \cref{fig:main_spectrum:a}), we write the single-particle eigenstates of TSTG as 
\begin{equation}
	\label{eqn:tripod_func}
	\begin{split}
		\ket{\Psi\left( \vec{k} \right)} = \sum_{\alpha} \left[ \sum_{i=0}^3 \left( \psi_{i,\alpha} \left( \vec{k} \right) \cre{c}{\vec{k},\vec{Q}_i,+,\alpha,s} \right)  \right. \\ + \left. \psi_{D,\alpha}\left( \vec{k} \right) \cre{b}{\vec{k},\vec{Q}_0,+,\alpha,s} \right] \ket{0},
	\end{split}
\end{equation}
with $\vec{Q}_i = \vec{q}_1 + \vec{q}_i$ for $i=1,2,3$ and $\vec{Q}_0= \vec{q}_1$. The first-quantized Hamiltonian acting on the ten-dimensional spinor 
\begin{equation}
    \Psi^T \hspace{-0.3 em } \left( \vec{k} \right) = \left(\psi_0^T\left( \vec{k} \right), \psi_1^T\left( \vec{k} \right), \psi_2^T\left( \vec{k} \right), \psi_3^T\left( \vec{k} \right), \psi_D^T\left( \vec{k} \right) \right)
\end{equation}
is given by
\begin{equation}
	\label{eqn:tripod_matrix}
	H_{\mathrm{Tri}} = \begin{pmatrix}
	\smallk \pauliVec & T'_1 & T'_2 & T'_3 & \frac{U}{2} \mathbb{1} \\ 
	T'_1 & h^{(1)} \left( \delta \vec{k} \right) & \mathbb{0} & \mathbb{0} & \mathbb{0} \\ 
	T'_2 & \mathbb{0} & h^{(2)} \left( \delta \vec{k} \right) & \mathbb{0} & \mathbb{0} \\ 
	T'_3 & \mathbb{0} & \mathbb{0} & h^{(3)} \left( \delta \vec{k} \right) & \mathbb{0} \\ 
	\frac{U}{2} \mathbb{1} & \mathbb{0} & \mathbb{0} & \mathbb{0} & \smallk \pauliVec
	\end{pmatrix},
\end{equation}
where we have introduced the shorthand notation $h^{(i)} \left( \delta \vec{k} \right) = \left( \smallk - \vec{q}_i \right) \pauliVec$ for $i=1,2,3$ and $\smallk = \vec{k} - \vec{q}_1$. In \cref{eqn:tripod_matrix}, we have denoted the two-dimensional Pauli vector by $\boldsymbol{\sigma} = \left( \sigma_x, \sigma_y \right)$ and defined the rescaled tunneling matrices $T'_i = T_i \sqrt{2}$ (for $i=1,2,3$), with $T_i$ being given in \cref{si:eqn:tunnelingMatrices}. 
The $10 \times 10$ Hamiltonian matrix in \cref{eqn:tripod_matrix} cannot be solved analytically. However, we are interested in the low-energy physics of TSTG near the $K_M$ point, for which $\delta k \sim E \ll 1$ (where $\delta k = \abs{\delta \vec{k}}$ and $E$ is the energy of the state at $\delta \vec{k}$), as can be seen in \cref{fig:main_spectrum}. Thanks to a series of justified approximations detailed in Appendix \ref{app:approx_single_part:tripod}, we can analytically obtain the low-energy dispersion relation near $K_M$
\begin{equation}
	\label{eqn:tripod_bands}
	E = \pm \frac{ \delta k \left(3 w_0^{\prime 2} + 2 \right) \pm \sqrt{9 \delta k^2 \left(w_0^{\prime 2} + 2 w_1^{\prime 2} \right)^2 + U^2 \Delta } }{2 \Delta},
\end{equation}
where $\Delta = \left(3 w_0^{\prime 2} + 3 w_1^{\prime 2} +1 \right)$ and $w'_{0,1} = w_{0,1} \sqrt{2}$. We plot the approximate dispersion relation of \cref{eqn:tripod_bands} in \cref{fig:main_spectrum}: the simplified tripod model qualitatively and quantitatively matches the low-energy spectrum of $\hat{H}_{0}$ obtained from numerical diagonalization with a large number of $\vec{Q}$ points. Note that a similar tripod model was derived in Ref.~\cite{LI19} but only for $U=0$. 

\subsection{The low-energy spectrum of TSTG}\label{sec:singleparticlespectrum:low}

The low-energy physics of TSTG with displacement field arises from the interplay between the almost-flat bands of $\hat{H}_{\mathrm{TBG}}$ and the Dirac cone bands of $\hat{H}_D$ with which they are coupled by $\hat{H}_U$. We are interested in a quantitative perturbation theory for the single-particle wave functions of TSTG in the presence of a non-zero displacement field. Ideally, we would also like to express the low-energy eigenstates of $\hat{H}_0$ \emph{only} in terms of the eigenstates of $\hat{H}_{\mathrm{TBG}}$: while they \emph{cannot} be analytically computed in the entire MBZ, their properties have been extensively studied in Refs.~\cite{SON19,BER20,BER20a,SON20b}.

In Appendix \ref{app:approx_single_part:spec}, we show that rather than starting from the \emph{full} TSTG Hamiltonian in \cref{eqn:singlePartTSTG} and \emph{then} projecting into its low energy states, an excellent approximation is to start from the TBG Hamiltonian projected into the active bands ($H_{\mathrm{TBG}}$) which is then hybridized with the Dirac cone fermions. For valley $\eta = +$ ($\eta=-$), the Hamiltonian $H_{\mathrm{TBG}}+\hat{H}_D + \hat{H}_U$ leads to close to the exact (\ie with an overlap higher than $99 \%$) eigenstates around the $K_M$ ($K_M'$) point within a radius $\Lambda$ for both the active and the Dirac cone bands. It also captures the correct eigenstates at $\Gamma_M$ for the active TBG bands, which are not changed much by the introduction of the displacement field. Note that around $\Gamma_M$, this method will \emph{not} give the correct eigenstates for the Dirac cone bands (which are, however, at high energy and do not contribute to the low-energy physics). Indeed, the high Fermi velocity of the Dirac cone bands implies that they hybridize with the higher energy (passive) bands of $\hat{H}_{\mathrm{TBG}}$ that we neglect in the projection (see \cref{fig:main_spectrum}). 

Using only three plane wave states (\ie three $\vec{Q}$ points) for the mirror-antisymmetric fermions (an approximation which was justified numerically in Appendix \ref{app:single_part_ham:a}), we can write the low-energy single-particle eigenstates of $\hat{H}_0$ for valley $\eta$, spin $s$, and band labeled by $m$ as 

\begin{equation}
	\label{eqn:8dimEigs}
	\begin{split}
		\ket{\Psi^{\eta,s,m} \left( \vec{k} \right)} = \left[ \sum_{i=1}^3 \sum_{\alpha} \left( \psi^{\eta,s,m}_{i,\alpha} \left( \vec{k} \right) \cre{b}{\vec{k},\eta \vec{q}_i, \eta ,\alpha,s} \right) \right. \\ \left. + \sum_{\abs{n} = 1} \phi^{\eta,s,m}_{n}\left( \vec{k} \right) \cre{c}{\vec{k},n,\eta,s} \right] \ket{0},
	\end{split}
\end{equation}
where we have defined the three two-component spinors on the sublattice space, $\psi^{\eta,s,m}_{i} \left( \vec{k} \right)$ (for $i=1,2,3$), and the two-component spinor in the space of the $n = \pm 1$ TBG active bands, $\phi^{\eta,s,m} \left( \vec{k} \right)$.
When acting on the eight-dimensional spinor 
\begin{equation}
	\label{eqn:8dimSpinor}
	\Psi^T \left( \vec{k} \right) = \left( \psi^T_{1} \left( \vec{k} \right), \psi^T_{2} \left( \vec{k} \right), \psi^T_{3} \left( \vec{k} \right), \phi^T \left( \vec{k} \right) \right),
\end{equation}
we obtain the following analytical expression for the low-energy the $8 \times 8$ first-quantized TSTG Hamiltonian 
\begin{equation}
	\label{eqn:8times8Ham}
	H_{8 \times 8} = \begin{pmatrix}
		\mathcal{E}^{\eta} & U^{\dagger \eta}_1 \left( \vec{k} \right) & U^{\dagger \eta}_2 \left( \vec{k} \right) & U^{\dagger \eta}_3 \left( \vec{k} \right) \\ 
		U^{\eta }_1 \left( \vec{k} \right) & h^{D,\eta}_{\vec{\eta \vec{q}_1}} \left( \vec{k} \right) & \mathbb{0} & \mathbb{0} \\ 
		U^{\eta }_2 \left( \vec{k} \right) & \mathbb{0} & h^{D,\eta}_{\vec{\eta \vec{q}_2}}  \left( \vec{k} \right)& \mathbb{0} \\ 
		U^{\eta }_3 \left( \vec{k} \right) & \mathbb{0} & \mathbb{0} & h^{D,\eta}_{\vec{\eta \vec{q}_3}} \left( \vec{k} \right)
	\end{pmatrix}.
\end{equation}
For the sake of brevity, in \cref{eqn:8dimSpinor}, we have suppressed the $\eta$, $s$, and $m$ indices. In addition, in \cref{eqn:8times8Ham} the $2 \times 2$ diagonal energy matrix for the TBG active bands in valley $\eta$ is given by
\begin{equation}
	\mathcal{E}^{\eta} \left( \vec{k} \right) = \begin{pmatrix}
		\epsilon^{\hat{c}}_{+1,\eta} \left( \vec{k} \right) & 0 \\
		0 & \epsilon^{\hat{c}}_{-1,\eta} \left( \vec{k} \right) \\
	\end{pmatrix}
\end{equation}
whereas the $2 \times 2$ displacement field perturbation matrices can be written in terms of the TBG wave functions defined in \cref{eqn:singlePartWavfTBG}
\begin{equation}
	\left[ U^\eta_i \left( \vec{k} \right) \right]_{\alpha,n} = \frac{U}{2} u^{\hat{c}}_{\eta\vec{q}_i \alpha; n \eta} \left( \vec{k} \right),
\end{equation}
for $i=1,2,3$. As anticipated in \cref{sec:singlehamiltonian:approximation}, there are two regions of interest in the BZ for the low-energy spectrum of $H_0$ and hence of $H_{8 \times 8}$: away and near the Dirac points of the MBZ. In deriving the single-particle projected TSTG Hamiltonian, we will now consider each of them individually.

\subsection{Single-particle projected TSTG Hamiltonian}\label{sec:singleparticlespectrum:proj}
We first provide the final expression of the single-particle projected TSTG Hamiltonian. We then sketch its derivation, with the detailed proof given in Appendix \ref{app:approx_single_part:spec}. The single-particle projected TSTG Hamiltonian reads
\begin{equation}
	\label{eqn:proj_single_part}
	H_{0} = H_{\mathrm{TBG}} + H_{D} + H^{\left( \hat{b}\hat{c} \right)}_{U} + H^{\left( \hat{c} \right)}_{U},
\end{equation}
where we have introduced the single-particle projected TBG~\cite{BER20a} and Dirac Hamiltonians, which are respectively given by
\begin{align}
	H_{\mathrm{TBG}} &= \sum_{\substack{\abs{n} = 1 \\ \eta,s \\ \vec{k} \in \mathrm{MBZ}}}  \epsilon_{n,\eta}^{\hat{c}} \left( \vec{k} \right) \cre{c}{\vec{k},n,\eta,s} \des{c}{\vec{k},n,\eta,s}, \label{eqn:proj_single_part_TBG} \\
	H_{D} &= \sum_{\substack{\abs{n} = 1 \\ \eta,s \\ \abs{\delta \vec{k}} \leq \Lambda}}  \epsilon_{n,\eta}^{\hat{b}} \left( \delta \vec{k} \right) \cre{b}{\delta \vec{k} + \eta \vec{q}_1,n,\eta,s} \des{b}{\delta \vec{k} + \eta \vec{q}_1,n,\eta,s}. \label{eqn:proj_single_part_Dirac}
\end{align}
Note that $H_{D}$ is only defined on a small region ($\abs{ \delta \vec{k} } \leq \Lambda \leq 0.2$) of the MBZ as a consequence of the high Fermi velocity of the Dirac cone bands for which
\begin{equation}
	\label{eqn:proj_single_part_Dirac_energy}
	\epsilon_{\pm 1,\eta}^{\hat{b}} \left( \delta \vec{k} \right) = \pm \abs{\delta \vec{k}}.
\end{equation} 
\Cref{eqn:proj_single_part} also incorporates the effects of a non-zero displacement field through the contributions $ H^{\left( \hat{b}\hat{c} \right)}_{U}$ (mixing the TBG and Dirac bands) and $H^{\left( \hat{c} \right)}_{U}$ (mixing the two active TBG bands within each valley and spin). These last two terms in \cref{eqn:proj_single_part} capture the effects of the perpendicular displacement field in the two regions of the MBZ and will be derived below.

\subsubsection{Perturbation theory away from the Dirac points}
Away from the Dirac points, \ie when $\vec{k} \in \mathcal{C}_\eta$, where
\begin{equation}
	\mathcal{C}_\eta = \left( \mathrm{MBZ} \setminus \bigcup_{i=1}^3 A_{\eta}^i \right),
\end{equation}
the hybridization between the eigenstates of $\hat{H}_D$ and the active bands of $\hat{H}_{\mathrm{TBG}}$ is suppressed by the difference in their energies. We can eliminate the $\psi_i$ spinors of \cref{eqn:8times8Ham} by writing them in terms of $\phi$
\begin{equation}
	\label{eqn:integratedPsi}
	\psi_i = \left(E - h_i \right)^{-1} U_i \phi,
\end{equation}
where we have suppressed the valley indices and made the $\vec{k}$-dependence implicit. In addition we have also introduced the shorthand notation $h_i \equiv h_{\eta \vec{q}_i} ^{D,\eta} \left( \vec{k} \right)$. \Cref{eqn:8times8Ham} can thus be cast as a non-linear eigenvalue equation
\begin{equation}
	\label{eqn:nonlinear_eig_8}
	\left[ \mathcal{E} + \sum_{i=1}^{3} U^{\dagger}_i \left(E - h_i \right)^{-1} U_i \right] \phi = E \phi.
\end{equation}
We expect the energy of the active bands to be only slightly changed by the hybridization with the Dirac cone Hamiltonian in the region $\mathcal{C}_{\eta}$ and have $\abs{E} \ll \abs{h_i}=\abs{\vec{k} - \eta \vec{q}_i}$. We can therefore ignore the $E$ dependence in the denominator of the second term of \cref{eqn:nonlinear_eig_8}~\footnote{Alternatively, we can expand $\left(E - h_i \right)^{-1}$ to linear order in $E$, and still end up with an analytically solvable equation. In this paper, we will ignore this linear contribution.}. This affords a major simplification as the Hamiltonians $h_i$ can be readily inverted to give a \emph{linear} eigenvalue equation 
\begin{equation}
	\label{eqn:linear_eig_8}
	\left[ \mathcal{E} + \sum_{i=1}^{3} U^{\dagger}_i h_i^{-1} U_i \right] \phi = E \phi.
\end{equation}
We show in Appendix \ref{app:approx_single_part:spec:a} that the amplitude of the mirror-antisymmetric operators is small enough in this region, validating an approximation even at large values of $U$: for $\vec{k} \in \mathcal{C}_{\eta}$, the displacement field only induces mixing between the active TBG bands. This contribution is captured by the effective Hamiltonian
\begin{equation}
	\label{eqn:singlePart:projUCeband}
	H^{\left( \hat{c} \right)}_{U} = \sum_{\substack{\abs{n}, \abs{m} = 1 \\ \eta,s}} \sum_{\substack{\vec{k} \in \mathcal{C}_{\eta}}}  \mathcal{B}^{\eta}_{nm} \left( \vec{k} \right) \cre{c}{\vec{k},n,\eta,s} \des{c}{\vec{k},m,\eta,s},
\end{equation}
where the matrix $\mathcal{B}^{\eta}_{nm}$ is given in \cref{si:eqn:BMatrix} of Appendix \ref{app:approx_single_part:spec:a} and represents a second-order contribution in $U$. For small enough displacement fields (\ie when $U^2 / \abs{\vec{k} - \eta \vec{q}_i}$ is much smaller than the bandwidth of the TBG flat bands), the active band states will not be significantly perturbed.

\subsubsection{Perturbation theory near the Dirac points}
Near any of the three Dirac points in the MBZ, the mixing between the TBG active bands and the Dirac cone Hamiltonian is significant. If $\vec{k}$ is near the $j$-th Dirac point in the MBZ (\ie $\vec{k} \in A_{\eta}^j$), we will have $\abs{h_j} \ll 1$, but $\abs{h_i} \approx \sqrt{3}$, for $i \neq j$. This implies that while the hybridization between the TBG active bands and the $j$-th Dirac Hamiltonian is relevant, there is little to no mixing with the Dirac cone bands stemming from the other two Dirac points of $\hat{H}_{D}$ in the MBZ. We can therefore approximate $\psi_i \approx 0$ for $i \neq j$ and write the single-particle TSTG wave functions for $\vec{k} \in A_{\eta}^j$ as
\begin{equation}
	\begin{split}
		\ket{\Psi^{\eta,s,m} \left( \vec{k} \right)} = \left[ \sum_{\alpha} \left( \psi^{\eta,s,m}_{j,\alpha} \left( \vec{k} \right) \cre{b}{\vec{k},\eta \vec{q}_j, \eta ,\alpha,s} \right) + \right. \\ 
		\left. \sum_{\abs{n} = 1} \phi^{\eta,s,m}_{n}\left( \vec{k} \right) \cre{c}{\vec{k},n,\eta,s} \right] \ket{0}.
	\end{split}
\end{equation}
In this region all four bands arising from the hybridization between the TBG active bands and the Dirac cone Hamiltonian are relevant for the low energy of TSTG. The corresponding first-quantized Hamiltonian reads
\begin{equation}
	H_{4 \times 4} = \begin{pmatrix}
		\mathcal{E}^{\eta} \left( \vec{k} \right) & U^{\dagger \eta}_j \left( \vec{k} \right) \\ 
		U^{\eta }_j \left( \vec{k} \right) & h^{D,\eta}_{\eta \vec{q}_j} \left( \vec{k} \right)
	\end{pmatrix}.
\end{equation}
In Appendix \ref{app:approx_single_part:spec:b}, we present a series of approximations which renders this $4 \times 4$ Hamiltonian exactly solvable in the (first) chiral limit. In the general case, we will write the projected displacement field Hamiltonian in this region of the MBZ in the energy band basis as
\begin{equation}
	\label{eqn:singlePart:projUAeband}
	H^{\left( \hat{b}\hat{c} \right)}_U \hspace{-0.4 em} =  \hspace{-1 em} \sum_{\substack{\eta,s \\ \abs{n},\abs{m} = 1 \\ \abs{\delta \vec{k}} \leq \Lambda}}  \hspace{-1.1 em}  \left[ N^{\eta}_{mn} \left( \delta \vec{k} \right) \cre{b}{\vec{k}_{\eta},m,\eta,s} \des{c}{\vec{k}_{\eta},n,\eta,s} + \mathrm{h.c.} \right],
\end{equation} 
where $\vec{k}_{\eta} \equiv \delta \vec{k} + \eta \vec{q}_1$ and the displacement field overlap matrix $N^{\eta}_{mn}$ is defined in \cref{si:eqn:singlePart:NMatrix} of Appendix \ref{app:approx_single_part:spec:b}.

\section{Many-body TSTG Hamiltonian}\label{sec:interaction}

This section introduces the interacting Hamiltonian for TSTG. We only quote the main results here; the complete derivations are relegated to Appendix \ref{app:interaction}. We start by writing the Coulomb repulsion Hamiltonian in terms of the moir\'e lattice fermion operators introduced in \cref{sec:singlehamiltonian}. Next, we show how the expression of the interaction Hamiltonian can be simplified by employing fermion operators corresponding to each mirror-symmetry sector. Using the energy band bases for the TBG and Dirac single-particle Hamiltonians defined in \cref{sec:singleparticlespectrum:ebasis}, we project the interaction Hamiltonian in the low-energy TSTG eigenstates. Finally, we write the expression for the fully-interacting TSTG Hamiltonian which is shown to have a spatial many-body charge-conjugation symmetry. 

\subsection{Coulomb interaction in TSTG}\label{sec:interaction:derivation}

The (unprojected) low-energy interaction Hamiltonian governing electron-electron repulsion in TSTG reads
\begin{equation}
	\label{eqn:interactionHamUnPro}
	\hat{H}_I = \frac{1}{2\Omega_{\mathrm{tot}}} \sum_{\substack{\vec{G} \in \mathcal{Q}_0 \\ \vec{q} \in \mathrm{MBZ} \\ l,l'}}  V^{l,l'}\left( \vec{q} + \vec{G} \right) \delta \rho^{l}_{\vec{G}+\vec{q}} \delta \rho^{l'}_{-\vec{G}-\vec{q}},
\end{equation}
where $\Omega_{\mathrm{tot}}$ is the total area of the TSTG sample and we have defined the Fourier transformation of the relative (to the single-layer graphene charge neutral point) electron density operators corresponding to layer $l$ to be
\begin{equation}
	\label{eqn:realtiveDensityOps}
	\begin{split}
		\delta \rho^{l}_{\vec{G}+\vec{q}}=\sum_{\eta, \alpha, s} \sum_{\substack{\vec{k} \in \mathrm{MBZ} \\ \vec{Q} \in \mathcal{Q}_{\eta, l}}} \left( \cre{a}{\vec{k},\vec{Q}, \eta,\alpha, s, l} \des{a}{\vec{k} - \vec{q},\vec{G} + \vec{Q}, \eta, \alpha, s, l} \right. \\ \left. - \frac{1}{2} \delta_{\vec{q},\vec{0}} \delta_{\vec{G},\vec{0}} \right).
	\end{split}
\end{equation}
In \cref{eqn:interactionHamUnPro}, $ V^{l,l'}\left( \vec{q} + \vec{G} \right)$ represents the Fourier transformation of the screened Coulomb potential $V^{l,l'} \left( \vec{r} \right)$ governing the repulsion between two electrons located respectively in layers $l$ and $l'$ and separated by a distance $\vec{r}$, measured in the plane of the single layer graphene. In the definition of the relative density operators from \cref{eqn:realtiveDensityOps}, we are effectively ignoring the inter-valley scattering processes, which are suppressed by the decay of the Coulomb potential in momentum space on a scale much smaller than the inter-valley separation of single layer graphene (see Appendix \ref{app:interaction:derivation:Moire}). 

For the typical gated arrangement used in experiments~\cite{HAO21,PAR21}, the interlayer distance (typically \SI{3}{\angstrom}) in TSTG is much smaller than the gate separation $\xi$ (usually \SI{10}{\nano\meter}) enabling us to neglect the dependence of $V^{l,l'}\left( \vec{q} + \vec{G} \right)$ on the layer indices $l$ and $l'$ (see Appendix \ref{app:interaction:derivation:Coulomb}) and write the screened Coulomb interaction as
\begin{equation}
	V^{l,l'}\left( \vec{q} \right) \approx V \left( \vec{q} \right) = \frac{2 \pi e^2}{\epsilon} \frac{\tanh \left( \xi \abs{\vec{q}}/2 \right)}{\abs{\vec{q}}}.
\end{equation}
This allows for a significant simplification, since the interaction Hamiltonian can now be written in terms of the relative density operators corresponding to the two mirror-symmetry sectors  
\begin{equation}
	\begin{split}
		\delta\rho^{\hat{c}}_{\vec{G} + \vec{q}}  &= \sum_{\substack{\vec{k} \in \mathrm{MBZ} \\ \vec{Q} \in \mathcal{Q}_{\pm} \\ \eta, \alpha, s}} \left( \cre{c}{\vec{k},\vec{Q},\eta ,\alpha, s} \des{c}{\vec{k} - \vec{q},\vec{G} + \vec{Q}, \eta, \alpha, s} - \frac{1}{2} \delta_{\vec{q},\vec{0}} \delta_{\vec{G},\vec{0}} \right) ,\\
	\delta\rho^{\hat{b}}_{\vec{G} + \vec{q}} &= \sum_{\substack{\vec{k} \in \mathrm{MBZ} \\ \vec{Q} \in \mathcal{Q}_{\eta} \\ \eta, \alpha, s}} \left( \cre{b}{\vec{k},\vec{Q},\eta ,\alpha, s} \des{b}{\vec{k} - \vec{q},\vec{G} + \vec{Q}, \eta, \alpha, s} - \frac{1}{2} \delta_{\vec{q},\vec{0}} \delta_{\vec{G},\vec{0}} \right),
	\end{split}
\end{equation} 
We can thus separate the interaction Hamiltonian from \cref{eqn:interactionHamUnPro} into three contributions
\begin{equation}
	\label{eqn:interactionHamUnProSeparated}
	\hat{H}_I = \hat{H}_{I,\mathrm{TBG}} + \hat{H}_{I,D} + \hat{H}_{I,\mathrm{TBG}-D}.
\end{equation}
The first and second terms in \cref{eqn:interactionHamUnProSeparated} respectively represent the interaction Hamiltonians for ordinary TBG and for Dirac cone fermions
\begin{align}
	\hat{H}_{I,\mathrm{TBG}} &= \frac{1}{2\Omega_{\mathrm{tot}}} \sum_{\substack{\vec{q} \in \mathrm{MBZ} \\ \vec{G} \in \mathcal{Q}_0}}  V\left( \vec{q} + \vec{G} \right) \delta \rho^{\hat{c}}_{\vec{G}+\vec{q}} \delta \rho^{\hat{c}}_{-\vec{G}-\vec{q}},\\
	\hat{H}_{I,D} &= \frac{1}{2\Omega_{\mathrm{tot}}} \sum_{\substack{\vec{q} \in \mathrm{MBZ} \\ \vec{G} \in \mathcal{Q}_0}}  V\left( \vec{q} + \vec{G} \right) \delta \rho^{\hat{b}}_{\vec{G}+\vec{q}} \delta \rho^{\hat{b}}_{-\vec{G}-\vec{q}}.
\end{align}
The third term corresponds to the Coulomb interaction between the TBG and Dirac cone fermions
\begin{equation}
	\begin{split}
		\hat{H}_{I,\mathrm{TBG}-D} = \frac{1}{2\Omega_{\mathrm{tot}}} \sum_{\substack{\vec{q} \in \mathrm{MBZ} \\ \vec{G} \in \mathcal{Q}_0}}  V\left( \vec{q} + \vec{G} \right) \\
		\times \left[ \delta \rho^{\hat{c}}_{\vec{G}+\vec{q}} \delta \rho^{\hat{b}}_{-\vec{G}-\vec{q}} + \mathrm{h.c.} \right].
	\end{split}
\end{equation}
Notice that the decomposition in \cref{eqn:interactionHamUnProSeparated} is valid even when the $m_z$ symmetry is broken in the presence of a perpendicular displacement field $U \neq 0$.

\subsection{Interaction projected Hamiltonian}\label{sec:coulomb}

Having derived the interaction Hamiltonian in the TSTG mirror-symmetry basis defined in \cref{eqn:ckQ,eqn:bkQ+}, we now turn our attention to projecting it in the low-energy TSTG single-particle eigenstates. As shown in Appendix \ref{app:interaction:projection}, the projected interaction Hamiltonian (henceforth denoted without a hat) reads
\begin{equation}
	\label{eqn:proj_int_hamiltonian}
	H_{I} = \frac{1}{2 \Omega_{\mathrm{tot}}} \hspace{-0.5 em} \sum_{\substack{\vec{G} \in \mathcal{Q}_0 \\ \vec{q} \in \mathrm{MBZ}}} \hspace{-0.5 em} \left( O^{\hat{c}}_{\vec{q},\vec{G}} + O^{\hat{b}}_{\vec{q},\vec{G}} \right)^\dagger  \left( O^{\hat{c}}_{\vec{q}, \vec{G}} + O^{\hat{b}}_{\vec{q}, \vec{G}} \right) 
\end{equation}
where we have introduced the operators
\begin{equation}	
	\begin{split}
	\label{eqn:proj_rho_c}
	O^{\hat{c}}_{\vec{q}, \vec{G}} = \sqrt{V \left( \vec{q} + \vec{G} \right)} \sum_{\substack{\vec{k} \in \mathrm{MBZ} \\ \abs{n},\abs{m} = 1 \\ \eta, s}} M^{\hat{c},\eta}_{mn} \left( \vec{k},\vec{q}+\vec{G} \right) \\
	\times \left(\cre{c}{\vec{k}+\vec{q},m,\eta,s} \des{c}{\vec{k},n,\eta,s} - \frac{1}{2} \delta_{\vec{q},\vec{0}} \delta_{m,n} \right).
	\end{split}
\end{equation}
and
\begin{equation}	
	\begin{split}
	\label{eqn:proj_rho_b}
	O^{\hat{b}}_{\vec{q}, \vec{G}} = \sqrt{V \left( \vec{q} + \vec{G} \right)} \sum_{\substack{\abs{\delta \vec{k}},\abs{\delta \vec{k} + \vec{q}} \leq \Lambda \\ \abs{n},\abs{m} = 1 \\ \eta, s}} \hspace{-1em} M^{\hat{b},\eta}_{mn} \left( \vec{k}_{\eta},\vec{q}+\vec{G} \right) \\
	\times \left(\cre{b}{\vec{k}_{\eta} + \vec{q},m,\eta,s} \des{b}{\vec{k}_{\eta},n,\eta,s} - \frac{1}{2} \delta_{\vec{q},\vec{0}} \delta_{m,n} \right),
	\end{split}
\end{equation}
with $\vec{k}_{\eta} \equiv \delta \vec{k} + \vec{q}_1$. Note that the expression of $O^{\hat{c}}_{\vec{q}, \vec{G}}$ is identical to the one corresponding to ordinary TBG derived in Ref.~\cite{BER20a}. Additionally, the operators $O^{\hat{f}}_{\vec{q}, \vec{G}}$ commute with each other, \ie $\left[ O^{\hat{b}}_{\vec{q}_1, \vec{G}_1}, O^{\hat{c}}_{\vec{q}_2, \vec{G}_2} \right]=0$, and obey $O^{\hat{f}}_{-\vec{q}, -\vec{G}} = O^{\dagger \hat{f}}_{\vec{q}, \vec{G}}$, for $\hat{f} = \hat{c}, \hat{b}$. In \cref{eqn:proj_rho_c,eqn:proj_rho_b}, the form factors $M^{\hat{c},\eta}_{mn}$ and $M^{\hat{b},\eta}_{mn}$ are defined in terms of the single-particle TBG and Dirac cone single-particle wave functions introduced in \cref{sec:singleparticlespectrum:ebasis} as
\begin{equation}
	\label{eqn:str_fact_def}
	M^{\hat{f},\eta}_{mn} \left( \vec{k},\vec{q}+\vec{G} \right) = \sum_{\substack{\alpha \\ \vec{Q} \in \mathcal{Q}_{\pm}}}  u^{* \hat{f}}_{\vec{Q}-\vec{G} \alpha; m \eta} \left( \vec{k} + \vec{q} \right) u^{\hat{f}}_{\vec{Q} \alpha; n \eta} \left( \vec{k} \right),
\end{equation}
for $\hat{f} = \hat{c}, \hat{b}$.

For the mirror-symmetric operators, the projection in the TSTG low-energy modes is equivalent to restricting the summation in \cref{eqn:proj_rho_c} to the active TBG bands. For the Dirac cone fermions, we additionally 
restrict the momenta in \cref{eqn:proj_rho_b} to lie near the Dirac points of $H_{D}$ located at $\eta \vec{q}_1$ for valley $\eta$. The TBG form-factors $M^{\hat{c},\eta}_{mn} \left( \vec{k},\vec{q}+\vec{G} \right)$ were shown to decay exponentially with $\abs{\vec{G}}$~\cite{BER20}. As such, only a few moir\'e reciprocal vectors $\vec{G}$ contribute to the summation in \cref{eqn:proj_int_hamiltonian}: the reciprocal vectors $\vec{G}$ for which $\abs{\vec{G}}=0,\sqrt{3}$. On the other hand, the Dirac cone form factors $M^{\hat{b},\eta}_{mn} \left( \vec{k},\vec{q}+\vec{G} \right)$ \emph{vanish completely} for any non-zero reciprocal vector $\vec{G}$, provided that the cutoff $\Lambda$ is small enough (as shown in Appendix \ref{app:interaction:projection}). 

Finally, we note that the projected interaction Hamiltonian in \cref{eqn:proj_int_hamiltonian} is a sum of positive semidefinite operators, and hence is itself positive semidefinite~\cite{KAN19}, similarly to the case of TBG~\cite{BUL20,BER20a}.
 
\subsection{Many-body projected TSTG Hamiltonian}

The expression of the interaction projected TSTG Hamiltonian \cref{eqn:proj_int_hamiltonian} can finally be combined with the projected single-particle Hamiltonian from \cref{eqn:proj_single_part} to yield the \emph{many-body} projected TSTG Hamiltonian  
\begin{equation}
	\label{eqn:proj_many_body}
	H = H_0 + H_I.
\end{equation}
Investigating the symmetries of $H$ under various different limits forms the object of \cref{sec:mbSym}. For now, we will only mention that $H$ features a spatial many-body charge conjugation symmetry $\mathcal{P}$ defined by the action of the single-particle anti-unitary transformation
\begin{equation}
	\mathcal{U} = m_z C_{2x} C_{2z} T P 
\end{equation}
followed by the interchange of the creation and annihilation fermion operators (see Appendix \ref{app:fullSym:chargeConj} for details). The many-body projected Hamiltonian $H$ is invariant under the action of $\mathcal{P}$, \ie
\begin{equation}
	\mathcal{P} H \mathcal{P}^{-1} = H
\end{equation}
In particular, $\mathcal{P}$ maps a many-body state with $N_e$ electrons to a state with $-N_e$ electrons, where number of electrons is measured with respect to the TSTG charge neutral point. As a consequence of the charge conjugation symmetry $\mathcal{P}$, the eigenspectrum of the fully-interacting projected TSTG Hamiltonian is symmetric about the charge neutral point. 

Finally, we note that the projected interaction Hamiltonian from \cref{eqn:proj_int_hamiltonian} is not normal ordered. The difference between $H_{I}$ and its normal-ordered form $\normord{H_I}$ is given by a quadratic contribution $\Delta H_I$, up to a constant term, \ie $H_{I} = \normord{H_I} + \Delta H_I + \mathrm{const.}$. By projecting the many-body TSTG Hamiltonian, we are effectively restricting ourselves to the $2N$ low-energy fermion modes distributed symmetrically around the charge neutral point. As shown in Appendix \ref{app:interaction:HF} and similarly to TBG~\cite{BER20a}, $\Delta H_I = \frac{1}{2} \left( H_{\mathrm{HF}}^N - H_{\mathrm{HF}}^{-N} \right)$, where $H_{\mathrm{HF}}^N$ represents the Hartree-Fock potential in the projected energy eigenstates contributed by the occupied eigenstates bellow the filling $N$. The quadratic contribution $\Delta H_I$ can therefore be thought as the effective potential arising in the projected many-body Hamiltonian from the energy-eigenstates which have been projected away. More importantly though, $\Delta H_I$ is essential for the existence of the experimentally observed $\mathcal{P}$ symmetry in TBG~\cite{NUC20}, as $\normord{H_I}$ alone lacks a spatial many-body charge conjugation symmetry.

\section{Exact symmetries of the many-body Hamiltonian}\label{sec:mbSym}

The single-particle TSTG Hamiltonian features a flavor-valley-spin $\left[\UN{2} \times \UN{2} \right]_{\hat{c}} \times \left[\UN{2} \times \UN{2} \right]_{\hat{b}}$ rotation symmetry in the $U=0$ case, which gets broken to a valley-spin $\UN{2} \times \UN{2}$ symmetry upon the introduction a perpendicular displacement field. Under various limits which will be discussed below, these symmetries are not only inherited by the many-body projected Hamiltonian, but also promoted to enlarged continuous groups of either the interaction Hamiltonian $H_I$, or of the full kinetic and interaction Hamiltonian, as a consequence of the discrete symmetries presented in \cref{sec:symmetries} and Appendix \ref{app:symmetries}. 

The aim of this section is to outline the symmetries of the many-body projected Hamiltonian from \cref{eqn:proj_many_body}. A more detailed exposition is given in Appendix \ref{app:fullSym}. As in \cref{sec:symmetries}, we will first consider the case without a perpendicular displacement field, and show that enlarged continuous symmetries arise for each individual mirror-symmetry sector. Finally, we will explore the effects of the perpendicular displacement field on the aforementioned continuous symmetries. 

Hereafter, we shall use $\zeta^{a}$, $\tau^a$, and $s^a$ to denote the identity matrix ($a=0$) and Pauli matrices ($a=x,y,z$) in the energy band $n=\pm 1$, valley $\eta = \pm$, and spin $s=\uparrow,\downarrow$ subspaces, respectively, for each mirror-symmetry sector. We will also rely on the results of Ref.~\cite{BER20a}, and make use of the gauge-fixing conventions detailed in Appendix \ref{app:gauge}, as well as on the resulting gauge-fixed forms of the single-particle (Appendix \ref{app:gauge_single_part}) and interaction (Appendix \ref{app:interaction:gauge}) projected Hamiltonians.

\subsection{Symmetries in the absence of displacement field}\label{sec:mbSym:noU}

In the absence of a perpendicular displacement field, the many-body projected Hamiltonian preserves the $C_{2z}$, $C_{3z}$, $m_z$ and $T$ symmetries of the single-particle TSTG Hamiltonian. Moreover, the two fermion flavors belonging to the two mirror-symmetry sectors remain uncoupled at the single-particle level and can (in principle) be individually rotated in the band, valley, and spin subspaces. We will therefore define two independent sets of generators corresponding respectively to the mirror-symmetric and mirror-antisymmetric fermion operators,
\begin{align}
	S_{\hat{c}}^{ a b}&=\sum_{\substack{\vec{k} \in \mathrm{MBZ} \\m,\eta,s\\n,\eta',s'}} \left(s_{\hat{c}}^{ab} \right)_{m \eta s, n \eta' s'} \cre{c}{\vec{k},m,\eta,s} \des{c}{\vec{k},n,\eta',s'}, \label{eqn:c_generator} \\
	S_{\hat{b}}^{ a b}&=\sum_{\substack{\abs{\delta \vec{k}} \leq \Lambda \\m,\eta,s\\n,\eta',s'}} \left(s_{\hat{b}}^{ab} \right)_{m \eta s, n \eta' s'} \cre{b}{\vec{k}_{\eta},m,\eta,s} \des{b}{\vec{k}_{\eta'},n,\eta',s'}, \label{eqn:b_generator}
\end{align}
where we have defined $\vec{k}_{\eta} \equiv \delta \vec{k} + \eta \vec{q}_1$ and $\vec{k}_{\eta'} \equiv \delta \vec{k} + \eta' \vec{q}_1$.
In \cref{eqn:c_generator,eqn:b_generator}, the $s^{ab}_{\hat{c}}$ ($s^{ab}_{\hat{b}}$) Hermitian matrices defined on the band, valley, and spin subspaces form a certain representation for the Lie algebra of the continuous symmetry group pertaining to the mirror-symmetric (mirror-antisymmetric) flavor. The two indices $a$ and $b$, indexing the generator $S_{\hat{c}}^{ a b}$ ($S_{\hat{b}}^{ a b}$) take different values depending on the continuous symmetry of the TSTG many-body Hamiltonian in the limit considered, but are unrelated to the band, valley, or spin Pauli matrix indices. We note that the generators $S_{\hat{c}}^{ a b}$ acting on the mirror-symmetric sector preserve momentum. On the other hand, $S_{\hat{b}}^{ a b}$ preserves momentum only if the matrix $s_{\hat{b}}^{ab}$ is diagonal in valley space. 

The generators from \cref{eqn:c_generator,eqn:b_generator} commute with the many-body TSTG Hamiltonian in different limits, and additionally commute with each other, \ie $\left[S_{\hat{c}}^{ a b},S_{\hat{b}}^{ c d} \right] = 0$. In what follows, we will analyze the various terms of the many-body TSTG Hamiltonian in the absence of displacement field  and determine the Lie algebra representation matrices $s^{ab}_{\hat{c}}$ and $s^{ab}_{\hat{b}}$ and the corresponding continuous symmetry groups. 
\subsubsection{Continuous symmetries of the mirror-antisymmetric sector}\label{sec:mbSym:noU:b}

The $\delta \vec{k}$-preserving symmetries of the single-particle Dirac Hamiltonian (where $\delta \vec{k}$ is the momentum measured from the Dirac points of $H_{D}$, located at $\eta \vec{q}_1$ in valley $\eta$) enforce certain relations between the single-particle eigenstates $u^{\hat{b}}_{\vec{Q} \alpha; n \eta} \left( \vec{k} \right)$. Using the gauge-fixing conventions of Appendix \ref{app:gauge:d}, it can be shown (see Appendix \ref{app:interaction:gauge:b}) that the $C_{2z}T$, $C_{2z}L$, and $C_{2z}T$ symmetries restrict the form factors $M^{\hat{b}} \left( \vec{k},\vec{q}+\vec{G} \right)$ to the following parameterization in the band and valley subspaces
\begin{equation}
	\label{eqn:form_factor_param_b}
	M^{ \hat{b},\eta}_{mn} \left(\vec{k}_\eta,\vec{q}+\vec{G} \right) = \sum_{j=0}^{1} \left( M_j \right)_{m\eta,n\eta} \alpha^{\hat{b}}_{j}\left(\delta \vec{k},\vec{q}+\vec{G} \right),
\end{equation}
where $\alpha^{\hat{b}}_{j}\left( \delta \vec{k},\vec{q}+\vec{G} \right)$ represent real scalar functions and we have defined $M_{0} = \zeta^0 \tau^0$ and $M_{1} = i \zeta^{y} \tau^{0}$. In Appendix \ref{app:fullSym:cont_sym_b_operators}, we show that \cref{eqn:form_factor_param_b} implies that the $O^{\hat{b}}_{\vec{q},\vec{G}}$ operators, governing the Coulomb interaction of the Dirac cone fermions in \cref{eqn:proj_int_hamiltonian}, have an enlarged $\left[\UN{4} \times \UN{4}\right]_{\hat{b}}$ symmetry. More specifically, we can define two sets of independent generators
\begin{equation}
	S_{\hat{b} \pm }^{ a b} = \sum_{\substack{\abs{\delta \vec{k}} \leq \Lambda \\m,\eta,s\\n,\eta',s'}} \left(s_{\hat{b} \pm }^{ab} \right)_{m \eta s, n \eta' s'} \cre{b}{\vec{k}_{\eta},m,\eta,s} \des{b}{\vec{k}_{\eta'},n,\eta',s'},
\end{equation}
which obey
\begin{equation}
	\label{eqn:generators_b_interaction}
	\left[S_{\hat{b} \pm }^{ a b},O^{\hat{b}}_{\vec{q},\vec{G}} \right] = 0,
\end{equation} 
for $a,b = 0,x,y,z$, with the corresponding representation matrices being given by 
\begin{equation}
	s^{ab}_{\hat{b} \pm } = \frac{1}{2} \left(\zeta^0 \pm \zeta^y \right) \tau^a s^b.
\end{equation}

As a consequence of its large Fermi velocity, the single-particle contribution $H_{D}$ cannot be ignored (\ie unlike the mirror-symmetric sector~\cite{BER20a}, there is no \emph{flat} limit for the mirror-antisymmetric one). Selecting only the subset of generators from \cref{eqn:generators_b_interaction} that additionally commute with $H_{D}$, we conclude that the mirror-antisymmetric sector enjoys a $\left[ \UN{4} \right]_{\hat{b}}$ symmetry whose generators obey 
\begin{equation}
	\label{eqn:generators_b}
	\left[S_{\hat{b} }^{ a b},O^{\hat{b}}_{\vec{q},\vec{G}} \right] = \left[S_{\hat{b} }^{ a b},H_{D} \right] = 0,
\end{equation}
for $a,b = 0,x,y,z$. The representation matrices of the $\left[ \UN{4} \right]_{\hat{b}}$ group are simply given by 
\begin{equation}
	s_{\hat{b} }^{ a b} = \zeta^0 \tau^a s^b.
\end{equation}
and correspond to full $\UN{4}$ rotations in the combined valley and spin subspaces.

\subsubsection{Continuous symmetries of the mirror-symmetric sector}\label{sec:mbSym:noU:c}

The continuous symmetries of the mirror-symmetric sector depend on the properties of the single-particle Hamiltonian $H_{\mathrm{TBG}}$ and the $O^{\hat{c}}_{\vec{q},\vec{G}}$ operators defined in \cref{eqn:proj_rho_c}. These have been derived and extensively discussed in Refs.~\cite{KAN19,SEO19,BUL20,BER20a}. As such, we will only enumerate these continuous symmetries pertaining to the mirror-symmetric sector of TSTG, with a more in-depth discussion being given in Appendix \ref{app:fullSym:cont_sym_c_operators}.  

The physically relevant limits of the projected TSTG Hamiltonian are the same as those arising in ordinary TBG~\cite{BER20a}: 

\begin{enumerate}
	\item \emph{The chiral-flat limit}. In the (first) chiral-flat limit, we neglect the single-particle dispersion of the TBG fermions. The many-body TSTG Hamiltonian then simply becomes $H = H_{D} + H_{I}$. As discussed in \cref{sec:mbSym:noU:b}, the dispersion of the high-velocity Dirac fermions implies that the contribution $H_D$ cannot be ignored.  Additionally, we take the chiral condition $w_0=0$ to hold exactly. It follows that the mirror-symmetric sector enjoys an enlarged $\left[ \UN{4} \times \UN{4} \right]_{\hat{c}}$ symmetry~\cite{BUL20,BER20a} generated by the 32 operators $S^{ab}_{\hat{c} \pm}$ (see Appendix \ref{app:fullSym:cont_sym_c_operators:cf}) for which the representation matrices read
	\begin{equation}
		\label{eqn:c_generator_matrix_cf}
		s^{ab}_{\hat{c} \pm} = \frac{1}{2} \left( \zeta^0 \pm \zeta^y \right) \tau^a s^b,
	\end{equation} 
	for $a,b=0,x,y,z$.
	\item \emph{The nonchiral-flat limit}. The nonchiral-flat limit is obtained by relaxing the chiral condition from the previous case, but still ignoring the dispersion of the TBG active bands, \ie $H = H_{D} + H_{I}$. As shown in Appendix \ref{app:fullSym:cont_sym_c_operators:ncf}, the mirror-symmetric sector has a $\left[ \UN{4} \right]_{\hat{c}}$ symmetry~\cite{KAN19,BER20a} generated by the operators in \cref{eqn:c_generator} for $a,b=0,x,y,z$. The corresponding representation matrices read
	\begin{equation}
		\begin{split}
			s^{0b}_{\hat{c}} = \zeta^0 \tau^0 s^b, &\quad
			s^{xb}_{\hat{c}} = \zeta^y \tau^x s^b, \\
			s^{yb}_{\hat{c}} = \zeta^y \tau^y s^b, &\quad
			s^{zb}_{\hat{c}} = \zeta^0 \tau^z s^b,
		\end{split}
	\end{equation}
	for $b=0,x,y,z$, and form a subset of the ones given in \cref{eqn:c_generator_matrix_cf} for the chiral-flat limit, but are different from either $s^{ab}_{\hat{c} +}$ or $s^{ab}_{\hat{c} -}$.
	\item \emph{The chiral-nonflat limit}. In the (first) chiral-nonflat limit, we assume the chiral condition $w_0=0$ to hold, but we no longer ignore the dispersion of the TBG active bands. As such, the full many-body TSTG Hamiltonian is restored, meaning that $H=H_{D} + H_{\mathrm{TBG}} + H_{I}$. In this case, the TBG fermions enjoy a $\left[ \UN{4} \right]_{\hat{c}}$ symmetry~\cite{BER20a} which is different from the one in the nonchiral-flat limit (see Appendix \ref{app:fullSym:cont_sym_c_operators:cnf}). The generators of this symmetry are given in \cref{eqn:c_generator} for $a,b=0,x,y,z$, with the representation matrices 
	\begin{equation}
		s^{ab}_{\hat{c}} = \zeta^0 \tau^a s^b
	\end{equation}
	corresponding to full $\UN{4}$ rotations in the combined valley and spin subspaces.
	\item \emph{The nonchiral-nonflat case}. Finally, moving away from the chiral condition and taking into consideration effects of the non-zero dispersion of the TBG active bands corresponds to the nonchiral-nonflat case. The many-body TSTG Hamiltonian given by $H=H_{D} + H_{\mathrm{TBG}} + H_{I}$ has only a $\left[\UN{2} \times \UN{2} \right]_{\hat{c}}$ valley-spin rotation symmetry (see Appendix \ref{app:fullSym:cont_sym_c_operators:ncnf}). The generators of this symmetry are also given in \cref{eqn:c_generator} for $a=0,z$ and $b=0,x,y,z$, and have the following representation matrices
	\begin{equation}
		s^{0b}_{\hat{c}} = \zeta^0 \tau^0 s^b, \quad
		s^{zb}_{\hat{c}} = \zeta^0 \tau^z s^b, 
	\end{equation}
	for $b=0,x,y,z$. They correspond to independent spin-charge rotations in the two valleys of the mirror-symmetric sector.
\end{enumerate}

\subsection{Exact symmetries in the presence of displacement field}
When $U \neq 0$, the TSTG many-body projected Hamiltonian is symmetric under the $C_{2z}$, $C_{3z}$ and $T$ symmetries. Additionally, the projected displacement field contribution $H^{\left( \hat{b} \hat{c} \right)}_{U}$ couples the two mirror-symmetry sector fermions, which can no longer be rotated independently in the band, valley, or spin subspaces. As such, we prove in Appendix \ref{app:fullSym:cont_sym_withU}, that the generators of continuous symmetries of $H$ in the presence of displacent field must have the form 
\begin{equation}
	\label{eqn:cont_generators}
	\begin{split}
		S^{ a b}=\sum_{\substack{m,\eta,s\\n,\eta',s'}} \left[ \sum_{\vec{k} \in \mathrm{MBZ} } \left(s^{ab} \right)_{m \eta s, n \eta' s'} \cre{c}{\vec{k},m,\eta,s} \des{c}{\vec{k},n,\eta',s'}
		 \right. \\ 
		+ \left. \sum_{\abs{\delta \vec{k}} \leq \Lambda} \left(s^{ab} \right)_{m \eta s, n \eta' s'} \cre{b}{\vec{k}_{\eta},m,\eta,s} \des{b}{\vec{k}_{\eta'},n,\eta',s'} \right],
	\end{split}
\end{equation} 
where the representation matrix $s^{ab}$ is diagonal in valley space. Note that the action of the generator in the two mirror-symmetry sectors is \emph{identical} (\ie they generate the same rotations in the valley and spin subspaces). 

Under any of the relevant limits of the many-body projected TSTG Hamiltonian, the generators from \cref{eqn:cont_generators} must, at the very least, obey the following commutation relations
\begin{equation}
	\left[S^{ a b},O^{\hat{b}}_{\vec{q},\vec{G}} \right] = \left[S^{ a b},O^{\hat{c}}_{\vec{q},\vec{G}} \right] = \left[S^{ a b}, H_D \right] = 0,
\end{equation}
in addition to commuting with the projected displacement field contributions $H^{\left( \hat{b} \hat{c} \right)}_{U}$ and $H^{\left( \hat{c} \right)}_{U}$. As a result, a non-zero displacement field breaks the symmetry of TSTG to the trivial $\UN{2} \times \UN{2}$ spin-valley rotation symmetry. The corresponding generators from \cref{eqn:cont_generators} are given simply by
\begin{equation}
	s^{0b}_{\hat{c}} = \zeta^0 \tau^0 s^b, \quad
	s^{zb}_{\hat{c}} = \zeta^0 \tau^z s^b, 
\end{equation}
for $b=0,x,y,z$.
\subsection{Summary}

\begin{table*}[!ht]
\begin{tabular}{l | c | c | r} 
	Operator / Hamiltonian & Flat band limit & Chiral limit ($w_0 = 0$) & Continuous symmetry \\ 
\hline \hline 
	$O^{\hat{b}}_{\vec{q},\vec{G}}$ & -- & -- & $\left[ \UN{4} \times \UN{4} \right]_{\hat{b}}$ \\
	$O^{\hat{c}}_{\vec{q},\vec{G}}$ & -- & yes & $\left[ \UN{4} \times \UN{4} \right]_{\hat{c}}$  \\ 
	$O^{\hat{c}}_{\vec{q},\vec{G}}$ & -- & no & $\left[ \UN{4} \right]_{\hat{c}}$  \\ 
\hline \hline
	$H_I$ & yes & yes & $\left[ \UN{4} \times \UN{4} \right]_{\hat{c}} \times \left[ \UN{4} \times \UN{4} \right]_{\hat{b}}$ \\
	$H_I$ & yes & no & $\left[ \UN{4} \right]_{\hat{c}} \times \left[ \UN{4} \times \UN{4} \right]_{\hat{b}}$ \\
	$H_D + H_I$ & yes & yes & $\left[ \UN{4} \times \UN{4} \right]_{\hat{c}} \times \left[ \UN{4} \right]_{\hat{b}}$ \\
	$H_D + H_I$ & yes & no & $\left[ \UN{4} \right]_{\hat{c}} \times \left[ \UN{4} \right]_{\hat{b}}$ \\
	$H_{D} + H_{\mathrm{TBG}} + H_I$ & no & yes & $\left[ \UN{4} \right]_{\hat{c}} \times \left[ \UN{4} \right]_{\hat{b}}$ \\
	$H_{D} + H_{\mathrm{TBG}} + H_I$ & no & no & $\left[ \UN{2} \times \UN{2} \right]_{\hat{c}} \times \left[ \UN{4} \right]_{\hat{b}}$ \\ 
	$H_{D} \left( + H_{\mathrm{TBG}} \right) + H^{\left( \hat{b} \hat{c} \right)}_{U}  + H^{\left( \hat{c} \right)}_{U} + H_I$ & no / yes & no / yes & $\UN{2} \times \UN{2}$   
\end{tabular} 
\caption{Continuous symmetries of the many-body projected TSTG Hamiltonian under different limits. We list the continuous symmetry groups corresponding to the two interaction operators $O^{\hat{b}}_{\vec{q},\vec{G}}$ and $O^{\hat{c}}_{\vec{q},\vec{G}}$, the projected interaction Hamiltonian $H_I$, as well as TSTG many-body Hamiltonians under different relevant limits. In the absence of displacement field, the two fermions flavors corresponding to different mirror-symmetry sectors can be independently rotated in the band, valley, and spin subspaces. As such, the continuous symmetry group for TSTG is the direct product of the continuous symmetry groups corresponding to each individual mirror-symmetry sector. Following to introduction of displacement field, the global TSTG symmetry is broken to the trivial $\UN{2} \times \UN{2}$ group.}
\label{tab:cont_sym_all_ham_int}
\end{table*}

In the absence of displacement field the TBG and Dirac cone fermions are uncoupled at the single-particle level. As a result, the many-body projected TSTG Hamiltonian inherits both the symmetries the many-body projected TBG Hamiltonian~\cite{BER20a} to which those of an interacting Dirac cone Hamiltonian are added, for a full symmetry of up to $\left[ \UN{4} \times \UN{4} \right]_{\hat{c}} \times \left[ \UN{4} \times \UN{4} \right]_{\hat{b}}$ of the projected interaction Hamiltonian $H_I$. The introduction of a perpendicular displacement field breaks the symmetries of the system to the trivial $\UN{2} \times \UN{2}$ symmetry, which corresponds to independent spin-charge rotations in the two TSTG valleys. For completeness, the enlarged band, valley, and spin rotation symmetries of TSTG under different physically relevant limits are presented in 
\cref{tab:cont_sym_all_ham_int}.

\section{Discussion}\label{sec:discussion}
The first part of this article was focused on the single-particle TSTG Hamiltonian. After reviewing a BM model for TSTG, we have derived the discrete crystalline symmetries of the system both with and without a perpendicular displacement field. In the absence of displacement field, we have uncovered a hidden anticommuting symmetry of the single-particle Hamiltonian, valid in the low-energy limit. The corresponding operator $L$ maps the high-velocity Dirac fermions from momentum $\delta \vec{k} + \eta \vec{q}_1$ to $-\delta \vec{k} + \eta \vec{q}_1$ in valley $\eta$, and hence denotes a non-local symmetry of the problem. We have also derived a series of approximations for the TSTG single-particle spectrum near charge neutrality, starting with a simplified tripod model which captured the essence of the TSTG band structure in the presence of displacement field. Finally, we provided more quantitiative perturbation schemes for the low-energy TSTG spectrum. They enabled us to obtain the TSTG eigenstates in the entire MBZ in terms of the TBG flat band wave functions, thus setting the stage for deriving the projected interaction Hamiltonian. 

In the second half of the paper, we introduced the Coulomb interaction Hamiltonian projected in the low-energy TSTG single-particle eigenstates. We showed that the electron-electron repulsion is comprised of three terms, corresponding to the interaction between the TBG fermions, the interaction between the Dirac electrons, and a term denoting the interaction between the TBG and high-velocity Dirac fermions. We then analyzed the symmetries of the many-body projected TSTG Hamiltonian. As a result of the local and non-local discrete symmetries at the single-particle level, we showed that the spin-valley $\UN{2} \times \UN{2}$ symmetry gets promoted to enlarged symmetry groups, up to a full $\left[ \UN{4} \times \UN{4} \right]_{\hat{c}} \times \left[ \UN{4} \right]_{\hat{b}}$ symmetry of the many-body projected Hamiltonian in the chiral-flat limit for $U=0$ (see \cref{tab:cont_sym_all_ham_int}). Moreover, we have shown that in the absence of displacement field, the enhanced rotation groups feature both local and non-local generators.

With the TSTG projected many-body Hamiltonian in hand, including its symmetries and derived gauge-fixing conditions, we have paved the way for understanding TSTG beyond the single-particle paradigm. Even in the absence of displacement field, the interaction naturally spoils the naive picture of decoupled TBG and high-velocity Dirac fermions~\cite{KHA20}. In light of the recent experiments~\cite{PAR21,HAO21}, this naturally raises questions about the fate of the insulating TBG phases, both with and without a perpendicularly applied displacement field. Such a study will be the core of our forthcoming work~\cite{TTG2}.

\begin{acknowledgments}
We thank Oskar Vafek, Pablo Jarillo-Herrero, and Dmitri Efetov for fruitful discussions. This work was supported by the DOE Grant No. DE-SC0016239, the Schmidt Fund for Innovative Research, Simons Investigator Grant No. 404513, the Packard Foundation, the Gordon and Betty Moore Foundation through Grant No. GBMF8685 towards the Princeton theory program, and a Guggenheim Fellowship from the John Simon Guggenheim Memorial Foundation. Further support was provided by the NSF-EAGER No. DMR 1643312, NSF-MRSEC No. DMR-1420541 and DMR-2011750, ONR No. N00014-20-1-2303, Gordon and Betty Moore Foundation through Grant GBMF8685 towards the Princeton theory program, BSF Israel US foundation No. 2018226, and the Princeton Global Network Funds.
\end{acknowledgments}

\bibliographystyle{apsrev4-2}
\bibliography{TTGbib,prep}

\begin{thebibliography}{139}%
\makeatletter
\providecommand \@ifxundefined [1]{%
 \@ifx{#1\undefined}
}%
\providecommand \@ifnum [1]{%
 \ifnum #1\expandafter \@firstoftwo
 \else \expandafter \@secondoftwo
 \fi
}%
\providecommand \@ifx [1]{%
 \ifx #1\expandafter \@firstoftwo
 \else \expandafter \@secondoftwo
 \fi
}%
\providecommand \natexlab [1]{#1}%
\providecommand \enquote  [1]{``#1''}%
\providecommand \bibnamefont  [1]{#1}%
\providecommand \bibfnamefont [1]{#1}%
\providecommand \citenamefont [1]{#1}%
\providecommand \href@noop [0]{\@secondoftwo}%
\providecommand \href [0]{\begingroup \@sanitize@url \@href}%
\providecommand \@href[1]{\@@startlink{#1}\@@href}%
\providecommand \@@href[1]{\endgroup#1\@@endlink}%
\providecommand \@sanitize@url [0]{\catcode `\\12\catcode `\$12\catcode
  `\&12\catcode `\#12\catcode `\^12\catcode `\_12\catcode `\%12\relax}%
\providecommand \@@startlink[1]{}%
\providecommand \@@endlink[0]{}%
\providecommand \url  [0]{\begingroup\@sanitize@url \@url }%
\providecommand \@url [1]{\endgroup\@href {#1}{\urlprefix }}%
\providecommand \urlprefix  [0]{URL }%
\providecommand \Eprint [0]{\href }%
\providecommand \doibase [0]{https://doi.org/}%
\providecommand \selectlanguage [0]{\@gobble}%
\providecommand \bibinfo  [0]{\@secondoftwo}%
\providecommand \bibfield  [0]{\@secondoftwo}%
\providecommand \translation [1]{[#1]}%
\providecommand \BibitemOpen [0]{}%
\providecommand \bibitemStop [0]{}%
\providecommand \bibitemNoStop [0]{.\EOS\space}%
\providecommand \EOS [0]{\spacefactor3000\relax}%
\providecommand \BibitemShut  [1]{\csname bibitem#1\endcsname}%
\let\auto@bib@innerbib\@empty
\bibitem [{\citenamefont {{Lopes dos Santos}}\ \emph
  {et~al.}(2007)\citenamefont {{Lopes dos Santos}}, \citenamefont {Peres},\
  and\ \citenamefont {Castro~Neto}}]{LOP07}%
  \BibitemOpen
  \bibfield  {author} {\bibinfo {author} {\bibfnamefont {J.~M.~B.}\
  \bibnamefont {{Lopes dos Santos}}}, \bibinfo {author} {\bibfnamefont
  {N.~M.~R.}\ \bibnamefont {Peres}},\ and\ \bibinfo {author} {\bibfnamefont
  {A.~H.}\ \bibnamefont {Castro~Neto}},\ }\href
  {https://doi.org/10.1103/PhysRevLett.99.256802} {\bibfield  {journal}
  {\bibinfo  {journal} {Phys. Rev. Lett.}\ }\textbf {\bibinfo {volume} {99}},\
  \bibinfo {pages} {256802} (\bibinfo {year} {2007})}\BibitemShut {NoStop}%
\bibitem [{\citenamefont {Su{\'a}rez~Morell}\ \emph {et~al.}(2010)\citenamefont
  {Su{\'a}rez~Morell}, \citenamefont {Correa}, \citenamefont {Vargas},
  \citenamefont {Pacheco},\ and\ \citenamefont {Barticevic}}]{SUA10}%
  \BibitemOpen
  \bibfield  {author} {\bibinfo {author} {\bibfnamefont {E.}~\bibnamefont
  {Su{\'a}rez~Morell}}, \bibinfo {author} {\bibfnamefont {J.~D.}\ \bibnamefont
  {Correa}}, \bibinfo {author} {\bibfnamefont {P.}~\bibnamefont {Vargas}},
  \bibinfo {author} {\bibfnamefont {M.}~\bibnamefont {Pacheco}},\ and\ \bibinfo
  {author} {\bibfnamefont {Z.}~\bibnamefont {Barticevic}},\ }\href
  {https://doi.org/10.1103/PhysRevB.82.121407} {\bibfield  {journal} {\bibinfo
  {journal} {Phys. Rev. B}\ }\textbf {\bibinfo {volume} {82}},\ \bibinfo
  {pages} {121407} (\bibinfo {year} {2010})}\BibitemShut {NoStop}%
\bibitem [{\citenamefont {Bistritzer}\ and\ \citenamefont
  {MacDonald}(2011)}]{BIS11}%
  \BibitemOpen
  \bibfield  {author} {\bibinfo {author} {\bibfnamefont {R.}~\bibnamefont
  {Bistritzer}}\ and\ \bibinfo {author} {\bibfnamefont {A.~H.}\ \bibnamefont
  {MacDonald}},\ }\href {https://doi.org/10.1073/pnas.1108174108} {\bibfield
  {journal} {\bibinfo  {journal} {PNAS}\ }\textbf {\bibinfo {volume} {108}},\
  \bibinfo {pages} {12233} (\bibinfo {year} {2011})}\BibitemShut {NoStop}%
\bibitem [{\citenamefont {Cao}\ \emph {et~al.}(2018)\citenamefont {Cao},
  \citenamefont {Fatemi}, \citenamefont {Demir}, \citenamefont {Fang},
  \citenamefont {Tomarken}, \citenamefont {Luo}, \citenamefont
  {{Sanchez-Yamagishi}}, \citenamefont {Watanabe}, \citenamefont {Taniguchi},
  \citenamefont {Kaxiras}, \citenamefont {Ashoori},\ and\ \citenamefont
  {{Jarillo-Herrero}}}]{CAO18}%
  \BibitemOpen
  \bibfield  {author} {\bibinfo {author} {\bibfnamefont {Y.}~\bibnamefont
  {Cao}}, \bibinfo {author} {\bibfnamefont {V.}~\bibnamefont {Fatemi}},
  \bibinfo {author} {\bibfnamefont {A.}~\bibnamefont {Demir}}, \bibinfo
  {author} {\bibfnamefont {S.}~\bibnamefont {Fang}}, \bibinfo {author}
  {\bibfnamefont {S.~L.}\ \bibnamefont {Tomarken}}, \bibinfo {author}
  {\bibfnamefont {J.~Y.}\ \bibnamefont {Luo}}, \bibinfo {author} {\bibfnamefont
  {J.~D.}\ \bibnamefont {{Sanchez-Yamagishi}}}, \bibinfo {author}
  {\bibfnamefont {K.}~\bibnamefont {Watanabe}}, \bibinfo {author}
  {\bibfnamefont {T.}~\bibnamefont {Taniguchi}}, \bibinfo {author}
  {\bibfnamefont {E.}~\bibnamefont {Kaxiras}}, \bibinfo {author} {\bibfnamefont
  {R.~C.}\ \bibnamefont {Ashoori}},\ and\ \bibinfo {author} {\bibfnamefont
  {P.}~\bibnamefont {{Jarillo-Herrero}}},\ }\href
  {https://doi.org/10.1038/nature26154} {\bibfield  {journal} {\bibinfo
  {journal} {Nature}\ }\textbf {\bibinfo {volume} {556}},\ \bibinfo {pages}
  {80} (\bibinfo {year} {2018})}\BibitemShut {NoStop}%
\bibitem [{\citenamefont {Cao}\ \emph {et~al.}(2020{\natexlab{a}})\citenamefont
  {Cao}, \citenamefont {Chowdhury}, \citenamefont {{Rodan-Legrain}},
  \citenamefont {{Rubies-Bigorda}}, \citenamefont {Watanabe}, \citenamefont
  {Taniguchi}, \citenamefont {Senthil},\ and\ \citenamefont
  {{Jarillo-Herrero}}}]{CAO20}%
  \BibitemOpen
  \bibfield  {author} {\bibinfo {author} {\bibfnamefont {Y.}~\bibnamefont
  {Cao}}, \bibinfo {author} {\bibfnamefont {D.}~\bibnamefont {Chowdhury}},
  \bibinfo {author} {\bibfnamefont {D.}~\bibnamefont {{Rodan-Legrain}}},
  \bibinfo {author} {\bibfnamefont {O.}~\bibnamefont {{Rubies-Bigorda}}},
  \bibinfo {author} {\bibfnamefont {K.}~\bibnamefont {Watanabe}}, \bibinfo
  {author} {\bibfnamefont {T.}~\bibnamefont {Taniguchi}}, \bibinfo {author}
  {\bibfnamefont {T.}~\bibnamefont {Senthil}},\ and\ \bibinfo {author}
  {\bibfnamefont {P.}~\bibnamefont {{Jarillo-Herrero}}},\ }\href
  {https://doi.org/10.1103/PhysRevLett.124.076801} {\bibfield  {journal}
  {\bibinfo  {journal} {Phys. Rev. Lett.}\ }\textbf {\bibinfo {volume} {124}},\
  \bibinfo {pages} {076801} (\bibinfo {year} {2020}{\natexlab{a}})}\BibitemShut
  {NoStop}%
\bibitem [{\citenamefont {Chen}\ \emph
  {et~al.}(2020{\natexlab{a}})\citenamefont {Chen}, \citenamefont {Sharpe},
  \citenamefont {Fox}, \citenamefont {Zhang}, \citenamefont {Wang},
  \citenamefont {Jiang}, \citenamefont {Lyu}, \citenamefont {Li}, \citenamefont
  {Watanabe}, \citenamefont {Taniguchi}, \citenamefont {Shi}, \citenamefont
  {Senthil}, \citenamefont {{Goldhaber-Gordon}}, \citenamefont {Zhang},\ and\
  \citenamefont {Wang}}]{CHE20a}%
  \BibitemOpen
  \bibfield  {author} {\bibinfo {author} {\bibfnamefont {G.}~\bibnamefont
  {Chen}}, \bibinfo {author} {\bibfnamefont {A.~L.}\ \bibnamefont {Sharpe}},
  \bibinfo {author} {\bibfnamefont {E.~J.}\ \bibnamefont {Fox}}, \bibinfo
  {author} {\bibfnamefont {Y.-H.}\ \bibnamefont {Zhang}}, \bibinfo {author}
  {\bibfnamefont {S.}~\bibnamefont {Wang}}, \bibinfo {author} {\bibfnamefont
  {L.}~\bibnamefont {Jiang}}, \bibinfo {author} {\bibfnamefont
  {B.}~\bibnamefont {Lyu}}, \bibinfo {author} {\bibfnamefont {H.}~\bibnamefont
  {Li}}, \bibinfo {author} {\bibfnamefont {K.}~\bibnamefont {Watanabe}},
  \bibinfo {author} {\bibfnamefont {T.}~\bibnamefont {Taniguchi}}, \bibinfo
  {author} {\bibfnamefont {Z.}~\bibnamefont {Shi}}, \bibinfo {author}
  {\bibfnamefont {T.}~\bibnamefont {Senthil}}, \bibinfo {author} {\bibfnamefont
  {D.}~\bibnamefont {{Goldhaber-Gordon}}}, \bibinfo {author} {\bibfnamefont
  {Y.}~\bibnamefont {Zhang}},\ and\ \bibinfo {author} {\bibfnamefont
  {F.}~\bibnamefont {Wang}},\ }\href
  {https://doi.org/10.1038/s41586-020-2049-7} {\bibfield  {journal} {\bibinfo
  {journal} {Nature}\ }\textbf {\bibinfo {volume} {579}},\ \bibinfo {pages}
  {56} (\bibinfo {year} {2020}{\natexlab{a}})}\BibitemShut {NoStop}%
\bibitem [{\citenamefont {Liu}\ \emph {et~al.}(2020{\natexlab{a}})\citenamefont
  {Liu}, \citenamefont {Wang}, \citenamefont {Watanabe}, \citenamefont
  {Taniguchi}, \citenamefont {Vafek},\ and\ \citenamefont {Li}}]{LIU20a}%
  \BibitemOpen
  \bibfield  {author} {\bibinfo {author} {\bibfnamefont {X.}~\bibnamefont
  {Liu}}, \bibinfo {author} {\bibfnamefont {Z.}~\bibnamefont {Wang}}, \bibinfo
  {author} {\bibfnamefont {K.}~\bibnamefont {Watanabe}}, \bibinfo {author}
  {\bibfnamefont {T.}~\bibnamefont {Taniguchi}}, \bibinfo {author}
  {\bibfnamefont {O.}~\bibnamefont {Vafek}},\ and\ \bibinfo {author}
  {\bibfnamefont {J.~I.~A.}\ \bibnamefont {Li}},\ }\href@noop {} {\bibfield
  {journal} {\bibinfo  {journal} {arXiv:2003.11072 [cond-mat]}\ } (\bibinfo
  {year} {2020}{\natexlab{a}})},\ \Eprint {https://arxiv.org/abs/2003.11072}
  {arXiv:2003.11072 [cond-mat]} \BibitemShut {NoStop}%
\bibitem [{\citenamefont {Lu}\ \emph {et~al.}(2019)\citenamefont {Lu},
  \citenamefont {Stepanov}, \citenamefont {Yang}, \citenamefont {Xie},
  \citenamefont {Aamir}, \citenamefont {Das}, \citenamefont {Urgell},
  \citenamefont {Watanabe}, \citenamefont {Taniguchi}, \citenamefont {Zhang},
  \citenamefont {Bachtold}, \citenamefont {MacDonald},\ and\ \citenamefont
  {Efetov}}]{LU19}%
  \BibitemOpen
  \bibfield  {author} {\bibinfo {author} {\bibfnamefont {X.}~\bibnamefont
  {Lu}}, \bibinfo {author} {\bibfnamefont {P.}~\bibnamefont {Stepanov}},
  \bibinfo {author} {\bibfnamefont {W.}~\bibnamefont {Yang}}, \bibinfo {author}
  {\bibfnamefont {M.}~\bibnamefont {Xie}}, \bibinfo {author} {\bibfnamefont
  {M.~A.}\ \bibnamefont {Aamir}}, \bibinfo {author} {\bibfnamefont
  {I.}~\bibnamefont {Das}}, \bibinfo {author} {\bibfnamefont {C.}~\bibnamefont
  {Urgell}}, \bibinfo {author} {\bibfnamefont {K.}~\bibnamefont {Watanabe}},
  \bibinfo {author} {\bibfnamefont {T.}~\bibnamefont {Taniguchi}}, \bibinfo
  {author} {\bibfnamefont {G.}~\bibnamefont {Zhang}}, \bibinfo {author}
  {\bibfnamefont {A.}~\bibnamefont {Bachtold}}, \bibinfo {author}
  {\bibfnamefont {A.~H.}\ \bibnamefont {MacDonald}},\ and\ \bibinfo {author}
  {\bibfnamefont {D.~K.}\ \bibnamefont {Efetov}},\ }\href
  {https://doi.org/10.1038/s41586-019-1695-0} {\bibfield  {journal} {\bibinfo
  {journal} {Nature}\ }\textbf {\bibinfo {volume} {574}},\ \bibinfo {pages}
  {653} (\bibinfo {year} {2019})}\BibitemShut {NoStop}%
\bibitem [{\citenamefont {Lu}\ \emph {et~al.}(2020)\citenamefont {Lu},
  \citenamefont {Lian}, \citenamefont {Chaudhary}, \citenamefont {Piot},
  \citenamefont {Romagnoli}, \citenamefont {Watanabe}, \citenamefont
  {Taniguchi}, \citenamefont {Poggio}, \citenamefont {MacDonald}, \citenamefont
  {Bernevig},\ and\ \citenamefont {Efetov}}]{LU20}%
  \BibitemOpen
  \bibfield  {author} {\bibinfo {author} {\bibfnamefont {X.}~\bibnamefont
  {Lu}}, \bibinfo {author} {\bibfnamefont {B.}~\bibnamefont {Lian}}, \bibinfo
  {author} {\bibfnamefont {G.}~\bibnamefont {Chaudhary}}, \bibinfo {author}
  {\bibfnamefont {B.~A.}\ \bibnamefont {Piot}}, \bibinfo {author}
  {\bibfnamefont {G.}~\bibnamefont {Romagnoli}}, \bibinfo {author}
  {\bibfnamefont {K.}~\bibnamefont {Watanabe}}, \bibinfo {author}
  {\bibfnamefont {T.}~\bibnamefont {Taniguchi}}, \bibinfo {author}
  {\bibfnamefont {M.}~\bibnamefont {Poggio}}, \bibinfo {author} {\bibfnamefont
  {A.~H.}\ \bibnamefont {MacDonald}}, \bibinfo {author} {\bibfnamefont {B.~A.}\
  \bibnamefont {Bernevig}},\ and\ \bibinfo {author} {\bibfnamefont {D.~K.}\
  \bibnamefont {Efetov}},\ }\href@noop {} {\bibfield  {journal} {\bibinfo
  {journal} {arXiv:2006.13963 [cond-mat]}\ } (\bibinfo {year} {2020})},\
  \Eprint {https://arxiv.org/abs/2006.13963} {arXiv:2006.13963 [cond-mat]}
  \BibitemShut {NoStop}%
\bibitem [{\citenamefont {Park}\ \emph
  {et~al.}(2020{\natexlab{a}})\citenamefont {Park}, \citenamefont {Cao},
  \citenamefont {Watanabe}, \citenamefont {Taniguchi},\ and\ \citenamefont
  {{Jarillo-Herrero}}}]{PAR20a}%
  \BibitemOpen
  \bibfield  {author} {\bibinfo {author} {\bibfnamefont {J.~M.}\ \bibnamefont
  {Park}}, \bibinfo {author} {\bibfnamefont {Y.}~\bibnamefont {Cao}}, \bibinfo
  {author} {\bibfnamefont {K.}~\bibnamefont {Watanabe}}, \bibinfo {author}
  {\bibfnamefont {T.}~\bibnamefont {Taniguchi}},\ and\ \bibinfo {author}
  {\bibfnamefont {P.}~\bibnamefont {{Jarillo-Herrero}}},\ }\href@noop {}
  {\bibfield  {journal} {\bibinfo  {journal} {arXiv:2008.12296 [cond-mat]}\ }
  (\bibinfo {year} {2020}{\natexlab{a}})},\ \Eprint
  {https://arxiv.org/abs/2008.12296} {arXiv:2008.12296 [cond-mat]} \BibitemShut
  {NoStop}%
\bibitem [{\citenamefont {Polshyn}\ \emph {et~al.}(2019)\citenamefont
  {Polshyn}, \citenamefont {Yankowitz}, \citenamefont {Chen}, \citenamefont
  {Zhang}, \citenamefont {Watanabe}, \citenamefont {Taniguchi}, \citenamefont
  {Dean},\ and\ \citenamefont {Young}}]{POL19}%
  \BibitemOpen
  \bibfield  {author} {\bibinfo {author} {\bibfnamefont {H.}~\bibnamefont
  {Polshyn}}, \bibinfo {author} {\bibfnamefont {M.}~\bibnamefont {Yankowitz}},
  \bibinfo {author} {\bibfnamefont {S.}~\bibnamefont {Chen}}, \bibinfo {author}
  {\bibfnamefont {Y.}~\bibnamefont {Zhang}}, \bibinfo {author} {\bibfnamefont
  {K.}~\bibnamefont {Watanabe}}, \bibinfo {author} {\bibfnamefont
  {T.}~\bibnamefont {Taniguchi}}, \bibinfo {author} {\bibfnamefont {C.~R.}\
  \bibnamefont {Dean}},\ and\ \bibinfo {author} {\bibfnamefont {A.~F.}\
  \bibnamefont {Young}},\ }\href {https://doi.org/10.1038/s41567-019-0596-3}
  {\bibfield  {journal} {\bibinfo  {journal} {Nat. Phys.}\ }\textbf {\bibinfo
  {volume} {15}},\ \bibinfo {pages} {1011} (\bibinfo {year}
  {2019})}\BibitemShut {NoStop}%
\bibitem [{\citenamefont {Saito}\ \emph {et~al.}(2020)\citenamefont {Saito},
  \citenamefont {Ge}, \citenamefont {Watanabe}, \citenamefont {Taniguchi},\
  and\ \citenamefont {Young}}]{SAI20}%
  \BibitemOpen
  \bibfield  {author} {\bibinfo {author} {\bibfnamefont {Y.}~\bibnamefont
  {Saito}}, \bibinfo {author} {\bibfnamefont {J.}~\bibnamefont {Ge}}, \bibinfo
  {author} {\bibfnamefont {K.}~\bibnamefont {Watanabe}}, \bibinfo {author}
  {\bibfnamefont {T.}~\bibnamefont {Taniguchi}},\ and\ \bibinfo {author}
  {\bibfnamefont {A.~F.}\ \bibnamefont {Young}},\ }\href
  {https://doi.org/10.1038/s41567-020-0928-3} {\bibfield  {journal} {\bibinfo
  {journal} {Nat. Phys.}\ }\textbf {\bibinfo {volume} {16}},\ \bibinfo {pages}
  {926} (\bibinfo {year} {2020})}\BibitemShut {NoStop}%
\bibitem [{\citenamefont {Saito}\ \emph {et~al.}(2021)\citenamefont {Saito},
  \citenamefont {Ge}, \citenamefont {Rademaker}, \citenamefont {Watanabe},
  \citenamefont {Taniguchi}, \citenamefont {Abanin},\ and\ \citenamefont
  {Young}}]{SAI21}%
  \BibitemOpen
  \bibfield  {author} {\bibinfo {author} {\bibfnamefont {Y.}~\bibnamefont
  {Saito}}, \bibinfo {author} {\bibfnamefont {J.}~\bibnamefont {Ge}}, \bibinfo
  {author} {\bibfnamefont {L.}~\bibnamefont {Rademaker}}, \bibinfo {author}
  {\bibfnamefont {K.}~\bibnamefont {Watanabe}}, \bibinfo {author}
  {\bibfnamefont {T.}~\bibnamefont {Taniguchi}}, \bibinfo {author}
  {\bibfnamefont {D.~A.}\ \bibnamefont {Abanin}},\ and\ \bibinfo {author}
  {\bibfnamefont {A.~F.}\ \bibnamefont {Young}},\ }\href
  {https://doi.org/10.1038/s41567-020-01129-4} {\bibfield  {journal} {\bibinfo
  {journal} {Nat. Phys.}\ ,\ \bibinfo {pages} {1}} (\bibinfo {year}
  {2021})}\BibitemShut {NoStop}%
\bibitem [{\citenamefont {Serlin}\ \emph {et~al.}(2020)\citenamefont {Serlin},
  \citenamefont {Tschirhart}, \citenamefont {Polshyn}, \citenamefont {Zhang},
  \citenamefont {Zhu}, \citenamefont {Watanabe}, \citenamefont {Taniguchi},
  \citenamefont {Balents},\ and\ \citenamefont {Young}}]{SER20}%
  \BibitemOpen
  \bibfield  {author} {\bibinfo {author} {\bibfnamefont {M.}~\bibnamefont
  {Serlin}}, \bibinfo {author} {\bibfnamefont {C.~L.}\ \bibnamefont
  {Tschirhart}}, \bibinfo {author} {\bibfnamefont {H.}~\bibnamefont {Polshyn}},
  \bibinfo {author} {\bibfnamefont {Y.}~\bibnamefont {Zhang}}, \bibinfo
  {author} {\bibfnamefont {J.}~\bibnamefont {Zhu}}, \bibinfo {author}
  {\bibfnamefont {K.}~\bibnamefont {Watanabe}}, \bibinfo {author}
  {\bibfnamefont {T.}~\bibnamefont {Taniguchi}}, \bibinfo {author}
  {\bibfnamefont {L.}~\bibnamefont {Balents}},\ and\ \bibinfo {author}
  {\bibfnamefont {A.~F.}\ \bibnamefont {Young}},\ }\href
  {https://doi.org/10.1126/science.aay5533} {\bibfield  {journal} {\bibinfo
  {journal} {Science}\ }\textbf {\bibinfo {volume} {367}},\ \bibinfo {pages}
  {900} (\bibinfo {year} {2020})}\BibitemShut {NoStop}%
\bibitem [{\citenamefont {Stepanov}\ \emph {et~al.}(2020)\citenamefont
  {Stepanov}, \citenamefont {Das}, \citenamefont {Lu}, \citenamefont
  {Fahimniya}, \citenamefont {Watanabe}, \citenamefont {Taniguchi},
  \citenamefont {Koppens}, \citenamefont {Lischner}, \citenamefont {Levitov},\
  and\ \citenamefont {Efetov}}]{STE20}%
  \BibitemOpen
  \bibfield  {author} {\bibinfo {author} {\bibfnamefont {P.}~\bibnamefont
  {Stepanov}}, \bibinfo {author} {\bibfnamefont {I.}~\bibnamefont {Das}},
  \bibinfo {author} {\bibfnamefont {X.}~\bibnamefont {Lu}}, \bibinfo {author}
  {\bibfnamefont {A.}~\bibnamefont {Fahimniya}}, \bibinfo {author}
  {\bibfnamefont {K.}~\bibnamefont {Watanabe}}, \bibinfo {author}
  {\bibfnamefont {T.}~\bibnamefont {Taniguchi}}, \bibinfo {author}
  {\bibfnamefont {F.~H.~L.}\ \bibnamefont {Koppens}}, \bibinfo {author}
  {\bibfnamefont {J.}~\bibnamefont {Lischner}}, \bibinfo {author}
  {\bibfnamefont {L.}~\bibnamefont {Levitov}},\ and\ \bibinfo {author}
  {\bibfnamefont {D.~K.}\ \bibnamefont {Efetov}},\ }\href
  {https://doi.org/10.1038/s41586-020-2459-6} {\bibfield  {journal} {\bibinfo
  {journal} {Nature}\ }\textbf {\bibinfo {volume} {583}},\ \bibinfo {pages}
  {375} (\bibinfo {year} {2020})}\BibitemShut {NoStop}%
\bibitem [{\citenamefont {Wu}\ \emph {et~al.}(2020{\natexlab{a}})\citenamefont
  {Wu}, \citenamefont {Zhang}, \citenamefont {Watanabe}, \citenamefont
  {Taniguchi},\ and\ \citenamefont {Andrei}}]{WU20a}%
  \BibitemOpen
  \bibfield  {author} {\bibinfo {author} {\bibfnamefont {S.}~\bibnamefont
  {Wu}}, \bibinfo {author} {\bibfnamefont {Z.}~\bibnamefont {Zhang}}, \bibinfo
  {author} {\bibfnamefont {K.}~\bibnamefont {Watanabe}}, \bibinfo {author}
  {\bibfnamefont {T.}~\bibnamefont {Taniguchi}},\ and\ \bibinfo {author}
  {\bibfnamefont {E.~Y.}\ \bibnamefont {Andrei}},\ }\href@noop {} {\bibfield
  {journal} {\bibinfo  {journal} {arXiv:2007.03735 [cond-mat]}\ } (\bibinfo
  {year} {2020}{\natexlab{a}})},\ \Eprint {https://arxiv.org/abs/2007.03735}
  {arXiv:2007.03735 [cond-mat]} \BibitemShut {NoStop}%
\bibitem [{\citenamefont {Yankowitz}\ \emph {et~al.}(2019)\citenamefont
  {Yankowitz}, \citenamefont {Chen}, \citenamefont {Polshyn}, \citenamefont
  {Zhang}, \citenamefont {Watanabe}, \citenamefont {Taniguchi}, \citenamefont
  {Graf}, \citenamefont {Young},\ and\ \citenamefont {Dean}}]{YAN19}%
  \BibitemOpen
  \bibfield  {author} {\bibinfo {author} {\bibfnamefont {M.}~\bibnamefont
  {Yankowitz}}, \bibinfo {author} {\bibfnamefont {S.}~\bibnamefont {Chen}},
  \bibinfo {author} {\bibfnamefont {H.}~\bibnamefont {Polshyn}}, \bibinfo
  {author} {\bibfnamefont {Y.}~\bibnamefont {Zhang}}, \bibinfo {author}
  {\bibfnamefont {K.}~\bibnamefont {Watanabe}}, \bibinfo {author}
  {\bibfnamefont {T.}~\bibnamefont {Taniguchi}}, \bibinfo {author}
  {\bibfnamefont {D.}~\bibnamefont {Graf}}, \bibinfo {author} {\bibfnamefont
  {A.~F.}\ \bibnamefont {Young}},\ and\ \bibinfo {author} {\bibfnamefont
  {C.~R.}\ \bibnamefont {Dean}},\ }\href
  {https://doi.org/10.1126/science.aav1910} {\bibfield  {journal} {\bibinfo
  {journal} {Science}\ }\textbf {\bibinfo {volume} {363}},\ \bibinfo {pages}
  {1059} (\bibinfo {year} {2019})}\BibitemShut {NoStop}%
\bibitem [{\citenamefont {Choi}\ \emph {et~al.}(2019)\citenamefont {Choi},
  \citenamefont {Kemmer}, \citenamefont {Peng}, \citenamefont {Thomson},
  \citenamefont {Arora}, \citenamefont {Polski}, \citenamefont {Zhang},
  \citenamefont {Ren}, \citenamefont {Alicea}, \citenamefont {Refael},
  \citenamefont {{von Oppen}}, \citenamefont {Watanabe}, \citenamefont
  {Taniguchi},\ and\ \citenamefont {{Nadj-Perge}}}]{CHO19}%
  \BibitemOpen
  \bibfield  {author} {\bibinfo {author} {\bibfnamefont {Y.}~\bibnamefont
  {Choi}}, \bibinfo {author} {\bibfnamefont {J.}~\bibnamefont {Kemmer}},
  \bibinfo {author} {\bibfnamefont {Y.}~\bibnamefont {Peng}}, \bibinfo {author}
  {\bibfnamefont {A.}~\bibnamefont {Thomson}}, \bibinfo {author} {\bibfnamefont
  {H.}~\bibnamefont {Arora}}, \bibinfo {author} {\bibfnamefont
  {R.}~\bibnamefont {Polski}}, \bibinfo {author} {\bibfnamefont
  {Y.}~\bibnamefont {Zhang}}, \bibinfo {author} {\bibfnamefont
  {H.}~\bibnamefont {Ren}}, \bibinfo {author} {\bibfnamefont {J.}~\bibnamefont
  {Alicea}}, \bibinfo {author} {\bibfnamefont {G.}~\bibnamefont {Refael}},
  \bibinfo {author} {\bibfnamefont {F.}~\bibnamefont {{von Oppen}}}, \bibinfo
  {author} {\bibfnamefont {K.}~\bibnamefont {Watanabe}}, \bibinfo {author}
  {\bibfnamefont {T.}~\bibnamefont {Taniguchi}},\ and\ \bibinfo {author}
  {\bibfnamefont {S.}~\bibnamefont {{Nadj-Perge}}},\ }\href
  {https://doi.org/10.1038/s41567-019-0606-5} {\bibfield  {journal} {\bibinfo
  {journal} {Nat. Phys.}\ }\textbf {\bibinfo {volume} {15}},\ \bibinfo {pages}
  {1174} (\bibinfo {year} {2019})}\BibitemShut {NoStop}%
\bibitem [{\citenamefont {Choi}\ \emph {et~al.}(2020)\citenamefont {Choi},
  \citenamefont {Kim}, \citenamefont {Peng}, \citenamefont {Thomson},
  \citenamefont {Lewandowski}, \citenamefont {Polski}, \citenamefont {Zhang},
  \citenamefont {Arora}, \citenamefont {Watanabe}, \citenamefont {Taniguchi},
  \citenamefont {Alicea},\ and\ \citenamefont {{Nadj-Perge}}}]{CHO20}%
  \BibitemOpen
  \bibfield  {author} {\bibinfo {author} {\bibfnamefont {Y.}~\bibnamefont
  {Choi}}, \bibinfo {author} {\bibfnamefont {H.}~\bibnamefont {Kim}}, \bibinfo
  {author} {\bibfnamefont {Y.}~\bibnamefont {Peng}}, \bibinfo {author}
  {\bibfnamefont {A.}~\bibnamefont {Thomson}}, \bibinfo {author} {\bibfnamefont
  {C.}~\bibnamefont {Lewandowski}}, \bibinfo {author} {\bibfnamefont
  {R.}~\bibnamefont {Polski}}, \bibinfo {author} {\bibfnamefont
  {Y.}~\bibnamefont {Zhang}}, \bibinfo {author} {\bibfnamefont {H.~S.}\
  \bibnamefont {Arora}}, \bibinfo {author} {\bibfnamefont {K.}~\bibnamefont
  {Watanabe}}, \bibinfo {author} {\bibfnamefont {T.}~\bibnamefont {Taniguchi}},
  \bibinfo {author} {\bibfnamefont {J.}~\bibnamefont {Alicea}},\ and\ \bibinfo
  {author} {\bibfnamefont {S.}~\bibnamefont {{Nadj-Perge}}},\ }\href@noop {}
  {\bibfield  {journal} {\bibinfo  {journal} {arXiv:2008.11746 [cond-mat]}\ }
  (\bibinfo {year} {2020})},\ \Eprint {https://arxiv.org/abs/2008.11746}
  {arXiv:2008.11746 [cond-mat]} \BibitemShut {NoStop}%
\bibitem [{\citenamefont {Kerelsky}\ \emph {et~al.}(2019)\citenamefont
  {Kerelsky}, \citenamefont {McGilly}, \citenamefont {Kennes}, \citenamefont
  {Xian}, \citenamefont {Yankowitz}, \citenamefont {Chen}, \citenamefont
  {Watanabe}, \citenamefont {Taniguchi}, \citenamefont {Hone}, \citenamefont
  {Dean}, \citenamefont {Rubio},\ and\ \citenamefont {Pasupathy}}]{KER19}%
  \BibitemOpen
  \bibfield  {author} {\bibinfo {author} {\bibfnamefont {A.}~\bibnamefont
  {Kerelsky}}, \bibinfo {author} {\bibfnamefont {L.~J.}\ \bibnamefont
  {McGilly}}, \bibinfo {author} {\bibfnamefont {D.~M.}\ \bibnamefont {Kennes}},
  \bibinfo {author} {\bibfnamefont {L.}~\bibnamefont {Xian}}, \bibinfo {author}
  {\bibfnamefont {M.}~\bibnamefont {Yankowitz}}, \bibinfo {author}
  {\bibfnamefont {S.}~\bibnamefont {Chen}}, \bibinfo {author} {\bibfnamefont
  {K.}~\bibnamefont {Watanabe}}, \bibinfo {author} {\bibfnamefont
  {T.}~\bibnamefont {Taniguchi}}, \bibinfo {author} {\bibfnamefont
  {J.}~\bibnamefont {Hone}}, \bibinfo {author} {\bibfnamefont {C.}~\bibnamefont
  {Dean}}, \bibinfo {author} {\bibfnamefont {A.}~\bibnamefont {Rubio}},\ and\
  \bibinfo {author} {\bibfnamefont {A.~N.}\ \bibnamefont {Pasupathy}},\ }\href
  {https://doi.org/10.1038/s41586-019-1431-9} {\bibfield  {journal} {\bibinfo
  {journal} {Nature}\ }\textbf {\bibinfo {volume} {572}},\ \bibinfo {pages}
  {95} (\bibinfo {year} {2019})}\BibitemShut {NoStop}%
\bibitem [{\citenamefont {Nuckolls}\ \emph {et~al.}(2020)\citenamefont
  {Nuckolls}, \citenamefont {Oh}, \citenamefont {Wong}, \citenamefont {Lian},
  \citenamefont {Watanabe}, \citenamefont {Taniguchi}, \citenamefont
  {Bernevig},\ and\ \citenamefont {Yazdani}}]{NUC20}%
  \BibitemOpen
  \bibfield  {author} {\bibinfo {author} {\bibfnamefont {K.~P.}\ \bibnamefont
  {Nuckolls}}, \bibinfo {author} {\bibfnamefont {M.}~\bibnamefont {Oh}},
  \bibinfo {author} {\bibfnamefont {D.}~\bibnamefont {Wong}}, \bibinfo {author}
  {\bibfnamefont {B.}~\bibnamefont {Lian}}, \bibinfo {author} {\bibfnamefont
  {K.}~\bibnamefont {Watanabe}}, \bibinfo {author} {\bibfnamefont
  {T.}~\bibnamefont {Taniguchi}}, \bibinfo {author} {\bibfnamefont {B.~A.}\
  \bibnamefont {Bernevig}},\ and\ \bibinfo {author} {\bibfnamefont
  {A.}~\bibnamefont {Yazdani}},\ }\href
  {https://doi.org/10.1038/s41586-020-3028-8} {\bibfield  {journal} {\bibinfo
  {journal} {Nature}\ }\textbf {\bibinfo {volume} {588}},\ \bibinfo {pages}
  {610} (\bibinfo {year} {2020})}\BibitemShut {NoStop}%
\bibitem [{\citenamefont {Wong}\ \emph {et~al.}(2020)\citenamefont {Wong},
  \citenamefont {Nuckolls}, \citenamefont {Oh}, \citenamefont {Lian},
  \citenamefont {Xie}, \citenamefont {Jeon}, \citenamefont {Watanabe},
  \citenamefont {Taniguchi}, \citenamefont {Bernevig},\ and\ \citenamefont
  {Yazdani}}]{WON20}%
  \BibitemOpen
  \bibfield  {author} {\bibinfo {author} {\bibfnamefont {D.}~\bibnamefont
  {Wong}}, \bibinfo {author} {\bibfnamefont {K.~P.}\ \bibnamefont {Nuckolls}},
  \bibinfo {author} {\bibfnamefont {M.}~\bibnamefont {Oh}}, \bibinfo {author}
  {\bibfnamefont {B.}~\bibnamefont {Lian}}, \bibinfo {author} {\bibfnamefont
  {Y.}~\bibnamefont {Xie}}, \bibinfo {author} {\bibfnamefont {S.}~\bibnamefont
  {Jeon}}, \bibinfo {author} {\bibfnamefont {K.}~\bibnamefont {Watanabe}},
  \bibinfo {author} {\bibfnamefont {T.}~\bibnamefont {Taniguchi}}, \bibinfo
  {author} {\bibfnamefont {B.~A.}\ \bibnamefont {Bernevig}},\ and\ \bibinfo
  {author} {\bibfnamefont {A.}~\bibnamefont {Yazdani}},\ }\href
  {https://doi.org/10.1038/s41586-020-2339-0} {\bibfield  {journal} {\bibinfo
  {journal} {Nature}\ }\textbf {\bibinfo {volume} {582}},\ \bibinfo {pages}
  {198} (\bibinfo {year} {2020})}\BibitemShut {NoStop}%
\bibitem [{\citenamefont {Xie}\ \emph {et~al.}(2019)\citenamefont {Xie},
  \citenamefont {Lian}, \citenamefont {J{\"a}ck}, \citenamefont {Liu},
  \citenamefont {Chiu}, \citenamefont {Watanabe}, \citenamefont {Taniguchi},
  \citenamefont {Bernevig},\ and\ \citenamefont {Yazdani}}]{XIE19}%
  \BibitemOpen
  \bibfield  {author} {\bibinfo {author} {\bibfnamefont {Y.}~\bibnamefont
  {Xie}}, \bibinfo {author} {\bibfnamefont {B.}~\bibnamefont {Lian}}, \bibinfo
  {author} {\bibfnamefont {B.}~\bibnamefont {J{\"a}ck}}, \bibinfo {author}
  {\bibfnamefont {X.}~\bibnamefont {Liu}}, \bibinfo {author} {\bibfnamefont
  {C.-L.}\ \bibnamefont {Chiu}}, \bibinfo {author} {\bibfnamefont
  {K.}~\bibnamefont {Watanabe}}, \bibinfo {author} {\bibfnamefont
  {T.}~\bibnamefont {Taniguchi}}, \bibinfo {author} {\bibfnamefont {B.~A.}\
  \bibnamefont {Bernevig}},\ and\ \bibinfo {author} {\bibfnamefont
  {A.}~\bibnamefont {Yazdani}},\ }\href
  {https://doi.org/10.1038/s41586-019-1422-x} {\bibfield  {journal} {\bibinfo
  {journal} {Nature}\ }\textbf {\bibinfo {volume} {572}},\ \bibinfo {pages}
  {101} (\bibinfo {year} {2019})}\BibitemShut {NoStop}%
\bibitem [{\citenamefont {Jiang}\ \emph {et~al.}(2019)\citenamefont {Jiang},
  \citenamefont {Lai}, \citenamefont {Watanabe}, \citenamefont {Taniguchi},
  \citenamefont {Haule}, \citenamefont {Mao},\ and\ \citenamefont
  {Andrei}}]{JIA19}%
  \BibitemOpen
  \bibfield  {author} {\bibinfo {author} {\bibfnamefont {Y.}~\bibnamefont
  {Jiang}}, \bibinfo {author} {\bibfnamefont {X.}~\bibnamefont {Lai}}, \bibinfo
  {author} {\bibfnamefont {K.}~\bibnamefont {Watanabe}}, \bibinfo {author}
  {\bibfnamefont {T.}~\bibnamefont {Taniguchi}}, \bibinfo {author}
  {\bibfnamefont {K.}~\bibnamefont {Haule}}, \bibinfo {author} {\bibfnamefont
  {J.}~\bibnamefont {Mao}},\ and\ \bibinfo {author} {\bibfnamefont {E.~Y.}\
  \bibnamefont {Andrei}},\ }\href {https://doi.org/10.1038/s41586-019-1460-4}
  {\bibfield  {journal} {\bibinfo  {journal} {Nature}\ }\textbf {\bibinfo
  {volume} {573}},\ \bibinfo {pages} {91} (\bibinfo {year} {2019})}\BibitemShut
  {NoStop}%
\bibitem [{\citenamefont {Choi}\ \emph {et~al.}(2021)\citenamefont {Choi},
  \citenamefont {Kim}, \citenamefont {Lewandowski}, \citenamefont {Peng},
  \citenamefont {Thomson}, \citenamefont {Polski}, \citenamefont {Zhang},
  \citenamefont {Watanabe}, \citenamefont {Taniguchi}, \citenamefont {Alicea},\
  and\ \citenamefont {{Nadj-Perge}}}]{CHO21}%
  \BibitemOpen
  \bibfield  {author} {\bibinfo {author} {\bibfnamefont {Y.}~\bibnamefont
  {Choi}}, \bibinfo {author} {\bibfnamefont {H.}~\bibnamefont {Kim}}, \bibinfo
  {author} {\bibfnamefont {C.}~\bibnamefont {Lewandowski}}, \bibinfo {author}
  {\bibfnamefont {Y.}~\bibnamefont {Peng}}, \bibinfo {author} {\bibfnamefont
  {A.}~\bibnamefont {Thomson}}, \bibinfo {author} {\bibfnamefont
  {R.}~\bibnamefont {Polski}}, \bibinfo {author} {\bibfnamefont
  {Y.}~\bibnamefont {Zhang}}, \bibinfo {author} {\bibfnamefont
  {K.}~\bibnamefont {Watanabe}}, \bibinfo {author} {\bibfnamefont
  {T.}~\bibnamefont {Taniguchi}}, \bibinfo {author} {\bibfnamefont
  {J.}~\bibnamefont {Alicea}},\ and\ \bibinfo {author} {\bibfnamefont
  {S.}~\bibnamefont {{Nadj-Perge}}},\ }\href@noop {} {\bibfield  {journal}
  {\bibinfo  {journal} {arXiv:2102.02209 [cond-mat]}\ } (\bibinfo {year}
  {2021})},\ \Eprint {https://arxiv.org/abs/2102.02209} {arXiv:2102.02209
  [cond-mat]} \BibitemShut {NoStop}%
\bibitem [{\citenamefont {Kang}\ and\ \citenamefont {Vafek}(2019)}]{KAN19}%
  \BibitemOpen
  \bibfield  {author} {\bibinfo {author} {\bibfnamefont {J.}~\bibnamefont
  {Kang}}\ and\ \bibinfo {author} {\bibfnamefont {O.}~\bibnamefont {Vafek}},\
  }\href {https://doi.org/10.1103/PhysRevLett.122.246401} {\bibfield  {journal}
  {\bibinfo  {journal} {Phys. Rev. Lett.}\ }\textbf {\bibinfo {volume} {122}},\
  \bibinfo {pages} {246401} (\bibinfo {year} {2019})}\BibitemShut {NoStop}%
\bibitem [{\citenamefont {Seo}\ \emph {et~al.}(2019)\citenamefont {Seo},
  \citenamefont {Kotov},\ and\ \citenamefont {Uchoa}}]{SEO19}%
  \BibitemOpen
  \bibfield  {author} {\bibinfo {author} {\bibfnamefont {K.}~\bibnamefont
  {Seo}}, \bibinfo {author} {\bibfnamefont {V.~N.}\ \bibnamefont {Kotov}},\
  and\ \bibinfo {author} {\bibfnamefont {B.}~\bibnamefont {Uchoa}},\ }\href
  {https://doi.org/10.1103/PhysRevLett.122.246402} {\bibfield  {journal}
  {\bibinfo  {journal} {Phys. Rev. Lett.}\ }\textbf {\bibinfo {volume} {122}},\
  \bibinfo {pages} {246402} (\bibinfo {year} {2019})}\BibitemShut {NoStop}%
\bibitem [{\citenamefont {Bultinck}\ \emph
  {et~al.}(2020{\natexlab{a}})\citenamefont {Bultinck}, \citenamefont {Khalaf},
  \citenamefont {Liu}, \citenamefont {Chatterjee}, \citenamefont {Vishwanath},\
  and\ \citenamefont {Zaletel}}]{BUL20}%
  \BibitemOpen
  \bibfield  {author} {\bibinfo {author} {\bibfnamefont {N.}~\bibnamefont
  {Bultinck}}, \bibinfo {author} {\bibfnamefont {E.}~\bibnamefont {Khalaf}},
  \bibinfo {author} {\bibfnamefont {S.}~\bibnamefont {Liu}}, \bibinfo {author}
  {\bibfnamefont {S.}~\bibnamefont {Chatterjee}}, \bibinfo {author}
  {\bibfnamefont {A.}~\bibnamefont {Vishwanath}},\ and\ \bibinfo {author}
  {\bibfnamefont {M.~P.}\ \bibnamefont {Zaletel}},\ }\href
  {https://doi.org/10.1103/PhysRevX.10.031034} {\bibfield  {journal} {\bibinfo
  {journal} {Phys. Rev. X}\ }\textbf {\bibinfo {volume} {10}},\ \bibinfo
  {pages} {031034} (\bibinfo {year} {2020}{\natexlab{a}})}\BibitemShut
  {NoStop}%
\bibitem [{\citenamefont {Hejazi}\ \emph {et~al.}(2020)\citenamefont {Hejazi},
  \citenamefont {Chen},\ and\ \citenamefont {Balents}}]{HEJ20}%
  \BibitemOpen
  \bibfield  {author} {\bibinfo {author} {\bibfnamefont {K.}~\bibnamefont
  {Hejazi}}, \bibinfo {author} {\bibfnamefont {X.}~\bibnamefont {Chen}},\ and\
  \bibinfo {author} {\bibfnamefont {L.}~\bibnamefont {Balents}},\ }\href@noop
  {} {\bibfield  {journal} {\bibinfo  {journal} {arXiv:2007.00134 [cond-mat]}\
  } (\bibinfo {year} {2020})},\ \Eprint {https://arxiv.org/abs/2007.00134}
  {arXiv:2007.00134 [cond-mat]} \BibitemShut {NoStop}%
\bibitem [{\citenamefont {Fernandes}\ and\ \citenamefont {Fu}(2021)}]{FER21}%
  \BibitemOpen
  \bibfield  {author} {\bibinfo {author} {\bibfnamefont {R.~M.}\ \bibnamefont
  {Fernandes}}\ and\ \bibinfo {author} {\bibfnamefont {L.}~\bibnamefont {Fu}},\
  }\href@noop {} {\bibfield  {journal} {\bibinfo  {journal} {arXiv:2101.07943
  [cond-mat]}\ } (\bibinfo {year} {2021})},\ \Eprint
  {https://arxiv.org/abs/2101.07943} {arXiv:2101.07943 [cond-mat]} \BibitemShut
  {NoStop}%
\bibitem [{\citenamefont {Fernandes}\ and\ \citenamefont
  {Venderbos}(2020)}]{FER20}%
  \BibitemOpen
  \bibfield  {author} {\bibinfo {author} {\bibfnamefont {R.~M.}\ \bibnamefont
  {Fernandes}}\ and\ \bibinfo {author} {\bibfnamefont {J.~W.~F.}\ \bibnamefont
  {Venderbos}},\ }\href {https://doi.org/10.1126/sciadv.aba8834} {\bibfield
  {journal} {\bibinfo  {journal} {Science Advances}\ }\textbf {\bibinfo
  {volume} {6}},\ \bibinfo {pages} {eaba8834} (\bibinfo {year}
  {2020})}\BibitemShut {NoStop}%
\bibitem [{\citenamefont {Venderbos}\ and\ \citenamefont
  {Fernandes}(2018)}]{VEN18}%
  \BibitemOpen
  \bibfield  {author} {\bibinfo {author} {\bibfnamefont {J.~W.~F.}\
  \bibnamefont {Venderbos}}\ and\ \bibinfo {author} {\bibfnamefont {R.~M.}\
  \bibnamefont {Fernandes}},\ }\href
  {https://doi.org/10.1103/PhysRevB.98.245103} {\bibfield  {journal} {\bibinfo
  {journal} {Phys. Rev. B}\ }\textbf {\bibinfo {volume} {98}},\ \bibinfo
  {pages} {245103} (\bibinfo {year} {2018})}\BibitemShut {NoStop}%
\bibitem [{\citenamefont {Potasz}\ \emph {et~al.}(2021)\citenamefont {Potasz},
  \citenamefont {Xie},\ and\ \citenamefont {MacDonald}}]{POT21}%
  \BibitemOpen
  \bibfield  {author} {\bibinfo {author} {\bibfnamefont {P.}~\bibnamefont
  {Potasz}}, \bibinfo {author} {\bibfnamefont {M.}~\bibnamefont {Xie}},\ and\
  \bibinfo {author} {\bibfnamefont {A.~H.}\ \bibnamefont {MacDonald}},\
  }\href@noop {} {\bibfield  {journal} {\bibinfo  {journal} {arXiv:2102.02256
  [cond-mat]}\ } (\bibinfo {year} {2021})},\ \Eprint
  {https://arxiv.org/abs/2102.02256} {arXiv:2102.02256 [cond-mat]} \BibitemShut
  {NoStop}%
\bibitem [{\citenamefont {Abouelkomsan}\ \emph {et~al.}(2020)\citenamefont
  {Abouelkomsan}, \citenamefont {Liu},\ and\ \citenamefont
  {Bergholtz}}]{ABO20}%
  \BibitemOpen
  \bibfield  {author} {\bibinfo {author} {\bibfnamefont {A.}~\bibnamefont
  {Abouelkomsan}}, \bibinfo {author} {\bibfnamefont {Z.}~\bibnamefont {Liu}},\
  and\ \bibinfo {author} {\bibfnamefont {E.~J.}\ \bibnamefont {Bergholtz}},\
  }\href {https://doi.org/10.1103/PhysRevLett.124.106803} {\bibfield  {journal}
  {\bibinfo  {journal} {Phys. Rev. Lett.}\ }\textbf {\bibinfo {volume} {124}},\
  \bibinfo {pages} {106803} (\bibinfo {year} {2020})}\BibitemShut {NoStop}%
\bibitem [{\citenamefont {Ahn}\ \emph {et~al.}(2019)\citenamefont {Ahn},
  \citenamefont {Park},\ and\ \citenamefont {Yang}}]{AHN19}%
  \BibitemOpen
  \bibfield  {author} {\bibinfo {author} {\bibfnamefont {J.}~\bibnamefont
  {Ahn}}, \bibinfo {author} {\bibfnamefont {S.}~\bibnamefont {Park}},\ and\
  \bibinfo {author} {\bibfnamefont {B.-J.}\ \bibnamefont {Yang}},\ }\href
  {https://doi.org/10.1103/PhysRevX.9.021013} {\bibfield  {journal} {\bibinfo
  {journal} {Phys. Rev. X}\ }\textbf {\bibinfo {volume} {9}},\ \bibinfo {pages}
  {021013} (\bibinfo {year} {2019})}\BibitemShut {NoStop}%
\bibitem [{\citenamefont {Bernevig}\ \emph
  {et~al.}(2020{\natexlab{a}})\citenamefont {Bernevig}, \citenamefont {Song},
  \citenamefont {Regnault},\ and\ \citenamefont {Lian}}]{BER20}%
  \BibitemOpen
  \bibfield  {author} {\bibinfo {author} {\bibfnamefont {B.~A.}\ \bibnamefont
  {Bernevig}}, \bibinfo {author} {\bibfnamefont {Z.-D.}\ \bibnamefont {Song}},
  \bibinfo {author} {\bibfnamefont {N.}~\bibnamefont {Regnault}},\ and\
  \bibinfo {author} {\bibfnamefont {B.}~\bibnamefont {Lian}},\ }\href@noop {}
  {\bibfield  {journal} {\bibinfo  {journal} {arXiv:2009.11301 [cond-mat]}\ }
  (\bibinfo {year} {2020}{\natexlab{a}})},\ \Eprint
  {https://arxiv.org/abs/2009.11301} {arXiv:2009.11301 [cond-mat]} \BibitemShut
  {NoStop}%
\bibitem [{\citenamefont {Bernevig}\ \emph
  {et~al.}(2020{\natexlab{b}})\citenamefont {Bernevig}, \citenamefont {Song},
  \citenamefont {Regnault},\ and\ \citenamefont {Lian}}]{BER20a}%
  \BibitemOpen
  \bibfield  {author} {\bibinfo {author} {\bibfnamefont {B.~A.}\ \bibnamefont
  {Bernevig}}, \bibinfo {author} {\bibfnamefont {Z.-D.}\ \bibnamefont {Song}},
  \bibinfo {author} {\bibfnamefont {N.}~\bibnamefont {Regnault}},\ and\
  \bibinfo {author} {\bibfnamefont {B.}~\bibnamefont {Lian}},\ }\href@noop {}
  {\bibfield  {journal} {\bibinfo  {journal} {arXiv:2009.12376 [cond-mat]}\ }
  (\bibinfo {year} {2020}{\natexlab{b}})},\ \Eprint
  {https://arxiv.org/abs/2009.12376} {arXiv:2009.12376 [cond-mat]} \BibitemShut
  {NoStop}%
\bibitem [{\citenamefont {Bernevig}\ \emph
  {et~al.}(2020{\natexlab{c}})\citenamefont {Bernevig}, \citenamefont {Lian},
  \citenamefont {Cowsik}, \citenamefont {Xie}, \citenamefont {Regnault},\ and\
  \citenamefont {Song}}]{BER20b}%
  \BibitemOpen
  \bibfield  {author} {\bibinfo {author} {\bibfnamefont {B.~A.}\ \bibnamefont
  {Bernevig}}, \bibinfo {author} {\bibfnamefont {B.}~\bibnamefont {Lian}},
  \bibinfo {author} {\bibfnamefont {A.}~\bibnamefont {Cowsik}}, \bibinfo
  {author} {\bibfnamefont {F.}~\bibnamefont {Xie}}, \bibinfo {author}
  {\bibfnamefont {N.}~\bibnamefont {Regnault}},\ and\ \bibinfo {author}
  {\bibfnamefont {Z.-D.}\ \bibnamefont {Song}},\ }\href@noop {} {\bibfield
  {journal} {\bibinfo  {journal} {arXiv:2009.14200 [cond-mat]}\ } (\bibinfo
  {year} {2020}{\natexlab{c}})},\ \Eprint {https://arxiv.org/abs/2009.14200}
  {arXiv:2009.14200 [cond-mat]} \BibitemShut {NoStop}%
\bibitem [{\citenamefont {Bultinck}\ \emph
  {et~al.}(2020{\natexlab{b}})\citenamefont {Bultinck}, \citenamefont
  {Chatterjee},\ and\ \citenamefont {Zaletel}}]{BUL20a}%
  \BibitemOpen
  \bibfield  {author} {\bibinfo {author} {\bibfnamefont {N.}~\bibnamefont
  {Bultinck}}, \bibinfo {author} {\bibfnamefont {S.}~\bibnamefont
  {Chatterjee}},\ and\ \bibinfo {author} {\bibfnamefont {M.~P.}\ \bibnamefont
  {Zaletel}},\ }\href {https://doi.org/10.1103/PhysRevLett.124.166601}
  {\bibfield  {journal} {\bibinfo  {journal} {Phys. Rev. Lett.}\ }\textbf
  {\bibinfo {volume} {124}},\ \bibinfo {pages} {166601} (\bibinfo {year}
  {2020}{\natexlab{b}})}\BibitemShut {NoStop}%
\bibitem [{\citenamefont {Cao}\ \emph {et~al.}(2020{\natexlab{b}})\citenamefont
  {Cao}, \citenamefont {Wang}, \citenamefont {Liu},\ and\ \citenamefont
  {Yao}}]{CAO20b}%
  \BibitemOpen
  \bibfield  {author} {\bibinfo {author} {\bibfnamefont {J.}~\bibnamefont
  {Cao}}, \bibinfo {author} {\bibfnamefont {M.}~\bibnamefont {Wang}}, \bibinfo
  {author} {\bibfnamefont {C.-C.}\ \bibnamefont {Liu}},\ and\ \bibinfo {author}
  {\bibfnamefont {Y.}~\bibnamefont {Yao}},\ }\href@noop {} {\bibfield
  {journal} {\bibinfo  {journal} {arXiv:2012.02575 [cond-mat]}\ } (\bibinfo
  {year} {2020}{\natexlab{b}})},\ \Eprint {https://arxiv.org/abs/2012.02575}
  {arXiv:2012.02575 [cond-mat]} \BibitemShut {NoStop}%
\bibitem [{\citenamefont {Cea}\ and\ \citenamefont {Guinea}(2020)}]{CEA20}%
  \BibitemOpen
  \bibfield  {author} {\bibinfo {author} {\bibfnamefont {T.}~\bibnamefont
  {Cea}}\ and\ \bibinfo {author} {\bibfnamefont {F.}~\bibnamefont {Guinea}},\
  }\href {https://doi.org/10.1103/PhysRevB.102.045107} {\bibfield  {journal}
  {\bibinfo  {journal} {Phys. Rev. B}\ }\textbf {\bibinfo {volume} {102}},\
  \bibinfo {pages} {045107} (\bibinfo {year} {2020})}\BibitemShut {NoStop}%
\bibitem [{\citenamefont {Christos}\ \emph {et~al.}(2020)\citenamefont
  {Christos}, \citenamefont {Sachdev},\ and\ \citenamefont {Scheurer}}]{CHR20}%
  \BibitemOpen
  \bibfield  {author} {\bibinfo {author} {\bibfnamefont {M.}~\bibnamefont
  {Christos}}, \bibinfo {author} {\bibfnamefont {S.}~\bibnamefont {Sachdev}},\
  and\ \bibinfo {author} {\bibfnamefont {M.~S.}\ \bibnamefont {Scheurer}},\
  }\href {https://doi.org/10.1073/pnas.2014691117} {\bibfield  {journal}
  {\bibinfo  {journal} {PNAS}\ }\textbf {\bibinfo {volume} {117}},\ \bibinfo
  {pages} {29543} (\bibinfo {year} {2020})}\BibitemShut {NoStop}%
\bibitem [{\citenamefont {Classen}\ \emph {et~al.}(2019)\citenamefont
  {Classen}, \citenamefont {Honerkamp},\ and\ \citenamefont {Scherer}}]{CLA19}%
  \BibitemOpen
  \bibfield  {author} {\bibinfo {author} {\bibfnamefont {L.}~\bibnamefont
  {Classen}}, \bibinfo {author} {\bibfnamefont {C.}~\bibnamefont {Honerkamp}},\
  and\ \bibinfo {author} {\bibfnamefont {M.~M.}\ \bibnamefont {Scherer}},\
  }\href {https://doi.org/10.1103/PhysRevB.99.195120} {\bibfield  {journal}
  {\bibinfo  {journal} {Phys. Rev. B}\ }\textbf {\bibinfo {volume} {99}},\
  \bibinfo {pages} {195120} (\bibinfo {year} {2019})}\BibitemShut {NoStop}%
\bibitem [{\citenamefont {Da~Liao}\ \emph {et~al.}(2019)\citenamefont
  {Da~Liao}, \citenamefont {Meng},\ and\ \citenamefont {Xu}}]{DA19}%
  \BibitemOpen
  \bibfield  {author} {\bibinfo {author} {\bibfnamefont {Y.}~\bibnamefont
  {Da~Liao}}, \bibinfo {author} {\bibfnamefont {Z.~Y.}\ \bibnamefont {Meng}},\
  and\ \bibinfo {author} {\bibfnamefont {X.~Y.}\ \bibnamefont {Xu}},\ }\href
  {https://doi.org/10.1103/PhysRevLett.123.157601} {\bibfield  {journal}
  {\bibinfo  {journal} {Phys. Rev. Lett.}\ }\textbf {\bibinfo {volume} {123}},\
  \bibinfo {pages} {157601} (\bibinfo {year} {2019})}\BibitemShut {NoStop}%
\bibitem [{\citenamefont {Da~Liao}\ \emph {et~al.}(2020)\citenamefont
  {Da~Liao}, \citenamefont {Kang}, \citenamefont {Brei{\o}}, \citenamefont
  {Xu}, \citenamefont {Wu}, \citenamefont {Andersen}, \citenamefont
  {Fernandes},\ and\ \citenamefont {Meng}}]{DA20}%
  \BibitemOpen
  \bibfield  {author} {\bibinfo {author} {\bibfnamefont {Y.}~\bibnamefont
  {Da~Liao}}, \bibinfo {author} {\bibfnamefont {J.}~\bibnamefont {Kang}},
  \bibinfo {author} {\bibfnamefont {C.~N.}\ \bibnamefont {Brei{\o}}}, \bibinfo
  {author} {\bibfnamefont {X.~Y.}\ \bibnamefont {Xu}}, \bibinfo {author}
  {\bibfnamefont {H.-Q.}\ \bibnamefont {Wu}}, \bibinfo {author} {\bibfnamefont
  {B.~M.}\ \bibnamefont {Andersen}}, \bibinfo {author} {\bibfnamefont {R.~M.}\
  \bibnamefont {Fernandes}},\ and\ \bibinfo {author} {\bibfnamefont {Z.~Y.}\
  \bibnamefont {Meng}},\ }\href@noop {} {\bibfield  {journal} {\bibinfo
  {journal} {arXiv:2004.12536 [cond-mat]}\ } (\bibinfo {year} {2020})},\
  \Eprint {https://arxiv.org/abs/2004.12536} {arXiv:2004.12536 [cond-mat]}
  \BibitemShut {NoStop}%
\bibitem [{\citenamefont {Dai}\ \emph {et~al.}(2016)\citenamefont {Dai},
  \citenamefont {Xiang},\ and\ \citenamefont {Srolovitz}}]{DAI16}%
  \BibitemOpen
  \bibfield  {author} {\bibinfo {author} {\bibfnamefont {S.}~\bibnamefont
  {Dai}}, \bibinfo {author} {\bibfnamefont {Y.}~\bibnamefont {Xiang}},\ and\
  \bibinfo {author} {\bibfnamefont {D.~J.}\ \bibnamefont {Srolovitz}},\ }\href
  {https://doi.org/10.1021/acs.nanolett.6b02870} {\bibfield  {journal}
  {\bibinfo  {journal} {Nano Lett.}\ }\textbf {\bibinfo {volume} {16}},\
  \bibinfo {pages} {5923} (\bibinfo {year} {2016})}\BibitemShut {NoStop}%
\bibitem [{\citenamefont {Dodaro}\ \emph {et~al.}(2018)\citenamefont {Dodaro},
  \citenamefont {Kivelson}, \citenamefont {Schattner}, \citenamefont {Sun},\
  and\ \citenamefont {Wang}}]{DOD18}%
  \BibitemOpen
  \bibfield  {author} {\bibinfo {author} {\bibfnamefont {J.~F.}\ \bibnamefont
  {Dodaro}}, \bibinfo {author} {\bibfnamefont {S.~A.}\ \bibnamefont
  {Kivelson}}, \bibinfo {author} {\bibfnamefont {Y.}~\bibnamefont {Schattner}},
  \bibinfo {author} {\bibfnamefont {X.~Q.}\ \bibnamefont {Sun}},\ and\ \bibinfo
  {author} {\bibfnamefont {C.}~\bibnamefont {Wang}},\ }\href
  {https://doi.org/10.1103/PhysRevB.98.075154} {\bibfield  {journal} {\bibinfo
  {journal} {Phys. Rev. B}\ }\textbf {\bibinfo {volume} {98}},\ \bibinfo
  {pages} {075154} (\bibinfo {year} {2018})}\BibitemShut {NoStop}%
\bibitem [{\citenamefont {Efimkin}\ and\ \citenamefont
  {MacDonald}(2018)}]{EFI18}%
  \BibitemOpen
  \bibfield  {author} {\bibinfo {author} {\bibfnamefont {D.~K.}\ \bibnamefont
  {Efimkin}}\ and\ \bibinfo {author} {\bibfnamefont {A.~H.}\ \bibnamefont
  {MacDonald}},\ }\href {https://doi.org/10.1103/PhysRevB.98.035404} {\bibfield
   {journal} {\bibinfo  {journal} {Phys. Rev. B}\ }\textbf {\bibinfo {volume}
  {98}},\ \bibinfo {pages} {035404} (\bibinfo {year} {2018})}\BibitemShut
  {NoStop}%
\bibitem [{\citenamefont {Eugenio}\ and\ \citenamefont {Dag}(2020)}]{EUG20}%
  \BibitemOpen
  \bibfield  {author} {\bibinfo {author} {\bibfnamefont {P.}~\bibnamefont
  {Eugenio}}\ and\ \bibinfo {author} {\bibfnamefont {C.}~\bibnamefont {Dag}},\
  }\href {https://doi.org/10.21468/SciPostPhysCore.3.2.015} {\bibfield
  {journal} {\bibinfo  {journal} {SciPost Physics Core}\ }\textbf {\bibinfo
  {volume} {3}},\ \bibinfo {pages} {015} (\bibinfo {year} {2020})}\BibitemShut
  {NoStop}%
\bibitem [{\citenamefont {Gonz{\'a}lez}\ and\ \citenamefont
  {Stauber}(2019)}]{GON19}%
  \BibitemOpen
  \bibfield  {author} {\bibinfo {author} {\bibfnamefont {J.}~\bibnamefont
  {Gonz{\'a}lez}}\ and\ \bibinfo {author} {\bibfnamefont {T.}~\bibnamefont
  {Stauber}},\ }\href {https://doi.org/10.1103/PhysRevLett.122.026801}
  {\bibfield  {journal} {\bibinfo  {journal} {Phys. Rev. Lett.}\ }\textbf
  {\bibinfo {volume} {122}},\ \bibinfo {pages} {026801} (\bibinfo {year}
  {2019})}\BibitemShut {NoStop}%
\bibitem [{\citenamefont {Guinea}\ and\ \citenamefont {Walet}(2018)}]{GUI18}%
  \BibitemOpen
  \bibfield  {author} {\bibinfo {author} {\bibfnamefont {F.}~\bibnamefont
  {Guinea}}\ and\ \bibinfo {author} {\bibfnamefont {N.~R.}\ \bibnamefont
  {Walet}},\ }\href {https://doi.org/10.1073/pnas.1810947115} {\bibfield
  {journal} {\bibinfo  {journal} {PNAS}\ }\textbf {\bibinfo {volume} {115}},\
  \bibinfo {pages} {13174} (\bibinfo {year} {2018})}\BibitemShut {NoStop}%
\bibitem [{\citenamefont {Guo}\ \emph {et~al.}(2018)\citenamefont {Guo},
  \citenamefont {Zhu}, \citenamefont {Feng},\ and\ \citenamefont
  {Scalettar}}]{GUO18}%
  \BibitemOpen
  \bibfield  {author} {\bibinfo {author} {\bibfnamefont {H.}~\bibnamefont
  {Guo}}, \bibinfo {author} {\bibfnamefont {X.}~\bibnamefont {Zhu}}, \bibinfo
  {author} {\bibfnamefont {S.}~\bibnamefont {Feng}},\ and\ \bibinfo {author}
  {\bibfnamefont {R.~T.}\ \bibnamefont {Scalettar}},\ }\href
  {https://doi.org/10.1103/PhysRevB.97.235453} {\bibfield  {journal} {\bibinfo
  {journal} {Phys. Rev. B}\ }\textbf {\bibinfo {volume} {97}},\ \bibinfo
  {pages} {235453} (\bibinfo {year} {2018})}\BibitemShut {NoStop}%
\bibitem [{\citenamefont {Hejazi}\ \emph
  {et~al.}(2019{\natexlab{a}})\citenamefont {Hejazi}, \citenamefont {Liu},
  \citenamefont {Shapourian}, \citenamefont {Chen},\ and\ \citenamefont
  {Balents}}]{HEJ19}%
  \BibitemOpen
  \bibfield  {author} {\bibinfo {author} {\bibfnamefont {K.}~\bibnamefont
  {Hejazi}}, \bibinfo {author} {\bibfnamefont {C.}~\bibnamefont {Liu}},
  \bibinfo {author} {\bibfnamefont {H.}~\bibnamefont {Shapourian}}, \bibinfo
  {author} {\bibfnamefont {X.}~\bibnamefont {Chen}},\ and\ \bibinfo {author}
  {\bibfnamefont {L.}~\bibnamefont {Balents}},\ }\href
  {https://doi.org/10.1103/PhysRevB.99.035111} {\bibfield  {journal} {\bibinfo
  {journal} {Phys. Rev. B}\ }\textbf {\bibinfo {volume} {99}},\ \bibinfo
  {pages} {035111} (\bibinfo {year} {2019}{\natexlab{a}})}\BibitemShut
  {NoStop}%
\bibitem [{\citenamefont {Hejazi}\ \emph
  {et~al.}(2019{\natexlab{b}})\citenamefont {Hejazi}, \citenamefont {Liu},\
  and\ \citenamefont {Balents}}]{HEJ19a}%
  \BibitemOpen
  \bibfield  {author} {\bibinfo {author} {\bibfnamefont {K.}~\bibnamefont
  {Hejazi}}, \bibinfo {author} {\bibfnamefont {C.}~\bibnamefont {Liu}},\ and\
  \bibinfo {author} {\bibfnamefont {L.}~\bibnamefont {Balents}},\ }\href
  {https://doi.org/10.1103/PhysRevB.100.035115} {\bibfield  {journal} {\bibinfo
   {journal} {Phys. Rev. B}\ }\textbf {\bibinfo {volume} {100}},\ \bibinfo
  {pages} {035115} (\bibinfo {year} {2019}{\natexlab{b}})}\BibitemShut
  {NoStop}%
\bibitem [{\citenamefont {Huang}\ \emph {et~al.}(2019)\citenamefont {Huang},
  \citenamefont {Zhang},\ and\ \citenamefont {Ma}}]{HUA19}%
  \BibitemOpen
  \bibfield  {author} {\bibinfo {author} {\bibfnamefont {T.}~\bibnamefont
  {Huang}}, \bibinfo {author} {\bibfnamefont {L.}~\bibnamefont {Zhang}},\ and\
  \bibinfo {author} {\bibfnamefont {T.}~\bibnamefont {Ma}},\ }\href
  {https://doi.org/10.1016/j.scib.2019.01.026} {\bibfield  {journal} {\bibinfo
  {journal} {Science Bulletin}\ }\textbf {\bibinfo {volume} {64}},\ \bibinfo
  {pages} {310} (\bibinfo {year} {2019})}\BibitemShut {NoStop}%
\bibitem [{\citenamefont {Huang}\ \emph {et~al.}(2020)\citenamefont {Huang},
  \citenamefont {Hosur},\ and\ \citenamefont {Pal}}]{HUA20a}%
  \BibitemOpen
  \bibfield  {author} {\bibinfo {author} {\bibfnamefont {Y.}~\bibnamefont
  {Huang}}, \bibinfo {author} {\bibfnamefont {P.}~\bibnamefont {Hosur}},\ and\
  \bibinfo {author} {\bibfnamefont {H.~K.}\ \bibnamefont {Pal}},\ }\href
  {https://doi.org/10.1103/PhysRevB.102.155429} {\bibfield  {journal} {\bibinfo
   {journal} {Phys. Rev. B}\ }\textbf {\bibinfo {volume} {102}},\ \bibinfo
  {pages} {155429} (\bibinfo {year} {2020})}\BibitemShut {NoStop}%
\bibitem [{\citenamefont {Isobe}\ \emph {et~al.}(2018)\citenamefont {Isobe},
  \citenamefont {Yuan},\ and\ \citenamefont {Fu}}]{ISO18}%
  \BibitemOpen
  \bibfield  {author} {\bibinfo {author} {\bibfnamefont {H.}~\bibnamefont
  {Isobe}}, \bibinfo {author} {\bibfnamefont {N.~F.~Q.}\ \bibnamefont {Yuan}},\
  and\ \bibinfo {author} {\bibfnamefont {L.}~\bibnamefont {Fu}},\ }\href
  {https://doi.org/10.1103/PhysRevX.8.041041} {\bibfield  {journal} {\bibinfo
  {journal} {Phys. Rev. X}\ }\textbf {\bibinfo {volume} {8}},\ \bibinfo {pages}
  {041041} (\bibinfo {year} {2018})}\BibitemShut {NoStop}%
\bibitem [{\citenamefont {Jain}\ \emph {et~al.}(2016)\citenamefont {Jain},
  \citenamefont {Juri{\v c}i{\'c}},\ and\ \citenamefont {Barkema}}]{JAI16}%
  \BibitemOpen
  \bibfield  {author} {\bibinfo {author} {\bibfnamefont {S.~K.}\ \bibnamefont
  {Jain}}, \bibinfo {author} {\bibfnamefont {V.}~\bibnamefont {Juri{\v
  c}i{\'c}}},\ and\ \bibinfo {author} {\bibfnamefont {G.~T.}\ \bibnamefont
  {Barkema}},\ }\href {https://doi.org/10.1088/2053-1583/4/1/015018} {\bibfield
   {journal} {\bibinfo  {journal} {2D Mater.}\ }\textbf {\bibinfo {volume}
  {4}},\ \bibinfo {pages} {015018} (\bibinfo {year} {2016})}\BibitemShut
  {NoStop}%
\bibitem [{\citenamefont {Julku}\ \emph {et~al.}(2020)\citenamefont {Julku},
  \citenamefont {Peltonen}, \citenamefont {Liang}, \citenamefont
  {Heikkil{\"a}},\ and\ \citenamefont {T{\"o}rm{\"a}}}]{JUL20}%
  \BibitemOpen
  \bibfield  {author} {\bibinfo {author} {\bibfnamefont {A.}~\bibnamefont
  {Julku}}, \bibinfo {author} {\bibfnamefont {T.~J.}\ \bibnamefont {Peltonen}},
  \bibinfo {author} {\bibfnamefont {L.}~\bibnamefont {Liang}}, \bibinfo
  {author} {\bibfnamefont {T.~T.}\ \bibnamefont {Heikkil{\"a}}},\ and\ \bibinfo
  {author} {\bibfnamefont {P.}~\bibnamefont {T{\"o}rm{\"a}}},\ }\href
  {https://doi.org/10.1103/PhysRevB.101.060505} {\bibfield  {journal} {\bibinfo
   {journal} {Phys. Rev. B}\ }\textbf {\bibinfo {volume} {101}},\ \bibinfo
  {pages} {060505} (\bibinfo {year} {2020})}\BibitemShut {NoStop}%
\bibitem [{\citenamefont {Kang}\ and\ \citenamefont {Vafek}(2018)}]{KAN18}%
  \BibitemOpen
  \bibfield  {author} {\bibinfo {author} {\bibfnamefont {J.}~\bibnamefont
  {Kang}}\ and\ \bibinfo {author} {\bibfnamefont {O.}~\bibnamefont {Vafek}},\
  }\href {https://doi.org/10.1103/PhysRevX.8.031088} {\bibfield  {journal}
  {\bibinfo  {journal} {Phys. Rev. X}\ }\textbf {\bibinfo {volume} {8}},\
  \bibinfo {pages} {031088} (\bibinfo {year} {2018})}\BibitemShut {NoStop}%
\bibitem [{\citenamefont {Kang}\ and\ \citenamefont {Vafek}(2020)}]{KAN20a}%
  \BibitemOpen
  \bibfield  {author} {\bibinfo {author} {\bibfnamefont {J.}~\bibnamefont
  {Kang}}\ and\ \bibinfo {author} {\bibfnamefont {O.}~\bibnamefont {Vafek}},\
  }\href {https://doi.org/10.1103/PhysRevB.102.035161} {\bibfield  {journal}
  {\bibinfo  {journal} {Phys. Rev. B}\ }\textbf {\bibinfo {volume} {102}},\
  \bibinfo {pages} {035161} (\bibinfo {year} {2020})}\BibitemShut {NoStop}%
\bibitem [{\citenamefont {Kennes}\ \emph {et~al.}(2018)\citenamefont {Kennes},
  \citenamefont {Lischner},\ and\ \citenamefont {Karrasch}}]{KEN18}%
  \BibitemOpen
  \bibfield  {author} {\bibinfo {author} {\bibfnamefont {D.~M.}\ \bibnamefont
  {Kennes}}, \bibinfo {author} {\bibfnamefont {J.}~\bibnamefont {Lischner}},\
  and\ \bibinfo {author} {\bibfnamefont {C.}~\bibnamefont {Karrasch}},\ }\href
  {https://doi.org/10.1103/PhysRevB.98.241407} {\bibfield  {journal} {\bibinfo
  {journal} {Phys. Rev. B}\ }\textbf {\bibinfo {volume} {98}},\ \bibinfo
  {pages} {241407} (\bibinfo {year} {2018})}\BibitemShut {NoStop}%
\bibitem [{\citenamefont {Khalaf}\ \emph {et~al.}(2020)\citenamefont {Khalaf},
  \citenamefont {Chatterjee}, \citenamefont {Bultinck}, \citenamefont
  {Zaletel},\ and\ \citenamefont {Vishwanath}}]{KHA20}%
  \BibitemOpen
  \bibfield  {author} {\bibinfo {author} {\bibfnamefont {E.}~\bibnamefont
  {Khalaf}}, \bibinfo {author} {\bibfnamefont {S.}~\bibnamefont {Chatterjee}},
  \bibinfo {author} {\bibfnamefont {N.}~\bibnamefont {Bultinck}}, \bibinfo
  {author} {\bibfnamefont {M.~P.}\ \bibnamefont {Zaletel}},\ and\ \bibinfo
  {author} {\bibfnamefont {A.}~\bibnamefont {Vishwanath}},\ }\href@noop {}
  {\bibfield  {journal} {\bibinfo  {journal} {arXiv:2004.00638 [cond-mat]}\ }
  (\bibinfo {year} {2020})},\ \Eprint {https://arxiv.org/abs/2004.00638}
  {arXiv:2004.00638 [cond-mat]} \BibitemShut {NoStop}%
\bibitem [{\citenamefont {K{\"o}nig}\ \emph {et~al.}(2020)\citenamefont
  {K{\"o}nig}, \citenamefont {Coleman},\ and\ \citenamefont {Tsvelik}}]{KON20}%
  \BibitemOpen
  \bibfield  {author} {\bibinfo {author} {\bibfnamefont {E.~J.}\ \bibnamefont
  {K{\"o}nig}}, \bibinfo {author} {\bibfnamefont {P.}~\bibnamefont {Coleman}},\
  and\ \bibinfo {author} {\bibfnamefont {A.~M.}\ \bibnamefont {Tsvelik}},\
  }\href {https://doi.org/10.1103/PhysRevB.102.104514} {\bibfield  {journal}
  {\bibinfo  {journal} {Phys. Rev. B}\ }\textbf {\bibinfo {volume} {102}},\
  \bibinfo {pages} {104514} (\bibinfo {year} {2020})}\BibitemShut {NoStop}%
\bibitem [{\citenamefont {Koshino}\ \emph {et~al.}(2018)\citenamefont
  {Koshino}, \citenamefont {Yuan}, \citenamefont {Koretsune}, \citenamefont
  {Ochi}, \citenamefont {Kuroki},\ and\ \citenamefont {Fu}}]{KOS18}%
  \BibitemOpen
  \bibfield  {author} {\bibinfo {author} {\bibfnamefont {M.}~\bibnamefont
  {Koshino}}, \bibinfo {author} {\bibfnamefont {N.~F.~Q.}\ \bibnamefont
  {Yuan}}, \bibinfo {author} {\bibfnamefont {T.}~\bibnamefont {Koretsune}},
  \bibinfo {author} {\bibfnamefont {M.}~\bibnamefont {Ochi}}, \bibinfo {author}
  {\bibfnamefont {K.}~\bibnamefont {Kuroki}},\ and\ \bibinfo {author}
  {\bibfnamefont {L.}~\bibnamefont {Fu}},\ }\href
  {https://doi.org/10.1103/PhysRevX.8.031087} {\bibfield  {journal} {\bibinfo
  {journal} {Phys. Rev. X}\ }\textbf {\bibinfo {volume} {8}},\ \bibinfo {pages}
  {031087} (\bibinfo {year} {2018})}\BibitemShut {NoStop}%
\bibitem [{\citenamefont {Ledwith}\ \emph {et~al.}(2020)\citenamefont
  {Ledwith}, \citenamefont {Tarnopolsky}, \citenamefont {Khalaf},\ and\
  \citenamefont {Vishwanath}}]{LED20}%
  \BibitemOpen
  \bibfield  {author} {\bibinfo {author} {\bibfnamefont {P.~J.}\ \bibnamefont
  {Ledwith}}, \bibinfo {author} {\bibfnamefont {G.}~\bibnamefont
  {Tarnopolsky}}, \bibinfo {author} {\bibfnamefont {E.}~\bibnamefont
  {Khalaf}},\ and\ \bibinfo {author} {\bibfnamefont {A.}~\bibnamefont
  {Vishwanath}},\ }\href {https://doi.org/10.1103/PhysRevResearch.2.023237}
  {\bibfield  {journal} {\bibinfo  {journal} {Phys. Rev. Research}\ }\textbf
  {\bibinfo {volume} {2}},\ \bibinfo {pages} {023237} (\bibinfo {year}
  {2020})}\BibitemShut {NoStop}%
\bibitem [{\citenamefont {Lewandowski}\ \emph {et~al.}(2020)\citenamefont
  {Lewandowski}, \citenamefont {Chowdhury},\ and\ \citenamefont
  {Ruhman}}]{LEW20}%
  \BibitemOpen
  \bibfield  {author} {\bibinfo {author} {\bibfnamefont {C.}~\bibnamefont
  {Lewandowski}}, \bibinfo {author} {\bibfnamefont {D.}~\bibnamefont
  {Chowdhury}},\ and\ \bibinfo {author} {\bibfnamefont {J.}~\bibnamefont
  {Ruhman}},\ }\href@noop {} {\bibfield  {journal} {\bibinfo  {journal}
  {arXiv:2007.15002 [cond-mat]}\ } (\bibinfo {year} {2020})},\ \Eprint
  {https://arxiv.org/abs/2007.15002} {arXiv:2007.15002 [cond-mat]} \BibitemShut
  {NoStop}%
\bibitem [{\citenamefont {Lian}\ \emph {et~al.}(2019)\citenamefont {Lian},
  \citenamefont {Wang},\ and\ \citenamefont {Bernevig}}]{LIA19}%
  \BibitemOpen
  \bibfield  {author} {\bibinfo {author} {\bibfnamefont {B.}~\bibnamefont
  {Lian}}, \bibinfo {author} {\bibfnamefont {Z.}~\bibnamefont {Wang}},\ and\
  \bibinfo {author} {\bibfnamefont {B.~A.}\ \bibnamefont {Bernevig}},\ }\href
  {https://doi.org/10.1103/PhysRevLett.122.257002} {\bibfield  {journal}
  {\bibinfo  {journal} {Phys. Rev. Lett.}\ }\textbf {\bibinfo {volume} {122}},\
  \bibinfo {pages} {257002} (\bibinfo {year} {2019})}\BibitemShut {NoStop}%
\bibitem [{\citenamefont {Lian}\ \emph
  {et~al.}(2020{\natexlab{a}})\citenamefont {Lian}, \citenamefont {Song},
  \citenamefont {Regnault}, \citenamefont {Efetov}, \citenamefont {Yazdani},\
  and\ \citenamefont {Bernevig}}]{LIA20}%
  \BibitemOpen
  \bibfield  {author} {\bibinfo {author} {\bibfnamefont {B.}~\bibnamefont
  {Lian}}, \bibinfo {author} {\bibfnamefont {Z.-D.}\ \bibnamefont {Song}},
  \bibinfo {author} {\bibfnamefont {N.}~\bibnamefont {Regnault}}, \bibinfo
  {author} {\bibfnamefont {D.~K.}\ \bibnamefont {Efetov}}, \bibinfo {author}
  {\bibfnamefont {A.}~\bibnamefont {Yazdani}},\ and\ \bibinfo {author}
  {\bibfnamefont {B.~A.}\ \bibnamefont {Bernevig}},\ }\href@noop {} {\bibfield
  {journal} {\bibinfo  {journal} {arXiv:2009.13530 [cond-mat]}\ } (\bibinfo
  {year} {2020}{\natexlab{a}})},\ \Eprint {https://arxiv.org/abs/2009.13530}
  {arXiv:2009.13530 [cond-mat]} \BibitemShut {NoStop}%
\bibitem [{\citenamefont {Lian}\ \emph
  {et~al.}(2020{\natexlab{b}})\citenamefont {Lian}, \citenamefont {Xie},\ and\
  \citenamefont {Bernevig}}]{LIA20a}%
  \BibitemOpen
  \bibfield  {author} {\bibinfo {author} {\bibfnamefont {B.}~\bibnamefont
  {Lian}}, \bibinfo {author} {\bibfnamefont {F.}~\bibnamefont {Xie}},\ and\
  \bibinfo {author} {\bibfnamefont {B.~A.}\ \bibnamefont {Bernevig}},\ }\href
  {https://doi.org/10.1103/PhysRevB.102.041402} {\bibfield  {journal} {\bibinfo
   {journal} {Phys. Rev. B}\ }\textbf {\bibinfo {volume} {102}},\ \bibinfo
  {pages} {041402} (\bibinfo {year} {2020}{\natexlab{b}})}\BibitemShut
  {NoStop}%
\bibitem [{\citenamefont {Liu}\ \emph {et~al.}(2012)\citenamefont {Liu},
  \citenamefont {Bergholtz}, \citenamefont {Fan},\ and\ \citenamefont
  {L{\"a}uchli}}]{LIU12}%
  \BibitemOpen
  \bibfield  {author} {\bibinfo {author} {\bibfnamefont {Z.}~\bibnamefont
  {Liu}}, \bibinfo {author} {\bibfnamefont {E.~J.}\ \bibnamefont {Bergholtz}},
  \bibinfo {author} {\bibfnamefont {H.}~\bibnamefont {Fan}},\ and\ \bibinfo
  {author} {\bibfnamefont {A.~M.}\ \bibnamefont {L{\"a}uchli}},\ }\href
  {https://doi.org/10.1103/PhysRevLett.109.186805} {\bibfield  {journal}
  {\bibinfo  {journal} {Phys. Rev. Lett.}\ }\textbf {\bibinfo {volume} {109}},\
  \bibinfo {pages} {186805} (\bibinfo {year} {2012})}\BibitemShut {NoStop}%
\bibitem [{\citenamefont {Liu}\ \emph {et~al.}(2018)\citenamefont {Liu},
  \citenamefont {Zhang}, \citenamefont {Chen},\ and\ \citenamefont
  {Yang}}]{LIU18}%
  \BibitemOpen
  \bibfield  {author} {\bibinfo {author} {\bibfnamefont {C.-C.}\ \bibnamefont
  {Liu}}, \bibinfo {author} {\bibfnamefont {L.-D.}\ \bibnamefont {Zhang}},
  \bibinfo {author} {\bibfnamefont {W.-Q.}\ \bibnamefont {Chen}},\ and\
  \bibinfo {author} {\bibfnamefont {F.}~\bibnamefont {Yang}},\ }\href
  {https://doi.org/10.1103/PhysRevLett.121.217001} {\bibfield  {journal}
  {\bibinfo  {journal} {Phys. Rev. Lett.}\ }\textbf {\bibinfo {volume} {121}},\
  \bibinfo {pages} {217001} (\bibinfo {year} {2018})}\BibitemShut {NoStop}%
\bibitem [{\citenamefont {Liu}\ \emph {et~al.}(2019{\natexlab{a}})\citenamefont
  {Liu}, \citenamefont {Liu},\ and\ \citenamefont {Dai}}]{LIU19}%
  \BibitemOpen
  \bibfield  {author} {\bibinfo {author} {\bibfnamefont {J.}~\bibnamefont
  {Liu}}, \bibinfo {author} {\bibfnamefont {J.}~\bibnamefont {Liu}},\ and\
  \bibinfo {author} {\bibfnamefont {X.}~\bibnamefont {Dai}},\ }\href
  {https://doi.org/10.1103/PhysRevB.99.155415} {\bibfield  {journal} {\bibinfo
  {journal} {Phys. Rev. B}\ }\textbf {\bibinfo {volume} {99}},\ \bibinfo
  {pages} {155415} (\bibinfo {year} {2019}{\natexlab{a}})}\BibitemShut
  {NoStop}%
\bibitem [{\citenamefont {Liu}\ and\ \citenamefont {Dai}(2021)}]{LIU21}%
  \BibitemOpen
  \bibfield  {author} {\bibinfo {author} {\bibfnamefont {J.}~\bibnamefont
  {Liu}}\ and\ \bibinfo {author} {\bibfnamefont {X.}~\bibnamefont {Dai}},\
  }\href@noop {} {\bibfield  {journal} {\bibinfo  {journal} {arXiv:1911.03760
  [cond-mat]}\ } (\bibinfo {year} {2021})},\ \Eprint
  {https://arxiv.org/abs/1911.03760} {arXiv:1911.03760 [cond-mat]} \BibitemShut
  {NoStop}%
\bibitem [{\citenamefont {Liu}\ \emph {et~al.}(2021)\citenamefont {Liu},
  \citenamefont {Khalaf}, \citenamefont {Lee},\ and\ \citenamefont
  {Vishwanath}}]{LIU21a}%
  \BibitemOpen
  \bibfield  {author} {\bibinfo {author} {\bibfnamefont {S.}~\bibnamefont
  {Liu}}, \bibinfo {author} {\bibfnamefont {E.}~\bibnamefont {Khalaf}},
  \bibinfo {author} {\bibfnamefont {J.~Y.}\ \bibnamefont {Lee}},\ and\ \bibinfo
  {author} {\bibfnamefont {A.}~\bibnamefont {Vishwanath}},\ }\href
  {https://doi.org/10.1103/PhysRevResearch.3.013033} {\bibfield  {journal}
  {\bibinfo  {journal} {Phys. Rev. Research}\ }\textbf {\bibinfo {volume}
  {3}},\ \bibinfo {pages} {013033} (\bibinfo {year} {2021})}\BibitemShut
  {NoStop}%
\bibitem [{\citenamefont {Ochi}\ \emph {et~al.}(2018)\citenamefont {Ochi},
  \citenamefont {Koshino},\ and\ \citenamefont {Kuroki}}]{OCH18}%
  \BibitemOpen
  \bibfield  {author} {\bibinfo {author} {\bibfnamefont {M.}~\bibnamefont
  {Ochi}}, \bibinfo {author} {\bibfnamefont {M.}~\bibnamefont {Koshino}},\ and\
  \bibinfo {author} {\bibfnamefont {K.}~\bibnamefont {Kuroki}},\ }\href
  {https://doi.org/10.1103/PhysRevB.98.081102} {\bibfield  {journal} {\bibinfo
  {journal} {Phys. Rev. B}\ }\textbf {\bibinfo {volume} {98}},\ \bibinfo
  {pages} {081102} (\bibinfo {year} {2018})}\BibitemShut {NoStop}%
\bibitem [{\citenamefont {Padhi}\ \emph {et~al.}(2020)\citenamefont {Padhi},
  \citenamefont {Tiwari}, \citenamefont {Neupert},\ and\ \citenamefont
  {Ryu}}]{PAD20}%
  \BibitemOpen
  \bibfield  {author} {\bibinfo {author} {\bibfnamefont {B.}~\bibnamefont
  {Padhi}}, \bibinfo {author} {\bibfnamefont {A.}~\bibnamefont {Tiwari}},
  \bibinfo {author} {\bibfnamefont {T.}~\bibnamefont {Neupert}},\ and\ \bibinfo
  {author} {\bibfnamefont {S.}~\bibnamefont {Ryu}},\ }\href
  {https://doi.org/10.1103/PhysRevResearch.2.033458} {\bibfield  {journal}
  {\bibinfo  {journal} {Phys. Rev. Research}\ }\textbf {\bibinfo {volume}
  {2}},\ \bibinfo {pages} {033458} (\bibinfo {year} {2020})}\BibitemShut
  {NoStop}%
\bibitem [{\citenamefont {Peltonen}\ \emph {et~al.}(2018)\citenamefont
  {Peltonen}, \citenamefont {Ojaj{\"a}rvi},\ and\ \citenamefont
  {Heikkil{\"a}}}]{PEL18}%
  \BibitemOpen
  \bibfield  {author} {\bibinfo {author} {\bibfnamefont {T.~J.}\ \bibnamefont
  {Peltonen}}, \bibinfo {author} {\bibfnamefont {R.}~\bibnamefont
  {Ojaj{\"a}rvi}},\ and\ \bibinfo {author} {\bibfnamefont {T.~T.}\ \bibnamefont
  {Heikkil{\"a}}},\ }\href {https://doi.org/10.1103/PhysRevB.98.220504}
  {\bibfield  {journal} {\bibinfo  {journal} {Phys. Rev. B}\ }\textbf {\bibinfo
  {volume} {98}},\ \bibinfo {pages} {220504} (\bibinfo {year}
  {2018})}\BibitemShut {NoStop}%
\bibitem [{\citenamefont {Po}\ \emph {et~al.}(2018)\citenamefont {Po},
  \citenamefont {Zou}, \citenamefont {Vishwanath},\ and\ \citenamefont
  {Senthil}}]{PO18}%
  \BibitemOpen
  \bibfield  {author} {\bibinfo {author} {\bibfnamefont {H.~C.}\ \bibnamefont
  {Po}}, \bibinfo {author} {\bibfnamefont {L.}~\bibnamefont {Zou}}, \bibinfo
  {author} {\bibfnamefont {A.}~\bibnamefont {Vishwanath}},\ and\ \bibinfo
  {author} {\bibfnamefont {T.}~\bibnamefont {Senthil}},\ }\href
  {https://doi.org/10.1103/PhysRevX.8.031089} {\bibfield  {journal} {\bibinfo
  {journal} {Phys. Rev. X}\ }\textbf {\bibinfo {volume} {8}},\ \bibinfo {pages}
  {031089} (\bibinfo {year} {2018})}\BibitemShut {NoStop}%
\bibitem [{\citenamefont {Po}\ \emph {et~al.}(2019)\citenamefont {Po},
  \citenamefont {Zou}, \citenamefont {Senthil},\ and\ \citenamefont
  {Vishwanath}}]{PO19}%
  \BibitemOpen
  \bibfield  {author} {\bibinfo {author} {\bibfnamefont {H.~C.}\ \bibnamefont
  {Po}}, \bibinfo {author} {\bibfnamefont {L.}~\bibnamefont {Zou}}, \bibinfo
  {author} {\bibfnamefont {T.}~\bibnamefont {Senthil}},\ and\ \bibinfo {author}
  {\bibfnamefont {A.}~\bibnamefont {Vishwanath}},\ }\href
  {https://doi.org/10.1103/PhysRevB.99.195455} {\bibfield  {journal} {\bibinfo
  {journal} {Phys. Rev. B}\ }\textbf {\bibinfo {volume} {99}},\ \bibinfo
  {pages} {195455} (\bibinfo {year} {2019})}\BibitemShut {NoStop}%
\bibitem [{\citenamefont {Repellin}\ \emph {et~al.}(2020)\citenamefont
  {Repellin}, \citenamefont {Dong}, \citenamefont {Zhang},\ and\ \citenamefont
  {Senthil}}]{REP20}%
  \BibitemOpen
  \bibfield  {author} {\bibinfo {author} {\bibfnamefont {C.}~\bibnamefont
  {Repellin}}, \bibinfo {author} {\bibfnamefont {Z.}~\bibnamefont {Dong}},
  \bibinfo {author} {\bibfnamefont {Y.-H.}\ \bibnamefont {Zhang}},\ and\
  \bibinfo {author} {\bibfnamefont {T.}~\bibnamefont {Senthil}},\ }\href
  {https://doi.org/10.1103/PhysRevLett.124.187601} {\bibfield  {journal}
  {\bibinfo  {journal} {Phys. Rev. Lett.}\ }\textbf {\bibinfo {volume} {124}},\
  \bibinfo {pages} {187601} (\bibinfo {year} {2020})}\BibitemShut {NoStop}%
\bibitem [{\citenamefont {Repellin}\ and\ \citenamefont
  {Senthil}(2020)}]{REP20a}%
  \BibitemOpen
  \bibfield  {author} {\bibinfo {author} {\bibfnamefont {C.}~\bibnamefont
  {Repellin}}\ and\ \bibinfo {author} {\bibfnamefont {T.}~\bibnamefont
  {Senthil}},\ }\href {https://doi.org/10.1103/PhysRevResearch.2.023238}
  {\bibfield  {journal} {\bibinfo  {journal} {Phys. Rev. Research}\ }\textbf
  {\bibinfo {volume} {2}},\ \bibinfo {pages} {023238} (\bibinfo {year}
  {2020})}\BibitemShut {NoStop}%
\bibitem [{\citenamefont {Roy}\ and\ \citenamefont {Juri{\v
  c}i{\'c}}(2019)}]{ROY19}%
  \BibitemOpen
  \bibfield  {author} {\bibinfo {author} {\bibfnamefont {B.}~\bibnamefont
  {Roy}}\ and\ \bibinfo {author} {\bibfnamefont {V.}~\bibnamefont {Juri{\v
  c}i{\'c}}},\ }\href {https://doi.org/10.1103/PhysRevB.99.121407} {\bibfield
  {journal} {\bibinfo  {journal} {Phys. Rev. B}\ }\textbf {\bibinfo {volume}
  {99}},\ \bibinfo {pages} {121407} (\bibinfo {year} {2019})}\BibitemShut
  {NoStop}%
\bibitem [{\citenamefont {Soejima}\ \emph {et~al.}(2020)\citenamefont
  {Soejima}, \citenamefont {Parker}, \citenamefont {Bultinck}, \citenamefont
  {Hauschild},\ and\ \citenamefont {Zaletel}}]{SOE20}%
  \BibitemOpen
  \bibfield  {author} {\bibinfo {author} {\bibfnamefont {T.}~\bibnamefont
  {Soejima}}, \bibinfo {author} {\bibfnamefont {D.~E.}\ \bibnamefont {Parker}},
  \bibinfo {author} {\bibfnamefont {N.}~\bibnamefont {Bultinck}}, \bibinfo
  {author} {\bibfnamefont {J.}~\bibnamefont {Hauschild}},\ and\ \bibinfo
  {author} {\bibfnamefont {M.~P.}\ \bibnamefont {Zaletel}},\ }\href
  {https://doi.org/10.1103/PhysRevB.102.205111} {\bibfield  {journal} {\bibinfo
   {journal} {Phys. Rev. B}\ }\textbf {\bibinfo {volume} {102}},\ \bibinfo
  {pages} {205111} (\bibinfo {year} {2020})}\BibitemShut {NoStop}%
\bibitem [{\citenamefont {Song}\ \emph {et~al.}(2019)\citenamefont {Song},
  \citenamefont {Wang}, \citenamefont {Shi}, \citenamefont {Li}, \citenamefont
  {Fang},\ and\ \citenamefont {Bernevig}}]{SON19}%
  \BibitemOpen
  \bibfield  {author} {\bibinfo {author} {\bibfnamefont {Z.}~\bibnamefont
  {Song}}, \bibinfo {author} {\bibfnamefont {Z.}~\bibnamefont {Wang}}, \bibinfo
  {author} {\bibfnamefont {W.}~\bibnamefont {Shi}}, \bibinfo {author}
  {\bibfnamefont {G.}~\bibnamefont {Li}}, \bibinfo {author} {\bibfnamefont
  {C.}~\bibnamefont {Fang}},\ and\ \bibinfo {author} {\bibfnamefont {B.~A.}\
  \bibnamefont {Bernevig}},\ }\href
  {https://doi.org/10.1103/PhysRevLett.123.036401} {\bibfield  {journal}
  {\bibinfo  {journal} {Phys. Rev. Lett.}\ }\textbf {\bibinfo {volume} {123}},\
  \bibinfo {pages} {036401} (\bibinfo {year} {2019})}\BibitemShut {NoStop}%
\bibitem [{\citenamefont {Song}\ \emph {et~al.}(2020)\citenamefont {Song},
  \citenamefont {Lian}, \citenamefont {Regnault},\ and\ \citenamefont
  {Bernevig}}]{SON20b}%
  \BibitemOpen
  \bibfield  {author} {\bibinfo {author} {\bibfnamefont {Z.-D.}\ \bibnamefont
  {Song}}, \bibinfo {author} {\bibfnamefont {B.}~\bibnamefont {Lian}}, \bibinfo
  {author} {\bibfnamefont {N.}~\bibnamefont {Regnault}},\ and\ \bibinfo
  {author} {\bibfnamefont {B.~A.}\ \bibnamefont {Bernevig}},\ }\href@noop {}
  {\bibfield  {journal} {\bibinfo  {journal} {arXiv:2009.11872 [cond-mat]}\ }
  (\bibinfo {year} {2020})},\ \Eprint {https://arxiv.org/abs/2009.11872}
  {arXiv:2009.11872 [cond-mat]} \BibitemShut {NoStop}%
\bibitem [{\citenamefont {Tarnopolsky}\ \emph {et~al.}(2019)\citenamefont
  {Tarnopolsky}, \citenamefont {Kruchkov},\ and\ \citenamefont
  {Vishwanath}}]{TAR19}%
  \BibitemOpen
  \bibfield  {author} {\bibinfo {author} {\bibfnamefont {G.}~\bibnamefont
  {Tarnopolsky}}, \bibinfo {author} {\bibfnamefont {A.~J.}\ \bibnamefont
  {Kruchkov}},\ and\ \bibinfo {author} {\bibfnamefont {A.}~\bibnamefont
  {Vishwanath}},\ }\href {https://doi.org/10.1103/PhysRevLett.122.106405}
  {\bibfield  {journal} {\bibinfo  {journal} {Phys. Rev. Lett.}\ }\textbf
  {\bibinfo {volume} {122}},\ \bibinfo {pages} {106405} (\bibinfo {year}
  {2019})}\BibitemShut {NoStop}%
\bibitem [{\citenamefont {Thomson}\ \emph {et~al.}(2018)\citenamefont
  {Thomson}, \citenamefont {Chatterjee}, \citenamefont {Sachdev},\ and\
  \citenamefont {Scheurer}}]{THO18}%
  \BibitemOpen
  \bibfield  {author} {\bibinfo {author} {\bibfnamefont {A.}~\bibnamefont
  {Thomson}}, \bibinfo {author} {\bibfnamefont {S.}~\bibnamefont {Chatterjee}},
  \bibinfo {author} {\bibfnamefont {S.}~\bibnamefont {Sachdev}},\ and\ \bibinfo
  {author} {\bibfnamefont {M.~S.}\ \bibnamefont {Scheurer}},\ }\href
  {https://doi.org/10.1103/PhysRevB.98.075109} {\bibfield  {journal} {\bibinfo
  {journal} {Phys. Rev. B}\ }\textbf {\bibinfo {volume} {98}},\ \bibinfo
  {pages} {075109} (\bibinfo {year} {2018})}\BibitemShut {NoStop}%
\bibitem [{\citenamefont {Uchida}\ \emph {et~al.}(2014)\citenamefont {Uchida},
  \citenamefont {Furuya}, \citenamefont {Iwata},\ and\ \citenamefont
  {Oshiyama}}]{UCH14}%
  \BibitemOpen
  \bibfield  {author} {\bibinfo {author} {\bibfnamefont {K.}~\bibnamefont
  {Uchida}}, \bibinfo {author} {\bibfnamefont {S.}~\bibnamefont {Furuya}},
  \bibinfo {author} {\bibfnamefont {J.-I.}\ \bibnamefont {Iwata}},\ and\
  \bibinfo {author} {\bibfnamefont {A.}~\bibnamefont {Oshiyama}},\ }\href
  {https://doi.org/10.1103/PhysRevB.90.155451} {\bibfield  {journal} {\bibinfo
  {journal} {Phys. Rev. B}\ }\textbf {\bibinfo {volume} {90}},\ \bibinfo
  {pages} {155451} (\bibinfo {year} {2014})}\BibitemShut {NoStop}%
\bibitem [{\citenamefont {Vafek}\ and\ \citenamefont {Kang}(2020)}]{VAF20}%
  \BibitemOpen
  \bibfield  {author} {\bibinfo {author} {\bibfnamefont {O.}~\bibnamefont
  {Vafek}}\ and\ \bibinfo {author} {\bibfnamefont {J.}~\bibnamefont {Kang}},\
  }\href {https://doi.org/10.1103/PhysRevLett.125.257602} {\bibfield  {journal}
  {\bibinfo  {journal} {Phys. Rev. Lett.}\ }\textbf {\bibinfo {volume} {125}},\
  \bibinfo {pages} {257602} (\bibinfo {year} {2020})}\BibitemShut {NoStop}%
\bibitem [{\citenamefont {Wang}\ \emph {et~al.}(2020)\citenamefont {Wang},
  \citenamefont {Zheng}, \citenamefont {Millis},\ and\ \citenamefont
  {Cano}}]{WAN20}%
  \BibitemOpen
  \bibfield  {author} {\bibinfo {author} {\bibfnamefont {J.}~\bibnamefont
  {Wang}}, \bibinfo {author} {\bibfnamefont {Y.}~\bibnamefont {Zheng}},
  \bibinfo {author} {\bibfnamefont {A.~J.}\ \bibnamefont {Millis}},\ and\
  \bibinfo {author} {\bibfnamefont {J.}~\bibnamefont {Cano}},\ }\href@noop {}
  {\bibfield  {journal} {\bibinfo  {journal} {arXiv:2010.03589 [cond-mat]}\ }
  (\bibinfo {year} {2020})},\ \Eprint {https://arxiv.org/abs/2010.03589}
  {arXiv:2010.03589 [cond-mat]} \BibitemShut {NoStop}%
\bibitem [{\citenamefont {van Wijk}\ \emph {et~al.}(2015)\citenamefont {van
  Wijk}, \citenamefont {Schuring}, \citenamefont {Katsnelson},\ and\
  \citenamefont {Fasolino}}]{WIJ15}%
  \BibitemOpen
  \bibfield  {author} {\bibinfo {author} {\bibfnamefont {M.~M.}\ \bibnamefont
  {van Wijk}}, \bibinfo {author} {\bibfnamefont {A.}~\bibnamefont {Schuring}},
  \bibinfo {author} {\bibfnamefont {M.~I.}\ \bibnamefont {Katsnelson}},\ and\
  \bibinfo {author} {\bibfnamefont {A.}~\bibnamefont {Fasolino}},\ }\href
  {https://doi.org/10.1088/2053-1583/2/3/034010} {\bibfield  {journal}
  {\bibinfo  {journal} {2D Mater.}\ }\textbf {\bibinfo {volume} {2}},\ \bibinfo
  {pages} {034010} (\bibinfo {year} {2015})}\BibitemShut {NoStop}%
\bibitem [{\citenamefont {Wilson}\ \emph {et~al.}(2020)\citenamefont {Wilson},
  \citenamefont {Fu}, \citenamefont {Das~Sarma},\ and\ \citenamefont
  {Pixley}}]{WIL20}%
  \BibitemOpen
  \bibfield  {author} {\bibinfo {author} {\bibfnamefont {J.~H.}\ \bibnamefont
  {Wilson}}, \bibinfo {author} {\bibfnamefont {Y.}~\bibnamefont {Fu}}, \bibinfo
  {author} {\bibfnamefont {S.}~\bibnamefont {Das~Sarma}},\ and\ \bibinfo
  {author} {\bibfnamefont {J.~H.}\ \bibnamefont {Pixley}},\ }\href
  {https://doi.org/10.1103/PhysRevResearch.2.023325} {\bibfield  {journal}
  {\bibinfo  {journal} {Phys. Rev. Research}\ }\textbf {\bibinfo {volume}
  {2}},\ \bibinfo {pages} {023325} (\bibinfo {year} {2020})}\BibitemShut
  {NoStop}%
\bibitem [{\citenamefont {Wu}\ \emph {et~al.}(2018)\citenamefont {Wu},
  \citenamefont {MacDonald},\ and\ \citenamefont {Martin}}]{WU18}%
  \BibitemOpen
  \bibfield  {author} {\bibinfo {author} {\bibfnamefont {F.}~\bibnamefont
  {Wu}}, \bibinfo {author} {\bibfnamefont {A.~H.}\ \bibnamefont {MacDonald}},\
  and\ \bibinfo {author} {\bibfnamefont {I.}~\bibnamefont {Martin}},\ }\href
  {https://doi.org/10.1103/PhysRevLett.121.257001} {\bibfield  {journal}
  {\bibinfo  {journal} {Phys. Rev. Lett.}\ }\textbf {\bibinfo {volume} {121}},\
  \bibinfo {pages} {257001} (\bibinfo {year} {2018})}\BibitemShut {NoStop}%
\bibitem [{\citenamefont {Wu}\ \emph {et~al.}(2019{\natexlab{a}})\citenamefont
  {Wu}, \citenamefont {Jian},\ and\ \citenamefont {Xu}}]{WU19}%
  \BibitemOpen
  \bibfield  {author} {\bibinfo {author} {\bibfnamefont {X.-C.}\ \bibnamefont
  {Wu}}, \bibinfo {author} {\bibfnamefont {C.-M.}\ \bibnamefont {Jian}},\ and\
  \bibinfo {author} {\bibfnamefont {C.}~\bibnamefont {Xu}},\ }\href
  {https://doi.org/10.1103/PhysRevB.99.161405} {\bibfield  {journal} {\bibinfo
  {journal} {Phys. Rev. B}\ }\textbf {\bibinfo {volume} {99}},\ \bibinfo
  {pages} {161405} (\bibinfo {year} {2019}{\natexlab{a}})}\BibitemShut
  {NoStop}%
\bibitem [{\citenamefont {Wu}\ \emph {et~al.}(2019{\natexlab{b}})\citenamefont
  {Wu}, \citenamefont {Hwang},\ and\ \citenamefont {Das~Sarma}}]{WU19a}%
  \BibitemOpen
  \bibfield  {author} {\bibinfo {author} {\bibfnamefont {F.}~\bibnamefont
  {Wu}}, \bibinfo {author} {\bibfnamefont {E.}~\bibnamefont {Hwang}},\ and\
  \bibinfo {author} {\bibfnamefont {S.}~\bibnamefont {Das~Sarma}},\ }\href
  {https://doi.org/10.1103/PhysRevB.99.165112} {\bibfield  {journal} {\bibinfo
  {journal} {Phys. Rev. B}\ }\textbf {\bibinfo {volume} {99}},\ \bibinfo
  {pages} {165112} (\bibinfo {year} {2019}{\natexlab{b}})}\BibitemShut
  {NoStop}%
\bibitem [{\citenamefont {Wu}\ and\ \citenamefont {Das~Sarma}(2020)}]{WU20b}%
  \BibitemOpen
  \bibfield  {author} {\bibinfo {author} {\bibfnamefont {F.}~\bibnamefont
  {Wu}}\ and\ \bibinfo {author} {\bibfnamefont {S.}~\bibnamefont {Das~Sarma}},\
  }\href {https://doi.org/10.1103/PhysRevLett.124.046403} {\bibfield  {journal}
  {\bibinfo  {journal} {Phys. Rev. Lett.}\ }\textbf {\bibinfo {volume} {124}},\
  \bibinfo {pages} {046403} (\bibinfo {year} {2020})}\BibitemShut {NoStop}%
\bibitem [{\citenamefont {Xie}\ \emph {et~al.}(2020{\natexlab{a}})\citenamefont
  {Xie}, \citenamefont {Song}, \citenamefont {Lian},\ and\ \citenamefont
  {Bernevig}}]{XIE20}%
  \BibitemOpen
  \bibfield  {author} {\bibinfo {author} {\bibfnamefont {F.}~\bibnamefont
  {Xie}}, \bibinfo {author} {\bibfnamefont {Z.}~\bibnamefont {Song}}, \bibinfo
  {author} {\bibfnamefont {B.}~\bibnamefont {Lian}},\ and\ \bibinfo {author}
  {\bibfnamefont {B.~A.}\ \bibnamefont {Bernevig}},\ }\href
  {https://doi.org/10.1103/PhysRevLett.124.167002} {\bibfield  {journal}
  {\bibinfo  {journal} {Phys. Rev. Lett.}\ }\textbf {\bibinfo {volume} {124}},\
  \bibinfo {pages} {167002} (\bibinfo {year} {2020}{\natexlab{a}})}\BibitemShut
  {NoStop}%
\bibitem [{\citenamefont {Xie}\ \emph {et~al.}(2020{\natexlab{b}})\citenamefont
  {Xie}, \citenamefont {Cowsik}, \citenamefont {Song}, \citenamefont {Lian},
  \citenamefont {Bernevig},\ and\ \citenamefont {Regnault}}]{XIE20a}%
  \BibitemOpen
  \bibfield  {author} {\bibinfo {author} {\bibfnamefont {F.}~\bibnamefont
  {Xie}}, \bibinfo {author} {\bibfnamefont {A.}~\bibnamefont {Cowsik}},
  \bibinfo {author} {\bibfnamefont {Z.-D.}\ \bibnamefont {Song}}, \bibinfo
  {author} {\bibfnamefont {B.}~\bibnamefont {Lian}}, \bibinfo {author}
  {\bibfnamefont {B.~A.}\ \bibnamefont {Bernevig}},\ and\ \bibinfo {author}
  {\bibfnamefont {N.}~\bibnamefont {Regnault}},\ }\href@noop {} {\bibfield
  {journal} {\bibinfo  {journal} {arXiv:2010.00588 [cond-mat]}\ } (\bibinfo
  {year} {2020}{\natexlab{b}})},\ \Eprint {https://arxiv.org/abs/2010.00588}
  {arXiv:2010.00588 [cond-mat]} \BibitemShut {NoStop}%
\bibitem [{\citenamefont {Xie}\ and\ \citenamefont
  {MacDonald}(2020{\natexlab{a}})}]{XIE20c}%
  \BibitemOpen
  \bibfield  {author} {\bibinfo {author} {\bibfnamefont {M.}~\bibnamefont
  {Xie}}\ and\ \bibinfo {author} {\bibfnamefont {A.~H.}\ \bibnamefont
  {MacDonald}},\ }\href {https://doi.org/10.1103/PhysRevLett.124.097601}
  {\bibfield  {journal} {\bibinfo  {journal} {Phys. Rev. Lett.}\ }\textbf
  {\bibinfo {volume} {124}},\ \bibinfo {pages} {097601} (\bibinfo {year}
  {2020}{\natexlab{a}})}\BibitemShut {NoStop}%
\bibitem [{\citenamefont {Xie}\ and\ \citenamefont
  {MacDonald}(2020{\natexlab{b}})}]{XIE20d}%
  \BibitemOpen
  \bibfield  {author} {\bibinfo {author} {\bibfnamefont {M.}~\bibnamefont
  {Xie}}\ and\ \bibinfo {author} {\bibfnamefont {A.~H.}\ \bibnamefont
  {MacDonald}},\ }\href@noop {} {\bibfield  {journal} {\bibinfo  {journal}
  {arXiv:2010.07928 [cond-mat]}\ } (\bibinfo {year} {2020}{\natexlab{b}})},\
  \Eprint {https://arxiv.org/abs/2010.07928} {arXiv:2010.07928 [cond-mat]}
  \BibitemShut {NoStop}%
\bibitem [{\citenamefont {Xu}\ and\ \citenamefont {Balents}(2018)}]{XU18}%
  \BibitemOpen
  \bibfield  {author} {\bibinfo {author} {\bibfnamefont {C.}~\bibnamefont
  {Xu}}\ and\ \bibinfo {author} {\bibfnamefont {L.}~\bibnamefont {Balents}},\
  }\href {https://doi.org/10.1103/PhysRevLett.121.087001} {\bibfield  {journal}
  {\bibinfo  {journal} {Phys. Rev. Lett.}\ }\textbf {\bibinfo {volume} {121}},\
  \bibinfo {pages} {087001} (\bibinfo {year} {2018})}\BibitemShut {NoStop}%
\bibitem [{\citenamefont {Xu}\ \emph {et~al.}(2018)\citenamefont {Xu},
  \citenamefont {Law},\ and\ \citenamefont {Lee}}]{XU18b}%
  \BibitemOpen
  \bibfield  {author} {\bibinfo {author} {\bibfnamefont {X.~Y.}\ \bibnamefont
  {Xu}}, \bibinfo {author} {\bibfnamefont {K.~T.}\ \bibnamefont {Law}},\ and\
  \bibinfo {author} {\bibfnamefont {P.~A.}\ \bibnamefont {Lee}},\ }\href
  {https://doi.org/10.1103/PhysRevB.98.121406} {\bibfield  {journal} {\bibinfo
  {journal} {Phys. Rev. B}\ }\textbf {\bibinfo {volume} {98}},\ \bibinfo
  {pages} {121406} (\bibinfo {year} {2018})}\BibitemShut {NoStop}%
\bibitem [{\citenamefont {You}\ and\ \citenamefont {Vishwanath}(2019)}]{YOU19}%
  \BibitemOpen
  \bibfield  {author} {\bibinfo {author} {\bibfnamefont {Y.-Z.}\ \bibnamefont
  {You}}\ and\ \bibinfo {author} {\bibfnamefont {A.}~\bibnamefont
  {Vishwanath}},\ }\href {https://doi.org/10.1038/s41535-019-0153-4} {\bibfield
   {journal} {\bibinfo  {journal} {npj Quantum Mater.}\ }\textbf {\bibinfo
  {volume} {4}},\ \bibinfo {pages} {1} (\bibinfo {year} {2019})}\BibitemShut
  {NoStop}%
\bibitem [{\citenamefont {Yuan}\ and\ \citenamefont {Fu}(2018)}]{YUA18}%
  \BibitemOpen
  \bibfield  {author} {\bibinfo {author} {\bibfnamefont {N.~F.~Q.}\
  \bibnamefont {Yuan}}\ and\ \bibinfo {author} {\bibfnamefont {L.}~\bibnamefont
  {Fu}},\ }\href {https://doi.org/10.1103/PhysRevB.98.045103} {\bibfield
  {journal} {\bibinfo  {journal} {Phys. Rev. B}\ }\textbf {\bibinfo {volume}
  {98}},\ \bibinfo {pages} {045103} (\bibinfo {year} {2018})}\BibitemShut
  {NoStop}%
\bibitem [{\citenamefont {Zhang}\ \emph
  {et~al.}(2020{\natexlab{a}})\citenamefont {Zhang}, \citenamefont {Jiang},
  \citenamefont {Wang},\ and\ \citenamefont {Zhang}}]{ZHA20}%
  \BibitemOpen
  \bibfield  {author} {\bibinfo {author} {\bibfnamefont {Y.}~\bibnamefont
  {Zhang}}, \bibinfo {author} {\bibfnamefont {K.}~\bibnamefont {Jiang}},
  \bibinfo {author} {\bibfnamefont {Z.}~\bibnamefont {Wang}},\ and\ \bibinfo
  {author} {\bibfnamefont {F.}~\bibnamefont {Zhang}},\ }\href
  {https://doi.org/10.1103/PhysRevB.102.035136} {\bibfield  {journal} {\bibinfo
   {journal} {Phys. Rev. B}\ }\textbf {\bibinfo {volume} {102}},\ \bibinfo
  {pages} {035136} (\bibinfo {year} {2020}{\natexlab{a}})}\BibitemShut
  {NoStop}%
\bibitem [{\citenamefont {Zou}\ \emph {et~al.}(2018)\citenamefont {Zou},
  \citenamefont {Po}, \citenamefont {Vishwanath},\ and\ \citenamefont
  {Senthil}}]{ZOU18}%
  \BibitemOpen
  \bibfield  {author} {\bibinfo {author} {\bibfnamefont {L.}~\bibnamefont
  {Zou}}, \bibinfo {author} {\bibfnamefont {H.~C.}\ \bibnamefont {Po}},
  \bibinfo {author} {\bibfnamefont {A.}~\bibnamefont {Vishwanath}},\ and\
  \bibinfo {author} {\bibfnamefont {T.}~\bibnamefont {Senthil}},\ }\href
  {https://doi.org/10.1103/PhysRevB.98.085435} {\bibfield  {journal} {\bibinfo
  {journal} {Phys. Rev. B}\ }\textbf {\bibinfo {volume} {98}},\ \bibinfo
  {pages} {085435} (\bibinfo {year} {2018})}\BibitemShut {NoStop}%
\bibitem [{\citenamefont {Kennes}\ \emph {et~al.}(2020)\citenamefont {Kennes},
  \citenamefont {Claassen}, \citenamefont {Xian}, \citenamefont {Georges},
  \citenamefont {Millis}, \citenamefont {Hone}, \citenamefont {Dean},
  \citenamefont {Basov}, \citenamefont {Pasupathy},\ and\ \citenamefont
  {Rubio}}]{KEN20}%
  \BibitemOpen
  \bibfield  {author} {\bibinfo {author} {\bibfnamefont {D.~M.}\ \bibnamefont
  {Kennes}}, \bibinfo {author} {\bibfnamefont {M.}~\bibnamefont {Claassen}},
  \bibinfo {author} {\bibfnamefont {L.}~\bibnamefont {Xian}}, \bibinfo {author}
  {\bibfnamefont {A.}~\bibnamefont {Georges}}, \bibinfo {author} {\bibfnamefont
  {A.~J.}\ \bibnamefont {Millis}}, \bibinfo {author} {\bibfnamefont
  {J.}~\bibnamefont {Hone}}, \bibinfo {author} {\bibfnamefont {C.~R.}\
  \bibnamefont {Dean}}, \bibinfo {author} {\bibfnamefont {D.~N.}\ \bibnamefont
  {Basov}}, \bibinfo {author} {\bibfnamefont {A.}~\bibnamefont {Pasupathy}},\
  and\ \bibinfo {author} {\bibfnamefont {A.}~\bibnamefont {Rubio}},\
  }\href@noop {} {\bibfield  {journal} {\bibinfo  {journal} {arXiv:2011.12638
  [cond-mat, physics:quant-ph]}\ } (\bibinfo {year} {2020})},\ \Eprint
  {https://arxiv.org/abs/2011.12638} {arXiv:2011.12638 [cond-mat,
  physics:quant-ph]} \BibitemShut {NoStop}%
\bibitem [{\citenamefont {Burg}\ \emph {et~al.}(2019)\citenamefont {Burg},
  \citenamefont {Zhu}, \citenamefont {Taniguchi}, \citenamefont {Watanabe},
  \citenamefont {MacDonald},\ and\ \citenamefont {Tutuc}}]{BUR19}%
  \BibitemOpen
  \bibfield  {author} {\bibinfo {author} {\bibfnamefont {G.~W.}\ \bibnamefont
  {Burg}}, \bibinfo {author} {\bibfnamefont {J.}~\bibnamefont {Zhu}}, \bibinfo
  {author} {\bibfnamefont {T.}~\bibnamefont {Taniguchi}}, \bibinfo {author}
  {\bibfnamefont {K.}~\bibnamefont {Watanabe}}, \bibinfo {author}
  {\bibfnamefont {A.~H.}\ \bibnamefont {MacDonald}},\ and\ \bibinfo {author}
  {\bibfnamefont {E.}~\bibnamefont {Tutuc}},\ }\href
  {https://doi.org/10.1103/PhysRevLett.123.197702} {\bibfield  {journal}
  {\bibinfo  {journal} {Phys. Rev. Lett.}\ }\textbf {\bibinfo {volume} {123}},\
  \bibinfo {pages} {197702} (\bibinfo {year} {2019})}\BibitemShut {NoStop}%
\bibitem [{\citenamefont {Carr}\ \emph
  {et~al.}(2020{\natexlab{a}})\citenamefont {Carr}, \citenamefont {Li},
  \citenamefont {Zhu}, \citenamefont {Kaxiras}, \citenamefont {Sachdev},\ and\
  \citenamefont {Kruchkov}}]{CAR20}%
  \BibitemOpen
  \bibfield  {author} {\bibinfo {author} {\bibfnamefont {S.}~\bibnamefont
  {Carr}}, \bibinfo {author} {\bibfnamefont {C.}~\bibnamefont {Li}}, \bibinfo
  {author} {\bibfnamefont {Z.}~\bibnamefont {Zhu}}, \bibinfo {author}
  {\bibfnamefont {E.}~\bibnamefont {Kaxiras}}, \bibinfo {author} {\bibfnamefont
  {S.}~\bibnamefont {Sachdev}},\ and\ \bibinfo {author} {\bibfnamefont
  {A.}~\bibnamefont {Kruchkov}},\ }\href
  {https://doi.org/10.1021/acs.nanolett.9b04979} {\bibfield  {journal}
  {\bibinfo  {journal} {Nano Lett.}\ }\textbf {\bibinfo {volume} {20}},\
  \bibinfo {pages} {3030} (\bibinfo {year} {2020}{\natexlab{a}})}\BibitemShut
  {NoStop}%
\bibitem [{\citenamefont {Carr}\ \emph
  {et~al.}(2020{\natexlab{b}})\citenamefont {Carr}, \citenamefont {Fang},\ and\
  \citenamefont {Kaxiras}}]{CAR20a}%
  \BibitemOpen
  \bibfield  {author} {\bibinfo {author} {\bibfnamefont {S.}~\bibnamefont
  {Carr}}, \bibinfo {author} {\bibfnamefont {S.}~\bibnamefont {Fang}},\ and\
  \bibinfo {author} {\bibfnamefont {E.}~\bibnamefont {Kaxiras}},\ }\href
  {https://doi.org/10.1038/s41578-020-0214-0} {\bibfield  {journal} {\bibinfo
  {journal} {Nat. Rev. Mater.}\ }\textbf {\bibinfo {volume} {5}},\ \bibinfo
  {pages} {748} (\bibinfo {year} {2020}{\natexlab{b}})}\BibitemShut {NoStop}%
\bibitem [{\citenamefont {Cea}\ \emph {et~al.}(2019)\citenamefont {Cea},
  \citenamefont {Walet},\ and\ \citenamefont {Guinea}}]{CEA19}%
  \BibitemOpen
  \bibfield  {author} {\bibinfo {author} {\bibfnamefont {T.}~\bibnamefont
  {Cea}}, \bibinfo {author} {\bibfnamefont {N.~R.}\ \bibnamefont {Walet}},\
  and\ \bibinfo {author} {\bibfnamefont {F.}~\bibnamefont {Guinea}},\ }\href
  {https://doi.org/10.1021/acs.nanolett.9b03335} {\bibfield  {journal}
  {\bibinfo  {journal} {Nano Lett.}\ }\textbf {\bibinfo {volume} {19}},\
  \bibinfo {pages} {8683} (\bibinfo {year} {2019})}\BibitemShut {NoStop}%
\bibitem [{\citenamefont {{Garc{\'i}a-Ruiz}}\ \emph {et~al.}(2020)\citenamefont
  {{Garc{\'i}a-Ruiz}}, \citenamefont {Thompson}, \citenamefont
  {{Mucha-Kruczy{\'n}ski}},\ and\ \citenamefont {Fal'ko}}]{GAR20}%
  \BibitemOpen
  \bibfield  {author} {\bibinfo {author} {\bibfnamefont {A.}~\bibnamefont
  {{Garc{\'i}a-Ruiz}}}, \bibinfo {author} {\bibfnamefont {J.~J.~P.}\
  \bibnamefont {Thompson}}, \bibinfo {author} {\bibfnamefont {M.}~\bibnamefont
  {{Mucha-Kruczy{\'n}ski}}},\ and\ \bibinfo {author} {\bibfnamefont {V.~I.}\
  \bibnamefont {Fal'ko}},\ }\href
  {https://doi.org/10.1103/PhysRevLett.125.197401} {\bibfield  {journal}
  {\bibinfo  {journal} {Phys. Rev. Lett.}\ }\textbf {\bibinfo {volume} {125}},\
  \bibinfo {pages} {197401} (\bibinfo {year} {2020})}\BibitemShut {NoStop}%
\bibitem [{\citenamefont {Haddadi}\ \emph {et~al.}(2020)\citenamefont
  {Haddadi}, \citenamefont {Wu}, \citenamefont {Kruchkov},\ and\ \citenamefont
  {Yazyev}}]{HAD20}%
  \BibitemOpen
  \bibfield  {author} {\bibinfo {author} {\bibfnamefont {F.}~\bibnamefont
  {Haddadi}}, \bibinfo {author} {\bibfnamefont {Q.}~\bibnamefont {Wu}},
  \bibinfo {author} {\bibfnamefont {A.~J.}\ \bibnamefont {Kruchkov}},\ and\
  \bibinfo {author} {\bibfnamefont {O.~V.}\ \bibnamefont {Yazyev}},\ }\href
  {https://doi.org/10.1021/acs.nanolett.9b05117} {\bibfield  {journal}
  {\bibinfo  {journal} {Nano Lett.}\ }\textbf {\bibinfo {volume} {20}},\
  \bibinfo {pages} {2410} (\bibinfo {year} {2020})}\BibitemShut {NoStop}%
\bibitem [{\citenamefont {Khalaf}\ \emph {et~al.}(2019)\citenamefont {Khalaf},
  \citenamefont {Kruchkov}, \citenamefont {Tarnopolsky},\ and\ \citenamefont
  {Vishwanath}}]{KHA19}%
  \BibitemOpen
  \bibfield  {author} {\bibinfo {author} {\bibfnamefont {E.}~\bibnamefont
  {Khalaf}}, \bibinfo {author} {\bibfnamefont {A.~J.}\ \bibnamefont
  {Kruchkov}}, \bibinfo {author} {\bibfnamefont {G.}~\bibnamefont
  {Tarnopolsky}},\ and\ \bibinfo {author} {\bibfnamefont {A.}~\bibnamefont
  {Vishwanath}},\ }\href {https://doi.org/10.1103/PhysRevB.100.085109}
  {\bibfield  {journal} {\bibinfo  {journal} {Phys. Rev. B}\ }\textbf {\bibinfo
  {volume} {100}},\ \bibinfo {pages} {085109} (\bibinfo {year}
  {2019})}\BibitemShut {NoStop}%
\bibitem [{\citenamefont {Lei}\ \emph {et~al.}(2020)\citenamefont {Lei},
  \citenamefont {Linhart}, \citenamefont {Qin}, \citenamefont {Libisch},\ and\
  \citenamefont {MacDonald}}]{LEI20}%
  \BibitemOpen
  \bibfield  {author} {\bibinfo {author} {\bibfnamefont {C.}~\bibnamefont
  {Lei}}, \bibinfo {author} {\bibfnamefont {L.}~\bibnamefont {Linhart}},
  \bibinfo {author} {\bibfnamefont {W.}~\bibnamefont {Qin}}, \bibinfo {author}
  {\bibfnamefont {F.}~\bibnamefont {Libisch}},\ and\ \bibinfo {author}
  {\bibfnamefont {A.~H.}\ \bibnamefont {MacDonald}},\ }\href@noop {} {\bibfield
   {journal} {\bibinfo  {journal} {arXiv:2010.05787 [cond-mat]}\ } (\bibinfo
  {year} {2020})},\ \Eprint {https://arxiv.org/abs/2010.05787}
  {arXiv:2010.05787 [cond-mat]} \BibitemShut {NoStop}%
\bibitem [{\citenamefont {Li}\ \emph {et~al.}(2019)\citenamefont {Li},
  \citenamefont {Wu},\ and\ \citenamefont {MacDonald}}]{LI19}%
  \BibitemOpen
  \bibfield  {author} {\bibinfo {author} {\bibfnamefont {X.}~\bibnamefont
  {Li}}, \bibinfo {author} {\bibfnamefont {F.}~\bibnamefont {Wu}},\ and\
  \bibinfo {author} {\bibfnamefont {A.~H.}\ \bibnamefont {MacDonald}},\
  }\href@noop {} {\bibfield  {journal} {\bibinfo  {journal} {arXiv:1907.12338
  [cond-mat]}\ } (\bibinfo {year} {2019})},\ \Eprint
  {https://arxiv.org/abs/1907.12338} {arXiv:1907.12338 [cond-mat]} \BibitemShut
  {NoStop}%
\bibitem [{\citenamefont {Liu}\ \emph {et~al.}(2019{\natexlab{b}})\citenamefont
  {Liu}, \citenamefont {Ma}, \citenamefont {Gao},\ and\ \citenamefont
  {Dai}}]{LIU19a}%
  \BibitemOpen
  \bibfield  {author} {\bibinfo {author} {\bibfnamefont {J.}~\bibnamefont
  {Liu}}, \bibinfo {author} {\bibfnamefont {Z.}~\bibnamefont {Ma}}, \bibinfo
  {author} {\bibfnamefont {J.}~\bibnamefont {Gao}},\ and\ \bibinfo {author}
  {\bibfnamefont {X.}~\bibnamefont {Dai}},\ }\href
  {https://doi.org/10.1103/PhysRevX.9.031021} {\bibfield  {journal} {\bibinfo
  {journal} {Phys. Rev. X}\ }\textbf {\bibinfo {volume} {9}},\ \bibinfo {pages}
  {031021} (\bibinfo {year} {2019}{\natexlab{b}})}\BibitemShut {NoStop}%
\bibitem [{\citenamefont {{Lopez-Bezanilla}}\ and\ \citenamefont
  {Lado}(2020)}]{LOP20}%
  \BibitemOpen
  \bibfield  {author} {\bibinfo {author} {\bibfnamefont {A.}~\bibnamefont
  {{Lopez-Bezanilla}}}\ and\ \bibinfo {author} {\bibfnamefont {J.~L.}\
  \bibnamefont {Lado}},\ }\href
  {https://doi.org/10.1103/PhysRevResearch.2.033357} {\bibfield  {journal}
  {\bibinfo  {journal} {Phys. Rev. Research}\ }\textbf {\bibinfo {volume}
  {2}},\ \bibinfo {pages} {033357} (\bibinfo {year} {2020})}\BibitemShut
  {NoStop}%
\bibitem [{\citenamefont {Mora}\ \emph {et~al.}(2019)\citenamefont {Mora},
  \citenamefont {Regnault},\ and\ \citenamefont {Bernevig}}]{MOR19}%
  \BibitemOpen
  \bibfield  {author} {\bibinfo {author} {\bibfnamefont {C.}~\bibnamefont
  {Mora}}, \bibinfo {author} {\bibfnamefont {N.}~\bibnamefont {Regnault}},\
  and\ \bibinfo {author} {\bibfnamefont {B.~A.}\ \bibnamefont {Bernevig}},\
  }\href {https://doi.org/10.1103/PhysRevLett.123.026402} {\bibfield  {journal}
  {\bibinfo  {journal} {Phys. Rev. Lett.}\ }\textbf {\bibinfo {volume} {123}},\
  \bibinfo {pages} {026402} (\bibinfo {year} {2019})}\BibitemShut {NoStop}%
\bibitem [{\citenamefont {Park}\ \emph
  {et~al.}(2020{\natexlab{b}})\citenamefont {Park}, \citenamefont {Chittari},\
  and\ \citenamefont {Jung}}]{PAR20b}%
  \BibitemOpen
  \bibfield  {author} {\bibinfo {author} {\bibfnamefont {Y.}~\bibnamefont
  {Park}}, \bibinfo {author} {\bibfnamefont {B.~L.}\ \bibnamefont {Chittari}},\
  and\ \bibinfo {author} {\bibfnamefont {J.}~\bibnamefont {Jung}},\ }\href
  {https://doi.org/10.1103/PhysRevB.102.035411} {\bibfield  {journal} {\bibinfo
   {journal} {Phys. Rev. B}\ }\textbf {\bibinfo {volume} {102}},\ \bibinfo
  {pages} {035411} (\bibinfo {year} {2020}{\natexlab{b}})}\BibitemShut
  {NoStop}%
\bibitem [{\citenamefont {Su{\'a}rez~Morell}\ \emph {et~al.}(2013)\citenamefont
  {Su{\'a}rez~Morell}, \citenamefont {Pacheco}, \citenamefont {Chico},\ and\
  \citenamefont {Brey}}]{SUA13}%
  \BibitemOpen
  \bibfield  {author} {\bibinfo {author} {\bibfnamefont {E.}~\bibnamefont
  {Su{\'a}rez~Morell}}, \bibinfo {author} {\bibfnamefont {M.}~\bibnamefont
  {Pacheco}}, \bibinfo {author} {\bibfnamefont {L.}~\bibnamefont {Chico}},\
  and\ \bibinfo {author} {\bibfnamefont {L.}~\bibnamefont {Brey}},\ }\href
  {https://doi.org/10.1103/PhysRevB.87.125414} {\bibfield  {journal} {\bibinfo
  {journal} {Phys. Rev. B}\ }\textbf {\bibinfo {volume} {87}},\ \bibinfo
  {pages} {125414} (\bibinfo {year} {2013})}\BibitemShut {NoStop}%
\bibitem [{\citenamefont {Wu}\ \emph {et~al.}(2020{\natexlab{b}})\citenamefont
  {Wu}, \citenamefont {Zhan},\ and\ \citenamefont {Yuan}}]{WU20c}%
  \BibitemOpen
  \bibfield  {author} {\bibinfo {author} {\bibfnamefont {Z.}~\bibnamefont
  {Wu}}, \bibinfo {author} {\bibfnamefont {Z.}~\bibnamefont {Zhan}},\ and\
  \bibinfo {author} {\bibfnamefont {S.}~\bibnamefont {Yuan}},\ }\href@noop {}
  {\bibfield  {journal} {\bibinfo  {journal} {arXiv:2012.13741 [cond-mat]}\ }
  (\bibinfo {year} {2020}{\natexlab{b}})},\ \Eprint
  {https://arxiv.org/abs/2012.13741} {arXiv:2012.13741 [cond-mat]} \BibitemShut
  {NoStop}%
\bibitem [{\citenamefont {Zhang}\ \emph
  {et~al.}(2020{\natexlab{b}})\citenamefont {Zhang}, \citenamefont {Xie},
  \citenamefont {Wu}, \citenamefont {Liu},\ and\ \citenamefont
  {Yazyev}}]{ZHA20a}%
  \BibitemOpen
  \bibfield  {author} {\bibinfo {author} {\bibfnamefont {S.}~\bibnamefont
  {Zhang}}, \bibinfo {author} {\bibfnamefont {B.}~\bibnamefont {Xie}}, \bibinfo
  {author} {\bibfnamefont {Q.}~\bibnamefont {Wu}}, \bibinfo {author}
  {\bibfnamefont {J.}~\bibnamefont {Liu}},\ and\ \bibinfo {author}
  {\bibfnamefont {O.~V.}\ \bibnamefont {Yazyev}},\ }\href@noop {} {\bibfield
  {journal} {\bibinfo  {journal} {arXiv:2012.11964 [cond-mat]}\ } (\bibinfo
  {year} {2020}{\natexlab{b}})},\ \Eprint {https://arxiv.org/abs/2012.11964}
  {arXiv:2012.11964 [cond-mat]} \BibitemShut {NoStop}%
\bibitem [{\citenamefont {Zhu}\ \emph {et~al.}(2020{\natexlab{a}})\citenamefont
  {Zhu}, \citenamefont {Carr}, \citenamefont {Massatt}, \citenamefont
  {Luskin},\ and\ \citenamefont {Kaxiras}}]{ZHU20}%
  \BibitemOpen
  \bibfield  {author} {\bibinfo {author} {\bibfnamefont {Z.}~\bibnamefont
  {Zhu}}, \bibinfo {author} {\bibfnamefont {S.}~\bibnamefont {Carr}}, \bibinfo
  {author} {\bibfnamefont {D.}~\bibnamefont {Massatt}}, \bibinfo {author}
  {\bibfnamefont {M.}~\bibnamefont {Luskin}},\ and\ \bibinfo {author}
  {\bibfnamefont {E.}~\bibnamefont {Kaxiras}},\ }\href
  {https://doi.org/10.1103/PhysRevLett.125.116404} {\bibfield  {journal}
  {\bibinfo  {journal} {Phys. Rev. Lett.}\ }\textbf {\bibinfo {volume} {125}},\
  \bibinfo {pages} {116404} (\bibinfo {year} {2020}{\natexlab{a}})}\BibitemShut
  {NoStop}%
\bibitem [{\citenamefont {Zhu}\ \emph {et~al.}(2020{\natexlab{b}})\citenamefont
  {Zhu}, \citenamefont {Cazeaux}, \citenamefont {Luskin},\ and\ \citenamefont
  {Kaxiras}}]{ZHU20a}%
  \BibitemOpen
  \bibfield  {author} {\bibinfo {author} {\bibfnamefont {Z.}~\bibnamefont
  {Zhu}}, \bibinfo {author} {\bibfnamefont {P.}~\bibnamefont {Cazeaux}},
  \bibinfo {author} {\bibfnamefont {M.}~\bibnamefont {Luskin}},\ and\ \bibinfo
  {author} {\bibfnamefont {E.}~\bibnamefont {Kaxiras}},\ }\href
  {https://doi.org/10.1103/PhysRevB.101.224107} {\bibfield  {journal} {\bibinfo
   {journal} {Phys. Rev. B}\ }\textbf {\bibinfo {volume} {101}},\ \bibinfo
  {pages} {224107} (\bibinfo {year} {2020}{\natexlab{b}})}\BibitemShut
  {NoStop}%
\bibitem [{\citenamefont {Chen}\ \emph
  {et~al.}(2019{\natexlab{a}})\citenamefont {Chen}, \citenamefont {Jiang},
  \citenamefont {Wu}, \citenamefont {Lyu}, \citenamefont {Li}, \citenamefont
  {Chittari}, \citenamefont {Watanabe}, \citenamefont {Taniguchi},
  \citenamefont {Shi}, \citenamefont {Jung}, \citenamefont {Zhang},\ and\
  \citenamefont {Wang}}]{CHE19}%
  \BibitemOpen
  \bibfield  {author} {\bibinfo {author} {\bibfnamefont {G.}~\bibnamefont
  {Chen}}, \bibinfo {author} {\bibfnamefont {L.}~\bibnamefont {Jiang}},
  \bibinfo {author} {\bibfnamefont {S.}~\bibnamefont {Wu}}, \bibinfo {author}
  {\bibfnamefont {B.}~\bibnamefont {Lyu}}, \bibinfo {author} {\bibfnamefont
  {H.}~\bibnamefont {Li}}, \bibinfo {author} {\bibfnamefont {B.~L.}\
  \bibnamefont {Chittari}}, \bibinfo {author} {\bibfnamefont {K.}~\bibnamefont
  {Watanabe}}, \bibinfo {author} {\bibfnamefont {T.}~\bibnamefont {Taniguchi}},
  \bibinfo {author} {\bibfnamefont {Z.}~\bibnamefont {Shi}}, \bibinfo {author}
  {\bibfnamefont {J.}~\bibnamefont {Jung}}, \bibinfo {author} {\bibfnamefont
  {Y.}~\bibnamefont {Zhang}},\ and\ \bibinfo {author} {\bibfnamefont
  {F.}~\bibnamefont {Wang}},\ }\href
  {https://doi.org/10.1038/s41567-018-0387-2} {\bibfield  {journal} {\bibinfo
  {journal} {Nat. Phys.}\ }\textbf {\bibinfo {volume} {15}},\ \bibinfo {pages}
  {237} (\bibinfo {year} {2019}{\natexlab{a}})}\BibitemShut {NoStop}%
\bibitem [{\citenamefont {Chen}\ \emph
  {et~al.}(2019{\natexlab{b}})\citenamefont {Chen}, \citenamefont {Sharpe},
  \citenamefont {Gallagher}, \citenamefont {Rosen}, \citenamefont {Fox},
  \citenamefont {Jiang}, \citenamefont {Lyu}, \citenamefont {Li}, \citenamefont
  {Watanabe}, \citenamefont {Taniguchi}, \citenamefont {Jung}, \citenamefont
  {Shi}, \citenamefont {{Goldhaber-Gordon}}, \citenamefont {Zhang},\ and\
  \citenamefont {Wang}}]{CHE19a}%
  \BibitemOpen
  \bibfield  {author} {\bibinfo {author} {\bibfnamefont {G.}~\bibnamefont
  {Chen}}, \bibinfo {author} {\bibfnamefont {A.~L.}\ \bibnamefont {Sharpe}},
  \bibinfo {author} {\bibfnamefont {P.}~\bibnamefont {Gallagher}}, \bibinfo
  {author} {\bibfnamefont {I.~T.}\ \bibnamefont {Rosen}}, \bibinfo {author}
  {\bibfnamefont {E.~J.}\ \bibnamefont {Fox}}, \bibinfo {author} {\bibfnamefont
  {L.}~\bibnamefont {Jiang}}, \bibinfo {author} {\bibfnamefont
  {B.}~\bibnamefont {Lyu}}, \bibinfo {author} {\bibfnamefont {H.}~\bibnamefont
  {Li}}, \bibinfo {author} {\bibfnamefont {K.}~\bibnamefont {Watanabe}},
  \bibinfo {author} {\bibfnamefont {T.}~\bibnamefont {Taniguchi}}, \bibinfo
  {author} {\bibfnamefont {J.}~\bibnamefont {Jung}}, \bibinfo {author}
  {\bibfnamefont {Z.}~\bibnamefont {Shi}}, \bibinfo {author} {\bibfnamefont
  {D.}~\bibnamefont {{Goldhaber-Gordon}}}, \bibinfo {author} {\bibfnamefont
  {Y.}~\bibnamefont {Zhang}},\ and\ \bibinfo {author} {\bibfnamefont
  {F.}~\bibnamefont {Wang}},\ }\href
  {https://doi.org/10.1038/s41586-019-1393-y} {\bibfield  {journal} {\bibinfo
  {journal} {Nature}\ }\textbf {\bibinfo {volume} {572}},\ \bibinfo {pages}
  {215} (\bibinfo {year} {2019}{\natexlab{b}})}\BibitemShut {NoStop}%
\bibitem [{\citenamefont {Chen}\ \emph
  {et~al.}(2020{\natexlab{b}})\citenamefont {Chen}, \citenamefont {He},
  \citenamefont {Zhang}, \citenamefont {Hsieh}, \citenamefont {Fei},
  \citenamefont {Watanabe}, \citenamefont {Taniguchi}, \citenamefont {Cobden},
  \citenamefont {Xu}, \citenamefont {Dean},\ and\ \citenamefont
  {Yankowitz}}]{CHE20b}%
  \BibitemOpen
  \bibfield  {author} {\bibinfo {author} {\bibfnamefont {S.}~\bibnamefont
  {Chen}}, \bibinfo {author} {\bibfnamefont {M.}~\bibnamefont {He}}, \bibinfo
  {author} {\bibfnamefont {Y.-H.}\ \bibnamefont {Zhang}}, \bibinfo {author}
  {\bibfnamefont {V.}~\bibnamefont {Hsieh}}, \bibinfo {author} {\bibfnamefont
  {Z.}~\bibnamefont {Fei}}, \bibinfo {author} {\bibfnamefont {K.}~\bibnamefont
  {Watanabe}}, \bibinfo {author} {\bibfnamefont {T.}~\bibnamefont {Taniguchi}},
  \bibinfo {author} {\bibfnamefont {D.~H.}\ \bibnamefont {Cobden}}, \bibinfo
  {author} {\bibfnamefont {X.}~\bibnamefont {Xu}}, \bibinfo {author}
  {\bibfnamefont {C.~R.}\ \bibnamefont {Dean}},\ and\ \bibinfo {author}
  {\bibfnamefont {M.}~\bibnamefont {Yankowitz}},\ }\href
  {https://doi.org/10.1038/s41567-020-01062-6} {\bibfield  {journal} {\bibinfo
  {journal} {Nat. Phys.}\ ,\ \bibinfo {pages} {1}} (\bibinfo {year}
  {2020}{\natexlab{b}})}\BibitemShut {NoStop}%
\bibitem [{\citenamefont {Hao}\ \emph {et~al.}(2021)\citenamefont {Hao},
  \citenamefont {Zimmerman}, \citenamefont {Ledwith}, \citenamefont {Khalaf},
  \citenamefont {Najafabadi}, \citenamefont {Watanabe}, \citenamefont
  {Taniguchi}, \citenamefont {Vishwanath},\ and\ \citenamefont {Kim}}]{HAO21}%
  \BibitemOpen
  \bibfield  {author} {\bibinfo {author} {\bibfnamefont {Z.}~\bibnamefont
  {Hao}}, \bibinfo {author} {\bibfnamefont {A.~M.}\ \bibnamefont {Zimmerman}},
  \bibinfo {author} {\bibfnamefont {P.}~\bibnamefont {Ledwith}}, \bibinfo
  {author} {\bibfnamefont {E.}~\bibnamefont {Khalaf}}, \bibinfo {author}
  {\bibfnamefont {D.~H.}\ \bibnamefont {Najafabadi}}, \bibinfo {author}
  {\bibfnamefont {K.}~\bibnamefont {Watanabe}}, \bibinfo {author}
  {\bibfnamefont {T.}~\bibnamefont {Taniguchi}}, \bibinfo {author}
  {\bibfnamefont {A.}~\bibnamefont {Vishwanath}},\ and\ \bibinfo {author}
  {\bibfnamefont {P.}~\bibnamefont {Kim}},\ }\bibfield  {journal} {\bibinfo
  {journal} {Science}\ }\href {https://doi.org/10.1126/science.abg0399}
  {10.1126/science.abg0399} (\bibinfo {year} {2021})\BibitemShut {NoStop}%
\bibitem [{\citenamefont {Park}\ \emph {et~al.}(2021)\citenamefont {Park},
  \citenamefont {Cao}, \citenamefont {Watanabe}, \citenamefont {Taniguchi},\
  and\ \citenamefont {{Jarillo-Herrero}}}]{PAR21}%
  \BibitemOpen
  \bibfield  {author} {\bibinfo {author} {\bibfnamefont {J.~M.}\ \bibnamefont
  {Park}}, \bibinfo {author} {\bibfnamefont {Y.}~\bibnamefont {Cao}}, \bibinfo
  {author} {\bibfnamefont {K.}~\bibnamefont {Watanabe}}, \bibinfo {author}
  {\bibfnamefont {T.}~\bibnamefont {Taniguchi}},\ and\ \bibinfo {author}
  {\bibfnamefont {P.}~\bibnamefont {{Jarillo-Herrero}}},\ }\href
  {https://doi.org/10.1038/s41586-021-03192-0} {\bibfield  {journal} {\bibinfo
  {journal} {Nature}\ }\textbf {\bibinfo {volume} {590}},\ \bibinfo {pages}
  {249} (\bibinfo {year} {2021})}\BibitemShut {NoStop}%
\bibitem [{\citenamefont {Polshyn}\ \emph {et~al.}(2020)\citenamefont
  {Polshyn}, \citenamefont {Zhu}, \citenamefont {Kumar}, \citenamefont {Zhang},
  \citenamefont {Yang}, \citenamefont {Tschirhart}, \citenamefont {Serlin},
  \citenamefont {Watanabe}, \citenamefont {Taniguchi}, \citenamefont
  {MacDonald},\ and\ \citenamefont {Young}}]{POL20}%
  \BibitemOpen
  \bibfield  {author} {\bibinfo {author} {\bibfnamefont {H.}~\bibnamefont
  {Polshyn}}, \bibinfo {author} {\bibfnamefont {J.}~\bibnamefont {Zhu}},
  \bibinfo {author} {\bibfnamefont {M.~A.}\ \bibnamefont {Kumar}}, \bibinfo
  {author} {\bibfnamefont {Y.}~\bibnamefont {Zhang}}, \bibinfo {author}
  {\bibfnamefont {F.}~\bibnamefont {Yang}}, \bibinfo {author} {\bibfnamefont
  {C.~L.}\ \bibnamefont {Tschirhart}}, \bibinfo {author} {\bibfnamefont
  {M.}~\bibnamefont {Serlin}}, \bibinfo {author} {\bibfnamefont
  {K.}~\bibnamefont {Watanabe}}, \bibinfo {author} {\bibfnamefont
  {T.}~\bibnamefont {Taniguchi}}, \bibinfo {author} {\bibfnamefont {A.~H.}\
  \bibnamefont {MacDonald}},\ and\ \bibinfo {author} {\bibfnamefont {A.~F.}\
  \bibnamefont {Young}},\ }\href {https://doi.org/10.1038/s41586-020-2963-8}
  {\bibfield  {journal} {\bibinfo  {journal} {Nature}\ }\textbf {\bibinfo
  {volume} {588}},\ \bibinfo {pages} {66} (\bibinfo {year} {2020})}\BibitemShut
  {NoStop}%
\bibitem [{\citenamefont {Shi}\ \emph {et~al.}(2020)\citenamefont {Shi},
  \citenamefont {Xu}, \citenamefont {Ezzi}, \citenamefont {Balakrishnan},
  \citenamefont {{Garcia-Ruiz}}, \citenamefont {Tsim}, \citenamefont {Mullan},
  \citenamefont {Barrier}, \citenamefont {Xin}, \citenamefont {Piot},
  \citenamefont {Taniguchi}, \citenamefont {Watanabe}, \citenamefont
  {Carvalho}, \citenamefont {Mishchenko}, \citenamefont {Geim}, \citenamefont
  {Fal'ko}, \citenamefont {Adam}, \citenamefont {Neto},\ and\ \citenamefont
  {Novoselov}}]{SHI20}%
  \BibitemOpen
  \bibfield  {author} {\bibinfo {author} {\bibfnamefont {Y.}~\bibnamefont
  {Shi}}, \bibinfo {author} {\bibfnamefont {S.}~\bibnamefont {Xu}}, \bibinfo
  {author} {\bibfnamefont {M.~M.~A.}\ \bibnamefont {Ezzi}}, \bibinfo {author}
  {\bibfnamefont {N.}~\bibnamefont {Balakrishnan}}, \bibinfo {author}
  {\bibfnamefont {A.}~\bibnamefont {{Garcia-Ruiz}}}, \bibinfo {author}
  {\bibfnamefont {B.}~\bibnamefont {Tsim}}, \bibinfo {author} {\bibfnamefont
  {C.}~\bibnamefont {Mullan}}, \bibinfo {author} {\bibfnamefont
  {J.}~\bibnamefont {Barrier}}, \bibinfo {author} {\bibfnamefont
  {N.}~\bibnamefont {Xin}}, \bibinfo {author} {\bibfnamefont {B.~A.}\
  \bibnamefont {Piot}}, \bibinfo {author} {\bibfnamefont {T.}~\bibnamefont
  {Taniguchi}}, \bibinfo {author} {\bibfnamefont {K.}~\bibnamefont {Watanabe}},
  \bibinfo {author} {\bibfnamefont {A.}~\bibnamefont {Carvalho}}, \bibinfo
  {author} {\bibfnamefont {A.}~\bibnamefont {Mishchenko}}, \bibinfo {author}
  {\bibfnamefont {A.~K.}\ \bibnamefont {Geim}}, \bibinfo {author}
  {\bibfnamefont {V.~I.}\ \bibnamefont {Fal'ko}}, \bibinfo {author}
  {\bibfnamefont {S.}~\bibnamefont {Adam}}, \bibinfo {author} {\bibfnamefont
  {A.~H.~C.}\ \bibnamefont {Neto}},\ and\ \bibinfo {author} {\bibfnamefont
  {K.~S.}\ \bibnamefont {Novoselov}},\ }\href@noop {} {\bibfield  {journal}
  {\bibinfo  {journal} {arXiv:2004.12414 [cond-mat]}\ } (\bibinfo {year}
  {2020})},\ \Eprint {https://arxiv.org/abs/2004.12414} {arXiv:2004.12414
  [cond-mat]} \BibitemShut {NoStop}%
\bibitem [{\citenamefont {Tsai}\ \emph {et~al.}(2020)\citenamefont {Tsai},
  \citenamefont {Zhang}, \citenamefont {Zhu}, \citenamefont {Luo},
  \citenamefont {Carr}, \citenamefont {Luskin}, \citenamefont {Kaxiras},\ and\
  \citenamefont {Wang}}]{TSA20}%
  \BibitemOpen
  \bibfield  {author} {\bibinfo {author} {\bibfnamefont {K.-T.}\ \bibnamefont
  {Tsai}}, \bibinfo {author} {\bibfnamefont {X.}~\bibnamefont {Zhang}},
  \bibinfo {author} {\bibfnamefont {Z.}~\bibnamefont {Zhu}}, \bibinfo {author}
  {\bibfnamefont {Y.}~\bibnamefont {Luo}}, \bibinfo {author} {\bibfnamefont
  {S.}~\bibnamefont {Carr}}, \bibinfo {author} {\bibfnamefont {M.}~\bibnamefont
  {Luskin}}, \bibinfo {author} {\bibfnamefont {E.}~\bibnamefont {Kaxiras}},\
  and\ \bibinfo {author} {\bibfnamefont {K.}~\bibnamefont {Wang}},\ }\href@noop
  {} {\bibfield  {journal} {\bibinfo  {journal} {arXiv:1912.03375 [cond-mat]}\
  } (\bibinfo {year} {2020})},\ \Eprint {https://arxiv.org/abs/1912.03375}
  {arXiv:1912.03375 [cond-mat]} \BibitemShut {NoStop}%
\bibitem [{\citenamefont {Burg}\ \emph {et~al.}(2020)\citenamefont {Burg},
  \citenamefont {Lian}, \citenamefont {Taniguchi}, \citenamefont {Watanabe},
  \citenamefont {Bernevig},\ and\ \citenamefont {Tutuc}}]{BUR20}%
  \BibitemOpen
  \bibfield  {author} {\bibinfo {author} {\bibfnamefont {G.~W.}\ \bibnamefont
  {Burg}}, \bibinfo {author} {\bibfnamefont {B.}~\bibnamefont {Lian}}, \bibinfo
  {author} {\bibfnamefont {T.}~\bibnamefont {Taniguchi}}, \bibinfo {author}
  {\bibfnamefont {K.}~\bibnamefont {Watanabe}}, \bibinfo {author}
  {\bibfnamefont {B.~A.}\ \bibnamefont {Bernevig}},\ and\ \bibinfo {author}
  {\bibfnamefont {E.}~\bibnamefont {Tutuc}},\ }\href@noop {} {\bibfield
  {journal} {\bibinfo  {journal} {arXiv:2006.14000 [cond-mat]}\ } (\bibinfo
  {year} {2020})},\ \Eprint {https://arxiv.org/abs/2006.14000}
  {arXiv:2006.14000 [cond-mat]} \BibitemShut {NoStop}%
\bibitem [{\citenamefont {Cao}\ \emph {et~al.}(2020{\natexlab{c}})\citenamefont
  {Cao}, \citenamefont {{Rodan-Legrain}}, \citenamefont {{Rubies-Bigorda}},
  \citenamefont {Park}, \citenamefont {Watanabe}, \citenamefont {Taniguchi},\
  and\ \citenamefont {{Jarillo-Herrero}}}]{CAO20a}%
  \BibitemOpen
  \bibfield  {author} {\bibinfo {author} {\bibfnamefont {Y.}~\bibnamefont
  {Cao}}, \bibinfo {author} {\bibfnamefont {D.}~\bibnamefont
  {{Rodan-Legrain}}}, \bibinfo {author} {\bibfnamefont {O.}~\bibnamefont
  {{Rubies-Bigorda}}}, \bibinfo {author} {\bibfnamefont {J.~M.}\ \bibnamefont
  {Park}}, \bibinfo {author} {\bibfnamefont {K.}~\bibnamefont {Watanabe}},
  \bibinfo {author} {\bibfnamefont {T.}~\bibnamefont {Taniguchi}},\ and\
  \bibinfo {author} {\bibfnamefont {P.}~\bibnamefont {{Jarillo-Herrero}}},\
  }\href {https://doi.org/10.1038/s41586-020-2260-6} {\bibfield  {journal}
  {\bibinfo  {journal} {Nature}\ }\textbf {\bibinfo {volume} {583}},\ \bibinfo
  {pages} {215} (\bibinfo {year} {2020}{\natexlab{c}})}\BibitemShut {NoStop}%
\bibitem [{\citenamefont {Liu}\ \emph {et~al.}(2020{\natexlab{b}})\citenamefont
  {Liu}, \citenamefont {Hao}, \citenamefont {Khalaf}, \citenamefont {Lee},
  \citenamefont {Ronen}, \citenamefont {Yoo}, \citenamefont {Haei~Najafabadi},
  \citenamefont {Watanabe}, \citenamefont {Taniguchi}, \citenamefont
  {Vishwanath},\ and\ \citenamefont {Kim}}]{LIU20b}%
  \BibitemOpen
  \bibfield  {author} {\bibinfo {author} {\bibfnamefont {X.}~\bibnamefont
  {Liu}}, \bibinfo {author} {\bibfnamefont {Z.}~\bibnamefont {Hao}}, \bibinfo
  {author} {\bibfnamefont {E.}~\bibnamefont {Khalaf}}, \bibinfo {author}
  {\bibfnamefont {J.~Y.}\ \bibnamefont {Lee}}, \bibinfo {author} {\bibfnamefont
  {Y.}~\bibnamefont {Ronen}}, \bibinfo {author} {\bibfnamefont
  {H.}~\bibnamefont {Yoo}}, \bibinfo {author} {\bibfnamefont {D.}~\bibnamefont
  {Haei~Najafabadi}}, \bibinfo {author} {\bibfnamefont {K.}~\bibnamefont
  {Watanabe}}, \bibinfo {author} {\bibfnamefont {T.}~\bibnamefont {Taniguchi}},
  \bibinfo {author} {\bibfnamefont {A.}~\bibnamefont {Vishwanath}},\ and\
  \bibinfo {author} {\bibfnamefont {P.}~\bibnamefont {Kim}},\ }\href
  {https://doi.org/10.1038/s41586-020-2458-7} {\bibfield  {journal} {\bibinfo
  {journal} {Nature}\ }\textbf {\bibinfo {volume} {583}},\ \bibinfo {pages}
  {221} (\bibinfo {year} {2020}{\natexlab{b}})}\BibitemShut {NoStop}%
\bibitem [{\citenamefont {Shen}\ \emph {et~al.}(2020)\citenamefont {Shen},
  \citenamefont {Chu}, \citenamefont {Wu}, \citenamefont {Li}, \citenamefont
  {Wang}, \citenamefont {Zhao}, \citenamefont {Tang}, \citenamefont {Liu},
  \citenamefont {Tian}, \citenamefont {Watanabe}, \citenamefont {Taniguchi},
  \citenamefont {Yang}, \citenamefont {Meng}, \citenamefont {Shi},
  \citenamefont {Yazyev},\ and\ \citenamefont {Zhang}}]{SHE20}%
  \BibitemOpen
  \bibfield  {author} {\bibinfo {author} {\bibfnamefont {C.}~\bibnamefont
  {Shen}}, \bibinfo {author} {\bibfnamefont {Y.}~\bibnamefont {Chu}}, \bibinfo
  {author} {\bibfnamefont {Q.}~\bibnamefont {Wu}}, \bibinfo {author}
  {\bibfnamefont {N.}~\bibnamefont {Li}}, \bibinfo {author} {\bibfnamefont
  {S.}~\bibnamefont {Wang}}, \bibinfo {author} {\bibfnamefont {Y.}~\bibnamefont
  {Zhao}}, \bibinfo {author} {\bibfnamefont {J.}~\bibnamefont {Tang}}, \bibinfo
  {author} {\bibfnamefont {J.}~\bibnamefont {Liu}}, \bibinfo {author}
  {\bibfnamefont {J.}~\bibnamefont {Tian}}, \bibinfo {author} {\bibfnamefont
  {K.}~\bibnamefont {Watanabe}}, \bibinfo {author} {\bibfnamefont
  {T.}~\bibnamefont {Taniguchi}}, \bibinfo {author} {\bibfnamefont
  {R.}~\bibnamefont {Yang}}, \bibinfo {author} {\bibfnamefont {Z.~Y.}\
  \bibnamefont {Meng}}, \bibinfo {author} {\bibfnamefont {D.}~\bibnamefont
  {Shi}}, \bibinfo {author} {\bibfnamefont {O.~V.}\ \bibnamefont {Yazyev}},\
  and\ \bibinfo {author} {\bibfnamefont {G.}~\bibnamefont {Zhang}},\ }\href
  {https://doi.org/10.1038/s41567-020-0825-9} {\bibfield  {journal} {\bibinfo
  {journal} {Nat. Phys.}\ }\textbf {\bibinfo {volume} {16}},\ \bibinfo {pages}
  {520} (\bibinfo {year} {2020})}\BibitemShut {NoStop}%
\bibitem [{TTG()}]{TTG2}%
  \BibitemOpen
  \href@noop {} {\bibinfo  {journal} {TSTG II, In preparation}\ }\BibitemShut
  {NoStop}%
\end{thebibliography}%

\appendix
\onecolumngrid
\newpage
\tableofcontents
\newpage
\crefalias{section}{appsec}
\crefalias{subsection}{appsec}
\section{Single-particle Hamiltonian}\label{app:single_part_ham}
In this appendix, we provide a detailed derivation of the TSTG single-particle Hamiltonian presented in \cref{sec:singlehamiltonian}. We explain how the TSTG Hamiltonian splits into a TBG-like contribution coupled to a high-velocity Dirac cone Hamiltonian by an externally applied displacement field. Finally, we introduce the energy-band basis which will be employed in writing the single-particle projected Hamiltonian in \cref{sec:singleparticlespectrum:proj}. 

\subsection{Derivation of the single-particle Hamiltonian}\label{app:single_part_ham:a}

Let $\cre{a}{\vec{p},\alpha,s,l}$ represent the fermion operator in the plane wave basis of graphene layer $l$. The momentum $\vec{p}$ is measured from the $\Gamma$ point of the monolayer graphene Brillouin Zone (BZ), $\alpha=A,B$ represents the sublattice index, $s=\uparrow,\downarrow$ is the spin index, and $l=1,2,3$ denotes the layer index (respectively corresponding to the lower, middle, and upper layers). Focusing on TSTG, we define $\vec{K}_{+}$ as the $K$ point in the top and bottom layer graphene BZ ($l=3,1$), and $\vec{K}_{-}$ as the $K$ point in the middle layer graphene BZ ($l=2$). $\vec{K}_{+}$ and $\vec{K}_{-}$ differ by a twist angle $\theta$. For concreteness, we assume $\mathbf{K}_{\pm}$ is along the direction with an angle $ \pm \theta/2$ to the $\hat{x}$ axis, as depicted in \cref{fig:mbz_qlattice:a}. Each graphene layer contains two valleys $K$ and $K'$, labeled by $\eta=\pm 1$ and located at momenta $\eta \vec{K}_{\pm}$, corresponding to two (decoupled) valleys of the moir\'e single-particle Hamiltonian.

For later use, we also introduce the 2D momenta
\begin{equation}
	\vec{q}_{1}=\left(\vec{K}_{+}-\vec{K}_{-}\right)=k_\theta \left( 0,1 \right)^T, \qquad 
	\vec{q}_{2}=C_{3z}\vec{q}_{1}=k_\theta \left(-\frac{\sqrt{3}}{2},-\frac{1}{2} \right)^T,\qquad
	\vec{q}_{3}=C_{3z}^2\vec{q}_{1}=k_\theta \left( \frac{\sqrt{3}}{2},-\frac{1}{2} \right)^T,
\end{equation}
whose coordinates are given in the $\left(k_x,k_y \right)$ basis and where $k_\theta=|\vec{K}_{-}-\vec{K}_{+}|=2|\mathbf{K}_{+}|\sin(\theta/2)$ corresponding to the twist angle $\theta$. We can then define the MBZ for the TSTG moir\'e lattice, which is generated by the reciprocal vectors
\begin{equation}
	\vec{b}_{M1}=\vec{q}_3-\vec{q}_1\ ,\qquad  \vec{b}_{M2}=\vec{q}_3-\vec{q}_2\ .
\end{equation}

To concentrate on the low energy physics of the two valleys, we define $\mathcal{Q}_{0}=\mathbb{Z}\mathbf{b}_{M1}+\mathbb{Z}\mathbf{b}_{M2}$ as the triangular moir\'e reciprocal lattice generated by the reciprocal basis vectors $\mathbf{b}_{M1}$ and $\mathbf{b}_{M2}$. We also define two shifted momentum lattices $\mathcal{Q}_{+}=\mathbf{q}_{1}+\mathcal{Q}_{0}$ and $\mathcal{Q}_{-}=-\mathbf{q}_{1}+\mathcal{Q}_{0}$, which together form a honeycomb lattice (as seen in \cref{fig:mbz_qlattice:b}). We then introduce the low-energy fermion operators $\cre{a}{\vec{k},\vec{Q},\eta,\alpha,s,l}$ defined as 
\begin{equation}	
		\label{si:eqn:low_en_fermions_a}
		\cre{a}{\vec{k},\vec{Q},\eta,\alpha,s,l} \equiv \cre{a}{\eta \vec{K}_l + \vec{k} - \vec{Q}, \alpha, s, l} \qquad \text{for} \qquad \vec{Q} \in \mathcal{Q}_{\eta,l}
\end{equation}
with $\vec{k} \in \text{MBZ}$ and $\vec{k}=\vec{0}$ representing the $\Gamma_{M}$ point. In addition, for a fixed valley $\eta$, we have introduced the notation
\begin{equation}
	\mathcal{Q}_{\eta,l} = \begin{cases}
		\mathcal{Q}_{\eta} &\qquad \text{for} \qquad l=1,3 \\
		\mathcal{Q}_{-\eta} &\qquad \text{for} \qquad l=2
	\end{cases},	
\end{equation}
and also denoted $\vec{K}_l = \vec{K}_+$ for $l=1,3$ and $\vec{K}_l = \vec{K}_-$ for $l=2$. Because of the staggered trilayer structure, there are twice as many fermion operators in the lattice $\mathcal{Q}_{\eta}$ ($\cre{a}{\vec{k},\vec{Q},\eta,\alpha,s,l}$, with $l=1,3$) than there are in lattice $\mathcal{Q}_{-\eta}$ ($\cre{a}{\vec{k},\vec{Q},\eta,\alpha,s,2}$). It is also worth noting that the low-energy fermions operators are not periodic in $\vec{k}$, but obey the Bloch periodicity property
\begin{equation}
	\label{si:eqn:shift_prop_a}
	\cre{a}{\vec{k},\vec{Q},\eta,\alpha,s}=\cre{a}{\vec{k}-\vec{G},\vec{Q}+\vec{G},\eta,\alpha,s},
\end{equation}
for any $\vec{G} \in \mathcal{Q}_0$.

Within each valley $\eta$, we introduce the first-quantized momentum space intra-layer Hamiltonian $h^{D,\eta}_{\vec{Q}} \left(\vec{k}\right)$ defined in sublattice space by
\begin{equation}
	\begin{split}
		h^{D,+}_{\vec{Q}} \left( \vec{k} \right) &= v_F \left( \vec{k} - \vec{Q} \right) \cdot \boldsymbol{\sigma}, \\
		h^{D,-}_{\vec{Q}} \left( \vec{k} \right) &= \sigma_x h^{D,+}_{-\vec{Q}} \left( -\vec{k} \right) \sigma_x,
	\end{split}
\end{equation}
where $v_F$ represents the Fermi velocity of the single graphene layer. $h^{D,\eta}_{\vec{Q}} \left(\vec{k}\right)$ represents a Dirac cone Hamiltonian that has been folded inside the first MBZ ($\vec{k} \in \mathrm{MBZ}$). In this paper, we employ dimensionless units, akin to the momentum and energy rescaling relation defined in \cref{eqn:nonDimRescale} of \cref{sec:singlehamiltonian:hamiltonian}, namely
\begin{equation}
	\vec{k} \to \frac{\vec{k}}{k_{\theta}}, \qquad
	E \to \frac{E}{v_F k_{\theta}}.
\end{equation}
We also define the first-quantized Hamiltonian $h^{I,\eta}_{\vec{Q},\vec{Q}'}$ describing the inter-layer tunneling between two adjacent graphene sheets as 
\begin{equation}
	\begin{split}
		h^{I,+}_{\vec{Q},\vec{Q}'} \left( \vec{k} \right) &= \sum_{j=1}^3 T_j \delta_{\vec{Q},\vec{Q}'+\vec{q}_j}, \\
		h^{I,-}_{\vec{Q},\vec{Q}'} \left( \vec{k} \right) &= \sigma_x h^{I,-}_{-\vec{Q},-\vec{Q}'} \left( -\vec{k} \right) \sigma_x,
	\end{split}
\end{equation}
where the tunneling matrices $T_j$ are given by
\begin{equation}
	\label{si:eqn:tunnelingMatrices}
	T_j = w_0 \sigma_0 + w_1 \left[ \sigma_{x} \cos \frac{2 \pi \left( j-1 \right)}{3} +  \sigma_{y} \sin \frac{2 \pi \left( j-1 \right)}{3} \right].
\end{equation}
Here $\sigma_0$ and $\boldsymbol{\sigma} = \left( \sigma_x, \sigma_y \right)$ represent the $2 \times 2$ identity matrix and Pauli matrices in the sublattice space, while $w_0 \geq 0$ and $w_1 \geq 0$ are the interlayer hoppings at the AA and AB stacking centers of two consecutive graphene sheets, respectively. Generically, in realistic systems $w_0 < w_1$ due to lattice relaxation and corrugation effects~\cite{UCH14,WIJ15,DAI16,JAI16,SON20b}. Note that $h^{I,\eta}_{\vec{Q},\vec{Q}'}$ vanishes unless $\vec{Q}$ and $\vec{Q}'$ belong to different shifted momentum lattices. We can now write the single-particle Hamiltonian for TSTG using the low-energy operators
\begin{align}
	\label{si:eqn:singlePart1}
	\hat{H}_{0} &= \sum_{\vec{k} \in \text{MBZ}} \sum_{\eta, \alpha, \beta, s} 
	\left( \sum_{l \in \setLayer} \sum_{\vec{Q} \in \mathcal{Q}_{\eta}} \left[ h^{D, \eta}_{\vec{Q}} \left( \vec{k} \right) \right]_{\alpha \beta} \cre{a}{\vec{k},\vec{Q},\eta,\alpha,s,l} \des{a}{\vec{k},\vec{Q},\eta,\beta,s,l}  
		+ \sum_{\vec{Q} \in \mathcal{Q}_{-\eta}} \left[ h^{D,\eta}_{\vec{Q}} \left( \vec{k} \right) \right]_{\alpha \beta} \cre{a}{\vec{k},\vec{Q},\eta,\alpha,s,2} \des{a}{\vec{k},\vec{Q},\eta,\beta,s,2} \nonumber \right. \\
		+& \sum_{l \in \setLayer} \sum_{\substack{\vec{Q} \in \mathcal{Q}_{-\eta} \\ \vec{Q}' \in \mathcal{Q}_{\eta}}} \left[ h^{I,\eta}_{\vec{Q},\vec{Q}'} \right]_{\alpha \beta} \cre{a}{\vec{k},\vec{Q},\eta,\alpha,s,2} \des{a}{\vec{k},\vec{Q}',\eta,\beta,s,l} 
		+ \sum_{l \in \setLayer} \sum_{\substack{\vec{Q} \in \mathcal{Q}_{\eta} \\ \vec{Q}' \in \mathcal{Q}_{-\eta}}} \left[ h^{I,\eta}_{\vec{Q},\vec{Q}'}  \right]_{\alpha \beta} \cre{a}{\vec{k},\vec{Q},\eta,\alpha,s,l} \des{a}{\vec{k},\vec{Q}',\eta,\beta,s,2} \nonumber \\
		+& \left.\frac{U}{2} \sum_{l \in \setLayer} (l-2) \,\delta_{\alpha,\beta} \sum_{\vec{Q} \in \mathcal{Q}_{\eta} }  \cre{a}{\vec{k},\vec{Q},\eta,\alpha,s,l} \des{a}{\vec{k},\vec{Q},\eta,\beta,s,l} \right).
\end{align}
In \cref{si:eqn:singlePart1}, we have introduced a perpendicular displacement field, which is equivalent to an onsite potential of $U/2$, $0$, $-U/2$ in the top, middle, and bottom layers, respectively. When $U=0$, the system is symmetric with respect to mirror reflections perpendicular to the $\hat{z}$ axis (to be defined later as a symmetry). Therefore, \cref{si:eqn:singlePart1} can be simplified significantly by working in the mirror-symmetric and mirror-antisymmetric bases. The mirror-symmetric operators are given by
\begin{equation}
	\label{si:eqn:ckQ+}
	\cre{c}{\vec{k},\vec{Q},\eta,\alpha,s}=\begin{cases}
		\frac{1}{\sqrt{2}} \left( \cre{a}{\vec{k},\vec{Q},\eta,\alpha,s,3} + \cre{a}{\vec{k},\vec{Q},\eta,\alpha,s,1}  \right) & \qquad \vec{Q} \in \mathcal{Q}_{\eta} \\
		\cre{a}{\vec{k},\vec{Q},\eta,\alpha,s,2} & \qquad \vec{Q} \in \mathcal{Q}_{-\eta}
	\end{cases},
\end{equation}
while the mirror-antisymmetric ones are given by
\begin{equation}
	\label{si:eqn:bkQ+}
	\cre{b}{\vec{k},\vec{Q},\eta,\alpha,s} = \frac{1}{\sqrt{2}} \left( \cre{a}{\vec{k},\vec{Q},\eta,\alpha,s,3} - \cre{a}{\vec{k},\vec{Q},\eta,\alpha,s,1}  \right)  \qquad \vec{Q} \in \mathcal{Q}_{\eta}.
\end{equation}
The low-energy operators corresponding to the two mirror-symmetry sector inherit the Bloch periodicity property from \cref{si:eqn:shift_prop_a} and obey
\begin{equation}
	\label{si:eqn:shift_prop_bc}
	\begin{split}
		\cre{b}{\vec{k},\vec{Q},\eta,\alpha,s}=\cre{b}{\vec{k}-\vec{G},\vec{Q}+\vec{G},\eta,\alpha,s}, \\
		\cre{c}{\vec{k},\vec{Q},\eta,\alpha,s}=\cre{c}{\vec{k}-\vec{G},\vec{Q}+\vec{G},\eta,\alpha,s},
	\end{split}
\end{equation}
for any $\vec{G} \in \mathcal{Q}_0$. When written in the mirror-symmetry sector basis, the Hamiltonian of TSTG splits into three terms
\begin{equation}
	\label{si:eqn:singlePart2}
		\hat{H}_{0} = \hat{H}_{\mathrm{TBG}} + \hat{H}_{D} + \hat{H}_{U}.
\end{equation}
In \cref{si:eqn:singlePart2}, the mirror-symmetric low-energy operators $\cre{c}{\vec{k},\vec{Q},\eta,\alpha,s}$ give rise to the term
\begin{equation}
	\label{si:eqn:singlePart2:TBG}
		\hat{H}_{\mathrm{TBG}} = \sum_{\vec{k} \in \text{MBZ}} \sum_{\eta, \alpha, \beta, s} \sum_{\vec{Q},\vec{Q}' \in \mathcal{Q}_{\pm}} \left[h^{\left(\eta\right)}_{\vec{Q},\vec{Q}'} \left( \vec{k} \right) \right]_{\alpha \beta} \cre{c}{\vec{k},\vec{Q},\eta,\alpha,s} \des{c}{\vec{k},\vec{Q}',\eta,\beta,s},
\end{equation}
which is similar to the ordinary twisted bilayer graphene (TBG) Hamiltonian, but with a rescaled tunneling amplitude, corresponding to the first-quantized Hamiltonian
\begin{equation}
	h^{\left(\eta\right)}_{\vec{Q},\vec{Q}'} \left( \vec{k} \right) = h^{D,\eta}_{\vec{Q}} \left( \vec{k} \right) \delta_{\vec{Q},\vec{Q}'} + \sqrt{2} h^{I,\eta}_{\vec{Q},\vec{Q}'} .
\end{equation}
At the same time, the mirror-symmetric $\cre{b}{\vec{k},\vec{Q},\eta,\alpha,s}$ operators, which are only defined for $\vec{Q} \in \mathcal{Q}_{\eta}$, give rise to a solitary Dirac cone contribution, folded inside the first MBZ
\begin{equation}
	\label{si:eqn:singlePart2:Dirac}
		\hat{H}_{D} = \sum_{\vec{k} \in \text{MBZ}} \sum_{\eta, \alpha, \beta, s} \sum_{\vec{Q} \in \mathcal{Q}_{\eta}} \left[h^{D,\eta}_{\vec{Q}} \left( \vec{k} \right) \right]_{\alpha \beta} \cre{b}{\vec{k},\vec{Q},\eta,\alpha,s} \des{b}{\vec{k},\vec{Q},\eta,\beta,s},
\end{equation}
while the third term in \cref{si:eqn:singlePart2} couples the TBG-like and the Dirac cone degrees of freedom
\begin{equation}
	\label{si:eqn:singlePart2:Displacement}
		\hat{H}_{U} = \sum_{\vec{k} \in \text{MBZ}} \sum_{\eta, \alpha, s} \sum_{\vec{Q} \in \mathcal{Q}_{\eta}} \frac{U}{2} \left( \cre{b}{\vec{k},\vec{Q},\eta,\alpha,s} \des{c}{\vec{k},\vec{Q},\eta,\alpha,s} + \cre{c}{\vec{k},\vec{Q},\eta,\alpha,s} \des{b}{\vec{k},\vec{Q},\eta,\alpha,s} \right),
\end{equation}
The Dirac cone and the TBG-like single-particle Hamiltonians are independent, unless the mirror symmetry is broken by the addition of a displacement field ($U \neq 0$).

\begin{figure}[!t]
	\twoFigureCBar{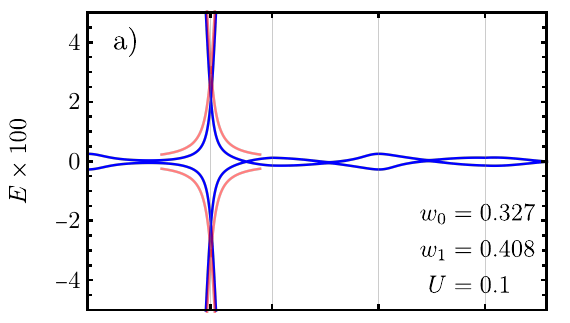}{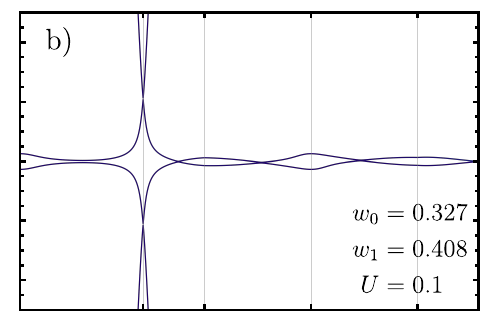}{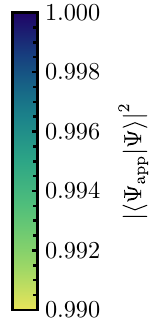}{\textwidth}{si:fig:threeQApproxNC1}\\[-\lineskip]
	\twoFigureCBar{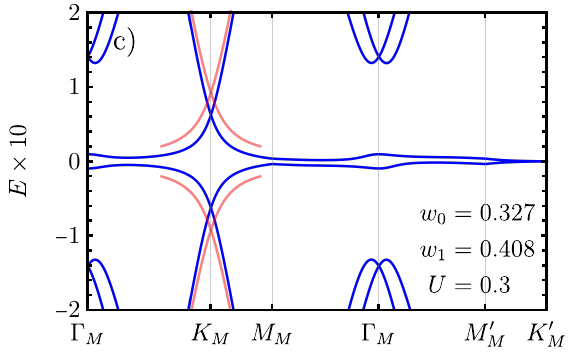}{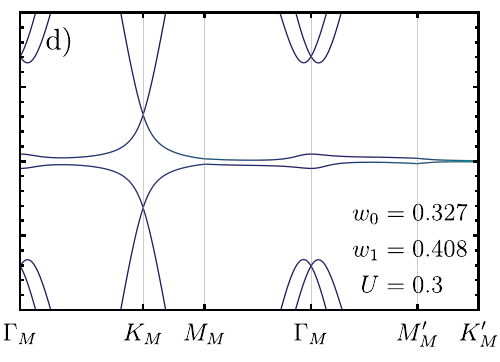}{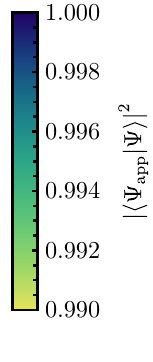}{\textwidth}{si:fig:threeQApproxNC2}
	\caption{The effects of the three-$\vec{Q}$ approximation on the single-particle spectrum of TSTG in the presence of displacement field for valley $\eta = +$ in the non-chiral limit. The blue line in panels (a) and (c) shows the low-energy spectrum of $\hat{H}_0$ (which we dub the unapproximated spectrum) obtained by employing $\abs{\mathcal{Q}_+}=\abs{\mathcal{Q}_-}=330$ points in the expression for $\hat{H}_{\mathrm{TBG}}$, $\hat{H}_D$, and $\hat{H}_U$ for two values of the displacement field $U$ in the non-chiral limit ($w_0/w_1 = 0.8$). In panels (b) and (d), we approximate the spectrum by reducing the number of $\vec{Q}$ points used in $\hat{H}_D$ and $\hat{H}_U$ to just three, as discussed in Appendix \ref{app:single_part_ham:a}. The bands in panels (b) and (d) are colored according to the overlap between the approximated ($\ket{\Psi_{\mathrm{app}}}$) and the unapproximated ($\ket{\Psi}$) single-particle wave functions: the overlap is always higher than $0.99$, thus justifying this approximation. In panels (a) and (c), the red lines denote the energy bands near the $K_M$ point obtained with the tripod model from Appendix \ref{app:approx_single_part:tripod}, which is seen to qualitatively predict the main features of the spectrum. The parameter values are indicated as in inset in the lower-left side of each plot.}
	\label{si:fig:threeQApproxNC}
\end{figure}

\begin{figure}[!t]
	\twoFigureCBar{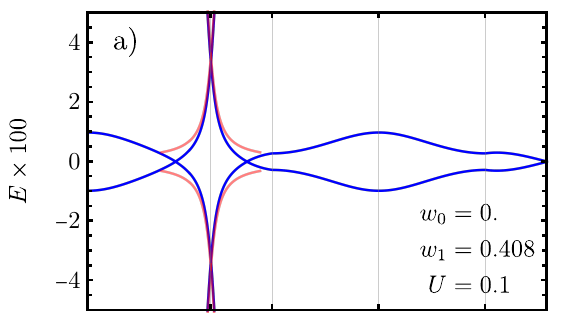}{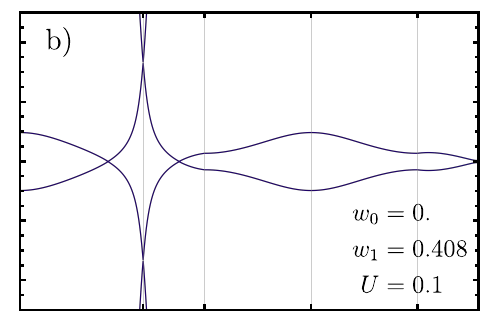}{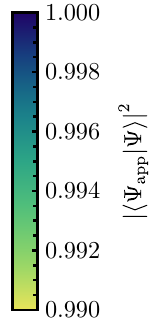}{\textwidth}{si:fig:threeQApproxC1}
	\twoFigureCBar{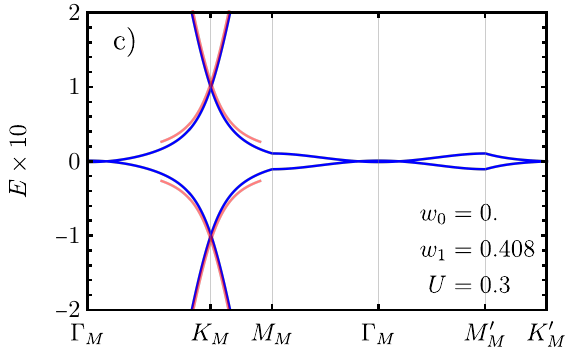}{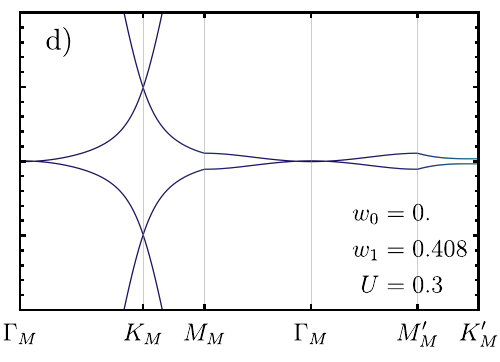}{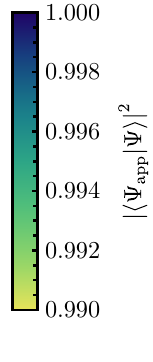}{\textwidth}{si:fig:threeQApproxC2}
	\caption{The effects of the three-$\vec{Q}$ approximation on the single-particle spectrum of TSTG in the presence of displacement field for valley $\eta = +$ in the chiral limit. The meaning of the panels is the same as \cref{si:fig:threeQApproxNC}. The overlap between the approximated and unapproximated single-particle wavefunction is always higher than $0.99$.}
	\label{si:fig:threeQApproxC}
\end{figure}

It is worth noting that in practice, we always take a finite number of lattice points inside the $\mathcal{Q}_{\pm}$ sublattices. As explained in Ref.~\cite{BER20}, we only consider the $\vec{Q}$ points with $\abs{\vec{Q}}$ smaller than a certain cutoff value, thus ensuring that all the discrete symmetries of the system are preserved. In what follows, we will denote the number of $\vec{Q}$ points in lattice $\mathcal{Q}_\eta$ by $\abs{\mathcal{Q}_\eta}$. The influence that the cutoff $\abs{\mathcal{Q}_\eta}$ has on the energy spectrum of the TBG Hamiltonian from \cref{si:eqn:singlePart2:TBG} was extensively discussed in Ref.~\cite{BER20}. In principle, while one could use the same cutoff in defining the Dirac Hamiltonian, a further approximation is justified in this case: we can restrict to considering only \emph{three} $\vec{Q}$ points in the Dirac Hamiltonian expression from \cref{si:eqn:singlePart2:Dirac}. This approximation (which we will henceforth call the \emph{three-$\vec{Q}$ approximation}) can be understood by remembering that we are interested in the low-energy physics of TSTG, which arises from the interplay between the almost-flat (\ie with a small bandwidth $\omega \ll v_F \vec{k}_{\theta}$) bands of TBG and the Dirac cone bands of $\hat{H}_D$. The flat bands of $\hat{H}_{\mathrm{TBG}}$ from \cref{si:eqn:singlePart2:TBG} have essentially zero energy with a small bandwidth $\omega$, hence the only eigenstates which can efficiently perturb the flat band modes of $\hat{H}_{\mathrm{TBG}}$ are the ones which have an energy significantly smaller than one. Since the MBZ forms a hexagon defined by the vertices $\pm \vec{q}_i$ (for $i=1,2,3$), the only possibility for $\abs{h_{\vec{Q}}^{D,\eta}\left( \vec{k} \right)} \ll 1$, with $\vec{k} \in \mathrm{MBZ}$ is for $\vec{Q}$ to be one of the $\eta \vec{q}_i$ points (for $i=1,2,3$) in each valley $\eta$. 

We explore the effects of the three-$\vec{Q}$ approximation on the one-particle energy spectrum in  \cref{si:fig:threeQApproxNC,si:fig:threeQApproxC} for both the non-chiral ($w_0 \neq 0$) and the chiral ($w_0 = 0$) limits, respectively. Taking the case when the same $\mathcal{Q}_{\pm}$ sublattice cutoff in employed for both $\hat{H}_{\mathrm{TBG}}$ and $\hat{H}_D$ as a reference, there is no discernible difference in the spectra when the three-$\vec{Q}$ approximation is employed. 

Moreover, even with $\vec{Q}\in \left\lbrace \eta \vec{q}_i \right\rbrace$, the low energy condition $\abs{h_{\vec{Q}}^{D,\eta}\left( \vec{k} \right)} \ll 1$ is only true for $\abs{\vec{k} - \vec{Q}} \leq \Lambda \ll 1$, where $\vec{k} \in \mathrm{MBZ}$. As depicted in \cref{fig:mbz_qlattice:b}, we will therefore introduce three zones $A^{i}_{\eta}$ (where $i=1,2,3$) inside the first MBZ for each valley $\eta$, which are defined as
\begin{equation}
	\label{si:eqn:Dirac_zones}
	A^{i}_{\eta} = \left\lbrace \vec{k} \in \mathrm{MBZ} \mid \abs{\vec{k} - \vec{\eta q_i}} \leq \Lambda  \right\rbrace.
\end{equation}
Typically, the cutoff $\Lambda$ will be much smaller than $1$, but bigger than the bandwidth $\omega$ of the flat bands of $\hat{H}_{\mathrm{TBG}}$. A physical cutoff is to take $\Lambda$ as the gap between the flat bands and the passive bands of $\hat{H}_{\mathrm{TBG}}$. With these approximations, we can write the Dirac cone Hamiltonian projected into the low-energy degrees of freedom as 
\begin{equation}
	\label{si:eqn:singlePart2:Dirac_approx}
		H_D = \sum_{\eta, \alpha, \beta, s} \sum_{i=1}^3 \sum_{\vec{k} \in A^{i}_{\eta}} \left[h^{D,\eta}_{\eta \vec{q}_i} \left( \vec{k} \right) \right]_{\alpha \beta} \cre{b}{\vec{k},\eta \vec{q}_i,\eta,\alpha,s} \des{b}{\vec{k},\eta \vec{q}_i,\eta,\beta,s}.
\end{equation} 
To emphasize that the Hamiltonian $H_D$ is projected into low-energy modes with the cutoff $\Lambda$, we have omitted the hat to differentiate it from the unprojected Dirac cone Hamiltonian $\hat{H}_D$. 

\subsection{Single-particle eigenstates}
\label{app:single_part_ham:b}

In the absence of a displacement field, the single-particle Hamiltonian $\hat{H}_0$ is a sum of two commuting terms, $\hat{H}_{\mathrm{TBG}}$ and $\hat{H}_D$, which can therefore be individually diagonalized. For this purpose, we introduce the energy band basis, which is defined according to 
\begin{equation}
	\begin{split}
		\cre{c}{\vec{k},n,\eta,s} = \sum_{\vec{Q}\in \mathcal{Q}_{\pm},\alpha} u^{\hat{c}}_{\vec{Q} \alpha; n \eta} \left( \vec{k} \right) \cre{c}{\vec{k},\vec{Q},\eta,\alpha,s},\\
		\cre{b}{\vec{k},n,\eta,s} = \sum_{\vec{Q}\in \mathcal{Q}_{\eta},\alpha} u^{\hat{b}}_{\vec{Q} \alpha; n \eta} \left( \vec{k} \right) \cre{b}{\vec{k},\vec{Q},\eta,\alpha,s},
	\end{split}	
\end{equation}
where $u^{\hat{c}}_{\vec{Q} \alpha; n \eta} \left( \vec{k} \right)$ and $u^{\hat{b}}_{\vec{Q} \alpha; n \eta} \left( \vec{k} \right)$ are the eigenstate wave functions of energy band $n$ of the first quantized single-particle Hamiltonians $h^{\left(\eta\right)}_{\vec{Q},\vec{Q}'} \left( \vec{k} \right)$ and $h^{D,\eta}_{\vec{Q}} \left( \vec{k} \right)$, respectively. For each valley and spin, we shall use the integer $n>0$ to denote the $n$-th conduction band and use the integer $n<0$ to label the $\abs{n}$-th valence band. They obey
\begin{equation}
	\begin{split}
		\sum_{\vec{Q}', \beta} \left[h^{\left(\eta\right)}_{\vec{Q},\vec{Q}'} \left( \vec{k} \right) \right]_{\alpha \beta} u^{\hat{c}}_{\vec{Q}' \beta; n \eta} \left( \vec{k} \right) &= \epsilon^{\hat{c}}_{n,\eta} \left( \vec{k} \right) u^{\hat{c}}_{\vec{Q} \alpha; n \eta} \left( \vec{k} \right),\\
		\sum_{\beta} \left[h^{D,\eta}_{\vec{Q}} \left( \vec{k} \right) \right]_{\alpha \beta} u^{\hat{b}}_{\vec{Q} \beta; n \eta} \left( \vec{k} \right) &= \epsilon^{\hat{b}}_{n,\eta} \left( \vec{k} \right) u^{\hat{b}}_{\vec{Q} \alpha; n \eta} \left( \vec{k} \right),
	\end{split}
\end{equation}
where $\epsilon^{\hat{c}}_{n,\eta} \left( \vec{k} \right)$ and $\epsilon^{\hat{b}}_{n,\eta} \left( \vec{k} \right)$ are the single-particle energies of the eigenstates $u^{\hat{c}}_{\vec{Q} \alpha; n \eta} \left( \vec{k} \right)$ and $u^{\hat{b}}_{\vec{Q} \alpha; n \eta} \left( \vec{k} \right)$, respectively. Owing to the Bloch periodicity property of \cref{si:eqn:shift_prop_bc}, we can generalize the eigenstate wave functions outside the first MBZ using the following embedding relations
\begin{equation}
	\label{si:eqn:embeddingWavF}
	u^{\hat{f}}_{\vec{Q} \alpha; n \eta} \left(\vec{k} + \vec{G}_0\right) = u^{\hat{f}}_{\vec{Q} - \vec{G}_0 \alpha; n \eta} \left( \vec{k} \right),
\end{equation}
ensuring that the energy band basis is defined periodically inside the MBZ, namely
\begin{equation}
	\label{si:eqn:periodicityEnB}
	\cre{f}{\vec{k},n,\eta,s} = \cre{f}{\vec{k}+\vec{G}_0,n,\eta,s},
\end{equation}
for $\cre{f}{} = \cre{b}{},\cre{c}{}$ and any MBZ reciprocal lattice vector $\vec{G}_0$ ($\vec{G}_0 \in \mathcal{Q}_0$).

\section{Symmetries of the single-particle Hamiltonian}\label{app:symmetries}
In this appendix, we extensively discuss the symmetries of the single-particle Hamiltonian from \cref{si:eqn:singlePart2} summarized in \cref{sec:symmetries}. It is instructive to consider the mirror-symmetric $U=0$ case first, as the Hamiltonian splits into two independent terms, namely $\hat{H}_{\mathrm{TBG}}$ and $\hat{H}_{D}$, which correspond respectively to the mirror-symmetric and mirror-antisymmetric sectors. For $\hat{H}_{\mathrm{TBG}}$, the various symmetries have been derived and discussed in Refs.~\cite{SON19, SON20b, BER20a, MOR19} whose notation and conventions we will follow. In addition to the crystalline symmetries, for $\hat{H}_{D}$, we also discuss the emergence of a low-energy effective symmetry, which is incompatible with a crystalline lattice $\mathcal{Q}_0$. 

In the presence of displacement field, $\hat{H}_{0}$ can no longer be split into commuting contributions; the symmetries must be discussed for the entire Hamiltonian. 

\subsection{Symmetries in the $U=0$ case}\label{app:symmetries:a}

\begin{enumerate}
	\item \emph{Discrete symmetries}. Since graphene has zero spin-orbit coupling (SOC), we can define a set of spinless symmetries for TSTG: the spinless unitary discrete symmetries $C_{2z}$, $C_{3z}$, $C_{2x}$, $m_z$, and the spinless anti-unitary time-reversal symmetry $T$. As discussed in \cref{sec:symmetries:a}, the mirror-symmetric term $\hat{H}_{\mathrm{TBG}}$ is symmetric under $C_{2z}$, $C_{3z}$, $C_{2x}$, $m_z$, and $T$, while the mirror-antisymmetric term has only the $C_{2z}$, $C_{3z}$, $m_z$, and $T$ symmetries (\ie it is not symmetric under $C_{2x}$). 
	
We denote the action of a spinless symmetry operator $g$ on the two flavors of fermions as
\begin{equation}
	\label{si:eqn:symmetry_action_bc}
	\begin{split}
		g \cre{c}{\vec{k},\vec{Q},\eta,\alpha,s} g^{-1} &= \sum_{\vec{Q}' \eta' \beta}\left[ D^{\hat{c}} \left( g \right) \right]_{\vec{Q}' \eta' \beta, \vec{Q} \eta \alpha}\cre{c}{g\vec{k},\vec{Q}',\eta',\beta,s},\\
		g \cre{b}{\vec{k},\vec{Q},\eta,\alpha,s} g^{-1} &= \sum_{\vec{Q}' \eta' \beta}\left[ D^{\hat{b}} \left( g \right) \right]_{\vec{Q}' \eta' \beta, \vec{Q} \eta \alpha}\cre{b}{g\vec{k},\vec{Q}',\eta',\beta,s}, 
	\end{split}
\end{equation}
where $D^{\hat{c}}(g)$ and $D^{\hat{b}}(g)$ are the representation matrices of the symmetry operator $g$ in the space of indices $\left\lbrace \vec{Q},\eta,\alpha \right\rbrace$ for each fermion operator. We denote $g \vec{k}$ to be the momentum obtained after acting the transformation $g$ on momentum $\vec{k}$. In particular, $C_{2z} \vec{k} = T \vec{k}=-\vec{k}$. The representation matrices for the discrete symmetries of TSTG are given by~\cite{SON19, SON20b, BER20a}
\begin{align}
	\left[D \left(C_{2z}\right)\right]_{\vec{Q}' \eta' \beta, \vec{Q} \eta \alpha} &= \delta_{\vec{Q}', -\vec{Q}} \delta_{\eta',-\eta} \left(\sigma_x\right)_{\beta\alpha}, \label{si:eqn:repSym_C2z} \\
	\left[D \left(C_{3z}\right)\right]_{\vec{Q}' \eta' \beta, \vec{Q} \eta \alpha} &= \delta_{\vec{Q}', C_{3z} \vec{Q}} \delta_{\eta', \eta} \left(e^{i\eta  \frac{2\pi}{3}\sigma_{z}} \right)_{\beta\alpha}, \label{si:eqn:repSym_C3z} \\ 
	\left[D \left(T\right)\right]_{\vec{Q}' \eta' \beta, \vec{Q} \eta \alpha} &= \delta_{\vec{Q}', -\vec{Q}} \delta_{\eta',-\eta} \delta_{\beta,\alpha}, \label{si:eqn:repSym_T} \\
	\left[D^{\hat{c}} \left(C_{2x}\right)\right]_{\vec{Q}' \eta' \beta, \vec{Q} \eta \alpha} &= \delta_{\vec{Q}', C_{2x} \vec{Q}} \delta_{\eta', \eta} \left(\sigma_x\right)_{\beta\alpha},  \label{si:eqn:repSym_C2x}
\end{align}
where $D(g)$ stands for both $D^{\hat{c}}(g)$ and $D^{\hat{b}}(g)$. The representation matrices for the mirror $m_z$ symmetry are different for the two fermion flavors
\begin{equation}
	\label{si:eqn:repSym_mz}
	\left[D^{\hat{c}} \left(m_z\right)\right]_{\vec{Q}' \eta' \beta, \vec{Q} \eta \alpha} = \delta_{\vec{Q}', \vec{Q}} \delta_{\eta', \eta} \delta_{\beta,\alpha},	\qquad
	\left[D^{\hat{b}} \left(m_z\right)\right]_{\vec{Q}' \eta' \beta, \vec{Q} \eta \alpha} = -\delta_{\vec{Q}', \vec{Q}} \delta_{\eta', \eta} \delta_{\beta,\alpha}.
\end{equation}
In particular, the combined symmetry $C_{2z}T$ does not change $\vec{k}$ ($C_{2z}T\vec{k}=\vec{k}$) and has the representation matrix
\begin{equation}
	\label{si:eqn:action_c2t}
	\left[ D(C_{2z} T) \right]_{\vec{Q}' \eta' \beta, \vec{Q} \eta \alpha}= \left[ D(C_{2z}) D(T) \right]_{\vec{Q}' \eta' \beta, \vec{Q} \eta \alpha} = \delta_{\vec{Q}',\vec{Q}} \delta_{\eta',\eta} \left( \sigma_x \right)_{\beta\alpha}. 
\end{equation}
Note that the $C_{2x}$ transformation exchanges the two $\mathcal{Q}_\pm$ sublattices, \ie it maps $\vec{Q} \in \mathcal{Q}_{\pm}$ to $C_{2x} \vec{Q} \in \mathcal{Q}_{\mp}$, without exchanging the valleys. Because the mirror-antisymmetric operators $\cre{b}{\vec{k},\vec{Q},\eta,\alpha,s}$ at a given valley $\eta$ only exist for $\vec{Q} \in \mathcal{Q}_{\eta}$, the action of $C_{2x}$ on them can not be defined. Therefore, $C_{2x}$ is not a symmetry of $\hat{H}_{D}$.  

\item  \emph{$\left[\UN{2} \times \UN{2} \right]_{\hat{c}} \times \left[\UN{2} \times \UN{2} \right]_{\hat{b}}$ spin-charge rotation symmetry}. In the single-particle Hamiltonian of TSTG for $U=0$, the two valleys $\eta=\pm$ and the two fermion flavors ($\cre{b}{}$ and $\cre{c}{}$) are decoupled. At the same time, monolayer graphene has zero (negligible) SOC, implying that in each valley, the $\SUN{2}$ spin for each fermion flavor can be freely rotated. Together with the charge $\UN{1}$ symmetry of each valley-flavor, this leads to a global $\left[\UN{2} \times \UN{2} \right]_{\hat{c}} \times \left[\UN{2} \times \UN{2} \right]_{\hat{b}}$  symmetry. The 16 generators of this symmetry are given by
\begin{align}
	\hat{S}_{\hat{c}}^{a b}&= \sum_{\substack{\alpha,\eta \\ s,s'}} \sum_{\substack{\vec{k} \in \text{MBZ} \\ \vec{Q} \in \mathcal{Q}_{0}}} (\tau^a)_{\eta\eta}(s^b)_{ss'} \cre{c}{\vec{k},\vec{Q},\eta,\alpha,s} \des{c}{\vec{k},\vec{Q},\eta,\alpha,s'}, \label{si:eqn:u2_generators_c} \\
	\hat{S}_{\hat{b}}^{a b}&= \sum_{\substack{\alpha,\eta \\ s,s'}} \sum_{\substack{\vec{k} \in \text{MBZ} \\ \vec{Q} \in \mathcal{Q}_{\eta}}} (\tau^a)_{\eta\eta}(s^b)_{ss'} \cre{b}{\vec{k},\vec{Q},\eta,\alpha,s} \des{b}{\vec{k},\vec{Q},\eta,\alpha,s'}, \label{si:eqn:u2_generators_b}
\end{align}
where $a=0,z$ and $b=0,x,y,z$. We have defined $\tau^a$ and $s^a$ ($a=0,x,y,z$) to be the $2\times2$ identity and Pauli matrices in the valley and spin spaces, respectively. 

\item \emph{Particle-hole transformations}. In addition to the above symmetries, one can also define a unitary particle-hole (PH) transformation $P$~\cite{SON19}. The action of the unitary PH transformation on the mirror-symmetric fermions is given by
\begin{equation}
	P \cre{c}{\vec{k},\vec{Q},\eta,\alpha,s} P^{-1} = \sum_{\vec{Q}' \eta' \beta}\left[ D^{\hat{c}} \left( P \right) \right]_{\vec{Q}' \eta' \beta, \vec{Q} \eta \alpha}\cre{c}{-\vec{k},\vec{Q}',\eta',\beta,s},
\end{equation}
with the representation matrix 
\begin{equation}
	\label{si:eqn:repSym_P}
	\left[D^{\hat{c}} \left( P \right) \right]_{\vec{Q}' \eta' \beta, \vec{Q} \eta \alpha} = \delta_{\vec{Q}', -\vec{Q}} \delta_{\eta', \eta} \delta_{\beta,\alpha} \zeta_{\vec{Q}},
\end{equation}
where 
\begin{equation}
	\label{si:eqn:zetaDef}
	\zeta_{\vec{Q}} = \begin{cases}
		+1 & \vec{Q} \in \mathcal{Q}_+ \\
		-1 & \vec{Q} \in \mathcal{Q}_-
	\end{cases}.
\end{equation} 
Note that $P$ transforms creation operators to creation operators (rather than annihilation operators), and exchanges the two $\mathcal{Q}_\pm$ sublattices, mapping $\vec{Q} \in \mathcal{Q}_\pm$ to $-\vec{Q} \in \mathcal{Q}_\mp$. In addition, the PH transformation obeys
\begin{equation}
	P^2=-1,\qquad
	\left[ P, C_{3z} \right] = 0,\qquad
	\left\lbrace P, C_{2x} \right\rbrace = 0,\qquad
	\left\lbrace P, C_{2z} \right\rbrace = 0,\qquad
	\left\lbrace P, T \right\rbrace = 0,\qquad
	\left[P, m_z \right]=0.
\end{equation}
 
The PH transformation anticommutes with $\hat{H}_{\mathrm{TBG}}$ defined in \cref{si:eqn:singlePart2:TBG}
\begin{equation}
	\left\lbrace P, \hat{H}_{\mathrm{TBG}} \right\rbrace = 0
\end{equation}
and hence does not represent a commuting symmetry of the Hamiltonian, but rather a relation between the positive and negative spectra of $\hat{H}_{\mathrm{TBG}}$. At the same time, because $P$ exchanges the two $\mathcal{Q}_\pm$ sublattices, without exchanging the valleys, its action cannot be defined on the $\cre{b}{}$ operators, and therefore, $\hat{H}_{D}$ is not PH-symmetric. Nevertheless, one can still introduce a combined transformation $C_{2x} P$, whose action
\begin{equation}
	C_{2x} P \cre{f}{\vec{k},\vec{Q},\eta,\alpha,s} \left( C_{2x} P \right)^{-1} = \sum_{\vec{Q}' \eta' \beta}\left[ D \left( C_{2x} P \right) \right]_{\vec{Q}' \eta' \beta, \vec{Q} \eta \alpha}\cre{f}{-C_{2x}\vec{k},\vec{Q}',\eta',\beta,s},
\end{equation}
can be defined for both fermion flavors $\cre{f}{} = \cre{b}{},\cre{c}{}$. Its representation matrix is the same for both symmetry sectors
\begin{equation}
	\left[D \left(C_{2x} P \right) \right]_{\vec{Q}' \eta' \beta, \vec{Q} \eta \alpha} = \delta_{\vec{Q}', - C_{2x} \vec{Q}} \delta_{\eta', \eta} \left(\sigma_x\right)_{\beta\alpha} \zeta_{\vec{Q}}	
\end{equation}
and is consistent with the representation matrices for the mirror-symmetric fermions of both $P$ and $C_{2x}$, defined in \cref{si:eqn:repSym_P,si:eqn:repSym_C2x}, respectively. The transformation $C_{2x} P$ represents an anticommuting symmetry of both $\hat{H}_{\mathrm{TBG}}$ and $\hat{H}_{D}$
\begin{equation}
	\left\lbrace C_{2x} P, \hat{H}_{\mathrm{TBG}} \right\rbrace = \left\lbrace C_{2x} P, \hat{H}_{D} \right\rbrace = 0
\end{equation}   
and satisfies $\left( C_{2x} P \right)^2 = 1$.

\item \emph{Chiral symmetries}. Besides PH, we can define two other anticommuting transformations $C$ and $C'$, which are known as the first and second chiral transformations~\cite{TAR19,BER20a}, respectively. Their action on the $\cre{b}{}$ and $\cre{c}{}$ operators is given by
\begin{equation}
	X \cre{f}{\vec{k},\vec{Q},\eta,\alpha,s} X^{-1} = \sum_{\vec{Q}' \eta' \beta}\left[ D \left( X \right) \right]_{\vec{Q}' \eta' \beta, \vec{Q} \eta \alpha}\cre{f}{\vec{k},\vec{Q}',\eta',\beta,s},
\end{equation}
where $\cre{f}{}=\cre{b}{},\cre{c}{}$ and $X = C, C'$. The representation matrices for the two chiral operators are given by
\begin{equation}
	\label{si:eqn:rep_mat_chiral}
	\begin{split}	
		\left[D \left( C \right) \right]_{\vec{Q}' \eta' \beta, \vec{Q} \eta \alpha} &= \delta_{\vec{Q}', \vec{Q}} \delta_{\eta', \eta} \left( \sigma_z \right)_{\beta\alpha}, \\	
		\left[D \left( C' \right) \right]_{\vec{Q}' \eta' \beta, \vec{Q} \eta \alpha} &= \delta_{\vec{Q}', \vec{Q}} \delta_{\eta', \eta} \left( \sigma_z \right)_{\beta\alpha} \zeta_{\vec{Q}},
	\end{split}
\end{equation}
where $\zeta_{\vec{Q}}$ is defined in \cref{si:eqn:zetaDef}. Similarly to the PH transformation, the chiral transformation reflects a relation between the positive and negative spectra of the Hamiltonian. For the TBG-like contribution, $\hat{H}_{\mathrm{TBG}}$ is symmetric under the chiral transformation only for specific parameter choices, namely
\begin{equation}
	\begin{split}
		\left\lbrace C, \hat{H}_{\mathrm{TBG}} \right\rbrace &= 0 \quad \text{if} \quad w_0 = 0, \\
		\left\lbrace C', \hat{H}_{\mathrm{TBG}} \right\rbrace &= 0 \quad \text{if} \quad w_1 = 0.
	\end{split}
\end{equation}  
The mirror-antisymmeric sector Hamiltonian $\hat{H}_{D}$ always has the chiral symmetry
\begin{equation}
	\left\lbrace C, \hat{H}_{D} \right\rbrace = \left\lbrace C', \hat{H}_{D} \right\rbrace = 0.
\end{equation}
Note however that in the case of $\hat{H}_D$, the first and second chiral transformations are equivalent up to a valley-charge rotation. To see this, consider the representation matrix for $C C'$ which is given by $\left[D \left(C C' \right) \right]_{\vec{Q}' \eta' \beta, \vec{Q} \eta \alpha} = \delta_{\vec{Q}', \vec{Q}} \delta_{\eta', \eta} \delta_{\beta,\alpha} \zeta_{\vec{Q}}$. Since $\hat{H}_D$ is defined in only one $\mathcal{Q}_{\pm}$ sublattice for each valley, $\left[D \left(C C' \right) \right]_{\vec{Q}' \eta' \beta, \vec{Q} \eta \alpha} = \delta_{\vec{Q}', \vec{Q}} \left( \tau_z \right)_{\eta' \eta} \delta_{\beta,\alpha}$, when acting on the $\cre{b}{}$ operators, implying that the two transformations are indeed equivalent up to a valley-charge rotation.

The two chiral symmetry operators satisfy
\begin{equation}
	\begin{split}
		C^2=1,\qquad
		\left\lbrace C, C_{2z} \right\rbrace = 0,\quad
		\left[C, T \right] = 0,\quad
		&\left[C, P \right]=0, \quad
		\left\lbrace C, C_{2z}T \right\rbrace=0,\quad
		\left\lbrace C,C_{2z}P \right\rbrace=0,\\
		C'^2=1, \quad
		\left[C', C_{2z} \right] = 0,\quad
		\left\lbrace C', T \right\rbrace = 0,\quad
		&\left\lbrace C', P \right\rbrace=0, \quad
		\left\lbrace C', C_{2z}T \right\rbrace=0,\quad
		\left\lbrace C',C_{2z}P \right\rbrace=0,
	\end{split}
\end{equation}
as well as $\left[C',C \right]=0$.

\item \emph{Effective low-energy symmetry $L$}. In this paper, we will be primarily interested in the low energy physics of $\hat{H}_{D}$. Consider therefore a simple $h \left( \delta \vec{k} \right) = \delta \vec{k} \pauliVec$ Dirac Hamiltonian. Letting $\delta \vec{k} = \vec{k} - \vec{q_1}$, we see that $h \left( \delta \vec{k} \right)$ is exactly equivalent to $h^{D,+}_{ \vec{q}_1} \left( \vec{k} \right)$. The Hamiltonian $h \left( \delta \vec{k} \right)$ has three distinct ``symmetries''
\begin{align}
	h \left( \delta \vec{k} \right) &= - h \left( - \delta \vec{k} \right), \label{si:eqn:dirac_simple_sym1} \\
	\sigma_z h \left( \delta \vec{k} \right) \sigma_z^{-1} &= -h \left( \delta \vec{k} \right), \label{si:eqn:dirac_simple_sym2} \\
	\sigma_x h \left( \delta \vec{k} \right) \sigma_x^{-1} &= h^{*} \left( \delta \vec{k} \right). \label{si:eqn:dirac_simple_sym3}
\end{align}
We first note that \cref{si:eqn:dirac_simple_sym2} is equivalent to the first chiral symmetry of $\hat{H}_{D}$ (given by the operator $C$), while \cref{si:eqn:dirac_simple_sym3} is equivalent with the $C_{2z}T$ symmetry of $\hat{H}_{D}$. \Cref{si:eqn:dirac_simple_sym1} however, represents a new emerging symmetry of $\hat{H}_{D}$ which we will discuss below. 

In \cref{si:fig:threeQApproxNC,si:fig:threeQApproxC} of Appendix \ref{app:single_part_ham:a}, we saw that an excellent approximation for $\hat{H}_{D}$ in the low-energy limit is given by \cref{si:eqn:singlePart2:Dirac_approx}, where for the $\cre{b}{\vec{k},\vec{Q},\eta,\alpha,s}$ operators we have considered only three $\vec{Q}$ points and the nearby $\vec{k}$ points in the MBZ. The Bloch periodicity property from \cref{si:eqn:shift_prop_bc} allows us to recast the projected Dirac cone Hamiltonian $H_{D}$ into a slightly simpler, albeit less symmetric form
\begin{equation}
	\label{si:eqn:singlePart2:Dirac_approx_nonSym}
	H_D = \sum_{\eta,\alpha,\beta,s} \sum_{\substack{\vec{k} \\ \abs{\vec{k} - \eta \vec{q}_1} \leq \Lambda}} \left[h^{D,\eta}_{\eta \vec{q}_1} \left( \vec{k} \right) \right]_{\alpha \beta} \cre{b}{\vec{k},\eta \vec{q}_1,\eta,\alpha,s} \des{b}{\vec{k},\eta \vec{q}_1,\eta,\beta,s}.
\end{equation}
Note that the price we payed for including only \emph{one} $\vec{Q}$ point is that $\vec{k}$ now takes values outside the first MBZ, unlike the displacement field Hamiltonian $\hat{H}_{U}$ given in \cref{si:eqn:singlePart2:Displacement}, which is defined inside the first MBZ.

We can now introduce the operator $L$ which implements the emerging low-energy symmetry of the Dirac Hamiltonian corresponding to \cref{si:eqn:dirac_simple_sym1}. Its action is only specified on the $\cre{b}{\delta \vec{k} + \eta \vec{q}_1, \eta \vec{q}_1, \eta, \alpha, s}$ operators from \cref{si:eqn:singlePart2:Dirac_approx_nonSym}, for $\abs{\delta \vec{k}} \leq \Lambda$ and can be written as 
\begin{equation}
	\label{si:eqn:L_operator_action}
	L \cre{b}{\delta \vec{k} + \eta \vec{q}_1, \vec{Q},\eta,\alpha,s} L^{-1} = \sum_{\vec{Q}' \eta' \beta} \left[D \left(L \right)\right]_{\vec{Q}' \eta' \beta, \vec{Q} \eta \alpha} \cre{b}{-\delta \vec{k} + \eta' \vec{q}_1, \vec{Q}',\eta',\beta,s},
\end{equation}
where the representation matrix is given by
\begin{equation}
	\left[D \left( L \right) \right]_{\vec{Q}' \eta' \beta, \vec{Q} \eta \alpha} = \delta_{\vec{Q}', \vec{Q}} \delta_{\eta', \eta} \delta_{\beta,\alpha}.
\end{equation}
We stress the fact that the action of this operator is only defined for $\cre{b}{\delta \vec{k} + \eta \vec{q}_1, \vec{Q},\eta,\alpha,s}$, where $\vec{Q} = \eta \vec{q}_1$, as shown schematically in \cref{si:fig:Laction:a}. If instead we chose to formulate our problem in terms of the  $\cre{b}{}$ operators of \cref{si:eqn:singlePart2:Dirac_approx}, $\delta \vec{k} + \eta \vec{q}_1 $ and $ - \delta \vec{k} + \eta \vec{q}_1$ need to be brought in the first MBZ by using \cref{si:eqn:shift_prop_bc}. This however results in a more complicated, yet equivalent form of the action of $L$ on the $\cre{b}{}$ operators, which, for completeness, we include below. Defining $\Theta$ to be the angle between $\delta \vec{k}$ and $\eta \vec{q}_1$ measured in the clockwise direction, we must have
\begin{align}
	L \cre{b}{\delta \vec{k} + \eta\vec{q}_2,\eta \vec{q}_2,\eta,\alpha,s} L^{-1} &= \cre{b}{-\delta \vec{k} + \eta\vec{q}_1, \eta\vec{q}_1,\eta,\alpha,s} \qquad 0 \leq \Theta < \pi/3, \nonumber \\
	L \cre{b}{\delta \vec{k} + \eta\vec{q}_2,\eta \vec{q}_2,\eta,\alpha,s} L^{-1} &= \cre{b}{-\delta \vec{k} + \eta\vec{q}_3, \eta\vec{q}_3,\eta,\alpha,s} \qquad \pi/3 \leq \Theta < 2\pi/3, \nonumber \\
	L \cre{b}{\delta \vec{k} + \eta\vec{q}_1,\eta \vec{q}_1,\eta,\alpha,s} L^{-1} &= \cre{b}{-\delta \vec{k} + \eta\vec{q}_3, \eta\vec{q}_3,\eta,\alpha,s} \qquad 2\pi/3 \leq \Theta < \pi, \nonumber \\
	L \cre{b}{\delta \vec{k} + \eta\vec{q}_1,\eta \vec{q}_1,\eta,\alpha,s} L^{-1} &= \cre{b}{-\delta \vec{k} + \eta\vec{q}_2, \eta\vec{q}_2,\eta,\alpha,s} \qquad \pi \leq \Theta < 4\pi/3, \nonumber \\
	L \cre{b}{\delta \vec{k} + \eta\vec{q}_3,\eta \vec{q}_3,\eta,\alpha,s} L^{-1} &= \cre{b}{-\delta \vec{k} + \eta\vec{q}_2, \eta\vec{q}_2,\eta,\alpha,s} \qquad 4\pi/3 \leq \Theta < 5\pi/3, \nonumber \\
	L \cre{b}{\delta \vec{k} + \eta\vec{q}_3,\eta \vec{q}_3,\eta,\alpha,s} L^{-1} &= \cre{b}{-\delta \vec{k} + \eta\vec{q}_1, \eta\vec{q}_1,\eta,\alpha,s} \qquad 5\pi/3 \leq \Theta < 2\pi. \label{si:eqn:L_operator_action_sym}
\end{align}
The transformations defined in \cref{si:eqn:L_operator_action_sym} are also illustrated schematically in \cref{si:fig:Laction:b}.

\begin{figure}[!t]
	\twoFigure{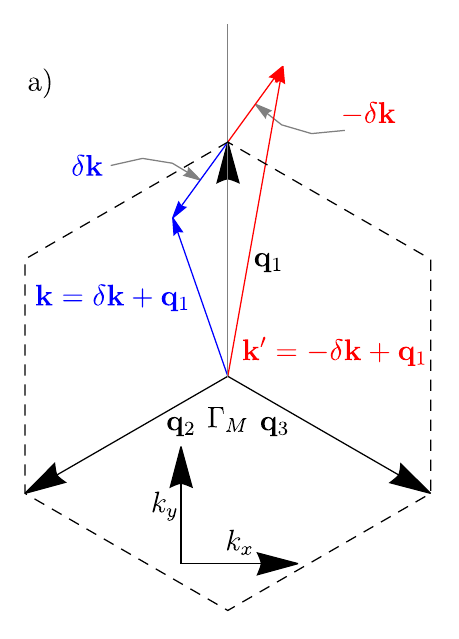}{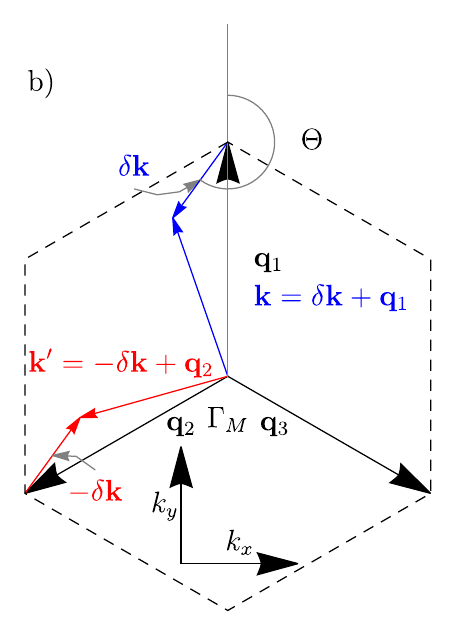}{0.5\columnwidth}{si:fig:Laction}
	\caption{The action of the low-energy symmetry of the projected Dirac cone Hamiltonian $H_D$. In panel (a), we illustrate the action of the $L$ transformation defined in \cref{si:eqn:L_operator_action} on the $\cre{b}{\delta \vec{k} + \vec{q}_1, \vec{q}_1, +, \alpha, s}$ operators from \cref{si:eqn:singlePart2:Dirac_approx_nonSym} (here, we focus on valley $\eta = +$). More specifically, $L$ maps the vector $\vec{k} = \delta \vec{k} + \vec{q}_1$ from inside the first MBZ (shown in dotted lines) to the vector $\vec{k}' = -\delta \vec{k} + \vec{q}_1$, which lies outside the first MBZ. Alternatively, in panel (b) we show that the action of $L$ can alternatively be defined on the $\cre{b}{\vec{k},\vec{Q},\eta,\alpha,s}$ operators from \cref{si:eqn:singlePart2:Dirac_approx}, for which $\vec{k}$ lies inside the first MBZ. According to \cref{si:eqn:L_operator_action_sym}, when the angle $\Theta$ between $\delta \vec{k}$ and $\vec{q}_1$ obeys $\pi \leq \Theta < 4\pi/3$, the operator $L$ maps the momentum $\vec{k} = \delta \vec{k} + \vec{q}_1$ to $\vec{k}' = -\delta \vec{k} + \vec{q}_2$, both of which lie in the first MBZ. We note that the resulting momenta $\vec{k}'$ in panels (a) and (b) are identical, up to a reciprocal moir\'e lattice vector. }
	\label{si:fig:Laction}
\end{figure}

In what follows, we will chose to use the $\cre{b}{}$ operators of \cref{si:eqn:singlePart2:Dirac_approx_nonSym} in discussing gauge-fixing in Appendix \ref{app:gauge:d}, as well as the form factors of the interaction Hamiltonian Appendix \ref{app:interaction:gauge:b}. Note however, that we will always be able to return to the more symmetrical $\cre{b}{}$ of \cref{si:eqn:singlePart2:Dirac_approx} by simply using Bloch periodicity in \cref{si:eqn:shift_prop_bc}.
 
The properties of the $L$ operator can be discussed from the action given in \cref{si:eqn:L_operator_action}. It represents an anticommuting symmetry of the projected Dirac Hamiltonian, obeying $\left\lbrace H_D,L \right \rbrace=0$. In addition, it also satisfies the following relations
\begin{equation}
	L^2=1,\qquad
	\left[ L, C_{2z} \right] = 0,\qquad
	\left[ L, T \right] = 0,\qquad
	\left[ L, C \right]=0.
\end{equation}
Finally, we note that $L$ maps $\delta \vec{k} + \eta \vec{q}_1$ to $-\delta \vec{k} + \eta \vec{q}_1$, two momentum points which are not related by any crystalline symmetry. Therefore, $L$ represents an emerging effective low-energy symmetry of $\hat{H}_D$ or of any low-energy Hamiltonian with a $\pi$ Berry phase (\ie which contains only odd terms in the low-energy momentum $\delta \vec{k}$). Moreover, the $L$ operator can be combined with the crystalline $C_{2z}$ symmetry to afford a $\delta \vec{k}$-preserving (non-crystalline) transformation whose action is defined on the operators $\cre{b}{\delta \vec{k} + \eta \vec{q}_1, \vec{Q},\eta,\alpha,s}$ for $\abs{\delta \vec{k}} \leq \Lambda$ and $\vec{Q} = \eta \vec{q}_1$ by
\begin{equation}
	\label{si:eqn:C2zL_operator_action}
	C_{2z} L  \cre{b}{\delta \vec{k} + \eta \vec{q}_1, \vec{Q},\eta,\alpha,s} \left( C_{2z} L \right)^{-1} = \sum_{\vec{Q}' \eta' \beta} \left[D \left(L \right)\right]_{\vec{Q}' \eta' \beta, \vec{Q} \eta \alpha} \cre{b}{\delta \vec{k} + \eta' \vec{q}_1, \vec{Q}',\eta',\beta,s},
\end{equation}
where the representation matrix is given by
\begin{equation}
	\left[D \left( C_{2z} L \right) \right]_{\vec{Q}' \eta' \beta, \vec{Q} \eta \alpha} = \delta_{\vec{Q}', - \vec{Q}} \delta_{\eta',- \eta} \left( \sigma_x \right)_{\beta\alpha}.
\end{equation}
The operator $C_{2z}L$ is an anticommuting symmetry of $H_D$, $\left\lbrace C_{2z} L, H_D \right\rbrace = 0$, and obeys $\left(C_{2z} L \right)=1$.  
\end{enumerate}

\subsection{Symmetries in the $U \neq 0$ case}\label{app:symmetries:b}

\begin{enumerate}
	\item \emph{Discrete symmetries}. The introduction of a perpendicular displacement field breaks the $C_{2x}$ and $m_z$ symmetries of TSTG. At the same time, $C_{2z}$, $C_{3z}$, and $T$ remain good symmetries of $\hat{H}_0$ and their representation matrices do not change.
	\item  \emph{$\UN{2} \times \UN{2}$ spin-charge rotation symmetry}. Following the introduction of $U \neq 0$, the two valleys of TSTG do remain decoupled. However, the two fermion flavors couple together breaking the initial global $\left[ \UN{2} \times \UN{2} \right]_{\hat{c}} \times \left[ \UN{2} \times \UN{2} \right]_{\hat{b}}$ symmetry into $\UN{2} \times \UN{2}$. The 8 generators of this symmetry are given by
\begin{equation}
	\hat{S}^{a b} = \sum_{\substack{\alpha,\eta \\ s,s'}} \sum_{\vec{k}} (\tau^a)_{\eta\eta}(s^b)_{ss'} \left( \sum_{\vec{Q} \in \mathcal{Q}_0} \cre{c}{\vec{k},\vec{Q},\eta,\alpha,s} \des{c}{\vec{k},\vec{Q},\eta,\alpha,s'} + \sum_{\vec{Q} \in \mathcal{Q}_\eta} \cre{b}{\vec{k},\vec{Q},\eta,\alpha,s} \des{b}{\vec{k},\vec{Q},\eta,\alpha,s'}  \right)
\end{equation}
where $a=0,z$ and $b=0,x,y,z$. In addition, we have defined $\tau^a$ and $s^a$ ($a=0,x,y,z$) to be the $2\times2$ identity and Pauli matrices in the valley and spin spaces, respectively.
	\item \emph{Combined particle-hole transformations}. Compared to the $U=0$ case, the introduction of a perpendicular displacement field breaks both the commuting mirror $m_z$ symmetry, as well as the anticommuting $C_{2x} P$ symmetry. The displacement field Hamiltonian defined in \cref{si:eqn:singlePart2:Displacement} anticommutes  with $m_z$, but commutes with $C_{2x} P$
\begin{equation}
	\left\lbrace m_z, \hat{H}_U \right\rbrace = \left[ C_{2x} P, \hat{H}_U \right] = 0,
\end{equation}
which represents the opposite situation to the TBG and Dirac cone Hamiltonians, for which 
\begin{equation}
\left\lbrace C_{2x} P, \hat{H}_{\mathrm{TBG}} + \hat{H}_{D} \right\rbrace = \left[ m_z, \hat{H}_{\mathrm{TBG}} + \hat{H}_{D} \right].
\end{equation}
However, combining $m_z$ with $C_{2x} P$ affords an anticommuting symmetry of $\hat{H}_0$ in the $U \neq 0$ case, obeying
\begin{equation}
	\left\lbrace m_z C_{2x} P,\hat{H}_0 \right\rbrace=0 .
\end{equation}
The action of $m_z C_{2x} P$ is given by
\begin{equation}
	\begin{split}
		\left( m_z C_{2x} P \right) \cre{c}{\vec{k},\vec{Q},\eta,\alpha,s} \left( m_z C_{2x} P \right)^{-1} &= \sum_{\vec{Q}' \eta' \beta}\left[ D^{\hat{c}} \left( m_z C_{2x} P \right) \right]_{\vec{Q}' \eta' \beta, \vec{Q} \eta \alpha}\cre{c}{-C_{2x}\vec{k},\vec{Q}',\eta',\beta,s},\\
		\left( m_z C_{2x} P \right) \cre{b}{\vec{k},\vec{Q},\eta,\alpha,s} \left( m_z C_{2x} P \right)^{-1} &= \sum_{\vec{Q}' \eta' \beta}\left[ D^{\hat{b}} \left( m_z C_{2x} P \right) \right]_{\vec{Q}' \eta' \beta, \vec{Q} \eta \alpha}\cre{b}{-C_{2x}\vec{k},\vec{Q}',\eta',\beta,s}, 
	\end{split}
\end{equation}
with the representation matrices
\begin{equation}
	\begin{split}
		\left[D^{\hat{c}} \left( m_z C_{2x} P \right) \right]_{\vec{Q}' \eta' \beta, \vec{Q} \eta \alpha} &= \delta_{\vec{Q}', - C_{2x} \vec{Q}} \delta_{\eta', \eta} \left(\sigma_x\right)_{\beta\alpha} \zeta_{\vec{Q}}, \\
		\left[D^{\hat{b}} \left( m_z C_{2x} P \right) \right]_{\vec{Q}' \eta' \beta, \vec{Q} \eta \alpha} &= -\delta_{\vec{Q}', - C_{2x} \vec{Q}} \delta_{\eta', \eta} \left(\sigma_x\right)_{\beta\alpha} \zeta_{\vec{Q}}.
	\end{split}
\end{equation}
where $\zeta_{\vec{Q}} = \pm 1$ for $\vec{Q} \in \mathcal{Q}_\pm$, respectively. As in the $U=0$ case, the combined PH transformation satisfies $\left( m_z C_{2x} P \right)^2=1$.

\item \emph{Chiral symmetries}. The TSTG Hamiltonian $\hat{H}_0$ defined in \cref{si:eqn:singlePart2} has chiral symmetry for the same parameter choices as $\hat{H}_{\mathrm{TBG}}$, namely
\begin{equation}
	\begin{split}
		\left\lbrace m_z C, \hat{H}_{0} \right\rbrace &= 0 \quad \text{if} \quad w_0 = 0, \\
		\left\lbrace m_z C', \hat{H}_{0} \right\rbrace &= 0 \quad \text{if} \quad w_1 = 0.
	\end{split}
\end{equation}  
\end{enumerate}

\section{Gauge-fixing of the single-particle spectrum}\label{app:gauge}
The symmetries presented in Appendix \ref{app:symmetries} yield certain relations between the eigenstates of the single-particle Hamiltonian $\hat{H}_0$. These relations are crucial in deriving the enhanced continuous symmetries of the interacting TSTG Hamiltonian in Appendix \ref{app:fullSym}. We here present the gauge-fixing conventions that will be used throughout the paper: they will prove instrumental for obtaining an explicit form of the projected interaction Hamiltonian in Appendix \ref{app:interaction}.
\subsection{Sewing matrices}\label{app:gauge:a}
To keep the discussion general, we will denote the wave functions $u^{\hat{f}}_{\mathbf{Q} \alpha; n \eta} \left( \vec{k} \right)$ to be the single-particle eigenstates of the Hamiltonian $\hat{H}_{\hat{f}}$, where $\cre{f}{} = \cre{b}{}$ for $\hat{H}_{\hat{f}} = \hat{H}_D$ and $\cre{f}{} = \cre{c}{}$ for $\hat{H}_{\hat{f}} = \hat{H}_{\mathrm{TBG}}$ (see Appendix \ref{app:single_part_ham:b}). Moreover, for the sake of brevity, we will consider the wave function $u^{\hat{f}}_{\mathbf{Q} \alpha; n \eta}\left( \vec{k} \right)$ as a column vector $u^{\hat{f}}_{n\eta}\left( \vec{k} \right)$ in the space of indices $\left\lbrace \vec{Q},\alpha \right\rbrace$. Furthermore, when a representation matrix $D \left(g \right)$ of an operation $g$ defined in \cref{si:eqn:symmetry_action_bc} acts on a wave function $u^{\hat{f}}_{n\eta'}\left( \vec{k} \right)$, we denote the resulting wave function in valley $\eta$ for short as $\sum_{\eta'} \left[D(g)\right]_{\eta\eta'} u^{\hat{f}}_{n\eta'}\left( \vec{k} \right)$, the components of which are given by $\sum_{\vec{Q}'\beta\eta'}\left[D\left(g\right)\right]_{\vec{Q}\alpha\eta,\vec{Q}'\beta\eta'}u^{\hat{f}}_{\vec{Q}'\beta;n\eta'}\left( \vec{k} \right)$. Namely, we suppress the indices $\left\lbrace \vec{Q},\alpha \right\rbrace$ of the representation matrix $D\left(g\right)$ to streamline notation.

When $g$ is a symmetry operator satisfying $\left[ \hat{H}_{\hat{f}},g \right]=0$ (or $\left\lbrace \hat{H}_{\hat{f}}, g \right\rbrace=0$), if $u^{\hat{f}}_{n\eta'}( \vec{k} )$ is an eigenstate wave function at momentum $\vec{k}$, the wave function $\sum_{\eta'} \left[ D^{\hat{f}} \left(g\right)\right]_{\eta\eta'} u^{\hat{f}}_{n\eta'}\left(\vec{k}\right)$ (an additional complex conjugation is needed if $g$ is anti-unitary) must also be an eigenstate wave function at momentum $g\vec{k}$ at the same (or opposite) single-particle energy. This allows us to define a sewing matrix corresponding to the symmetry operator $g$ and the eigenstates $u^{\hat{f}}_{n\eta'}( \vec{k} )$
\begin{equation}
	\label{si:eqn:symmetry_sewing_definition}
	\sum_{\eta'}  \left[ D^{\hat{f}} \left(g\right)\right]_{\eta\eta'} u^{\hat{f}}_{n\eta'}\left(\vec{k}\right) = \sum_{m \eta'} \left[ B^g_{\hat{f}} \left( \vec{k} \right) \right]_{m \eta', n\eta} u^{\hat{f}}_{m\eta'} \left(g \vec{k} \right).
\end{equation}

In the absence of a displacement field the single-particle Hamiltonian can be decoupled into two commuting terms. Therefore, we can gauge-fix the wave functions of $\hat{H}_{\mathrm{TBG}}$ and $\hat{H}_{D}$ separately. In the energy band basis of $\hat{H}_{\mathrm{TBG}}$ and $\hat{H}_D$, a symmetry $g$ acts as 
\begin{equation}
	\label{si:eqn:symmetry_action_sewing}
	\begin{split}
		g \cre{c}{\vec{k}, n, \eta', s} g^{-1} &= \sum_{m,\eta} \left[ B^g_{\hat{c}} \left( \vec{k} \right) \right]_{m \eta, n\eta'}\cre{c}{g\vec{k}, m, \eta, s},\\
		g \cre{b}{\vec{k}, n, \eta', s} g^{-1} &= \sum_{m,\eta} \left[ B^g_{\hat{b}} \left( \vec{k} \right) \right]_{m \eta, n\eta'}\cre{b}{g\vec{k}, m, \eta, s}.
	\end{split}
\end{equation}
\subsection{Gauge-fixing the mirror-symmetric operators}\label{app:gauge:b}
For the mirror-symmetric operators $\cre{c}{\vec{k}, n, \eta, s}$, the gauge-fixing was discussed at length in Refs.~\cite{SON20b, BER20a}. We will only summarize the results here and refer the reader to Refs.~\cite{SON20b, BER20a} for complete proofs. All sewing matrices are closed within each pair of bands $n= \pm n_B$ for any $n_B \geq 1$. Therefore, within each pair of PH-symmetric bands with band indices $n= \pm n_B$, we will use $\zeta^a$ and $\tau^a$ ($a=0,x,y,z$) to denote the identity and Pauli matrices in the energy band $n=\pm n_B$ and the valley spaces, respectively. For all the symmetries that leave $\vec{k}$ invariant, the following $\vec{k}$-independent gauge-fixings will be adopted in this paper
\begin{equation}
	\label{si:eqn:sewing_mats_TBG}
	B_{\hat{c}}^{C_{2z} T} \left( \vec{k} \right) = \zeta^0 \tau^0 \qquad
	B_{\hat{c}}^{C_{2z} P} \left( \vec{k} \right) = \zeta^y \tau^y \qquad
	B_{\hat{c}}^{C} \left( \vec{k} \right) = \zeta^y \tau^z \qquad
	B_{\hat{c}}^{m_z} \left( \vec{k} \right) = \zeta^0 \tau^0
\end{equation}
where the sewing matrix of the chiral symmetry operator is only applicable in the first chiral limit, when $w_0 = 0$. Additionally, we can further fix the relative gauge between wave functions at momenta $\vec{k}$ and $-\vec{k}$ by fixing the sewing matrices of $C_{2z}$ and $P$. 
\begin{equation}
	\label{si:eqn:sewing_mats_TBG_kDep}
	B_{\hat{c}}^{C_{2z}}\left(\vec{k}\right)=\begin{cases}
		\zeta^0\tau^x &\vec{k} \neq \vec{k}_{M_M} \\
		-\zeta^0\tau^x &\vec{k} = \vec{k}_{M_M} \\ 
	\end{cases} \quad
	B_{\hat{c}}^{T}\left(\vec{k}\right)=\begin{cases}
		\zeta^0\tau^x &\vec{k} \neq \vec{k}_{M_M} \\
		-\zeta^0\tau^x &\vec{k} = \vec{k}_{M_M} \\ 
	\end{cases} \quad
	B_{\hat{c}}^{P}\left(\vec{k}\right)=\begin{cases}
		-i\zeta^y\tau^z &\vec{k} \neq \vec{k}_{M_M} \\
		i\zeta^y\tau^z &\vec{k} = \vec{k}_{M_M} \\ 
	\end{cases},
\end{equation}
where $\vec{k}_{M_M}$ denotes one of the three equivalent $M_M$ points in the MBZ (as shown in \cref{fig:mbz_qlattice:b}). The reason for the additional minus sign of the sewing matrix $B_{\hat{c}}^{P}\left(\vec{k}\right)$ at $\vec{k} = \vec{k}_{M_M}$ was explain in Ref.~\cite{BER20a}: the sewing matrix $B_{\hat{c}}^{P}\left(\vec{k}\right)$ must have additional minus signs at an odd number of the four $P$-invariant momenta due to the odd topological winding number of the $n = \pm 1 $ bands protected by $C_{2z}T$. Because the transformations $C_{2z}T$ and $C_{2z}P$ have been gauge-fixed in a $\vec{k}$-independent manner in \cref{si:eqn:sewing_mats_TBG}, the sewing matrices $B_{\hat{c}}^{C_{2z}}$ and $B_{\hat{c}}^{T}$ also have additional minus signs at $\vec{k} = \vec{k}_{M_M}$.

In addition to the gauge-fixing conditions given above, we fix the relative sign between the single-particle wave functions $u^{\hat{c}}_{+, \eta}\left( \vec{k} \right)$ and $u^{\hat{c}}_{-, \eta}\left( \vec{k} \right)$ imposing~\cite{BER20a}
\begin{equation}
	\label{si:eqn:c_continuous}
	\lim_{\vec{q} \rightarrow \mathbf{0}}\abs{u^{\dagger \hat{c}}_{n \eta} \left( \vec{k} + \vec{q} \right) u^{\hat{c}}_{n \eta}\left( \vec{k} \right)- u^{\dagger \hat{c}}_{-n \eta} \left(\vec{k} +\vec{q} \right) u_{-n \eta}^{\hat{c}} \left(\vec{k} \right)}=0.
\end{equation} 

\subsection{The Chern band basis for the mirror-symmetric operators}\label{app:gauge:c}

For the future discussion of the many-body states, we also introduce the Chern band basis~\cite{HEJ20,BUL20,SON20b,BER20a} within the lowest two bands in each valley-spin flavor $\eta = \pm $, as defined in Refs.~\cite{SON20b,BER20a}. Under the gauge-fixings of \cref{si:eqn:c_continuous,si:eqn:sewing_mats_TBG}, the Chern band basis operators are defined by 
\begin{equation}
	\label{si:eqn:chern_band_TBG}
	\cre{d}{\vec{k},e_Y,\eta,s} = \frac{1}{\sqrt{2}} \left( \cre{c}{\vec{k},+1,\eta,s} + i e_Y \cre{c}{\vec{k},-1,\eta,s} \right)
\end{equation}
where $e_Y=\pm1$. As proven in Refs.~\cite{SON20b,BER20a}, the operators $\cre{d}{\vec{k},e_Y,\eta,s}$ for $\vec{k} \in \mathrm{MBZ}$ and fixed $e_Y$, $\eta$, and $s$ corresponds to a Chern band carrying Chern number $e_Y$.

\subsection{Gauge-fixing the mirror-antisymmetric operators}\label{app:gauge:d}
For the mirror-antisymmetric operators $\cre{b}{\vec{k},n,\eta,s}$, we will focus on the $C_{2z}$, $T$, and $C$ symmetries, which are compatible with a crystalline lattice, as well as on the low-energy emerging symmetry $L$, which is not. Note that we will not consider the second chiral transformation $C'$, as it is equivalent to $C$ up to a valley-charge rotation, as shown in Appendix \ref{app:symmetries:a}. For each valley $\eta$, we will restrict ourselves to the projected bands corresponding to $n=\pm 1$, as well as to $\vec{k} \in A_{\eta}^i$, as defined in \cref{si:eqn:Dirac_zones}. Similarly to the discussion surrounding \cref{si:eqn:L_operator_action} however, we will temporarily allow $\vec{k}$ to be outside of the first MBZ, and instead consider the points $\vec{k} = \delta\vec{k} + \eta \vec{q}_1$, where $\abs{\delta \vec{k}} \leq \Lambda$.

The action of the symmetry operation $g$ on the momentum $\vec{k}$ can be defined straight-forwardly in the case of the crystalline symmetries: for $g=C_{2z}$ or $g=T$, $g \vec{k} = -\vec{k}$, while for $g=C$, $g \vec{k} = \vec{k}$. In the case of the emerging symmetry $g=L$, we must have that $g \left( \delta \vec{k} + \eta \vec{q}_1 \right) = - \delta \vec{k} + \eta \vec{q}_1 $, where $\abs{\delta \vec{k}} \leq \Lambda$. In what follows, we will parameterize the sewing matrices according to $\delta\vec{k}$ and adopt the following shorthand notation
\begin{equation}
	\left[B_{\hat{b}}^{g} \left( \delta \vec{k} \right) \right]_{m\eta,n\eta'} \equiv \left[B_{\hat{b}}^{g} \left( \delta \vec{k} + \eta \vec{q}_1 \right) \right]_{m\eta,n\eta'}
\end{equation}
for a given transformation $g$. For example, using the shorthand notation, the action of the $L$ transformation reads
\begin{equation}
	\label{si:eqn:symmetry_action_L}
	L \cre{b}{\delta \vec{k} + \eta' \vec{q}_1, n, \eta', s} L^{-1} = \sum_{m,\eta} \left[ B^g_{\hat{b}} \left( \delta \vec{k} \right) \right]_{m \eta, n\eta'}\cre{b}{-\delta \vec{k} + \eta \vec{q}_1, m, \eta, s}.
\end{equation}
As in the case of the $\cre{c}{\vec{k},n,\eta,s}$ operators, the sewing matrices are closed within the $n=\pm n_B$ bands subspace (for any integer $n_B > 0$). We will therefore use $\zeta^a$ and $\tau^a$ ($a=0,x,y,z$) to denote the identity and Pauli matrices in the energy band $n=\pm n_B$ and the valley spaces, respectively.

We start by fixing the sewing matrices for the $\delta \vec{k}$-preserving transformations in a $\delta \vec{k}$-independent way. The sewing matrix for $C_{2z}T$ can be chosen to be 
\begin{equation}
	\label{si:eqn:gauge_cond_real}
	B_{\hat{b}}^{C_{2z} T} \left( \delta \vec{k} \right) = \zeta^0 \tau^0.
\end{equation}
At the same time, the sewing matrix for the first chiral symmetry must have the form
\begin{equation}
	\left[B_{\hat{b}}^{C} \left( \delta \vec{k} \right) \right]_{m\eta,n\eta'} = \delta_{\eta,\eta'} \delta_{-m,n} e^{i \phi^{C}_{n,\eta'}},
\end{equation}
where $\phi^{C}_{n,\eta'}$ represents a phase dependent on the band and valley. Because $\left\lbrace C,C_{2z} T \right\rbrace = 0$ and $C^2 = 1$, the sewing matrix for $C$ must satisfy the requirements 
\begin{equation}
	B_{\hat{b}}^{C} \left( \delta \vec{k} \right) B_{\hat{c}}^{C_{2z} T} \left( \delta \vec{k} \right) = - B_{\hat{c}}^{C_{2z} T} \left(\delta \vec{k} \right) B_{\hat{b}}^{*C} \left(\delta \vec{k} \right) 
	\quad \text{and} \quad 
	B_{\hat{b}}^{C} \left(\delta \vec{k} \right) B_{\hat{b}}^{C} \left(\delta \vec{k} \right) = 1.
\end{equation} 
We are therefore free to choose $B_{\hat{b}}^{C} \left( \delta \vec{k} \right) = \zeta^y \tau^z$, as in the mirror-symmetric sector. Finally, the transformation $C_{2z}L$ must have a sewing matrix of the form
\begin{equation}
	\left[B_{\hat{b}}^{C_{2z}L} \left( \delta \vec{k} \right) \right]_{m\eta,n\eta'} = \delta_{-\eta,\eta'} \delta_{-m,n} e^{i \phi^{C_{2z}L}_{n,\eta'}},
\end{equation}
which owing to $\left\lbrace C_{2z} L, C \right\rbrace = 0$, $\left[ C_{2z} L, C_{2z} T \right] = 0$, and $\left(C_{2z} L \right)^2 = 1$ must obey 
\begin{align}
	B_{\hat{b}}^{C} \left( \delta \vec{k} \right) B_{\hat{b}}^{C_{2z}L} \left( \delta \vec{k} \right) &= - B_{\hat{b}}^{C_{2z}L} \left( \delta \vec{k} \right) B_{\hat{b}}^{C} \left( \delta \vec{k} \right) \\
	B_{\hat{b}}^{C_{2z}T} \left( \delta \vec{k} \right) B_{\hat{b}}^{* C_{2z}L} \left( \delta \vec{k} \right) &=  B_{\hat{b}}^{C_{2z}L} \left( \delta \vec{k} \right) B_{\hat{b}}^{C_{2z}T} \left( \delta \vec{k} \right) \\
	B_{\hat{b}}^{C_{2z}L} \left( \delta \vec{k} \right) B_{\hat{b}}^{C_{2z}L} \left( \delta \vec{k} \right) &= 1,
\end{align} 
implying that $B_{\hat{b}}^{C_{2z}L} \left( \delta \vec{k} \right) = \zeta^y \tau^y$, as in the mirror-symmetric case for $C_{2z}P$.  

Having fixed the sewing matrices for the $\delta \vec{k}$-preserving transformation, we now consider whether $\delta \vec{k}$-independent sewing matrices can be found for any of the non-$\delta \vec{k}$-preserving symmetries, $C_{2z}$, $T$, and $L$
\begin{equation}
	\begin{split}
		\left[B_{\hat{b}}^{C_{2z}} \left( \delta \vec{k} \right) \right]_{m\eta,n\eta'} = \delta_{-\eta,\eta'} \delta_{m,n} e^{i \phi^{C_{2z}}_{n,\eta'}}, \\
		\left[B_{\hat{b}}^{T} \left( \delta \vec{k} \right) \right]_{m\eta,n\eta'} = \delta_{-\eta,\eta'} \delta_{m,n} e^{i \phi^{T}_{n,\eta'}}, \\
		\left[B_{\hat{b}}^{L} \left( \delta \vec{k} \right) \right]_{m\eta,n\eta'} = \delta_{\eta,\eta'} \delta_{-m,n} e^{i \phi^{L}_{n,\eta'}}. 
	\end{split}
\end{equation}
As we are interested in a $\delta \vec{k}$-independent gauge-fixing, we will temporarily suppress the $\delta \vec{k}$ parameter of the sewing matrices. To be compatible with the gauge-fixing of the sewing matrices $B_{\hat{b}}^{C_{2z}L} \left( \delta \vec{k} \right)$ and $B_{\hat{b}}^{C_{2z}T} \left( \delta \vec{k} \right)$, we must have that
\begin{align}
		B_{\hat{b}}^{C_{2z}} B_{\hat{b}}^{T} &= B_{\hat{b}}^{T} B_{\hat{b}}^{*C_{2z}} = \zeta^0 \tau^0, \label{si:eqn:gauge_cond_c2zt}\\
		B_{\hat{b}}^{C_{2z}} B_{\hat{b}}^{L} &= \zeta^y \tau^y \label{si:eqn:gauge_cond_c2zl}.
\end{align}
We first try to fix the sewing matrices corresponding to $T$ and $C_{2z}$ which satisfy 
\begin{equation}
	\label{si:eqn:gauge_cond_square}
	B_{\hat{b}}^{T} B_{\hat{b}}^{*T} = B_{\hat{b}}^{C_{2z}} B_{\hat{b}}^{C_{2z}} = 1,
\end{equation}
because $T^2 = C_{2z}^2 = 1$. Additionally, the commutation relations $\left[C_{2z} T, T  \right] = \left[C_{2z} T, C_{2z}  \right] = 0$ together with \cref{si:eqn:gauge_cond_real} imply that the sewing matrices of $T$ and $C_{2z}$ must be real. On the other hand, $\left[ T, C  \right] = \left\lbrace C_{2z}, C  \right\rbrace=0$ and so 
\begin{equation}	
	\label{si:eqn:gauge_cond_t_c2_c}	
	B_{\hat{b}}^{T} B_{\hat{b}}^{*C} - B_{\hat{b}}^{C} B_{\hat{b}}^{T} = B_{\hat{b}}^{C_{2z}} B_{\hat{b}}^{C} + B_{\hat{b}}^{C} B_{\hat{b}}^{C_{2z}} = 0.
\end{equation}
The only way \cref{si:eqn:gauge_cond_c2zt,si:eqn:gauge_cond_t_c2_c,si:eqn:gauge_cond_square} can be satisfied is if $B_{\hat{b}}^{C_{2z}} = B_{\hat{b}}^{T} = \zeta^0 \tau^x$. However, this choice is incompatible with the commutation relations $\left[C_{2z} L, T  \right] = \left[C_{2z} L, C_{2z}  \right] = 0$, which would require the $\delta \vec{k}$-independent sewing matrices of $C_{2z}$
and $T$ to commute with $B_{\hat{b}}^{C_{2z}L} = \zeta^y \tau^y$. Therefore, we conclude that the sewing matrices $B_{\hat{b}}^{C_{2z}} \left( \delta \vec{k} \right)$ and $B_{\hat{b}}^{T} \left( \delta \vec{k} \right)$ have to be $\delta \vec{k}$-dependent. 

Alternatively, it would have been impossible to gauge-fix $B_{\hat{b}}^{L} \left( \delta \vec{k} \right)$ first in a $\delta \vec{k}$-independent manner, as that would have implied a $\delta \vec{k}$-independent sewing matrix for $C_{2z}$, leading to another contradiction. Hence, the only symmetry transformations for which we can choose $\delta \vec{k}$-independent sewing matrices are $C_{2z} T$, $C_{2z}L$, $C$, and $m_z$ for which the sewing matrices are given by
\begin{equation}
	\label{si:eqn:sewing_mats_Dirac}
	B_{\hat{b}}^{C_{2z} T} \left( \delta \vec{k} \right) = \zeta^0 \tau^0 \qquad
	B_{\hat{b}}^{C_{2z} L} \left( \delta \vec{k} \right) = \zeta^y \tau^y \qquad
	B_{\hat{b}}^{C} \left( \delta \vec{k} \right) = \zeta^y \tau^z \qquad
	B_{\hat{b}}^{m_z} \left( \delta \vec{k} \right) = - \zeta^0 \tau^0,
\end{equation}
where $C$ is an anticommuting symmetry of the Dirac Hamiltonian for all interlayer hoppings $w_0$, $w_1$.

\section{Approximations of the single-particle spectrum}\label{app:approx_single_part}
This appendix details the various approximations for the single-particle spectrum of TSTG that are mentioned in \cref{sec:singleparticlespectrum}. We first introduce a modified tripod model (similar to the one derived in Ref.~\cite{BIS11}) in order to qualitatively understand the band structure of TSTG near the $K_M$ points for $U \neq 0$. This simplified tripod model paves the road toward deriving various quantitative $\vec{k}$-dependent perturbation schemes for the low-energy single-particle Hamiltonian of TSTG in the presence of displacement field. These will ultimately allow us to obtain the single-particle eigenstates of TSTG analytically in terms of the single-particle eigenstates of $\hat{H}_{\mathrm{TBG}}$.
\subsection{A tripod model of TSTG with displacement field}\label{app:approx_single_part:tripod}
To gain a better understanding of how the TBG Hamiltonian $\hat{H}_{\mathrm{TBG}}$ is influenced by the coupling with the Dirac cone Hamiltonian $\hat{H}_{D}$ in the presence of a displacement field ($U \neq 0$), it is instructive to consider a simple model in conjunction with a series of analytically tractable approximations. For this purpose, we focus on the $\eta=+$ valley near the $K_M$ point (located at $\vec{k} = \vec{q}_1$) and employ a modified tripod model~\cite{BIS11}. This is equivalent to considering only four $\vec{Q}$-points in the $\mathcal{Q}_{\pm}$ sublattices, namely $\mathcal{Q}_{+} = \left\lbrace \vec{q}_1 \right\rbrace$ and $\mathcal{Q}_{-}= \left\lbrace 2\vec{q}_1, \vec{q}_1 + \vec{q}_2, \vec{q}_1 + \vec{q}_3 \right\rbrace$. We write the single-particle eigenstates as 
\begin{equation}
	\ket{\Psi\left( \vec{k} \right)} = \sum_{\alpha} \left[ \sum_{i=0}^3 \left( \psi_{i,\alpha} \left( \vec{k} \right) \cre{c}{\vec{k},\vec{Q}_i,+,\alpha,s} \right) + \psi_{D,\alpha}\left( \vec{k} \right) \cre{b}{\vec{k},\vec{Q}_0,+,\alpha,s} \right] \ket{0},
\end{equation}
where we have denoted $\vec{Q}_i = \vec{q}_1 + \vec{q}_i$ for $i=1,2,3$ and $\vec{Q}_0= \vec{q}_1$. The first-quantized Hamiltonian acting on the ten-dimensional spinor $\Psi^T \left( \vec{k} \right) = \left( \psi_0^T\left( \vec{k} \right), \psi_1^T\left( \vec{k} \right), \psi_2^T\left( \vec{k} \right), \psi_3^T\left( \vec{k} \right), \psi_D^T\left( \vec{k} \right) \right)$ is given by
\begin{equation}
	H_{\mathrm{Tri}} = \begin{pmatrix}
	\smallk \pauliVec & T'_1 & T'_2 & T'_3 & \frac{U}{2} \mathbb{1} \\ 
	T'_1 & \left( \smallk - \vec{q}_1 \right) \pauliVec & \mathbb{0} & \mathbb{0} & \mathbb{0} \\ 
	T'_2 & \mathbb{0} & \left( \smallk - \vec{q}_2 \right) \pauliVec & \mathbb{0} & \mathbb{0} \\ 
	T'_3 & \mathbb{0} & \mathbb{0} & \left( \smallk - \vec{q}_3 \right) \pauliVec & \mathbb{0} \\ 
	\frac{U}{2} \mathbb{1} & \mathbb{0} & \mathbb{0} & \mathbb{0} & \smallk \pauliVec
	\end{pmatrix},
\end{equation}
with $\smallk = \vec{k} - \vec{q}_1$ and $T'_i = T_i \sqrt{2}$ (for $i=1,2,3$). As required from \cref{si:eqn:singlePart2:Displacement}, the displacement field only couples the Dirac cone and the TBG fermions at $\vec{Q}_0$. In what follows, we will suppress the momentum $\vec{k}$ variable of the two- and ten-dimensional spinors. For an eigenstate $\Psi$ of energy $E$, we must have
\begin{align}
	\smallk \pauliVec \psi_0 + \sum_{j=1}^3 T'_j \psi_{j} + \frac{U}{2} \psi_D &= E \psi_0 \label{si:eqb:tripodEval:1} \\
	T'_i \psi_0 + \left( \delta \vec{k} - \vec{q}_i \right) \pauliVec \psi_i &= E \psi_i \label{si:eqb:tripodEval:2} \\
	\frac{U}{2} \psi_0 + \smallk \pauliVec \psi_D &= E \psi_D \label{si:eqb:tripodEval:3} 
\end{align}
Using \cref{si:eqb:tripodEval:2,si:eqb:tripodEval:3}, we can eliminate $\psi_D$ and $\psi_i$ by writing them in terms of $\psi_0$ as
\begin{equation}
	\begin{split}
		\psi_D &= \frac{E + \smallk \pauliVec}{E^2 - \delta k^2} \frac{U}{2} \psi_0 \\
		\psi_i &= \frac{E + \left( \smallk - \vec{q}_i \right) \pauliVec}{E^2 - \left( \smallk - \vec{q}_i \right)^2} T'_i \psi_0 
	\end{split}
\end{equation}
and cast \cref{si:eqb:tripodEval:1} in the form
\begin{equation}
	\label{si:eqn:tripod_model_eval}
	\smallk \pauliVec \psi_0 + \sum_{j=1}^3 T'_j \frac{\left[E + \left( \smallk - \vec{q}_j \right) \pauliVec\right]}{E^2 - \left( \smallk - \vec{q}_j \right)^2} T'_j \psi_0 + \frac{U^2}{4} \frac{E + \smallk \pauliVec}{E^2 - \delta k^2} \psi_0 = E \psi_0.
\end{equation}
We are interested in the low-energy solutions of \cref{si:eqn:tripod_model_eval} near the $K_M$ point and therefore we must have $\abs{\smallk} \sim \abs{E} \ll 1$. The denominator of the second term in \cref{si:eqn:tripod_model_eval} can thus be expanded as 
\begin{equation}
	\frac{1}{E^2 - \left( \smallk - \vec{q}_i \right)^2} = 
	\frac{-1}{1 - \left( E^2 - \delta k^2 + 2 \smallk \cdot \vec{q}_i \right)} = -1 - 2\smallk \cdot \vec{q}_i + \bigOrd{\abs{\smallk}^2},
\end{equation}
leading to
\begin{equation}
	\begin{split}
		\sum_{j=1}^3 T'_j \frac{\left[E + \left( \smallk - \vec{q}_j \right) \pauliVec\right]}{E^2 - \left( \smallk - \vec{q}_j \right)^2} T'_j &= -\sum_{j=1}^3 T'_j \left[E +  \smallk \pauliVec - \vec{q}_j \pauliVec - \left(\vec{q}_i \pauliVec\right) \left( \smallk \cdot \vec{q}_i \right) \right] T'_j + \bigOrd{\abs{\smallk}^2} \\
		&= - 3\left( w_0^{\prime 2} + w_1^{\prime 2} \right) E -3 w_0^{\prime 2}  \smallk \pauliVec + 3 \left( w_0^{\prime 2} - w_1^{\prime 2} \right) \smallk \pauliVec + \bigOrd{\abs{\smallk}^2},
	\end{split}
\end{equation}
where for simplicity we have defined the rescaled hopping parameters $w_0' = w_0 \sqrt{2}$ and $w'_1 = w_1 \sqrt{2}$. This allows us to simplify \cref{si:eqn:tripod_model_eval} into
\begin{equation}
	\begin{split}
		\left[ \left(1 - 3 w_1^{\prime 2} \right) \smallk \pauliVec - E \left(3 w_0^{\prime 2} + 3 w_1^{\prime 2} + 1 \right) +  \frac{U^2}{4} \frac{E + \smallk \pauliVec}{E^2 - \delta k^2} + \bigOrd{\abs{\smallk}^2} \right] \psi_0 = 0, \\
		\left\lbrace \left[ \frac{U^2}{4} + \left(1 - 3 w_1^{\prime 2} \right) \left( E^2 - \delta k^2 \right) \right] \smallk \pauliVec - E \left[ \left(3 w_0^{\prime 2} + 3 w_1^{\prime 2} + 1 \right) \left( E^2 - \delta k^2 \right) - \frac{U^2}{4}\right] + \bigOrd{\abs{\smallk}^4} \right\rbrace \psi_0 = 0 \\
	\end{split}
\end{equation}
where $E^2 - \delta k^2 \neq 0$. This eigenvalue equation has non-trivial solutions for $\psi_{0}$ only if
\begin{equation}
	\left[\frac{U^2}{4} + \left(1 - 3 w_1^{\prime 2} \right) \left( E^2 - \delta k^2 \right) \right] \delta k = \pm E \left[ \left(3 w_0^{\prime 2} + 3 w_1^{\prime 2} + 1 \right) \left( E^2 - \delta k^2 \right) - \frac{U^2}{4} \right],
\end{equation}
which leads to the following four-band dispersion relation:
\begin{equation}
	\label{si:eqn:tripod_bands}
	E = \pm \frac{ \delta k \left(3 w_0^{\prime 2} + 2 \right) \pm \sqrt{9 \delta k^2 \left(w_0^{\prime 2} + 2 w_1^{\prime 2} \right)^2 + U^2 \left(3
   w_0^{\prime 2} + 3 w_1^{\prime 2} +1 \right)} }{2\left( 3 w_0^{\prime 2} + 3 w_1^{\prime 2} + 1 \right)}.
\end{equation}
By expanding \cref{si:eqn:tripod_bands} in $\delta k$ to linear order, one can see that the displacement field splits the two Dirac cones (one stemming from $\hat{H}_{\mathrm{TBG}}$ and the other one, from $\hat{H}_{D}$) away from zero
\begin{equation}
	E = \pm \frac{ \delta k \left(3 w_0^{\prime 2} + 2 \right)}{2\left( 3 w_0^{\prime 2} + 3 w_1^{\prime 2} + 1 \right)} \pm \frac{U}{2\sqrt{ 3 w_0^{\prime 2} + 3 w_1^{\prime 2} +1 }}.
\end{equation}

In \cref{si:fig:threeQApproxNC,si:fig:threeQApproxC}, we compare the low-energy spectrum obtained from \cref{si:eqn:tripod_bands} with the one computed by numerical diagonalizing $\hat{H}_0$ with a large number of $\vec{Q}$ points. This simplified tripod model is seen to predict the appropriate qualitative features of the single-particle energy spectrum near the $K_M$ point. Moreover, in the limit $U=0$, \cref{si:eqn:tripod_bands} reduces to the high-velocity Dirac cone spectrum ($E = \pm \delta k$) and the TBG tripod approximation spectrum of Ref.~\cite{BIS11} (with the tunneling amplitudes rescalled by a factor of $\sqrt{2}$)
\begin{equation}
	E = \pm \frac{ \delta k \left(1 - 3 w^{\prime 2}_1 \right) }{\left( 3 w_0^{\prime 2} + 3 w_1^{\prime 2} + 1 \right)}.
\end{equation}

Finally, we note that a similar tripod model exists at the $K_M'$ point in valley $\eta = +$, but because the Dirac cone bands are much higher in energy in this region of the MBZ, the effects of $H_D$ can be ignored for the low-energy physics. Therefore, a tripod model for TSTG near the $K_M'$ point in valley $\eta = +$ is entirely equivalent to the tripod model introduced for ordinary TBG~\cite{BIS11}.

\subsection{Single-particle spectrum in the presence of displacement field}\label{app:approx_single_part:spec}

\begin{figure}[!th]
	\centering
	\fiveFigureCBar{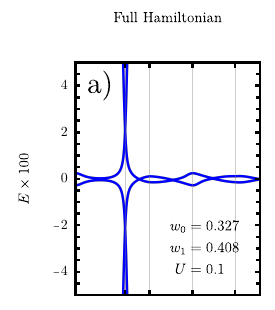}{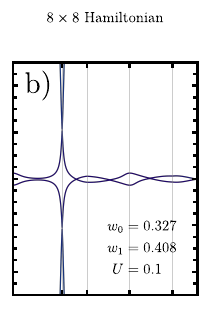}{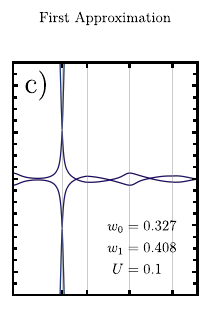}{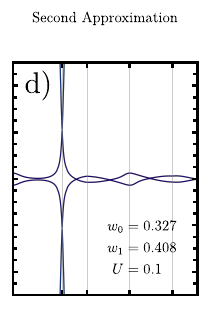}{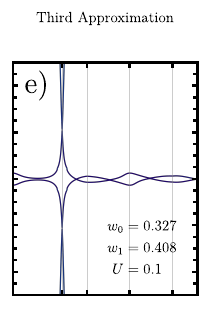}{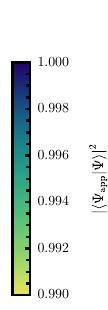}{\columnwidth}{fig:pertCompNC:1}
	\fiveFigureCBar{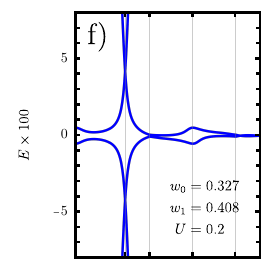}{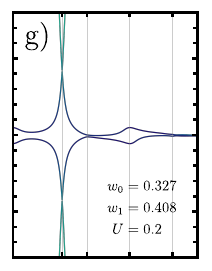}{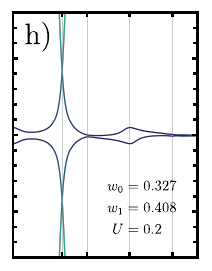}{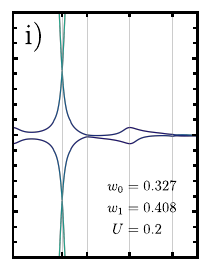}{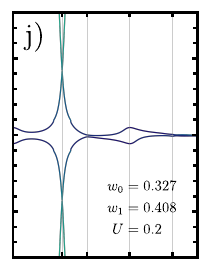}{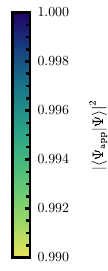}{\columnwidth}{fig:pertCompNC:2}%
	\fiveFigureCBar{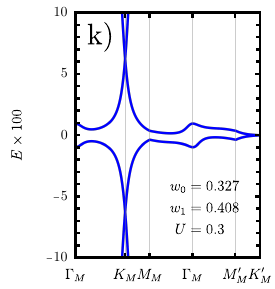}{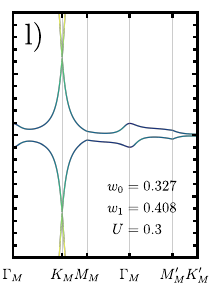}{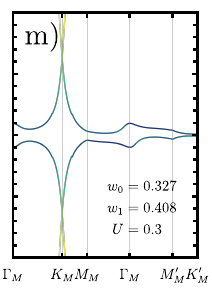}{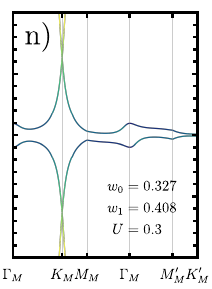}{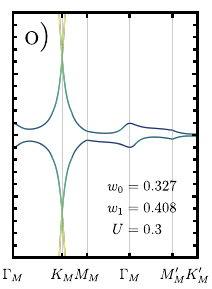}{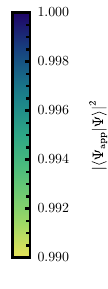}{\columnwidth}{fig:pertCompNC:3}
	\caption{Various approximations used to compute the single-particle low-energy spectrum of TSTG in the presence of displacement field in the non-chiral limit. For each row, the first panel denotes the unapproximated spectrum (computed as in \cref{si:fig:threeQApproxNC1:a,si:fig:threeQApproxNC2:a}), while the second panel denotes the spectrum obtained numerically from the $8 \times 8$ Hamiltonian in \cref{si:eqn:8x8_hamiltonian}. In the third to fifth panels, we always approximate the spectrum away from the Dirac points of the MBZ using \cref{si:eqn:approx_single_part_away_Dirac}. For the energy spectrum near the Dirac points, we employ the first, second, and third approximations of Appendix \ref{app:approx_single_part:spec:b} in the third, fourth, and fifth panels, respectively. We use a different cutoff $\Lambda$ for the each value of the displacement field $U$, namely $\Lambda = 0.1$, $0.175$, and $0.2$ for $U=0.1$, $0.2$, and $0.3$, respectively. For each approximation, the bands are colored according to the overlap between the corresponding approximated and unapproximated single-particle wave functions. This overlap is always greater than $0.99$. The values of the TSTG parameters are given as an inset for each plot.}
	\label{fig:pertCompNC}
\end{figure}

\begin{figure}[!th]
	\centering
	\fiveFigureCBar{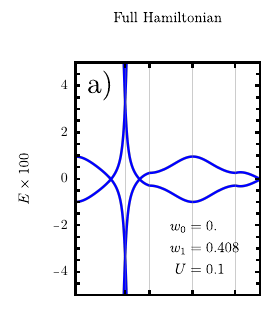}{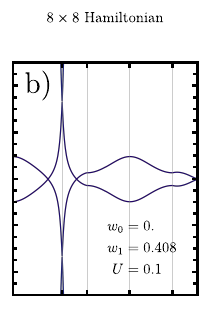}{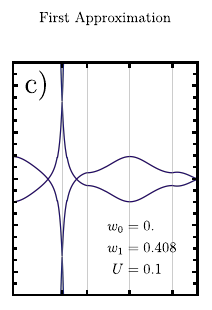}{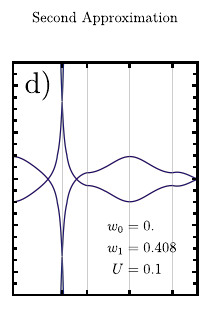}{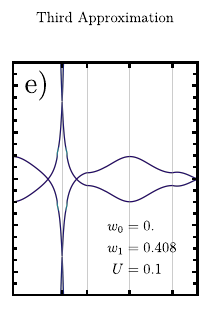}{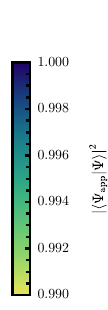}{\columnwidth}{fig:pertCompC:1}
	\fiveFigureCBar{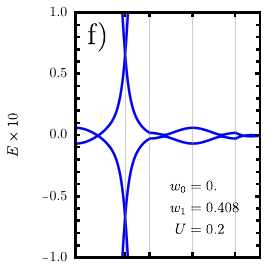}{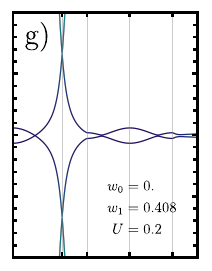}{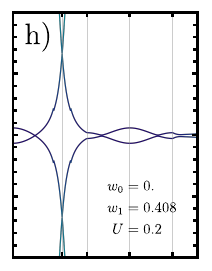}{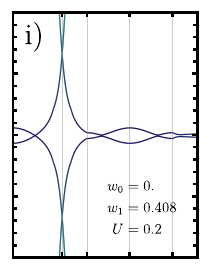}{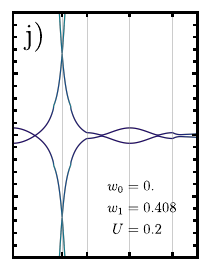}{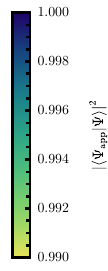}{\columnwidth}{fig:pertCompC:2}%
	\fiveFigureCBar{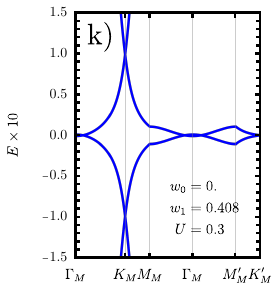}{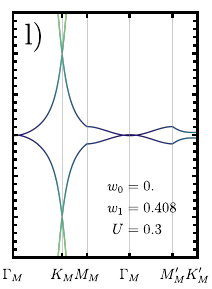}{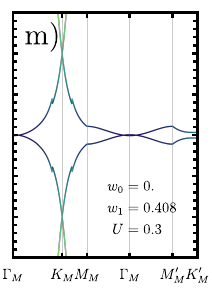}{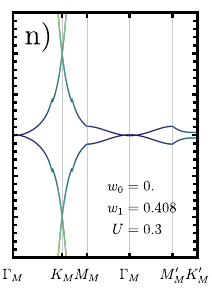}{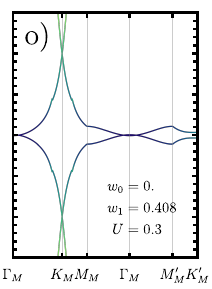}{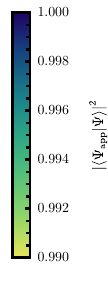}{\columnwidth}{fig:pertCompC:3}
	\caption{Various approximations used to compute the single-particle low-energy spectrum of TSTG in the presence of displacement field in the chiral limit. The meaning of the panels is the same as in \cref{fig:pertCompNC}. All approximate models perform remarkably well, with no differences from the exact Hamiltonian visible by eye.}
	\label{fig:pertCompC}
\end{figure}

Having developed an intuition for the effect of the displacement field coupling on the TSTG single-particle spectrum, we now turn toward more quantitative perturbation schemes. As we are ultimately interested in the low-energy physics of TSTG, one approach would be to \emph{first} consider the single-particle spectrum of the \emph{full} TSTG Hamiltonian from \cref{si:eqn:singlePart2} and \emph{then} project into its low-energy eigenstates. In our case, however, a better approach is to consider the TBG Hamiltonian of \cref{si:eqn:singlePart2:TBG} \emph{projected} into its flat bands and \emph{then} hybridize it with the high-velocity Dirac cone in the presence of displacement field.

\subsubsection{Perturbative effect of the displacement field}

We assume that $w_0 \leq 0.8 w_1$, which implies the existence of a sizable gap between the active bands ($n = \pm 1$) and the passive bands of the TBG Hamiltonian: in particular, the active band bandwidth $\omega$ is much smaller that the gap~\cite{BER20}. The approach of starting from the projected TBG single-particle Hamiltonian (henceforth denoted without a hat)
\begin{equation}
	\label{si:eqn:projTBG}
	H_{\mathrm{TBG}} = \sum_{\abs{n} = 1} \sum_{\eta,s} \sum_{\vec{k} \in \mathrm{MBZ}} \epsilon_{n,\eta}^{\hat{c}} \left( \vec{k} \right) \cre{c}{\vec{k},n,\eta,s} \des{c}{\vec{k},n,\eta,s},
\end{equation}
rather than the unprojected one is justified by using perturbation theory arguments. For the experimentally relevant values of the displacement field~\cite{HAO21} (corresponding to $U<0.3$), we can develop a perturbation theory in $U$ for the hybridization between the two mirror-symmetry sectors. The hybridization between the TBG active bands and the Dirac cone Hamiltonian happens already at first order in $U$, while mixing between the TBG active and passive bands happens only as a second order virtual process in $U$.  

The low-energy physics of TSTG is therefore governed by the Hamiltonian 
\begin{equation}
	\label{si:eqn:full_projected_Ham_with_U}
	H_{0} = H_{\mathrm{TBG}} + \hat{H}_D + \hat{H}_U,
\end{equation}
where the projected TBG Hamiltonian was given by \cref{si:eqn:projTBG}, while the Dirac cone and displacement field contributions are respectively given by
\begin{equation}
	\hat{H}_D = \sum_{\substack{\vec{k} \in \mathrm{MBZ} \\ \eta,s}} \sum_{\alpha, \beta} \sum_{i=1}^3 \left[h^{D,\eta}_{\vec{\eta \vec{q}_i}} \left( \vec{k} \right) \right]_{\alpha \beta} \cre{b}{\vec{k},\eta \vec{q}_i,\eta,\alpha,s} \des{b}{\vec{k},\eta \vec{q}_i,\eta,\beta,s},
\end{equation}
and
\begin{equation}
	\hat{H}_U = \frac{U}{2} \sum_{\substack{\vec{k} \in \mathrm{MBZ} \\ \eta,s}} \sum_{\alpha} \sum_{i=1}^3 \sum_{\abs{n} = 1} \left( u^{\hat{c}}_{\eta \vec{q}_i \alpha; n \eta}\left( \vec{k} \right) \cre{b}{\vec{k},\eta \vec{q}_i,\eta,\alpha,s}  \des{c}{\vec{k},n,\eta,s} + u^{*\hat{c}}_{\eta \vec{q}_i \alpha; n \eta}\left( \vec{k} \right)  \cre{c}{\vec{k},n,\eta,s} \des{b}{\vec{k},\eta \vec{q}_i,\eta,\alpha,s} \right).
\end{equation}
Among other things, we checked numerically the validity of the TSTG Hamiltonian from \cref{si:eqn:full_projected_Ham_with_U} in the first two columns of \cref{fig:pertCompNC,fig:pertCompC}. The low-energy TSTG spectrum obtained from \cref{si:eqn:full_projected_Ham_with_U} and the one obtained from $\hat{H}_{0} = \hat{H}_{\mathrm{TBG}} + \hat{H}_D + \hat{H}_U$ (\ie starting from the unprojected TBG Hamiltonian) are consistent both in energy and eigenstates to an error smaller than $1 \%$. Moreover, as discussed in Appendix \ref{app:single_part_ham:a}, we have restricted to only three plane-wave states (\ie $\vec{Q}$ points) in the expression of $\hat{H}_D$. However, the eigenstates of the TBG active bands are evaluated in all generality on the $\mathcal{Q}_{\pm}$ sublattice using the approximations discussed in Ref.~\cite{BER20}. 

For the sake of making this appendix self-contained, we briefly review the notation provided in \cref{sec:singleparticlespectrum:low}. The single-particle eigenstates of $H_{0}$ for valley $\eta$ and spin $s$ labeled by $m$ are given by  
\begin{equation}
	\label{si:eqn:8x8_eigenstates}
	\ket{\Psi^{\eta,s,m} \left( \vec{k} \right)} = \left[ \sum_{i=1}^3 \sum_{\alpha} \left( \psi^{\eta,s,m}_{i,\alpha} \left( \vec{k} \right) \cre{b}{\vec{k},\eta \vec{q}_i, \eta ,\alpha,s} \right) + \sum_{\abs{n} = 1} \phi^{\eta,s,m}_{n}\left( \vec{k} \right) \cre{c}{\vec{k},n,\eta,s} \right] \ket{0}.
\end{equation}
In \cref{si:eqn:8x8_eigenstates}, we have defined three two-component spinors on the sublattice space, $\psi^{\eta,s,m}_{i} \left( \vec{k} \right)$ (for $i=1,2,3$), corresponding to the three Dirac points in the MBZ, and the two-component spinor in the space of the $n = \pm 1$ active TBG bands, $\phi^{\eta,s,m} \left( \vec{k} \right)$.
We have also employed $m$ to label the different bands of $H_0$. The single-particle eigenvalue equation 
\begin{equation}
	H_0\ket{\Psi^{\eta,s,m} \left( \vec{k} \right)} = E^{\eta,m} \left( \vec{k} \right) \ket{\Psi^{\eta,s,m} \left( \vec{k} \right)}
\end{equation} 
can be written in matrix form as

\begin{equation}
	\label{si:eqn:8x8_hamiltonian}
	\renewcommand{\arraystretch}{1.6}
	\begin{pmatrix}
		\mathcal{E}^{\eta} \left( \vec{k} \right) & U^{\dagger \eta}_1 \left( \vec{k} \right) & U^{\dagger \eta}_2 \left( \vec{k} \right) & U^{\dagger \eta}_3 \left( \vec{k} \right) \\ 
		U^{\eta }_1 \left( \vec{k} \right) & h^{D,\eta}_{\vec{\eta \vec{q}_1}} \left( \vec{k} \right) & \mathbb{0} & \mathbb{0} \\ 
		U^{\eta }_2 \left( \vec{k} \right) & \mathbb{0} & h^{D,\eta}_{\vec{\eta \vec{q}_2}}  \left( \vec{k} \right)& \mathbb{0} \\ 
		U^{\eta }_3 \left( \vec{k} \right) & \mathbb{0} & \mathbb{0} & h^{D,\eta}_{\vec{\eta \vec{q}_3}} \left( \vec{k} \right)
	\end{pmatrix}
	\begin{pmatrix}
		\phi^{\eta,s,m} \left( \vec{k} \right) \\ \psi^{\eta,s,m}_{1} \left( \vec{k} \right) \\ \psi^{\eta,s,m}_{2} \left( \vec{k} \right) \\ \psi^{\eta,s,m}_{3} \left( \vec{k} \right)		
	\end{pmatrix}
	=
	E^{\eta,m} \left( \vec{k} \right)
	\begin{pmatrix}
		\phi^{\eta,s,m} \left( \vec{k} \right) \\ \psi^{\eta,s,m}_{1} \left( \vec{k} \right) \\ \psi^{\eta,s,m}_{2} \left( \vec{k} \right) \\ \psi^{\eta,s,m}_{3} \left( \vec{k} \right)		
	\end{pmatrix}
\end{equation}
where we have defined the $2 \times 2$ diagonal energy matrix for the TBG active bands in valley $\eta$
\begin{equation}
	\mathcal{E}^{\eta} \left( \vec{k} \right) = \begin{pmatrix}
		\epsilon^{\hat{c}}_{+1,\eta} \left( \vec{k} \right) & 0 \\
		0 & \epsilon^{\hat{c}}_{-1,\eta} \left( \vec{k} \right) \\
	\end{pmatrix}
\end{equation}
as well as the displacement field $2 \times 2$ perturbation matrices
\begin{equation}
	\left[ U^\eta_i \left( \vec{k} \right) \right]_{\alpha,n} = \frac{U}{2} u^{\hat{c}}_{\eta\vec{q}_i \alpha; n \eta} \left( \vec{k} \right),
\end{equation}
for $i=1,2,3$. In what follows, we will temporarily suppress the $m$, $\eta$, and $s$ indices as well as the momentum $\vec{k}$ parameter. In addition, we will introduce the following shorthand notation 
\begin{equation}
	h_i \equiv h_{\eta \vec{q}_i} ^{D,\eta} \left( \vec{k} \right),
\end{equation}
for $i=1,2,3$. 

There are two regions of interest pertaining to the low-energy eigenstates of $H_{0}$. When $\vec{k}$ is away from the Dirac points, \ie $\vec{k} \in \mathcal{C}_\eta$, where we have defined the region
\begin{equation}
	 \mathcal{C}_\eta = \mathrm{MBZ} \setminus \bigcup_{i=1}^3 A_{\eta}^i
\end{equation}
in terms of the regions $A_{\eta}^i$ introduced in \cref{si:eqn:Dirac_zones}, the hybridization between the eigenstates of $\hat{H}_D$ and the active bands of $\hat{H}_{\mathrm{TBG}}$ is suppressed by the difference in energy. One can therefore avoid solving the $8 \times 8$ Hamiltonian in \cref{si:eqn:8x8_hamiltonian} and employ perturbation theory to find the effect of the displacement on the active bands in this region. When $\vec{k}$ is near any of the three Dirac point of $\hat{H}_{D}$ in the MBZ, \ie $\vec{k} \in A_{\eta}^i$, we can no longer ignore the effects of the Dirac cone bands and more refined approximation methods needs to be developed. We will now explore these two cases and attempt to solve the Hamiltonian in \cref{si:eqn:8x8_hamiltonian}.

\subsubsection{Perturbation theory away from the Dirac points}\label{app:approx_single_part:spec:a}
When $\vec{k} \in \mathcal{C}_\eta$, the hybridization between the active TBG bands and the Dirac cone bands is suppressed by the difference in energy. We can therefore eliminate the $\psi_i$ spinors of \cref{si:eqn:8x8_hamiltonian} by writing them in terms of the $\phi$ spinors. 
\begin{equation}
	\label{si:eqn:eigenstateReduction1}
	\psi_i = \left(E - h_i \right)^{-1} U_i \phi.
\end{equation}
This allows us to formulate our problem as a non-linear eigenvalue equation for $\phi$ which simply reads
\begin{equation}
	\label{si:eqn:nonlinear_eig_8}
	\left[ \mathcal{E} + \sum_{i=1}^{3} U^{\dagger}_i \left(E - h_i \right)^{-1} U_i \right] \phi = E \phi.
\end{equation}
We expect the energy of the active bands to be only slightly changed by the hybridization with the Dirac cone Hamiltonian and have $\abs{E} \ll \abs{h_i}=\abs{\vec{k} - \eta \vec{q}_i}$. For the low-energy bands of $H_0$ we can thus ignore the $E$ dependence in the denominator of the second term of \cref{si:eqn:nonlinear_eig_8}. This affords a major simplification as the Hamiltonians $h_i$ can be readily inverted. Introducing the notation $\boldsymbol{\sigma}^{+} = \left(\sigma_x,\sigma_y \right)$ and $\boldsymbol{\sigma}^{-} = \left(-\sigma_x,\sigma_y \right)$ to denote the Pauli vector corresponding to the two valleys $\eta = \pm$, the eigenvalue equation becomes

\begin{equation}
	\label{si:eqn:approx_single_part_away_Dirac}
	\begin{pmatrix}
		\epsilon^{\hat{c}}_{+1,\eta} \left( \vec{k} \right) + \mathcal{B}_{+1,+1}^\eta \left( \vec{k} \right) & \mathcal{B}_{+1, -1}^\eta \left( \vec{k} \right) \\
		\mathcal{B}_{-1, +1}^\eta \left( \vec{k} \right) & \epsilon^{\hat{c}}_{-1,\eta} \left( \vec{k} \right) + \mathcal{B}_{-1, -1}^\eta \left( \vec{k} \right) \\
	\end{pmatrix}
	\begin{pmatrix}
		\phi^{\eta,s,m}_{+1} \left( \vec{k} \right) \\ \phi^{\eta,s,m}_{-1} \left( \vec{k} \right)	
	\end{pmatrix}
	=E^{\eta,m} \left( \vec{k} \right)
	\begin{pmatrix}
		\phi^{\eta,s,m}_{+1} \left( \vec{k} \right) \\ \phi^{\eta,s,m}_{-1} \left( \vec{k} \right)	
	\end{pmatrix},
\end{equation}
where we have defined the displacement field perturbation matrix
\begin{equation}
	\label{si:eqn:BMatrix}
	\mathcal{B}^{\eta}_{n m} \left( \vec{k} \right) = \frac{U^2}{4} \sum_{i=1}^3 \sum_{\alpha,\beta} \frac{u^{*\hat{c}}_{\eta\vec{q}_i \alpha; n \eta} \left( \vec{k} \right) \left[ \left( \vec{k} - \eta \vec{q}_i \right) \pauliVecEta \right]_{\alpha \beta} u^{\hat{c}}_{\eta\vec{q}_i \beta; m \eta} \left( \vec{k} \right) }{\abs{\vec{k} - \eta \vec{q}_i}^2}.
\end{equation}
As \cref{si:eqn:approx_single_part_away_Dirac} is only a $2 \times 2$ matrix, it can be readily diagonalized to obtain the low-energy band dispersion
\begin{equation}
	E^{\eta,\pm 1 } = \frac{\epsilon^{\hat{c}}_{+1,\eta} + \epsilon^{\hat{c}}_{-1,\eta} + \mathcal{B}_{+1,+1}^\eta + \mathcal{B}_{-1,-1}^\eta \pm \sqrt{\left[\epsilon^{\hat{c}}_{+1,\eta} - \epsilon^{\hat{c}}_{-1,\eta} + \mathcal{B}_{+1,+1}^\eta - \mathcal{B}_{-1,-1}^\eta \right]^2 + \mathcal{B}_{-1,+1}^\eta \mathcal{B}_{+1,-1}^\eta } }{2}.
\end{equation}
\begin{figure}[!t]
	\centering
	\threeFigureCBar{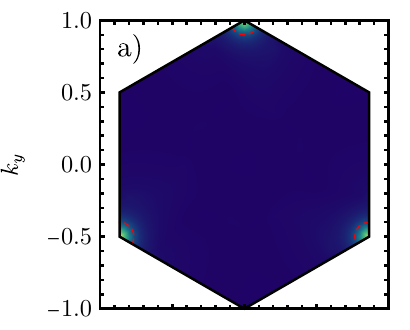}{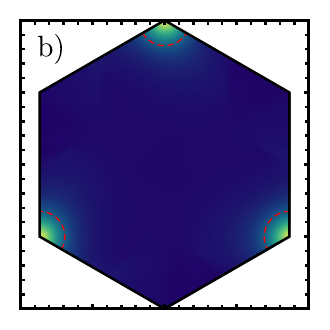}{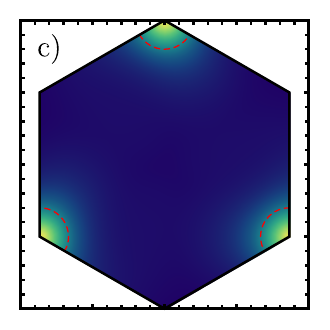}{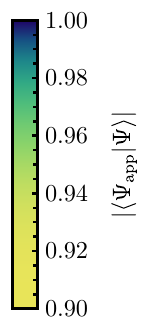}{\columnwidth}{fig:eigenstateSupprt:1}
	\threeFigureCBar{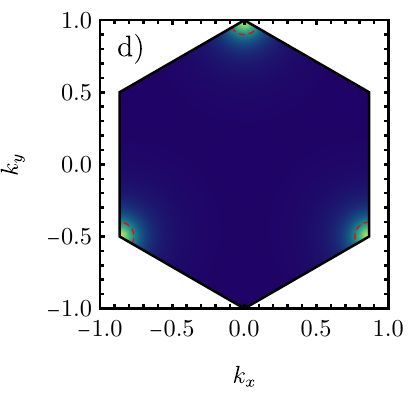}{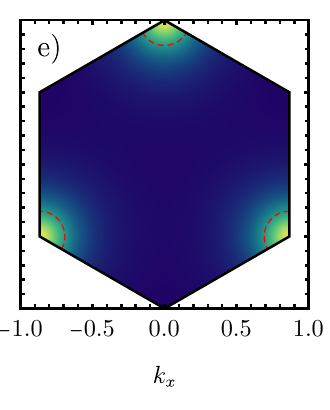}{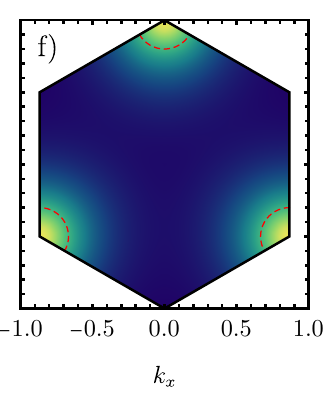}{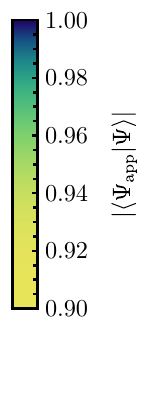}{\columnwidth}{fig:eigenstateSupprt:2}
	\caption{The amplitude of the mirror-symmetric operators in the lowest-energy single-particle eigenstates of TSTG. We consider the overlap between the unapproximated ($\ket{\psi}$) and approximated ($\ket{\psi_{\mathrm{app}}}$) wave functions corresponding to the lowest-energy conduction band for valley $\eta=+$ in the MBZ of TSTG. The approximated wave function $\ket{\psi_{\mathrm{app}}}$ is obtained directly from $\ket{\psi}$ by setting the amplitudes of all mirror-antisymmetric creation operators to zero (\ie $\ket{\psi_{\mathrm{app}}} = \frac{1 + m_z}{2} \ket{\psi}$). The boundaries of the $A_{+}^i$ zones (for $i=1,2,3$) are shown with red dashed lines. We consider the nonchiral limit ($w_0 / w_1=0.8$, $w_1 = 0.408$) in panels a)-c) and the chiral limit ($w_0 = 0$, $w_1 = 0.408$) in panels d)-f). The values of the displacement field are $U=0.1$ in panels a) and d), $U=0.2$ in panels b) and e), and $U=0.3$ panels in c) and f). Away from the high-velocity Dirac points in the TSTG MBZ, the weight of the mirror-antisymmetric operators in the low-energy eigenstates is negligible.}
	\label{fig:eigenstateSupprt}
\end{figure}%
The corresponding eigenstates can be found from the $\phi$ spinors and \cref{si:eqn:eigenstateReduction1}. We prove the validity of the approximation from \cref{si:eqn:approx_single_part_away_Dirac} numerically in the third, fourth and fifth columns of \cref{fig:pertCompNC,fig:pertCompC} in both the non-chiral and (first) chiral limits, respectively. However, a further approximation can be used: as shown in \cref{fig:eigenstateSupprt}, the weight of the mirror-antisymmetric operators is small enough in this region of the MBZ to approximate $\psi_i \approx 0$. The effects of the displacement field for $\vec{k} \in \mathcal{C}_{\eta}$ can then be captured by the following effective Hamiltonian which is second order in the displacement field
\begin{equation}
	\label{si:eqn:singlePart:projUCeband}
	H^{\left(\hat{c} \right)}_{U} = \sum_{\substack{\abs{n}, \abs{m} = 1 \\ \eta,s}} \sum_{\substack{\vec{k} \in \mathcal{C}_{\eta}}}  \mathcal{B}^{\eta}_{nm} \left( \vec{k} \right) \cre{c}{\vec{k},n,\eta,s} \des{c}{\vec{k},n',\eta,s}.
\end{equation}
Finally, we note that if $U^2 / \abs{\vec{k} - \eta \vec{q}_i} \ll \omega$ for $i=1,2,3$, then the active TBG band states will not be significantly perturbed by the displacement field.

\subsubsection{Perturbation theory near the Dirac points}\label{app:approx_single_part:spec:b}
Near any of the three Dirac points in the MBZ, the hybridization between the TBG active bands and the folded Dirac cone Hamiltonian is significant. If $\vec{k}$ is near the $j$-th Dirac point in the MBZ (\ie $\vec{k} \in A_{\eta}^j$), we will have $\abs{h_j} \ll 1$, but $\abs{h_i} \approx \sqrt{3}$, for $i \neq j$. This implies that while the hybridization between the TBG active bands and the $j$-th Dirac Hamiltonian will be relevant, there will be little to no mixing with the Dirac cone bands stemming from the other two Dirac points of $\hat{H}_{D}$ in the MBZ. We can therefore approximate $\psi_i \approx 0$ for $i \neq j$ and write the single-particle TSTG wave functions as
\begin{equation}
	\ket{\Psi^{\eta,s,m} \left( \vec{k} \right)} = \left[ \sum_{\alpha} \left( \psi^{\eta,s,m}_{j,\alpha} \left( \vec{k} \right) \cre{b}{\vec{k},\eta \vec{q}_j, \eta ,\alpha,s} \right) + \sum_{n = \pm 1} \phi^{\eta,s,m}_{n}\left( \vec{k} \right) \cre{c}{\vec{k},n,\eta,s} \right] \ket{0}.
\end{equation}
reflecting the four bands per spin per valley which are relevant for the low-energy of TSTG, namely the two TBG active bands and two Dirac cone bands. The eigenvalue equation $H_0 \ket{\Psi^{\eta,s,m} \left( \vec{k} \right)} = E^{\eta,m} \left( \vec{k} \right) \ket{\Psi^{\eta,s,m} \left( \vec{k} \right)}$ can then be written as a $4 \times 4$ matrix eigenvalue equation
\begin{equation}
	\label{si:eqn:4x4_hamiltonian}
	\renewcommand{\arraystretch}{1.6}
	\begin{pmatrix}
		\mathcal{E}^{\eta} \left( \vec{k} \right) & U^{\dagger \eta}_j \left( \vec{k} \right) \\ 
		U^{\eta }_j \left( \vec{k} \right) & h^{D,\eta}_{\eta \vec{q}_j} \left( \vec{k} \right) \\ 
	\end{pmatrix}
	\begin{pmatrix}
		\phi^{\eta,s,m} \left( \vec{k} \right) \\ \psi^{\eta,s,m}_{j} \left( \vec{k} \right)		
	\end{pmatrix}
	=
	E^{\eta,m} \left( \vec{k} \right)
	\begin{pmatrix}
		\phi^{\eta,s,m} \left( \vec{k} \right) \\ \psi^{\eta,s,m}_{j} \left( \vec{k} \right) 	
	\end{pmatrix}.
\end{equation}
We will call this the \emph{first approximation}. While this Hamiltonian cannot be solved analytically, we can employ a series of approximations to make the computations tractable. We begin by eliminating the $\phi$ spinor to obtain a nonlinear eigenvalue equation for $\psi_j$
\begin{equation}
	\left[h_j + U_j \frac{E - \frac{\epsilon^{\hat{c}}_+ + \epsilon^{\hat{c}}_-}{2} + \frac{\epsilon^{\hat{c}}_+ - \epsilon^{\hat{c}}_-}{2} \sigma_z}{\left(E - \frac{\epsilon^{\hat{c}}_+ + \epsilon^{\hat{c}}_-}{2} \right)^2 - \left( \frac{\epsilon^{\hat{c}}_+ - \epsilon^{\hat{c}}_-}{2} \right)^2} U^{\dagger}_j \right] \psi_j = E \psi_j,
\end{equation} 
where we have suppressed the valley index and made the dependence on the $\vec{k}$ parameter implicit. We then ignore the term $\frac{\epsilon^{\hat{c}}_+ + \epsilon^{\hat{c}}_-}{2}$ to first order in $E$ (an approximation which is exact in the chiral limit, where $\epsilon^{\hat{c}}_- = - \epsilon^{\hat{c}}_+$), obtaining
\begin{equation}
	\label{si:eqn:nonlinear_eig_first_to_second}
	\left[h_j + U_j \frac{E + \frac{\epsilon^{\hat{c}}_+ - \epsilon^{\hat{c}}_-}{2} \sigma_z}{E^2 - \left( \frac{\epsilon^{\hat{c}}_+ - \epsilon^{\hat{c}}_-}{2} \right)^2} U^{\dagger}_j \right] \psi_j = E \psi_j.
\end{equation} 
The solution to \cref{si:eqn:nonlinear_eig_first_to_second} can be found as an asymptotic series in the small parameter $\frac{\epsilon^{\hat{c}}_+ - \epsilon^{\hat{c}}_-}{2}$: to first order in $\frac{\epsilon^{\hat{c}}_+ - \epsilon^{\hat{c}}_-}{2}$, the nonlinear eigenvalue equation becomes
\begin{equation}
	\left[h_j + U_j \frac{E + \frac{\epsilon^{\hat{c}}_+ - \epsilon^{\hat{c}}_-}{2} \sigma_z}{E^2} U^{\dagger}_j \right] \psi_j = E \psi_j,
\end{equation}
an approximation which we call the \emph{second approximation}. Solving this equation is still impossible to do analytically, so we take one further simplification and ignore the $\frac{\epsilon^{\hat{c}}_+ - \epsilon^{\hat{c}}_-}{2}$ term altogether to afford the \emph{third approximation}
\begin{equation}
	\left[h_j +  \frac{U_j U^{\dagger}_j}{E}  \right] \psi_j = E \psi_j.
\end{equation}
Since $h_j$ is a $2 \times 2$ matrix, finding the energies of this equation essentially amounts to solving for the roots of a fourth-order polynomial, a cumbersome task to do analytically. In the chiral limit however, under the gauge-fixing of \cref{si:eqn:sewing_mats_TBG}, we have that 
\begin{equation}
	\sum_{\beta} \left( \sigma_{z} \right)_{\alpha \beta} \left[U_j \right]_{\beta,n} = \eta \sum_{n'} \left[U_j \right]_{\alpha,n'} \left( \sigma_{y} \right)_{n' n},
\end{equation}
which translates to $\sigma_z U_j U_j^\dagger \sigma_z =U_j U_j^\dagger$, implying that the term $U_j U_j^\dagger$ is essentially a $2 \times 2$ diagonal matrix
\begin{equation}
	U_j U^{\dagger}_j = 
	\begin{pmatrix}
		A & 0 \\
		0 & D	
	\end{pmatrix},
\end{equation}  
where $A$ and $D$ are real functions of $\vec{k}$, obtained from the moir\'e eigenstates. The characteristic equation of the third approximation 
\begin{equation}
	\det\left[E^2 - E h_j - 
		\begin{pmatrix}
			A & 0 \\
			0 & D
		\end{pmatrix}
	 \right] = 0
\end{equation}
is equivalent to finding the roots of a second order polynomial in $E^2$, an equation which can be readily solved analytically.

We note that the validity of \cref{si:eqn:4x4_hamiltonian} (\ie the first approximation), as well as of the second and third approximations was checked numerically in \cref{fig:pertCompNC,fig:pertCompC} in the third, fourth and fifth columns, respectively. In what follows, we will restrict to the first approximation from \cref{si:eqn:4x4_hamiltonian} and write the corresponding projected displacement field Hamiltonian as 
\begin{equation}
	\label{si:eqn:singlePart:projUA}
		H^{\left( \hat{b}\hat{c} \right)}_U = \frac{U}{2} \sum_{i=1}^3 \sum_{\vec{k} \in A^{i}_{\eta}} \sum_{\substack{\abs{n} = 1 \\ \alpha}} \left( u^{\hat{c}}_{\eta \vec{q}_i \alpha; n \eta}\left( \vec{k} \right) \cre{b}{\vec{k},\eta \vec{q}_i,\eta,\alpha,s}  \des{c}{\vec{k},n,\eta,s} + \mathrm{h.c.} \right).
\end{equation} 
In terms of the energy band basis, the projected displacement field Hamiltonian reads
\begin{equation}
	\label{si:eqn:singlePart:projUAeband}
		H^{\left( \hat{b}\hat{c} \right)}_U = \sum_{\substack{\abs{n},\abs{m} = 1 \\ \eta,s}} \sum_{\abs{\delta \vec{k}} \leq \Lambda} \left[ N^{\eta}_{mn} \left( \delta \vec{k} \right) \cre{b}{\delta \vec{k} + \eta \vec{q}_1,m,\eta,s} \des{c}{\delta \vec{k} + \eta \vec{q}_1,n,\eta,s} + \mathrm{h.c.} \right],
\end{equation} 
where we have defined the displacement field overlap matrix
\begin{equation}
	\label{si:eqn:singlePart:NMatrix}
	N^{\eta}_{mn} \left( \delta \vec{k} \right) = \sum_{\alpha} u^{*\hat{b}}_{\eta \vec{q}_1 \alpha; m \eta} \left( \delta \vec{k} + \eta \vec{q}_1 \right) u^{\hat{c}}_{\eta \vec{q}_1 \alpha; n \eta} \left( \delta \vec{k} + \eta \vec{q}_1 \right).
\end{equation}
For simplicity, in \cref{si:eqn:singlePart:projUAeband} we have used the periodicity of the energy band fermion operators \cref{si:eqn:periodicityEnB}, as well as the embedding relation of the single-particle wave functions \cref{si:eqn:embeddingWavF} to bring together the zones $A_{\eta}^i$ (for $i=1,2,3$) into a full circular region around the $\eta \vec{q}_1$ point.

\section{Gauge-fixing the single-particle projected Hamiltonian}\label{app:gauge_single_part}
The discrete symmetries of the single-particle Hamiltonian from Appendix \ref{app:symmetries} together with the gauge-fixing conditions in Appendix \ref{app:gauge} directly determine the explicit forms of the various terms of the single-particle projected TSTG Hamiltonian. The goal of this appendix is to parameterize the various terms appearing in the single-particle projected TSTG Hamiltonian from \cref{eqn:proj_single_part} of \cref{sec:singleparticlespectrum:proj}. The resulting parameterizations will be used for the unambiguous identification of the continuous symmetry groups of the projected many-body TSTG Hamiltonian in Appendix \ref{app:fullSym}, together with their corresponding generators. 

In what follows, we will find it useful to employ a series of conventions pertaining to the energy band operators $\cre{c}{\vec{k},n,\eta,s}$ and $\cre{b}{\vec{k},n,\eta,s}$. For both mirror-symmetry sector operators, we will use $\zeta^a$ (with $a=0,x,y,z$) to denote the identity and Pauli matrices in the energy band subspace (restricting to the pair of bands $n = \pm 1$), $\tau^a$ (with $a=0,x,y,z$) for the valley subspace, and $s^a$ (with $a=0,x,y,z$) for the spin subspace. 


\subsection{Parameterized forms of the single-particle projected TBG and Dirac Hamiltonian}\label{app:gauge_single_part:energy}
Independent of the gauge chosen in Appendix \ref{app:gauge}, the form of the Hamiltonians $H_{\mathrm{TBG}}$ and $H_{D}$ in the energy band basis is constrained by the symmetries discussed in Appendix \ref{app:symmetries}. For the sake of completeness, we will list the resulting parameterizations here.
\subsubsection{Parameterization of $H_{\mathrm{TBG}}$}

The energies of the TBG active bands $\epsilon_{n,\eta}^{\hat{c}} \left( \vec{k} \right)$ can be viewed as the elements of a $\vec{k}$-dependent matrix which we dub the TBG energy band matrix and which is diagonal in the valley and energy band subspaces. The (unprojected) TBG Hamiltonian from \cref{si:eqn:singlePart2:TBG} is symmetric under the combined $C_{2z}P$ transformation
\begin{equation}
	\left \lbrace \hat{H}_{\mathrm{TBG}},C_{2z} P \right \rbrace = 0.
\end{equation}
This implies that $\epsilon_{n,\eta}^{\hat{c}} \left( \vec{k} \right) = -\epsilon_{-n,-\eta}^{\hat{c}} \left( \vec{k} \right)$ and so the TBG energy band matrix can be parameterized as 
\begin{equation}
	\epsilon_{n,\eta}^{\hat{c}} \left( \vec{k} \right) = 
	\left[\zeta^z \tau^0 \right]_{nn,\eta\eta} \beta^{\hat{c}}_0 \left( \vec{k} \right) +
	\left[\zeta^0 \tau^z \right]_{nn,\eta\eta} \beta^{\hat{c}}_1 \left( \vec{k} \right),
\end{equation}
or in matrix notation 
\begin{equation}
	\label{si:eqn:param_e_c}
	\epsilon^{\hat{c}} \left( \vec{k} \right) = 
	\zeta^z \tau^0 \beta^{\hat{c}}_0 \left( \vec{k} \right) +
	\zeta^0 \tau^z \beta^{\hat{c}}_1 \left( \vec{k} \right).
\end{equation}
The functions $\beta^{\hat{c}}_{0,1} \left( \vec{k} \right)$ in \cref{si:eqn:param_e_c} represent real function whose exact form depends on the TBG band energies
\begin{equation}
	\begin{split}
		\beta^{\hat{c}}_0 \left( \vec{k} \right) &= \frac{\epsilon_{+1,+}^{\hat{c}} \left( \vec{k} \right) + \epsilon_{+1,-}^{\hat{c}} \left( \vec{k} \right) - \epsilon_{-1,+}^{\hat{c}} \left( \vec{k} \right) - \epsilon_{-1,-}^{\hat{c}} \left( \vec{k} \right)}{4} \\
		\beta^{\hat{c}}_1 \left( \vec{k} \right) &= \frac{\epsilon_{+1,+}^{\hat{c}} \left( \vec{k} \right) - \epsilon_{+1,-}^{\hat{c}} \left( \vec{k} \right) + \epsilon_{-1,+}^{\hat{c}} \left( \vec{k} \right) - \epsilon_{-1,-}^{\hat{c}} \left( \vec{k} \right)}{4} 
	\end{split}
\end{equation}

Additionally, as a consequence of the $C_{2z}$ commuting symmetry of $\hat{H}_{\mathrm{TBG}}$, the TBG energy bands obey $\epsilon_{n,\eta}^{\hat{c}} \left( \vec{k} \right) = - \epsilon_{n,-\eta}^{\hat{c}} \left(-\vec{k} \right)$, which further imposes that
\begin{equation}
	\beta^{\hat{c}}_{0} \left( \vec{k} \right) = \beta^{\hat{c}}_{0} \left(-\vec{k} \right) \qquad \text{and} \qquad
	\beta^{\hat{c}}_{1} \left( \vec{k} \right) = -\beta^{\hat{c}}_{1} \left(-\vec{k} \right).
\end{equation}

In the (first) chiral limit ($w_0 = 0$), the projected TBG Hamiltonian is also symmetric under the chiral transformation 
\begin{equation}
	\left \lbrace H_{\mathrm{TBG}},C \right \rbrace = 0.
\end{equation}
In this case, the band energies additionally satisfy $\epsilon_{n,\eta}^{\hat{c}} \left( \vec{k} \right) = \epsilon_{-n,\eta}^{\hat{c}} \left(\vec{k} \right)$, and thus $\beta^{\hat{c}}_{1} \left( \vec{k} \right)=0$. In the chiral limit, the TBG energy band matrix can be written in the simple form 
\begin{equation}
	\label{si:eqn:param_e_c_chiral}
	\epsilon^{\hat{c}} \left( \vec{k} \right) = 
	\zeta^z \tau^0 \beta^{\hat{c}}_0 \left( \vec{k} \right).
\end{equation}

\subsubsection{Parameterization of $H_{D}$}
Similar to the active bands of $H_{\mathrm{TBG}}$, we can also view the band energies of the projected Dirac Hamiltonian from \cref{sec:singleparticlespectrum:proj} as the elements of a matrix diagonal in the energy band and valley subspaces. The exact form of the Dirac energy bands from \cref{eqn:proj_single_part_Dirac_energy} allows the following straight-forward parameterization 
\begin{equation}
	\label{si:eqn:param_e_b}
	\epsilon_{n,\eta}^{\hat{b}} \left( \delta \vec{k} \right) = \left[ \zeta^z \tau^0 \right]_{n n,\eta \eta} \abs{\delta \vec{k}}.
\end{equation}

\subsection{Parameterization of the projected displacement field Hamiltonian}\label{app:gauge_single_part:displacement}
As shown in Appendix \ref{app:approx_single_part}, the projected displacement field Hamiltonian has two contributions, $H_{U}^{\left( \hat{b}\hat{c} \right)}$ and $H_{U}^{\left( \hat{c} \right)}$, which are defined in disjoint regions of the MBZ, one close and one away from the Dirac points, and which are linear and quadratic in the displacement field parameter $U$, respectively. We will now discuss the parameterized forms for each contribution.
\subsubsection{Parameterized form of $H_{U}^{\left( \hat{b}\hat{c} \right)}$}
The first contribution to the displacement field Hamiltonian from \cref{si:eqn:singlePart:projUAeband} is written in terms of the energy band basis operators of the TBG and Dirac cone sectors and the displacement field overlap matrix $N \left(\delta \vec{k} \right)$, whose exact form was given in \cref{si:eqn:singlePart:NMatrix}. Without explicitly solving for $N \left(\delta \vec{k} \right)$, we will now analyze the constraints imposed on its form by the symmetries of TSTG presented in Appendix \ref{app:symmetries} in conjunction with the $\vec{k}$-independent sewing matrices fixed in Appendix \ref{app:gauge}. 

\begin{enumerate}
	\item The antiunitary $C_{2z} T$ symmetry of TSTG has the same $\vec{k}$-independent sewing matrix for both the TBG and Dirac cone eigenstates, \ie $B_{\hat{b}}^{C_{2z}T} = B_{\hat{c}}^{C_{2z}T} = \zeta^0 \tau^0$. Using the shorthand notation from \cref{si:eqn:symmetry_sewing_definition}, we obtain
	\begin{align}
		&N^{\eta}_{mn} \left( \delta \vec{k} \right) = \frac{U}{2} \sum_{\alpha} u^{*\hat{b}}_{\eta \vec{q}_1 \alpha; m \eta} \left( \delta \vec{k} + \eta \vec{q}_1 \right) 
		u^{\hat{c}}_{\eta \vec{q}_1 \alpha; n \eta} \left( \delta \vec{k} + \eta \vec{q}_1 \right) \nonumber \\
		=& \frac{U}{2} \sum_{\alpha,\eta_1,\eta_2} \left[ u^{\dagger \hat{b}}_{m \eta_1} \left( \delta \vec{k} + \eta_1 \vec{q}_1 \right) \left[D^\dagger \left(C_{2z} T \right) \right]_{\eta \eta_1} \right]_{\eta \vec{q}_1 \alpha} 
		\left[ \left[D \left(C_{2z} T \right) \right]_{\eta \eta_2} u^{\hat{c}}_{n \eta_2} \left( \delta \vec{k} + \eta_2 \vec{q}_1 \right) \right]_{\eta \vec{q}_1 \alpha} \nonumber \\
		=& \frac{U}{2} \sum_{\substack{\alpha,\eta_1,\eta_2 \\ m',n'}} \left[B_{\hat{b}}^{\dagger C_{2z} T} \right]_{m \eta, m' \eta_1} u^{\hat{b}}_{\eta_1 \vec{q}_1 \alpha; m' \eta_1} \left( \delta \vec{k} + \eta_1 \vec{q}_1 \right) 
		 u^{*\hat{c}}_{\eta_2 \vec{q}_1 \alpha; n' \eta_2} \left( \delta \vec{k} + \eta_2 \vec{q}_1 \right) \left[B_{\hat{c}}^{ C_{2z} T} \right]_{n' \eta_2,n \eta}, \label{si:eqn:singl_part_N_gauge_c2t} 
	\end{align}		
	Equivalently, \cref{si:eqn:singl_part_N_gauge_c2t} can be written in matrix form as 
	\begin{equation}
		\label{si:eqn:singl_part_N_gauge_c2t_mat} 
		N \left( \delta \vec{k} \right) = \left(\zeta^0 \tau^0 \right) N^{*} \left( \delta \vec{k} \right) \left(\zeta^0 \tau^0 \right) = N^{*} \left( \delta \vec{k} \right).
	\end{equation}
 	proving that the displacement field overlap matrix is real.
	\item In the (first) chiral limit, both the TBG and the Dirac cone Hamiltonians are symmetric under the chiral transformation $C$, which has the $\vec{k}$-independent sewing matrices $B_{\hat{b}}^{C} = B_{\hat{c}}^{C} = \zeta^y \tau^z$. Correspondingly, the displacement field overlap matrix obeys
	\begin{align}
		&N^{\eta}_{mn} \left( \delta \vec{k} \right) = \frac{U}{2} \sum_{\alpha} u^{\hat{b}*}_{\eta \vec{q}_1 \alpha; m \eta} \left( \delta \vec{k} + \eta \vec{q}_1 \right) 
		u^{\hat{c}}_{\eta \vec{q}_1 \alpha; n \eta} \left( \delta \vec{k} + \eta \vec{q}_1 \right) \nonumber \\
		=& \frac{U}{2} \sum_{\alpha,\eta_1,\eta_2} \left[ u^{\dagger \hat{b}}_{m \eta_1} \left( \delta \vec{k} + \eta_1 \vec{q}_1 \right) \left[D^\dagger \left(C \right) \right]_{\eta \eta_1} \right]_{\eta \vec{q}_1 \alpha} 
		\left[ \left[D \left(C \right) \right]_{\eta \eta_2} u^{\hat{c}}_{n \eta_2} \left( \delta \vec{k} + \eta_2 \vec{q}_1 \right) \right]_{\eta \vec{q}_1 \alpha} \nonumber \\
		=& \frac{U}{2} \sum_{\substack{\alpha,\eta_1,\eta_2 \\ m',n'}} \left[B_{\hat{b}}^{\dagger C } \right]_{m \eta, m' \eta_1} u^{*\hat{b}}_{\eta_1 \vec{q}_1 \alpha; m' \eta_1} \left( \delta \vec{k} + \eta_1 \vec{q}_1 \right) 
		 u^{\hat{c}}_{\eta_2 \vec{q}_1 \alpha; n' \eta_2} \left( \delta \vec{k} + \eta_2 \vec{q}_1 \right) \left[B_{\hat{c}}^{ C} \right]_{n' \eta_2,n \eta}, \label{si:eqn:singl_part_N_gauge_c} 
	\end{align}
	which in matrix form is equivalent to 
	\begin{equation}
		\label{si:eqn:singl_part_N_gauge_c_mat}
		N \left( \delta \vec{k} \right) = \left(\zeta^y \tau^z \right) N \left( \delta \vec{k} \right) \left(\zeta^y \tau^z \right).
	\end{equation}
\end{enumerate} 

The displacement field overlap matrix is diagonal in valley subspace and, as a consequence of \cref{si:eqn:singl_part_N_gauge_c2t_mat}, its elements are real. It can therefore always be parameterized as 
\begin{equation}
	\label{si:eqn:parameter:NMatrix}
	N \left(\delta \vec{k} \right) = \sum_{b \in \left\lbrace 0,z \right\rbrace} \zeta^0\tau^b \lambda^{\left( \hat{b}\hat{c} \right)}_{0b} \left( \delta \vec{k} \right) +
	\zeta^x\tau^b \lambda^{\left( \hat{b}\hat{c} \right)}_{xb} \left( \delta \vec{k} \right) +
	i \zeta^y\tau^b \lambda^{\left( \hat{b}\hat{c} \right)}_{yb} \left( \delta \vec{k} \right) +
	\zeta^z\tau^b \lambda^{\left( \hat{b}\hat{c} \right)}_{zb} \left( \delta \vec{k} \right),
\end{equation}
where $\lambda^{\left( \hat{b}\hat{c} \right)}_{ab} \left( \delta \vec{k} \right)$ (for $a=0,x,y,z$ and $b=0,z$) represent generic real functions, whose exact form depends on the single-particle wave functions of the TBG and Dirac Hamiltonians. Additionally, in the (first) chiral limit, due to the constraint imposed by \cref{si:eqn:singl_part_N_gauge_c_mat}, $\lambda^{\left( \hat{b}\hat{c} \right)}_{ab} \left( \delta \vec{k} \right) = 0$ for $a=x,z$ and $b=0,z$ and so the displacement field overlap matrix must be given by 
\begin{equation}
	\label{si:eqn:parameter:NMatrix_c}
	N \left(\delta \vec{k} \right) = \zeta^0\tau^0 \lambda^{\left( \hat{b}\hat{c} \right)}_{00} \left( \delta \vec{k} \right) + 
	i\zeta^y\tau^0 \lambda^{\left( \hat{b}\hat{c} \right)}_{y0} \left( \delta \vec{k} \right) +
	\zeta^0\tau^z \lambda^{\left( \hat{b}\hat{c} \right)}_{0z} \left( \delta \vec{k} \right) +
	i\zeta^y\tau^z \lambda^{\left( \hat{b}\hat{c} \right)}_{yz} \left( \delta \vec{k} \right).
\end{equation}

\subsubsection{Parameterized form of $H_{U}^{\left( \hat{c} \right)}$}
The matrix $\mathcal{B} \left( \vec{k} \right)$ given in \cref{si:eqn:BMatrix} governs the second-order contribution to the displacement field projected Hamiltonian, away from the Dirac points. Here, we derive its parameterized form in a similar fashion to $N \left( \delta \vec{k} \right)$. 

\begin{enumerate}
	\item Owing to the antiunitary $C_{2z} T$ symmetry of the TBG Hamiltonian with the $\vec{k}$-independent sewing matrix $B_{\hat{c}}^{C_{2z}T} = \zeta^0 \tau^0$, the matrix $\mathcal{B} \left( \vec{k} \right)$ obeys
	\begin{align}
		&\mathcal{B}^{\eta}_{n m} \left( \vec{k} \right) = \frac{U^2}{4} \sum_{i=1}^3 \sum_{\alpha,\beta} \frac{u^{*\hat{c}}_{\eta\vec{q}_i \alpha; n \eta} \left( \vec{k} \right) \left[ \left( \vec{k} - \eta \vec{q}_i \right) \pauliVecEta \right]_{\alpha \beta} u^{\hat{c}}_{\eta\vec{q}_i \beta; m \eta} \left( \vec{k} \right) }{\abs{\vec{k} - \eta \vec{q}_i}^2} \nonumber \\
		=& \frac{U^2}{4} \sum_{i=1}^3 \sum_{\substack{\alpha,\beta \\ \alpha',\beta'}} \frac{u^{*\hat{c}}_{\eta\vec{q}_i \alpha'; n \eta} \left( \vec{k} \right) \left( \sigma_x \right)_{\alpha' \alpha} \left[ \left( \vec{k} - \eta \vec{q}_i \right) \pauliVecEta \right]^{*}_{\alpha \beta} \left( \sigma_x \right)_{\beta \beta'} u^{\hat{c}}_{\eta\vec{q}_i \beta'; m \eta} \left( \vec{k} \right) }{\abs{\vec{k} - \eta \vec{q}_i}^2} \nonumber \\
		=& \frac{U^2}{4} \sum_{i=1}^3 \sum_{\substack{\alpha,\beta \\ \eta_1,\eta_2}} \frac{ \left[ u^{\dagger \hat{c}}_{n \eta_1} \left( \vec{k} \right) \left[D^\dagger \left(C_{2z} T \right) \right]_{\eta \eta_1} \right]_{\eta \vec{q}_i \alpha} 
		\left[ \left( \vec{k} - \eta \vec{q}_i \right) \pauliVecEta \right]^{*}_{\alpha \beta}
		\left[ \left[D \left(C_{2z} T \right) \right]_{\eta \eta_2} u^{\hat{c}}_{m \eta_2} \left( \vec{k} \right) \right]_{\eta \vec{q}_i \beta} }{\abs{\vec{k} - \eta \vec{q}_i}^2} \nonumber \\
		=& \frac{U^2}{4} \sum_{i=1}^3 \sum_{\substack{\alpha,\beta \\ \eta_1,\eta_2 \\ m',n' }} \left[B_{\hat{c}}^{ \dagger C_{2z} T } \right]_{m \eta, m' \eta_1} 
		\frac{
		u^{\hat{c}}_{\eta_1 \vec{q}_i \alpha; m' \eta_1} \left( \vec{k} \right) 
		\left[ \left( \vec{k} - \eta \vec{q}_i \right) \pauliVecEta \right]^{*}_{\alpha \beta} 
		u^{*\hat{c}}_{\eta_2 \vec{q}_i \beta; n' \eta_2} \left( \vec{k} \right)}
		{\abs{\vec{k} - \eta \vec{q}_i}^2}
		\left[B_{\hat{c}}^{ C_{2z} T} \right]_{n' \eta_2,n \eta}
	 \label{si:eqn:singl_part_B_gauge_c2t} 
	\end{align}		
	Equivalently, \cref{si:eqn:singl_part_B_gauge_c2t} can be written in matrix form as 
	\begin{equation}
		\label{si:eqn:singl_part_B_gauge_c2t_mat} 
		\mathcal{B} \left( \vec{k} \right) = \left(\zeta^0 \tau^0 \right) \mathcal{B}^{*} \left( \vec{k} \right) \left(\zeta^0 \tau^0 \right) = \mathcal{B}^{*} \left( \vec{k} \right).
	\end{equation}
 	proving that the $\mathcal{B} \left( \vec{k} \right)$ matrix is real.
	\item In the (first) chiral limit, the TBG Hamiltonian is symmetric under the chiral transformation $C$, which has the $\vec{k}$-independent sewing matrix $B_{\hat{c}}^{C} = \zeta^y \tau^z$. Correspondingly, the displacement field perturbation matrix satisfies
	\begin{align}
		&\mathcal{B}^{\eta}_{n m} \left( \vec{k} \right) = \frac{U^2}{4} \sum_{i=1}^3 \sum_{\alpha,\beta} \frac{u^{*\hat{c}}_{\eta\vec{q}_i \alpha; n \eta} \left( \vec{k} \right) \left[ \left( \vec{k} - \eta \vec{q}_i \right) \pauliVecEta \right]_{\alpha \beta} u^{\hat{c}}_{\eta\vec{q}_i \beta; m \eta} \left( \vec{k} \right) }{\abs{\vec{k} - \eta \vec{q}_i}^2} \nonumber \\
		=& - \frac{U^2}{4} \sum_{i=1}^3 \sum_{\substack{\alpha,\beta \\ \alpha',\beta'}} \frac{u^{*\hat{c}}_{\eta\vec{q}_i \alpha'; n \eta} \left( \vec{k} \right) \left( \sigma_z \right)_{\alpha' \alpha} \left[ \left( \vec{k} - \eta \vec{q}_i \right) \pauliVecEta \right]_{\alpha \beta} \left( \sigma_z \right)_{\beta \beta'} u^{\hat{c}}_{\eta\vec{q}_i \beta'; m \eta} \left( \vec{k} \right) }{\abs{\vec{k} - \eta \vec{q}_i}^2} \nonumber \\
		=& - \frac{U^2}{4} \sum_{i=1}^3 \sum_{\substack{\alpha,\beta \\ \eta_1,\eta_2}} \frac{ \left[ u^{\dagger \hat{c}}_{n \eta_1} \left( \vec{k} \right) \left[D^\dagger \left(C \right) \right]_{\eta \eta_1} \right]_{\eta \vec{q}_i \alpha} 
		\left[ \left( \vec{k} - \eta \vec{q}_i \right) \pauliVecEta \right]_{\alpha \beta}
		\left[ \left[D \left(C \right) \right]_{\eta \eta_2} u^{\hat{c}}_{m \eta_2} \left( \vec{k} \right) \right]_{\eta \vec{q}_i \beta} }{\abs{\vec{k} - \eta \vec{q}_i}^2} \nonumber \\
		=& - \frac{U^2}{4} \sum_{i=1}^3 \sum_{\substack{\alpha,\beta \\ \eta_1,\eta_2 \\ m',n' }} \left[B_{\hat{c}}^{ \dagger C } \right]_{m \eta, m' \eta_1} 
		\frac{
		u^{*\hat{c}}_{\eta_1 \vec{q}_i \alpha; m' \eta_1} \left( \vec{k} \right) 
		\left[ \left( \vec{k} - \eta \vec{q}_i \right) \pauliVecEta \right]_{\alpha \beta} 
		u^{\hat{c}}_{\eta_2 \vec{q}_i \beta; n' \eta_2} \left( \vec{k} \right)}
		{\abs{\vec{k} - \eta \vec{q}_i}^2}
		\left[B_{\hat{c}}^{ C } \right]_{n' \eta_2,n \eta}
	 \label{si:eqn:singl_part_B_gauge_c} 
	\end{align}
	where we have used \cref{si:eqn:rep_mat_chiral}. We can rewrite \cref{si:eqn:singl_part_B_gauge_c} in an equivalent matrix form 
	\begin{equation}
		\label{si:eqn:singl_part_B_gauge_c_mat}
		\mathcal{B} \left( \vec{k} \right) = - \left(\zeta^y \tau^z \right) \mathcal{B} \left( \vec{k} \right) \left(\zeta^y \tau^z \right).
	\end{equation}
\end{enumerate} 

The displacement field perturbation matrix is diagonal in valley subspace and, as a consequence of \cref{si:eqn:singl_part_B_gauge_c2t_mat}, its elements are real. It can therefore always be parameterized as 
\begin{equation}
	\label{si:eqn:parameter:BMatrix}
	\mathcal{B} \left( \vec{k} \right) = \sum_{b \in \left\lbrace 0,z \right\rbrace} \zeta^0\tau^b \lambda^{\left( \hat{c} \right)}_{0b} \left( \vec{k} \right) +
	\zeta^x\tau^b \lambda^{\left( \hat{c} \right)}_{xb} \left( \vec{k} \right) +
	i \zeta^y\tau^b \lambda^{\left( \hat{c} \right)}_{yb} \left( \vec{k} \right) +
	\zeta^z\tau^b \lambda^{\left( \hat{c} \right)}_{zb} \left( \vec{k} \right),
\end{equation}
where $\lambda^{\left( \hat{c} \right)}_{ab} \left( \vec{k} \right)$ (for $a=0,x,y,z$ and $b=0,z$) represent generic real functions, whose exact form depends on the single-particle wave functions of the TBG Hamiltonian. Additionally, in the (first) chiral limit, due to the constraint imposed by \cref{si:eqn:singl_part_B_gauge_c_mat}, $\lambda^{\left( \hat{c} \right)}_{ab} \left( \vec{k} \right) = 0$ for $a=0,y$ and $b=0,z$ and so the displacement field perturbation matrix must be given by 
\begin{equation}
	\label{si:eqn:parameter:BMatrix_c}
	\mathcal{B} \left( \vec{k} \right) = \zeta^x\tau^0 \lambda^{\left( \hat{c} \right)}_{x0} \left( \vec{k} \right) + 
	\zeta^z\tau^0 \lambda^{\left( \hat{c} \right)}_{z0} \left( \vec{k} \right) +
	\zeta^x\tau^z \lambda^{\left( \hat{c} \right)}_{xz} \left( \vec{k} \right) +
	\zeta^z\tau^z \lambda^{\left( \hat{c} \right)}_{zz} \left( \vec{k} \right).
\end{equation}

\section{Interaction Hamiltonian}\label{app:interaction}
In this appendix, we derive the TSTG interaction Hamiltonian. First, we show how the electron-electron repulsion Hamiltonian can be written using the fermion operators defined on the moir\'e lattice in \cref{si:eqn:ckQ+,si:eqn:bkQ+} of Appendix \ref{app:single_part_ham}. We then project the TSTG interaction Hamiltonian in the eigenstates of the single-particle projected Hamiltonian from \cref{eqn:proj_single_part}, namely the active TBG bands and the low-energy Dirac cone modes. We also show that the projected TSTG Hamiltonian includes an effective Hartree-Fock potential arising from the TSTG bands that have been projected away. Finally, we gauge-fix the terms of the projected interaction Hamiltonian according to the symmetries of Appendix \ref{app:symmetries} and the gauge-fixing conditions of Appendix \ref{app:gauge}.

\subsection{Derivation of the interaction Hamiltonian}\label{app:interaction:derivation}
Here, we derive the low-energy interaction Hamiltonian governing electron-electron repulsion in TSTG~\cite{BER20a}. We start by writing the Fourier transformation of the electron density operators in terms of the low-energy fermion operators from \cref{si:eqn:low_en_fermions_a} defined on the moir\'e lattice. Finally, we simplify the expression of the interaction Hamiltonian by employing the mirror-symmetric and antisymmetric fermion operators.
\subsubsection{Interaction Hamiltonian in the moir\'e lattice}\label{app:interaction:derivation:Moire}
For each graphene layer $l$ in TSTG, we define the real-space electron operators,
\begin{equation}
	\label{si:eqn:fourierOverGBZ}
	\cre{a}{\vec{R}, \alpha, s , l} = \frac{1}{\sqrt{N}} \sum_{\vec{p} \in \mathrm{BZ}_l} e^{-i \vec{p} \cdot R_{\theta,l} \left(\vec{R} + \vec{t}_\alpha \right)} \cre{a}{\vec{p}, \alpha, s, l},
\end{equation}
where $\vec{R}$ denotes the single-layer graphene unit cell coordinates, $\alpha$ represents the sublattice index, $s$ is the electron spin, and $\vec{t}_\alpha$ is the displacement of the atoms belonging to sublattice $\alpha$ from the origin of the unit cell. Moreover, in \cref{si:eqn:fourierOverGBZ}, $\mathrm{BZ}_l$ is the BZ of the graphene layer $l$, while $R_{\theta,l}$ denotes the rotation matrix corresponding to the twist in layer $l$ relative to the coordinates chosen in \cref{fig:mbz_qlattice}. Note that in this notation, an atom belonging to layer $l$ having the unit cell coordinate $\vec{R}$ and belonging to the sublattice $\alpha$ is located at position $R_{\theta,l} \left(\vec{R} + \vec{t}_\alpha \right)$. As discussed in \cref{sec:singlehamiltonian} and Appendix \ref{app:single_part_ham}, the low-energy physics is dominated by the electron states near the Dirac points $\pm \vec{K}_{l}$, allowing us to approximate
\begin{equation}
	\label{si:eqn:fourierOverMBZApprox}
	\cre{a}{\vec{R}, \alpha, s , l} \approx \frac{1}{\sqrt{N}} \sum_{\eta} \sum_{\vec{k} \in \mathrm{MBZ}} \sum_{\vec{Q} \in \mathcal{Q}_{\eta, l}} e^{-i \left( \eta \vec{K}_{l} + \vec{k} - \vec{Q} \right) \cdot R_{\theta,l} \left(\vec{R} + \vec{t}_\alpha \right)} \cre{a}{\vec{k}, \vec{Q}, \eta ,\alpha, s, l},
\end{equation}
where $N$ represents the number of single-layer graphene unit cells and we have used the same notation as in Appendix \ref{app:single_part_ham:a}. The approximation in \cref{si:eqn:fourierOverMBZApprox} consists in imposing a finite cut-off for the number of points in the $\mathcal{Q}_{\eta, l}$ sublattice, such that we always have $\abs{\vec{Q}} \ll \vec{K}_l$. Using the real-space operators $\cre{a}{\vec{R}, \alpha, s , l}$, we can write the interaction Hamiltonian as 
\begin{equation}
	\label{si:eqn:realSpaceInteraction}
	\hat{H}_I = \frac{1}{2} \sum_{\vec{R},\vec{R}'} \sum_{\substack{\alpha, s, l \\ \alpha', s', l'}} V^{l,l'} \left[ R_{\theta,l} \left(\vec{R} + \vec{t}_\alpha \right) - R_{\theta,l'} \left(\vec{R}' + \vec{t}_{\alpha'} \right) \right] \normord{\cre{a}{\vec{R},\alpha,s,l}\des{a}{\vec{R},\alpha,s,l} \cre{a}{\vec{R}',\alpha',s',l'}\des{a}{\vec{R}',\alpha',s',l'}},
\end{equation}
where $V^{l,l'}\left( \vec{r} \right)$ represents the screened Coulomb interaction potential between two fermions located in layers $l$ and $l'$ which are separated by the vector $\vec{r}$ in the plane of the single-layer graphene. In \cref{si:eqn:realSpaceInteraction}, we have used $\normord{\left( \dots \right)}$ to denote normal-ordering for the fermion operators. The Coulomb interaction potential obeys the symmetry condition
\begin{equation}
	\label{si:eqn:gen_pot_symmetry}
	V^{l,l'} \left( \vec{r} \right) = V^{l',l} \left( \vec{r} \right).
\end{equation}
Defining the Fourier transformation of $V^{l,l'}\left( \vec{r} \right)$ over the MBZ, 
\begin{equation}
	V^{l,l'}\left( \vec{r} \right) = \frac{1}{\Omega_{\mathrm{tot}}} \sum_{\vec{G} \in \mathcal{Q}_0} \sum_{\vec{q} \in \mathrm{MBZ}} e^{-\left( \vec{q} + \vec{G} \right) \cdot \vec{r}} V^{l,l'}\left( \vec{q} + \vec{G} \right),
\end{equation}
where $\Omega_{\mathrm{tot}}$ represents the total area of the TSTG sample, we will only require that $V^{l,l'} \left( \vec{q} + \vec{G} \right)$ decays with $\abs{\vec{G}}$ and becomes negligible when $\abs{\vec{G}} \sim \abs{\vec{K}_l}$ (so that the interaction is diagonal in valley index), but will otherwise leave it unspecified for the moment, to keep the discussion general. 

We now introduce the interaction Hamiltonian in momentum space
\begin{equation}
	\label{si:eqn:interaction_density_normord}
	\hat{H}_I = \frac{1}{2\Omega_{\mathrm{tot}}} \sum_{\vec{G} \in \mathcal{Q}_0} \sum_{\substack{\vec{q} \in \mathrm{MBZ}\\l,l'}}  V^{l,l'}\left( \vec{q} + \vec{G} \right) \normord{ \rho^{l}_{\vec{G}+\vec{q}} \rho^{l'}_{-\vec{G}-\vec{q}}}.
\end{equation}
In \cref{si:eqn:interaction_density_normord} the density operator for layer $l$ is defined as
\begin{equation}
	\rho^{l}_{\vec{G} + \vec{q}} = \sum_{\vec{R}, \alpha, s} e^{i\left( \vec{G} + \vec{q} \right) \cdot R_{\theta,l} \left( \vec{R} +\vec{t}_{\alpha} \right) }  \cre{a}{\vec{R},\alpha,s,l} \des{a}{\vec{R},\alpha,s,l},
\end{equation}
and can be re-expressed with the aid of \cref{si:eqn:fourierOverMBZApprox} as
\begin{equation}
	\label{si:eqn:density1}
	\begin{split}
		\rho^{l}_{\vec{G} + \vec{q}} 
		=& \frac{1}{N} \sum_{\vec{R}, \alpha, s} e^{i\left( \vec{G} + \vec{q} \right) \cdot R_{\theta,l} \left( \vec{R} +\vec{t}_{\alpha} \right) } \sum_{\eta, \eta'} \sum_{\substack{\vec{k},\vec{k}' \in \mathrm{MBZ} \\ \vec{Q} \in \mathcal{Q}_{\eta, l}  \\ \vec{Q}' \in \mathcal{Q}_{\eta', l} }} e^{-i \left[ \left( \eta-\eta' \right) \vec{K}_l + \vec{k} - \vec{k}' - \vec{Q} + \vec{Q}' \right] \cdot  R_{\theta,l} \left( \vec{R} +\vec{t}_{\alpha} \right)} \cre{a}{\vec{k}, \vec{Q}, \eta, \alpha, s, l} \des{a}{\vec{k}', \vec{Q}', \eta', \alpha, s, l} = \\
		=& \sum_{\substack{\eta,\eta' \\ \alpha, s}} \sum_{\substack{\vec{k},\vec{k}' \in \mathrm{MBZ} \\ \vec{Q} \in \mathcal{Q}_{\eta, l}  \\ \vec{Q}' \in \mathcal{Q}_{\eta', l} }} \sum_{\vec{P}} \cre{a}{\vec{k},\vec{Q}, \eta, \alpha, s, l} \des{a}{\vec{k}',\vec{Q}', \eta', \alpha, s, l'} e^{i \vec{P} \cdot \vec{t}_{\alpha}} \delta_{\left( \eta'-\eta \right) \vec{K}_l - \vec{k} + \vec{k}' + \vec{Q} - \vec{Q}' + \vec{G} + \vec{q},R_{\theta,l} \vec{P}} ,
	\end{split} 
\end{equation}
where the sum indexed by $\vec{P}$ is over the reciprocal lattice of single-layer graphene. In evaluating the summation over the single-layer graphene real-space lattice vectors $\vec{R}$, we have employed the Poisson resummation formula
\begin{equation}
	\frac{1}{N} \sum_{\vec{R}} e^{i \vec{k} \cdot R_{\theta, l} \vec{R}} = \sum_{\vec{P}} \delta_{\vec{k}, R_{\theta, l} \vec{P}}.  
\end{equation}
We now turn our attention toward simplifying the summations in \cref{si:eqn:density1}. For this, we consider two possibilities:
\begin{enumerate}
	\item $\eta = \eta'$ (intra-valley scattering). In this case, the momentum-conserving $\delta$-function reads 
\begin{equation}
	\delta_{- \vec{k} + \vec{k}' + \vec{Q} - \vec{Q}' + \vec{G} + \vec{q},R_{\theta,l} \vec{P}}.
\end{equation} 
Since the interaction potential $V_{\vec{q} + \vec{G}}$ only contributes for $\abs{\vec{G}} \ll \abs{\vec{K}_{l}}$, we have that $\abs{- \vec{k} + \vec{k}' + \vec{Q} - \vec{Q}' + \vec{G} + \vec{q}} \ll \abs{\vec{K}_l}$, and so the only non-vanishing terms in the sum from \cref{si:eqn:density1} corresponds to $\vec{P}= \vec{0}$.
	\item $\eta=-\eta'$ (inter-valley scattering). In this case, the momentum-conserving $\delta$-function becomes 
\begin{equation}
	\delta_{- \vec{k} + \vec{k}' + \vec{Q} - \vec{Q}' + \vec{G} + \vec{q},R_{\theta,l} \vec{P} - 2 \eta \vec{K}_{l}}.
\end{equation} 
However, because $2 \eta \vec{K}_{l}$ is not a reciprocal vector of the graphene layer $l$ (whereas $R_{\theta,l}\vec{P}$ is), $\abs{\vec{P} -  2 \eta \vec{K}_{l}} \sim \abs{\vec{K}_l}$, and so the $\delta$-function always vanishes.
\end{enumerate} 
Imposing $\eta = \eta'$ and $\vec{P}=\vec{0}$ in \cref{si:eqn:density1}, we find that the density operators simplify
\begin{equation}
	\label{si:eqn:density2}
	\begin{split}
		\rho^{l}_{\vec{G} + \vec{q}} =& \sum_{\eta, \alpha, s} \sum_{\substack{\vec{k},\vec{k}' \in \mathrm{MBZ} \\ \vec{Q},\vec{Q}' \in \mathcal{Q}_{\eta, l}}} \cre{a}{\vec{k},\vec{Q}, \eta, \alpha, s, l} \des{a}{\vec{k}',\vec{Q}', \eta, \alpha, s, l} \delta_{- \vec{k} + \vec{k}' + \vec{Q} - \vec{Q}' + \vec{G} + \vec{q}, \vec{0}} \\
		=& \sum_{\eta, \alpha, s} \sum_{\substack{\vec{k} \in \mathrm{MBZ} \\ \vec{Q} \in \mathcal{Q}_{\eta, l}}} \cre{a}{\vec{k},\vec{Q}, \eta,\alpha, s, l} \des{a}{\vec{k} - \vec{q},\vec{G} + \vec{Q}, \eta, \alpha, s, l}, 
	\end{split}
\end{equation}
where we have employed the Bloch periodicity from \cref{si:eqn:shift_prop_a}.

In what follows, we will find it easier to recast the interaction Hamiltonian in \cref{si:eqn:interaction_density_normord} into a more symmetrical form. We will therefore introduce the Fourier transformation of the electron density relative to the filling of single-layer graphene at the charge neutral point, for which $\expec{\cre{a}{\vec{k},\vec{Q}, \eta, \alpha, s, l} \des{a}{\vec{k} - \vec{q},\vec{G} + \vec{Q}, \eta, \alpha, s, l}} = \frac{1}{2} \delta_{\vec{q},\vec{0}} \delta_{\vec{G},\vec{0}}$. The interaction Hamiltonian then becomes
\begin{equation}
	\label{si:eqn:interaction_density_relative}
	\hat{H}_I = \frac{1}{2\Omega_{\mathrm{tot}}} \sum_{\vec{G} \in \mathcal{Q}_0} \sum_{\substack{\vec{q} \in \mathrm{MBZ}\\l,l'}}  V^{l,l'}\left( \vec{q} + \vec{G} \right) \delta \rho^{l}_{\vec{G}+\vec{q}} \delta \rho^{l'}_{-\vec{G}-\vec{q}},
\end{equation}
where the relative electron density operators are defined as
\begin{equation}
	\delta \rho^{l}_{\vec{G}+\vec{q}}=\sum_{\eta, \alpha, s} \sum_{\substack{\vec{k} \in \mathrm{MBZ} \\ \vec{Q} \in \mathcal{Q}_{\eta, l}}} \left( \cre{a}{\vec{k},\vec{Q}, \eta,\alpha, s, l} \des{a}{\vec{k} - \vec{q},\vec{G} + \vec{Q}, \eta, \alpha, s, l} - \frac{1}{2} \delta_{\vec{q},\vec{0}} \delta_{\vec{G},\vec{0}} \right).
\end{equation}
Note that the expressions in \cref{si:eqn:interaction_density_normord,si:eqn:interaction_density_relative} are equivalent up to a redefinition of the chemical potential.

\subsubsection{Coulomb repulsion potential}\label{app:interaction:derivation:Coulomb}
\begin{figure}[!t]
	\centering
	\includegraphics[width=0.3\columnwidth]{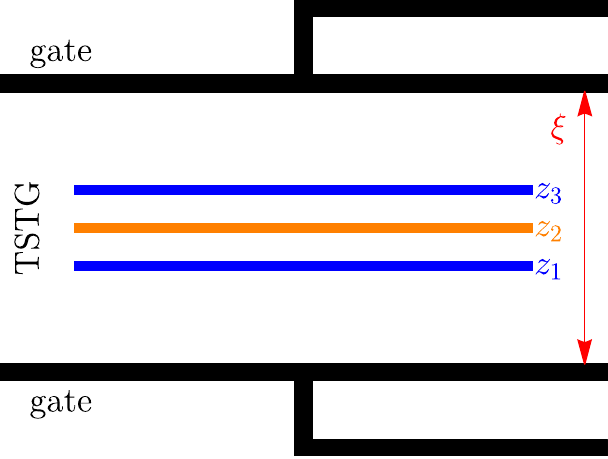}
	\caption{Gated TSTG experimental setup. We assume that the TSTG sample is located midway between two gate plates which are separated by a distance $\xi$. The three graphene monolayers are colored according to the twist angle (see also \cref{fig:mbz_qlattice}) and are located at heights $z_l$ (for $l=1,2,3$) measured from the middle between the two gates (for this geometry, $z_2 = 0$). For the typical experimental setups, the distance between adjacent graphene monolayers ($\abs{z_3 - z_2} = \abs{z_2 - z_1} \sim 3\,\si{\angstrom} $) is much smaller than the gate separation ($\xi \sim 10\,\si{\nano\metre}$).}
	\label{si:fig:expSet}
\end{figure}

To make further approximations and simplify the expression of $\hat{H}_I$, we need to discuss the exact form of the interaction potential between the electrons of TSTG. Here we assume that the TSTG sample is situated midway between a top gate plate and a bottom gate plate which are a distance $\xi$ away from each other in the $\hat{z}$ direction (see \cref{si:fig:expSet}). We also assume that the height of the graphene layer $l$ is given by $z_l$ (as measured from the middle between the two gates). The potential between two electrons in two layers $l$ and $l'$, separated by a distance $\vec{r}$ in the plane of the single layer graphene is given by summing over an infinite series of image charges
\begin{equation}
	V^{l,l'} \left( \vec{r} \right) = \frac{e^2}{\epsilon} \sum_{n=-\infty}^{\infty} \frac{\left( -1 \right)^n}{\sqrt{r^2 + \left[n \xi + z_l - z_{l'} \left( -1 \right)^n \right]^2}},
\end{equation} 
with $\epsilon$ being the dielectric constant, $e$ denoting the charge of an electron, and $r = \abs{\vec{r}}$. Note that the potential obeys the symmetry condition from \cref{si:eqn:gen_pot_symmetry}, as it can be seen by changing the dummy summation variable $n \to n \left( -1 \right)^n$. The separation between the top and bottom plates $\xi$ is usually around \SI{10}{\nano\metre}. On the other hand the $z_l$ is of the same order as the inter-layer separation, which is approximately \SI{3}{\angstrom}. It is therefore justified to ignore the $z_l$ and $z_{l'}$ dependence of the interaction potential and approximate
\begin{equation}
		V^{l,l'} \left( \vec{r} \right) \approx \frac{e^2}{\epsilon} \sum_{n=-\infty}^{\infty} \frac{\left( -1 \right)^n}{\sqrt{r^2 + \left(n \xi \right)^2}}.
\end{equation}
which affords a significant simplification, since the interaction potential is now independent on the layer indices. We will henceforth suppress the layer indices and write the interaction potential as $V^{l,l'} \left( \vec{r} \right) = V \left( \vec{r} \right)$, whose Fourier transformation reads~\cite{BER20a}
\begin{equation}
	V \left( \vec{q} \right) = \frac{2 \pi e^2}{\epsilon} \frac{\tanh \left( \xi q/2 \right)}{q}, 
\end{equation}
where $q=\abs{\vec{q}}$.

\subsubsection{Decoupling the interaction Hamiltonian in mirror symmetry sectors}\label{app:interaction:derivation:Mirror}
As a consequence of the $m_z$ symmetry of TSTG in the absence of displacement field, it is useful to construct relative density operators corresponding to the mirror-symmetric and mirror-antisymmetric operators of TSTG, which are respectively given by 
\begin{equation}
	\begin{split}
		\delta\rho^{\hat{c}}_{\vec{G} + \vec{q}}  &= \sum_{\eta, \alpha, s} \sum_{\substack{\vec{k} \in \mathrm{MBZ} \\ \vec{Q} \in \mathcal{Q}_{\pm}}} \left( \cre{c}{\vec{k},\vec{Q},\eta ,\alpha, s} \des{c}{\vec{k} - \vec{q},\vec{G} + \vec{Q}, \eta, \alpha, s} - \frac{1}{2} \delta_{\vec{q},\vec{0}} \delta_{\vec{G},\vec{0}} \right) ,\\
	\delta\rho^{\hat{b}}_{\vec{G} + \vec{q}} &= \sum_{\eta, \alpha, s} \sum_{\substack{\vec{k} \in \mathrm{MBZ} \\ \vec{Q} \in \mathcal{Q}_{\eta}}} \left( \cre{b}{\vec{k},\vec{Q},\eta ,\alpha, s} \des{b}{\vec{k} - \vec{q},\vec{G} + \vec{Q}, \eta, \alpha, s} - \frac{1}{2} \delta_{\vec{q},\vec{0}} \delta_{\vec{G},\vec{0}} \right).	
	\end{split}
\end{equation} 
Because 
\begin{equation}
	\delta\rho^{\hat{c}}_{\vec{G} + \vec{q}} + \delta\rho^{\hat{b}}_{\vec{G} + \vec{q}} = \sum_{l=1}^3 \delta\rho^{l}_{\vec{G} + \vec{q}},
\end{equation}
we can use the independence of the interaction potential $V^{l,l'} \left( \vec{r} \right)$ on the layer indices $l$ and $l'$ in \cref{si:eqn:interaction_density_relative} and show that, in a similar fashion to the one-particle Hamiltonian, it also decouples into three different terms
\begin{equation}
	\label{si:eqn:interaction_density}
	\begin{split}
		\hat{H}_I &= \frac{1}{2\Omega_{\mathrm{tot}}} \sum_{\vec{G} \in \mathcal{Q}_0} \sum_{\vec{q} \in \mathrm{MBZ}}  V\left( \vec{q} + \vec{G} \right) \left[ \delta \rho^{\hat{b}}_{\vec{G}+\vec{q}} + \delta \rho^{\hat{c}}_{\vec{G}+\vec{q}} \right] \left[ \delta \rho^{\hat{b}}_{-\vec{G}-\vec{q}} +  \delta \rho^{\hat{c}}_{-\vec{G}-\vec{q}} \right] \\
		&= \hat{H}_{I,\mathrm{TBG}} + \hat{H}_{I,D} + \hat{H}_{I,\mathrm{TBG}-D}.
	\end{split}
\end{equation}
The first term in \cref{si:eqn:interaction_density}
\begin{equation}
	\hat{H}_{I,\mathrm{TBG}} = \frac{1}{2\Omega_{\mathrm{tot}}} \sum_{\vec{G} \in \mathcal{Q}_0} \sum_{\vec{q} \in \mathrm{MBZ}}  V\left( \vec{q} + \vec{G} \right) \delta \rho^{\hat{c}}_{\vec{G}+\vec{q}} \delta \rho^{\hat{c}}_{-\vec{G}-\vec{q}}
\end{equation}
represents the interaction Hamiltonian that appears in ordinary TBG (see for example Ref.~\cite{BER20a}),
\begin{equation}
	\hat{H}_{I,D} = \frac{1}{2\Omega_{\mathrm{tot}}} \sum_{\vec{G} \in \mathcal{Q}_0} \sum_{\vec{q} \in \mathrm{MBZ}}  V\left( \vec{q} + \vec{G} \right) \delta \rho^{\hat{b}}_{\vec{G}+\vec{q}} \delta \rho^{\hat{b}}_{-\vec{G}-\vec{q}}
\end{equation}
denotes the interaction between the Dirac cone fermions, while
\begin{equation}
	\hat{H}_{I,\mathrm{TBG}-D} = \frac{1}{2\Omega_{\mathrm{tot}}} \sum_{\vec{G} \in \mathcal{Q}_0} \sum_{\vec{q} \in \mathrm{MBZ}}  V\left( \vec{q} + \vec{G} \right) \left[ \delta \rho^{\hat{c}}_{\vec{G}+\vec{q}} \delta \rho^{\hat{b}}_{-\vec{G}-\vec{q}} + \delta \rho^{\hat{b}}_{\vec{G}+\vec{q}} \delta \rho^{\hat{c}}_{-\vec{G}-\vec{q}} \right]
\end{equation}
is the interaction between the Dirac cone fermions and the TBG electrons.

\subsection{Projecting the interaction Hamiltonian}\label{app:interaction:projection}
Having derived the TSTG interaction Hamiltonian, we now turn our attention towards projecting it in the low-energy modes of the single-particle Hamiltonian $\hat{H}_0$. This is done by writing the relative density operators in \cref{si:eqn:interaction_density} in the energy band basis defined in Appendix \ref{app:single_part_ham:b} and then restricting the summation to the active TBG bands and the Dirac cone eigenstates with an energy lower than the gap between the TBG active and passive bands. 
\subsubsection{Projected density operators} 
To keep the discussion general, we will consider a generic energy band basis given by the operators $\cre{f}{\vec{k},n,\eta,s}$ with the corresponding single-particle wave function $u^{\hat{f}}_{\vec{Q} \alpha; n \eta} \left( \vec{k} \right)$. The wave functions $u^{\hat{f}}_{\vec{Q} \alpha; n \eta} \left( \vec{k} \right)$ are defined on a certain $\mathcal{Q}_{\hat{f},\eta}$ sublattice (which might depend on the valley $\eta$) and obey the following completeness relation 
\begin{equation}
	\label{si:eqn:completeness}
	\delta_{\vec{G},\vec{0}} \delta_{\alpha,\beta} = \sum_{\substack{\vec{Q} \in \mathcal{Q}_{\hat{f},\eta} \\ n}} u^{* \hat{f}}_{\vec{Q}-\vec{G} \beta; n \eta} \left( \vec{k} \right) u^{\hat{f}}_{\vec{Q}\alpha; n \eta} \left( \vec{k} \right),
\end{equation}
for $\vec{G} \in \mathcal{Q}_0$. \Cref{si:eqn:completeness} can be used to obtain the operators $\cre{f}{\vec{k},\vec{Q},\alpha,\eta,s}$ in terms of the energy band basis as
\begin{equation}
	\label{si:eqn:eband_inverse}
	\cre{f}{\vec{k},\vec{Q},\alpha,\eta,s} = \sum_{n} u^{* \hat{f}}_{\vec{Q} \alpha; n \eta} \cre{f}{\vec{k},n,\eta,s}.
\end{equation}
Using \cref{si:eqn:eband_inverse}, the relative density operators corresponding to species $\cre{f}{\vec{k},\vec{Q},\alpha,\eta,s}$ become
\begin{equation}
	\label{si:eqn:rel_density_eband_1}
	\delta \rho^{\hat{f}}_{\vec{G} + \vec{q}} = \sum_{\eta, \alpha, s} \sum_{\vec{k} \in \mathrm{MBZ}} \sum_{\vec{Q} \in \mathcal{Q}_{\hat{f},\eta}} \sum_{m,n} u^{* \hat{f}}_{\vec{Q}-\vec{G} \alpha; m \eta} \left( \vec{k} + \vec{q} \right) u^{\hat{f}}_{\vec{Q} \alpha; n \eta} \left( \vec{k} \right) \left(\cre{f}{\vec{k}+\vec{q},m,\eta,s} \des{f}{\vec{k},n,\eta,s} - \frac{1}{2} \delta_{\vec{q},\vec{0}} \delta_{m,n} \right).
\end{equation}
Defining the form factor matrix
\begin{equation}
	\label{si:eqn:str_fact_def}
	M^{\hat{f},\eta}_{mn} \left( \vec{k},\vec{q}+\vec{G} \right) = \sum_{\alpha} \sum_{\vec{Q} \in \mathcal{Q}_{\hat{f},\eta}} u^{* \hat{f}}_{\vec{Q}-\vec{G} \alpha; m \eta} \left( \vec{k} + \vec{q} \right) u^{\hat{f}}_{\vec{Q} \alpha; n \eta} \left( \vec{k} \right) = \sum_{\alpha} \sum_{\vec{Q} \in \mathcal{Q}_{\hat{f},\eta}} u^{* \hat{f}}_{\vec{Q} \alpha; m \eta} \left( \vec{k} + \vec{q} + \vec{G} \right) u^{\hat{f}}_{\vec{Q} \alpha; n \eta} \left( \vec{k} \right),
\end{equation}
the expression in \cref{si:eqn:rel_density_eband_1} can be further simplified into
\begin{equation}
	\label{si:eqn:rel_density_eband_2}
	\delta \rho^{\hat{f}}_{\vec{G} + \vec{q}} = \sum_{\substack{\eta, s \\ m,n}} \sum_{\vec{k} \in \mathrm{MBZ}} M^{\hat{f},\eta}_{mn} \left( \vec{k},\vec{q}+\vec{G} \right) \left(\cre{f}{\vec{k}+\vec{q},m,\eta,s} \des{f}{\vec{k},n,\eta,s} - \frac{1}{2} \delta_{\vec{q},\vec{0}} \delta_{m,n} \right).
\end{equation}
It is worth mentioning that owing to the embedding relation \cref{si:eqn:embeddingWavF}, the form factor matrix is periodic in the first argument, \ie
\begin{equation}
	\label{si:eqn:str_fact_per}
	M^{\hat{f},\eta}_{mn} \left( \vec{k},\vec{q}+\vec{G} \right) = M^{\hat{f},\eta}_{mn} \left( \vec{k} + \vec{G}_0,\vec{q}+\vec{G} \right),
\end{equation} 
for any $\vec{G}_0 \in \mathcal{Q}_{0}$. Also, following straight-forwardly from their definition, the form factors obey the Hermiticity condition~\cite{BER20a}
\begin{equation}
	\label{si:eqn:formFactHerm}
	M^{\hat{f},\eta}_{mn} \left( \vec{k},\vec{q}+\vec{G} \right) = M^{*\hat{f},\eta}_{nm} \left( \vec{k} - \vec{q},-\vec{q}-\vec{G} \right).
\end{equation}

For the mirror-symmetric operators $\cre{c}{\vec{k},n,\eta,s}$, the projected density operators are obtained by restricting to the active bands (\ie $\abs{n},\abs{m} = 1$), yielding
\begin{equation}
	\label{si:eqn:proj_rho_c}
	\overline{\delta \rho^{\hat{c}}}_{\vec{G}+\vec{q}} = \sum_{\eta, s} \sum_{\vec{k} \in \mathrm{MBZ} } \sum_{\abs{n},\abs{m} = 1} M^{\hat{c},\eta}_{mn} \left( \vec{k},\vec{q}+\vec{G} \right) \left(\cre{c}{\vec{k}+\vec{q},m,\eta,s} \des{c}{\vec{k},n,\eta,s} - \frac{1}{2} \delta_{\vec{q},\vec{0}} \delta_{m,n} \right).
\end{equation}
The overbar used in the definition of $\overline{\delta \rho^{\hat{c}}}_{\vec{G}+\vec{q}}$ emphasizes the fact that it represents the \emph{projected}, rather the unprojected relative density operator.

On the other hand, for the mirror-antisymmetric operators $\cre{b}{\vec{k},n,\eta,s}$, the summation is also constrained to include only the energy band basis operators with momenta $\vec{k} \in \mathcal{A}_{\eta}$, where 
\begin{equation}
	\mathcal{A}_{\eta} = \bigcup_{i=1}^3 A_{\eta}^i,
\end{equation}
in addition to requiring $\abs{n},\abs{m} = 1$. The corresponding projected density operators are given by 
\begin{equation}
	\overline{\delta \rho^{\hat{b}}}_{\vec{G}+\vec{q}} = \sum_{\eta, s} \sum_{\substack{\vec{k} \\ \vec{k},\vec{k}+\vec{q} \in \mathcal{A}_{\eta}}} \sum_{\abs{n},\abs{m} = 1} M^{\hat{b},\eta}_{mn} \left( \vec{k},\vec{q}+\vec{G} \right) \left(\cre{b}{\vec{k}+\vec{q},m,\eta,s} \des{b}{\vec{k},n,\eta,s} - \frac{1}{2} \delta_{\vec{q},\vec{0}} \delta_{m,n} \right)
\end{equation}
However, due to the periodicity of the energy band operators from \cref{si:eqn:periodicityEnB} and the periodicity of the form factors from \cref{si:eqn:str_fact_per}, we can change the disconnected region of summation for the $\vec{k}$ momenta to an equivalent connected region which is defined by the conditions $\abs{\vec{k}-\eta \vec{q}_1},\abs{\vec{k} + \vec{q} - \eta \vec{q}_1} \leq \Lambda$. To make this restricted summation more apparent, we can rewrite the density operators in terms of $\delta \vec{k} = \vec{k} - \eta \vec{q}_1$ as
\begin{equation}
	\label{si:eqn:proj_rho_b}
	\overline{\delta \rho^{\hat{b}}}_{\vec{G}+\vec{q}} = \sum_{\eta, s} \sum_{\delta \vec{k}} \sum_{\abs{n},\abs{m} = 1} M^{\hat{b},\eta}_{mn} \left(\delta \vec{k}+\eta \vec{q}_1,\vec{q}+\vec{G} \right) \left(\cre{b}{\delta \vec{k}+\eta \vec{q}_1+\vec{q},m,\eta,s} \des{b}{\delta \vec{k}+\eta \vec{q}_1,n,\eta,s} - \frac{1}{2} \delta_{\vec{q},\vec{0}} \delta_{m,n} \right).
\end{equation}
In \cref{si:eqn:proj_rho_b} and in the following equations involving the form factors of the mirror-antisymmetryic operators, the constraint $\abs{\delta\vec{k}},\abs{\delta\vec{k} + \vec{q}} \leq \Lambda$ is implicit. 

\subsubsection{Projected interaction Hamiltonian}

Following the notation of Ref.~\cite{BER20a}, we define a set of new operators
\begin{equation}
	\label{si:eqn:oqg_definition}
	O^{\hat{f}}_{\vec{q},\vec{G}} = \sqrt{V\left( \vec{q} + \vec{G} \right)} \, \overline{\delta \rho^{\hat{f}}}_{\vec{G}+\vec{q}},
\end{equation}
for $\cre{f}{}=\cre{b}{},\cre{c}{}$. This allows us to write the projected interaction Hamiltonian $H_{I}$ in terms of the $O^{\hat{c}}_{\vec{q},\vec{G}}$ and $O^{\hat{b}}_{\vec{q},\vec{G}}$ operators, corresponding to the original mirror-symmetric and mirror-anitsymmetric operators. It simply reads
\begin{equation}
	\label{si:eqn:proj_int_hamiltonian}
	H_{I} = \frac{1}{2 \Omega_{\mathrm{tot}}} \sum_{\vec{q} \in \mathrm{MBZ}} \sum_{\vec{G} \in \mathcal{Q}_0} \left( O^{\hat{c}}_{-\vec{q},-\vec{G}} + O^{\hat{b}}_{-\vec{q},-\vec{G}} \right) \left( O^{\hat{c}}_{\vec{q}, \vec{G}} + O^{\hat{b}}_{\vec{q}, \vec{G}} \right).
\end{equation}
Because $O^{\dagger \hat{f}}_{\vec{q},\vec{G}} =O^{\hat{f}}_{-\vec{q},-\vec{G}}$, for any $ \hat{f} = \hat{b}, \hat{c}$, the projected interaction Hamiltonian from \cref{si:eqn:proj_int_hamiltonian} is a positive semidefinite operator.

It is important to note that only the Dirac fermions in a fraction of the MBZ (\ie $\cre{b}{\vec{k},n,\eta,s}$ with $n = \pm 1$ and $\vec{k} \in \mathcal{A}_{\eta} $) contribute to the interaction Hamiltonian. Using this fact, the expression for $H_{I}$ given in \cref{si:eqn:proj_int_hamiltonian} can be further simplified. To see this, one must first note that the \emph{projected} relative density operators corresponding to the mirror-antisymmetric sector can be equivalently written as 
\begin{equation}
	\label{si:eqn:proj_density_dirac_alternative}
	\overline{\delta\rho^{\hat{b}}}_{\vec{G} + \vec{q}} = \sum_{\eta, \alpha, s} \sum_{i,j}\sum_{\substack{\vec{k} \in A^i_{\eta} \\ \left(\vec{k} - \vec{q} \right) \in A^j_{\eta}}} \left( \cre{b}{\vec{k},\eta \vec{q}_i ,\alpha, s, l} \des{b}{\vec{k} - \vec{q}, \eta \vec{q}_j ,\alpha, s, l} \delta_{\vec{G},\eta \left( \vec{q}_j - \vec{q}_i \right)} - \frac{1}{2} \delta_{\vec{q},\vec{0}} \delta_{\vec{G},\vec{0}} \right),
\end{equation}
where we are summing over those values of $\vec{k}$ in MBZ where both $\vec{k}$ and $\vec{k}- \vec{q}$ lie inside the zone $\mathcal{A}_{\eta}$. For $\vec{q}$ inside the first MBZ, for a small enough momentum cutoff $\Lambda$, it follows that $\overline{\delta\rho^{\hat{b}}}_{\vec{G} + \vec{q}}$ vanishes, unless $\vec{G} = \vec{0}$, a statement which we prove below. 

We start with the conditions $\vec{k} \in A^i_{\eta}$ and $\left(\vec{k} - \vec{q} \right) \in A^j_{\eta}$ which imply that 
\begin{equation}
	\abs{\vec{k}-\eta \vec{q}_i} \leq \Lambda \qquad \text{and} \quad
	\abs{\vec{k} + \vec{q} - \eta \vec{q}_j} \leq \Lambda.
\end{equation}
Writing $\vec{q}$ as 
\begin{equation}
	\vec{q} = \left(\vec{k} + \vec{q} - \eta \vec{q}_j \right) - \left( \vec{k}-\eta \vec{q}_i \right)+\eta \left( \vec{q}_j - \vec{q}_i \right) = \left(\vec{k} + \vec{q} - \eta \vec{q}_j \right) - \left( \vec{k}-\eta \vec{q}_i \right)+ \vec{G},
\end{equation} 
we can find that its modulus squared is given by 
\begin{equation}
	q^2 = \left(\vec{k} + \vec{q} - \eta \vec{q}_j \right)^2 + \left( \vec{k}-\eta \vec{q}_i \right)^2 + G^2 + 2 \vec{G} \cdot \left(\vec{k} + \vec{q} - \eta \vec{q}_j \right) - 2 \vec{G} \cdot \left( \vec{k}-\eta \vec{q}_i \right) - 2\left( \vec{k}-\eta \vec{q}_i \right) \cdot \left(\vec{k} + \vec{q} - \eta \vec{q}_j \right). 
\end{equation}
Using the following inequalities 
\begin{equation}
	\begin{split}
		2 \vec{G} \cdot \left(\vec{k} + \vec{q} - \eta \vec{q}_j \right) &\geq - 2 G \Lambda, \\
		- 2 \vec{G} \cdot \left( \vec{k}-\eta \vec{q}_i \right) &\geq - 2 G \Lambda, \\
		- 2\left( \vec{k}-\eta \vec{q}_i \right) \cdot \left(\vec{k} + \vec{q} - \eta \vec{q}_j \right) &\geq - 2 \Lambda^2, \\
	\end{split}
\end{equation}
we deduce that 
\begin{equation}
	\label{si:eqn:ineq_rec_vec}
	q^2 \geq G^2 - 4 G \Lambda - 2 \Lambda^2.
\end{equation}
Given that $\vec{q} \in \mathrm{MBZ}$, $q^2 \leq 1$ and so \cref{si:eqn:ineq_rec_vec} can only be satisfied if $G^2 - 4 G \Lambda - 2 \Lambda^2 \leq 1$, which is trivially satisfied if $\vec{G} = \vec{0}$. However, if $\vec{G}$ is the smallest non-zero reciprocal lattice vector, $G = \sqrt{3}$, the inequality $G^2 - 4 G \Lambda - 2 \Lambda^2 \leq 1$ can only be satisfied for $\Lambda \geq 2 - \sqrt{3}$. Since we assume the cutoff $\Lambda$ to be smaller than 0.2 (for typical values of $\Lambda$, see \cref{fig:eigenstateSupprt}), this leads to a contradiction. We thus find that the projected density operators $\overline{\delta\rho^{\hat{b}}}_{\vec{G} + \vec{q}}$ vanish unless $\vec{G} = 0$. In addition, the condition $\vec{G} = 0$ implies that we must have $\abs{\vec{q}} \lesssim \Lambda$ for $\overline{\delta\rho^{\hat{b}}}_{\vec{G} + \vec{q}}$ to be non-vanishing. Consequently, we find that  
\begin{equation}
	O^{\hat{b}}_{\vec{q},\vec{G}} = O^{\hat{b}}_{\vec{q},\vec{0}} \delta_{\vec{G},\vec{0}}.
\end{equation}

The projected interaction Hamiltonian can thus be written as 
\begin{equation}
	H_{I} = H_{I,\mathrm{TBG}} + H_{I,D} + H_{I,\mathrm{TBG}-D}, 
\end{equation}
where the first term denotes the same projected interaction Hamiltonian that appears in ordinary TBG~\cite{BER20a},
\begin{equation}
	 H_{I,\mathrm{TBG}} = \frac{1}{2\Omega_{\mathrm{tot}}} \sum_{\vec{G} \in \mathcal{Q}_0} \sum_{\vec{q} \in \mathrm{MBZ}} O^{\hat{c}}_{-\vec{q},-\vec{G}} O^{\hat{c}}_{\vec{q},\vec{G}},
\end{equation}
while 
\begin{equation}
	 H_{I,D} = \frac{1}{2\Omega_{\mathrm{tot}}} \sum_{\vec{q} \in \mathrm{MBZ}} O^{\hat{b}}_{-\vec{q},\vec{0}} O^{\hat{b}}_{\vec{q},\vec{0}}
\end{equation}
represents the projected interaction Hamiltonian for the Dirac fermions. Additionally, 
\begin{equation}
	H_{I,\mathrm{TBG}-D} = \frac{1}{2\Omega_{\mathrm{tot}}} \sum_{\vec{q} \in \mathrm{MBZ}} \left( O^{\hat{c}}_{-\vec{q},\vec{0}} O^{\hat{b}}_{\vec{q},\vec{0}} + O^{\hat{b}}_{-\vec{q},\vec{0}} O^{\hat{c}}_{\vec{q},\vec{0}} \right)
\end{equation} 
denotes the projected interaction between the TBG and Dirac electrons.

\subsection{Gauge-Fixing the $O^{\hat{c}}_{\vec{q},\vec{G}}$ and $O^{\hat{b}}_{\vec{q},\vec{G}}$ Operators}
\label{app:interaction:gauge}
The operators $O^{\hat{c}}_{\vec{q},\vec{G}}$ have been introduced in Ref.~\cite{BER20a} for TBG, where the procedure used to gauge-fix their form (through the form factors $M^{\hat{c}}$) was also thoroughly explained.  Here, we will focus on gauge-fixing the $O^{\hat{b}}_{\vec{q},\vec{G}}$ operators (through the form factors $M^{\hat{b}}$) and briefly summarize the results of the gauge-fixing procedure used in Ref.~\cite{BER20a} for the $O^{\hat{c}}_{\vec{q},\vec{G}}$ operators.

\subsubsection{Gauge-fixing the $M^{\hat{b}} \left( \vec{k},\vec{q}+\vec{Q} \right)$ form factors} \label{app:interaction:gauge:b}

To fix the exact form of the coefficients $M^{\hat{b},\eta}_{mn} \left( \vec{k},\vec{q}+\vec{G} \right)$, for $\abs{m}, \abs{n} = 1$, we impose a series of constraints arising from the symmetries defined in Appendix \ref{app:symmetries}, as well as from the specific gauge choices of Appendix \ref{app:gauge}. More precisely, for each $\delta \vec{k}$-preserving symmetry with a $\delta \vec{k}$-independent sewing matrix, the form factors will satisfy certain commutation relations, similarly to the single-particle terms in Appendix \ref{app:gauge_single_part:displacement}. The resulting parameterizations of the form factors will prove instrumental in deriving the continuous symmetries of the many-body TSTG Hamiltonian in Appendix \ref{app:fullSym}.
\begin{enumerate}
	\item The $C_{2z} T$ symmetry, which has the $\delta \vec{k}$-independent sewing matrix $B_{\hat{b}}^{C_{2z}T} = \zeta^0 \tau^0$ imposes the real condition. Namely, we must have that
	\begin{align}
		&M_{mn}^{\hat{b},\eta}\left(\delta \vec{k} + \eta \vec{q}_1,\vec{q}+\vec{G}\right) 
= \sum_{\alpha} \sum_{\vec{Q} \in \mathcal{Q}_{\eta}} u^{*\hat{b}}_{\vec{Q}-\vec{G} \alpha; m \eta} \left( \delta \vec{k} +\eta \vec{q}_1 + \vec{q} \right) u^{\hat{b}}_{\vec{Q} \alpha; n \eta} \left( \delta \vec{k} +\eta \vec{q}_1 \right) \nonumber \\
=& \sum_{\mathbf{Q}\in\mathcal{Q}_{\eta}} \sum_{\alpha,\eta_1,\eta_2} 
\left[u^{\dagger\hat{b}}_{m\eta_1} \left(\delta \vec{k} +\eta_1 \vec{q}_1 + \vec{q} \right) \left[D^\dagger \left( C_{2z} T \right) \right]_{\eta \eta_1} \right]_{\vec{Q}-\vec{G} \alpha} 
\left[\left[D \left( C_{2z} T \right) \right]_{\eta \eta_2} u^{\hat{b}}_{n\eta_2}\left(\delta \vec{k} +\eta_2 \vec{q}_1\right)\right]_{\vec{Q} \alpha} \nonumber\\
=& \sum_{\substack{\alpha,\eta_1,\eta_2 \\ m',n'}} \left[B^{\dagger C_{2z} T}_{\hat{b}} \right]_{m \eta,m' \eta_1} \sum_{\vec{Q} \in \mathcal{Q}_{\eta_1}} u^{\hat{b}}_{\vec{Q}-\vec{G} \alpha; m' \eta_1} \left( \delta \vec{k} +\eta_1 \vec{q}_1 + \vec{q} \right) u^{*\hat{b}}_{\vec{Q} \alpha; n' \eta_2} \left( \delta \vec{k} +\eta_2 \vec{q}_1 \right) \left[B^{C_{2z} T}_{\hat{b}} \right]_{n' \eta_2,n \eta}. \label{si:eqn:b_form_gauge_C2T}
	\end{align}
	It then follows that the form factor matrix elements are real 
	\begin{equation}
		\label{si:eqn:b_form_matrix_gauge_C2T}
		\begin{split}
			M_{mn}^{\hat{b},\eta}\left(\delta \vec{k} + \eta \vec{q}_1,\vec{q}+\vec{G}\right) &= \sum_{m', n', \eta' }
			\left(\zeta^0 \tau^0 \right)_{m \eta, m' \eta'}
			M_{m'n'}^{*\hat{b},\eta'}\left(\delta \vec{k} + \eta' \vec{q}_1,\vec{q}+\vec{G}\right) 
			\left(\zeta^0 \tau^0 \right)_{n' \eta', n \eta} \\
			&= 	M_{mn}^{*\hat{b},\eta}\left(\delta \vec{k} + \eta \vec{q}_1,\vec{q}+\vec{G}\right).
		\end{split}  
	\end{equation}
	\item Due to the chiral transformation $C$, which has the $\delta \vec{k}$-independent sewing matrix $B_{\hat{b}}^C = \zeta^y \tau^z $ in the pair of bands $n = \pm 1$, we must have
	\begin{align}
		&M_{mn}^{\hat{b},\eta}\left(\delta \vec{k} + \eta \vec{q}_1,\vec{q}+\vec{G}\right) 
= \sum_{\alpha} \sum_{\vec{Q} \in \mathcal{Q}_{\eta}} u^{*\hat{b}}_{\vec{Q}-\vec{G} \alpha; m \eta} \left( \delta \vec{k} +\eta \vec{q}_1 + \vec{q} \right) u^{\hat{b}}_{\vec{Q} \alpha; n \eta} \left( \delta \vec{k} +\eta \vec{q}_1 \right) \nonumber \\
=& \sum_{\mathbf{Q}\in\mathcal{Q}_{\eta}} \sum_{\alpha,\eta_1,\eta_2} 
\left[u^{\dagger\hat{b}}_{m\eta_1} \left(\delta \vec{k} +\eta_1 \vec{q}_1 + \vec{q} \right) \left[D^\dagger \left( C \right) \right]_{\eta \eta_1} \right]_{\vec{Q}-\vec{G} \alpha} 
\left[\left[D \left( C \right) \right]_{\eta \eta_2} u^{\hat{b}}_{n\eta_2}\left(\delta \vec{k} +\eta_2 \vec{q}_1\right)\right]_{\vec{Q} \alpha} \nonumber\\
=& \sum_{\substack{\alpha,\eta_1,\eta_2 \\ m',n'}} \left[B^{\dagger C}_{\hat{b}} \right]_{m \eta,m' \eta_1}\sum_{\alpha} \sum_{\vec{Q} \in \mathcal{Q}_{\eta_1}} u^{*\hat{b}}_{\vec{Q}-\vec{G} \alpha; m' \eta_1} \left( \delta \vec{k} +\eta_1 \vec{q}_1 + \vec{q} \right) u^{\hat{b}}_{\vec{Q} \alpha; n' \eta_2} \left( \delta \vec{k} +\eta_2 \vec{q}_1 \right) \left[B^C_{\hat{b}} \right]_{n' \eta_2,n \eta}, 
	\end{align}
	which can be rewritten as 
	\begin{equation}
		\label{si:eqn:b_form_gauge_C}
		M_{mn}^{\hat{b},\eta}\left(\delta \vec{k} + \eta \vec{q}_1,\vec{q}+\vec{G}\right) = \sum_{m', n', \eta' }
		\left( \zeta^y \tau^z \right)_{m \eta, m' \eta'}
		M_{m'n'}^{\hat{b},\eta'}\left(\delta \vec{k} + \eta' \vec{q}_1,\vec{q}+\vec{G}\right) 
		\left( \zeta^y \tau^z \right)_{n' \eta', n \eta}.
	\end{equation}	
	\item Finally, the combination transformation $C_{2z} L$ which has a $\delta \vec{k}$-independent sewing matrix $B_{\hat{b}}^{C_{2z}L}=\zeta^y \tau^y$ imposes the following condition on the form factor matrices
	\begin{align}
		&M_{mn}^{\hat{b},\eta}\left(\delta \vec{k}+ \eta \vec{q}_1,\vec{q}+\vec{G}\right) 
= \sum_{\alpha} \sum_{\vec{Q} \in \mathcal{Q}_{\eta}} u^{*\hat{b}}_{\vec{Q}-\vec{G} \alpha; m \eta} \left( \delta \vec{k} +\eta \vec{q}_1 + \vec{q} \right) u^{\hat{b}}_{\vec{Q} \alpha; n \eta} \left( \delta \vec{k} +\eta \vec{q}_1 \right) \nonumber \\
=& \sum_{\mathbf{Q}\in\mathcal{Q}_{\eta}} \sum_{\alpha,\eta_1,\eta_2} 
\left[u^{\dagger\hat{b}}_{m\eta_1} \left(\delta \vec{k} +\eta_1 \vec{q}_1 + \vec{q} \right) \left[D^\dagger \left(C_{2z}L \right) \right]_{\eta \eta_1} \right]_{\vec{Q}-\vec{G} \alpha} 
\left[\left[D \left(C_{2z}L \right) \right]_{\eta \eta_2} u^{\hat{b}}_{n\eta_2}\left(\delta \vec{k} +\eta_2 \vec{q}_1\right)\right]_{\vec{Q} \alpha} \nonumber\\
=& \sum_{\substack{\alpha,\eta_1,\eta_2 \\ m',n'}} \left[B^{\dagger C_{2z}L }_{\hat{b}} \right]_{m \eta,m' \eta_1} \sum_{\vec{Q} \in \mathcal{Q}_{\eta_1}} u^{*\hat{b}}_{\vec{Q}-\vec{G} \alpha; m' \eta_1} \left( \delta \vec{k} +\eta_1 \vec{q}_1 + \vec{q} \right) u^{\hat{b}}_{\vec{Q} \alpha; n' \eta_2} \left( \delta \vec{k} +\eta_2 \vec{q}_1 \right) \left[B^{C_{2z}L}_{\hat{b}} \right]_{n' \eta_2,n \eta},\label{si:eqn:b_form_gauge_C2zL} 
	\end{align}
	which requires that
	\begin{equation}
		\label{si:eqn:b_form_matrix_gauge_C2zL}
		M_{mn}^{\hat{b},\eta}\left(\delta \vec{k} + \eta \vec{q}_1,\vec{q}+\vec{G}\right) = \sum_{m', n', \eta' }
		\left( \zeta^y \tau^y \right)_{m \eta, m' \eta'}
		M_{m'n'}^{\hat{b},\eta'}\left(\delta \vec{k} + \eta' \vec{q}_1,\vec{q}+\vec{G}\right) 
		\left( \zeta^y \tau^y \right)_{n' \eta', n \eta}.
	\end{equation}	
\end{enumerate} 
As a direct product of $2 \times 2$ matrices in valley and spin space, we can generically parameterize the form factors as 
\begin{equation}
	\label{si:eqn:form_factor_exp}
	M^{ \hat{b},\eta}_{mn} \left(\delta \vec{k} + \eta \vec{q}_1,\vec{q}+\vec{G} \right) = \sum_{\substack{a \in \left\lbrace 0,x,y,z \right\rbrace \\ d \in \left\lbrace 0,z \right\rbrace}}\left( \zeta^a \tau^d \right)_{m\eta,n\eta} \alpha^{\hat{b}}_{ad}\left(\delta \vec{k},\vec{q}+\vec{G} \right),
\end{equation}
where only $d=0,z$ are allowed since $M^{\hat{b},\eta}_{mn}\left(\delta \vec{k} + \eta \vec{q}_1,\vec{q}+\vec{G} \right)$ is diagonal in valley space, and $\alpha_{cd} \left(\delta \vec{k},\vec{q}+\vec{G} \right)$ represent generic complex functions. At the same time, \cref{si:eqn:b_form_matrix_gauge_C2T,si:eqn:b_form_gauge_C,si:eqn:b_form_matrix_gauge_C2zL} impose a series of constraints on the form factors which restrict the number of terms allowed in the parameterization from \cref{si:eqn:form_factor_exp}. More precisely, the matrix $M^{\hat{b}} \left(\delta \vec{k} + \eta \vec{q}_1,\vec{q}+\vec{G}\right)$ turns out to be a sum of only two terms 
\begin{equation}
	\label{si:eqn:gen_form_b_M}
	M^{ \hat{b},\eta}_{mn} \left(\delta \vec{k} + \eta \vec{q}_1,\vec{q}+\vec{G} \right) = \left( \zeta^0 \tau^0 \right)_{m\eta,n\eta} \alpha^{\hat{b}}_0 \left(\delta \vec{k},\vec{q}+\vec{G}\right) + i \left( \zeta^y \tau^0 \right)_{m\eta,n\eta} \alpha^{\hat{b}}_1 \left(\delta \vec{k},\vec{q}+\vec{G}\right),
\end{equation}
where $\alpha_j\left(\delta \vec{k},\vec{q}+\vec{G}\right)$ (for $j=0,1$) represent two real functions.

\subsubsection{Gauge-fixing the $M^{\hat{c}} \left( \vec{k},\vec{q}+\vec{Q} \right)$ form factors} \label{app:interaction:gauge:c}

The gauge-fixing of the form factors for to the mirror-symmetric sector was fully detailed in Ref.~\cite{BER20a}. The single-particle wave functions $u^{\hat{c}}_{\vec{Q} \alpha; n \eta} \left( \vec{k} \right)$ at a given $\vec{k}$ point can be related by the $\vec{k}$-independent sewing matrices of the combined symmetry operators $C_{2z} P$ and $C_{2z} T$. The gauge-fixing from \cref{si:eqn:sewing_mats_TBG} restrict the form factors to the following parameterization in the band and valley subspaces~\cite{BER20a}
\begin{equation}
	\label{si:eqn:gen_form_c_M_nonchiral}
	M^{\hat{c}} \left(\vec{k},\vec{q}+\vec{G}\right) = \zeta^0\tau^0 \alpha^{\hat{c}}_0 \left(\vec{k},\vec{q}+\vec{G}\right) + \zeta^x\tau^z \alpha^{\hat{c}}_1 \left(\vec{k},\vec{q}+\vec{G}\right) + i\zeta^y\tau^0 \alpha^{\hat{c}}_2 \left(\vec{k},\vec{q}+\vec{G}\right) + \zeta^z\tau^z \alpha^{\hat{c}}_3 \left(\vec{k},\vec{q}+\vec{G}\right),
\end{equation}
where $\alpha^{\hat{c}}_j \left( \vec{k},\vec{q}+\vec{G} \right)$ (for $j=0,1,2,3$) are all real function. 

Furthermore, in the (first) chiral limit $w_0=0$, the single-particle wave functions $u^{\hat{c}}_{\vec{Q} \alpha; n \eta} \left( \vec{k} \right)$ at a given $\vec{k}$ can additionally be related by the $\vec{k}$-independent sewing matrix of the chiral symmetry operator $C$. This implies that the form factors will be further restricted to the parameterization
\begin{equation}
	\label{si:eqn:gen_form_c_M_chiral}
	M^{\hat{c}} \left(\vec{k},\vec{q}+\vec{G}\right) = \zeta^0\tau^0 \alpha^{\hat{c}}_0 \left(\vec{k},\vec{q}+\vec{G}\right) + i\zeta^y\tau^0 \alpha^{\hat{c}}_2 \left(\vec{k},\vec{q}+\vec{G}\right).
\end{equation}

\subsection{Hartree-Fock Potential in the projected interaction Hamiltonian}\label{app:interaction:HF}
In Appendix \ref{app:interaction:projection}, we have derived the projected interaction Hamiltonian starting from \cref{si:eqn:interaction_density} without first normal-ordering. We also noted in Appendix \ref{app:interaction:derivation} that the normal-ordered and the non-normal-ordered forms of the \emph{unprojected} interaction Hamiltonian only differ by a chemical potential term. Here, we show that the \emph{projected} interaction Hamiltonian $H_I$ is \emph{not} equivalent to its normal ordered form up to a redefinitation of the chemical potential. Instead, the projected interaction Hamiltonian is the sum between its normal-ordered form $\normord{H_I}$ and a single-particle Hamiltonian $H_{\mathrm{HF}}$, which can be understood as the electron potential from the remote bands which are projected away 
\begin{equation}
	\label{si:eqn:HF:normOrdDif}
	H_I = \normord{H_I} + H_{\mathrm{HF}} + \mathrm{const.}
\end{equation}

We will now show that $H_{\mathrm{HF}}$ can be thought of as an effective background Hartree-Fock potential. The proof is similar to the one given in Ref.~\cite{BER20a} for TBG. However, because we are dealing with two fermion flavors, we will find it easier to employ a different notation that treats the single-particle Dirac cone and TBG eigenstates on equal footing. We let $\cre{f}{i}$ denote the creation operator for some TBG or Dirac cone energy band eigenstate ($\cre{c}{\vec{k}, n, \eta, s}$ or $\cre{b}{\vec{k}, n, \eta, s}$) with corresponding single-particle energy $\epsilon_i$. The momentum, band, valley, spin and fermion flavor are encoded in the index $i$, which is chosen such that $\epsilon_i \leq \epsilon_j$ for any $i<j$. Projecting the TSTG Hamiltonian becomes equivalent to only considering those fermion operators whose single-particle energy lies within a certain interval.

When written in terms of the new operators, the projected interaction Hamiltonian is given by 
\begin{equation}
	\label{si:eqn:HF:projInteraction}
	H_I = \frac{1}{2 \Omega_{\mathrm{tot}}}\sum_{\substack{i,j,m,n \\ N_- \leq i,j,m,n \leq N_+ \\ \vec{G} \in \mathcal{Q}_0}} \left[ \mathcal{M}^{*\vec{G}}_{ij} \left( \cre{f}{j} \des{f}{i} - \frac{1}{2} \delta_{ij} \right) \mathcal{M}^{\vec{G}}_{mn} \left( \cre{f}{m} \des{f}{n} - \frac{1}{2} \delta_{mn} \right) \right]. 
\end{equation} 
where the matrix elements $\mathcal{M}^{\vec{G}}_{ij}$ are defined as 
\begin{equation}
	\mathcal{M}^{\vec{G}}_{ij} = \sqrt{V \left( \vec{q} + \vec{G} \right)} M ^{\hat{f},\eta}_{mn} \left( \vec{k}, \vec{q} + \vec{G} \right) \delta_{s_1,s_2} \delta_{\eta,\eta'},
\end{equation}
where the indices $i$ and $j$ are such that $\cre{f}{i} = \cre{f}{\vec{k}+\vec{q},m,\eta,s_1}$ and $\des{f}{j} = \des{f}{\vec{k},n,\eta',s_2}$, $\vec{G}$ is a reciprocal vector. The projection in \cref{si:eqn:HF:projInteraction} is implemented by restricting the fermion indices to lie between $N_-$ and $N_+$, which respectively denote the index of the lowest- and highest-energy single-particle eigenstates included in the projection. We note that the \emph{unprojected} interaction Hamiltonian has the same form as \cref{si:eqn:HF:projInteraction}, but without imposing any restrictions on the fermion indices
\begin{equation}
	\label{si:eqn:HF:Interaction}
	\hat{H}_I = \frac{1}{2 \Omega_{\mathrm{tot}}} \sum_{\substack{ i,j,m,n  \\ \vec{G} \in \mathcal{Q}_0} } \left[ \mathcal{M}^{*\vec{G}}_{ij} \left( \cre{f}{j} \des{f}{i} - \frac{1}{2} \delta_{ij} \right) \mathcal{M}^{\vec{G}}_{mn} \left( \cre{f}{m} \des{f}{n} - \frac{1}{2} \delta_{mn} \right) \right]. 
\end{equation}

We will now derive the Hartree-Fock contribution arising from $\hat{H}_I$ by fully filling all the energy eigenstates indexed by $i$, with $i \leq N$. The filled states give rise to a mean-field
\begin{equation}
	\label{si:eqn:HF:meanfield}
	\expec{\cre{f}{i} \des{f}{j}} = \heaviside{N-i} \delta_{ij},
\end{equation}
where $\heaviside{x}$ is the Heaviside step function. The Hartree term arising from this filling is simply given by
\begin{equation}
	\hat{H}_{\mathrm{H}}^{N} = \frac{1}{2 \Omega_{\mathrm{tot}}} \sum_{\substack{i,m,n \\ \vec{G} \in \mathcal{Q}_0}} \left( \mathcal{M}^{*\vec{G}}_{ii} \mathcal{M}^{\vec{G}}_{mn} \cre{f}{m} \des{f}{n} + \mathcal{M}^{*\vec{G}}_{mn} \mathcal{M}^{\vec{G}}_{ii} \cre{f}{n} \des{f}{m} \right) \heaviside{N-i} .
\end{equation} 
The mean-field from \cref{si:eqn:HF:meanfield} also gives rise to a quadratic Fock contribution
\begin{equation}
	\hat{H}_{\mathrm{F}}^{N} =  - \frac{1}{2 \Omega_{\mathrm{tot}}} \sum_{\substack{i,m,n  \\ \vec{G} \in \mathcal{Q}_0}} \left( \mathcal{M}^{*\vec{G}}_{im} \mathcal{M}^{\vec{G}}_{in} \cre{f}{m} \des{f}{n} + \mathcal{M}^{*\vec{G}}_{ni} \mathcal{M}^{\vec{G}}_{mi} \cre{f}{m} \des{f}{n} \right)\heaviside{N-i}.
\end{equation}
We now project the Hartree-Fock contributions arising from the partial filling from \cref{si:eqn:HF:meanfield} to the active energy modes with fermion indices between $N_-$ and $N_+$ (\ie the TSTG eigenstates in which we project the interaction Hamiltonian). The resulting Hartree and Fock potentials read
\begin{align}
	H_{\mathrm{H}}^{N} &= \frac{1}{2 \Omega_{\mathrm{tot}}} \sum_{\substack{i,m,n \\ N_- \leq m,n \leq N_+ \\ \vec{G} \in \mathcal{Q}_0}} \left( \mathcal{M}^{*\vec{G}}_{ii} \mathcal{M}^{\vec{G}}_{mn} \cre{f}{m} \des{f}{n} + \mathcal{M}^{*\vec{G}}_{mn} \mathcal{M}^{\vec{G}}_{ii} \cre{f}{n} \des{f}{m} \right) \heaviside{N-i}, \label{si:eqn:HF:projectedPotentialsH} \\
	H_{\mathrm{F}}^{N} &=  - \frac{1}{2 \Omega_{\mathrm{tot}}} \sum_{\substack{i,m,n \\ N_- \leq m,n \leq N_+  \\ \vec{G} \in \mathcal{Q}_0}} \left( \mathcal{M}^{*\vec{G}}_{im} \mathcal{M}^{\vec{G}}_{in} \cre{f}{m} \des{f}{n} + \mathcal{M}^{*\vec{G}}_{mi} \mathcal{M}^{\vec{G}}_{ni} \cre{f}{n} \des{f}{m} \right)\heaviside{N-i}. \label{si:eqn:HF:projectedPotentialsF}	
\end{align}

We now turn our attention to the projected interaction Hamiltonian $H_I$, which can be written as the sum between its normal-ordered form, a quadratic part and a constant, as seen in \cref{si:eqn:HF:normOrdDif}. The quadratic part (which has been denoted $H_{\mathrm{HF}} = H_I - \normord{H_I}$ in anticipation of the results of this section) can be written (up to a constant term) as 
\begin{equation}
		H_{\mathrm{HF}} = \frac{1}{4 \Omega_{\mathrm{tot}}} \sum_{\substack{i,m,n \\ N_- \leq i,m,n \leq N_+ \\ \vec{G} \in \mathcal{Q}_0}} \left[ \left( \mathcal{M}^{*\vec{G}}_{ii} \mathcal{M}^{\vec{G}}_{mn} - \mathcal{M}^{*\vec{G}}_{im} \mathcal{M}^{\vec{G}}_{in} \right) \cre{f}{m} \des{f}{n} + \left( \mathcal{M}^{*\vec{G}}_{mn} \mathcal{M}^{\vec{G}}_{ii} - \mathcal{M}^{*\vec{G}}_{mi} \mathcal{M}^{\vec{G}}_{ni} \right) \cre{f}{n} \des{f}{m} \right].
\end{equation}
Employing \cref{si:eqn:HF:projectedPotentialsH,si:eqn:HF:projectedPotentialsF}, we find that the quadratic part is indeed an effective background Hartree-Fock potential arising from the bands that have been projected away, \ie 
\begin{equation}
	H_{\mathrm{HF}} = \frac{1}{2} \left[ H_{\mathrm{H}}^{N_+} + H_{\mathrm{F}}^{N_+} - \left( H_{\mathrm{H}}^{N_-} + H_{\mathrm{F}}^{N_-} \right) \right],
\end{equation}
thus completing the proof. Finally, we note that the effective Hartree-Fock potential $H_{\mathrm{HF}}$ is crucial in proving the charge-conjugation symmetry of the projected many-body Hamiltonian in Appendix \ref{app:fullSym:chargeConj}. 

\section{Symmetries of the projected many-body Hamiltonian}\label{app:fullSym}

In this appendix, we discuss the symmetries of the many-body TSTG projected Hamiltonian $H$ defined in \cref{eqn:proj_many_body} in different physically-relevant limits. We commence by showing that $H$ inherits a spatial many-body charge conjugation symmetry from the single-graphene layers. Next, we show that the many-body TSTG projected Hamiltonian enjoys enlarged continuous symmetries in various limits of interest. In the absence of displacement field, the fermion flavors corresponding to the two different mirror-symmetry sectors are decoupled at the single-particle level allowing us to discuss the continuous symmetries of each flavor independently. We conclude this appendix by showing that a nonzero displacement field breaks the symmetries of the system to the trivial $\UN{2} \times \UN{2}$ spin-valley rotation symmetry.

\subsection{Spatial many-body charge conjugation symmetry}\label{app:fullSym:chargeConj}

The full projected Hamiltonian $H=H_0+H_I$ has a spatial many-body charge-conjugation symmetry $\mathcal{P}$, which ensures that all the physical phenomena are particle-hole symmetric about the charge neutral point. Here, we define this spatial many-body charge-conjugation symmetry transformation and prove explicitly that, up to a constant, it leaves the projected TSTG many-body Hamiltonian invariant.

\subsubsection{Definition}
We define the spatial many-body charge conjugation operation $\mathcal{P}$ as the combined antiunitary single-particle transformation 
\begin{equation}
	\mathcal{U} \equiv m_z C_{2x} C_{2z} T P
\end{equation}
followed by an interchange between fermion creation and annihilation operators. Its action on the energy band operators $\hat{f}=\hat{b},\hat{c}$
\begin{equation}
	\label{si:eqn:action_MB_PH}
	\mathcal{P} \cre{f}{\vec{k},n,\eta,s} \mathcal{P}^{-1} = \sum_{n',\eta'} \left[ B^{\mathcal{U}}_{\hat{f}} \left( \vec{k} \right) \right]_{n' \eta', n \eta} \des{f}{-C_{2x} \vec{k},n',\eta',s}, \quad
	\mathcal{P} \des{f}{\vec{k},n,\eta,s} \mathcal{P}^{-1} = \sum_{n',\eta'} \left[ B^{\mathcal{U}}_{\hat{f}} \left( \vec{k} \right) \right]^{*}_{n' \eta', n \eta} \cre{f}{-C_{2x} \vec{k},n',\eta',s}.
\end{equation}
The representation matrices for the combined single-particle transformation $\mathcal{U}$ obey
\begin{equation}
	D^{\hat{f}} \left( \mathcal{U} \right) = D^{\hat{f}} \left( C_{2x}P \right) D^{\hat{f}} \left( m_z \right) D^{\hat{f}} \left( C_{2z}T \right),
\end{equation}
and, as required by \cref{si:eqn:repSym_C2z,si:eqn:repSym_C2x,si:eqn:repSym_T,si:eqn:repSym_mz}, are given explicitly by  
\begin{equation}
	\begin{split}
		\left[D^{\hat{c}} \left( \mathcal{U} \right) \right]_{\vec{Q}' \eta' \beta, \vec{Q} \eta \alpha} &= \delta_{\vec{Q}', -C_{2x} \vec{Q}} \delta_{\eta',\eta} \delta_{\beta,\alpha} \zeta_{\vec{Q}}, \\
		\left[D^{\hat{b}} \left( \mathcal{U} \right) \right]_{\vec{Q}' \eta' \beta, \vec{Q} \eta \alpha} &= - \delta_{\vec{Q}', -C_{2x} \vec{Q}} \delta_{\eta',\eta} \delta_{\beta,\alpha} \zeta_{\vec{Q}}.
	\end{split}
\end{equation}
At the same time, the corresponding sewing matrices can be found from the relation 
\begin{equation}
	B^{\mathcal{U}}_{\hat{f}} \left( \vec{k} \right) = B^{C_{2x}P}_{\hat{f}} \left( \vec{k} \right) B^{m_z}_{\hat{f}} \left( \vec{k} \right) B^{C_{2z}T}_{\hat{f}} \left( \vec{k} \right),
\end{equation}
which, under the gauge-fixing of Appendix \ref{app:gauge}, can be simplified into
\begin{equation}
	\begin{split}
		B^{\mathcal{U}}_{\hat{c}} \left( \vec{k} \right) &= B^{C_{2x}P}_{\hat{c}} \left( \vec{k} \right), \\
		B^{\mathcal{U}}_{\hat{b}} \left( \vec{k} \right) &= -B^{C_{2x}P}_{\hat{b}} \left( \vec{k} \right).
	\end{split}
\end{equation}
In what follows, we will not explicitly fix the sewing matrices for the $\mathcal{U}$ transformation, but note that since $C_{2x} P$ anticommutes with the single-particle Hamiltonians $\hat{H}_{\mathrm{TBG}}$ and $\hat{H}_{D}$ and preserves the valley, the sewing matrices must have the form 
\begin{equation}
    \label{si:eqn:gauge_p_implicit}
	\left[ B^{\mathcal{U}}_{\hat{f}} \left( \vec{k} \right) \right]_{n' \eta',n \eta} = \delta_{\eta,\eta'} \delta_{-n,n'} e^{i \phi_{n' \eta'}^{\hat{f},\mathcal{U}} \left( \vec{k} \right)},
\end{equation}
where $\phi_{n' \eta'}^{\hat{f},\mathcal{U}} \left( \vec{k} \right)$ are gauge-dependent phases, which we will leave unspecified. We now proceed to show that the various terms of the many-body projected TSTG Hamiltonian are symmetric under the spatial many-body charge conjugation symmetry $\mathcal{P}$.
\subsubsection{Spatial many-body charge conjugation symmetry of $H_{\mathrm{TBG}}$ and $H_{D}$}

The single-particle TBG and Dirac cone Hamiltonians anticommute with the antiunitary transformation $\mathcal{U}$, namely 
\begin{equation}
	\label{si:eqn:anticommSingPartU}
	\left \lbrace \mathcal{U}, \hat{H}_{\hat{f}} \right \rbrace = 0,
\end{equation}
for $\hat{f} = \hat{c},\hat{b}$. For the sake of brevity, in \cref{si:eqn:anticommSingPartU}, we have introduced the notation $\hat{H}_{\hat{f}} = \hat{H}_{\mathrm{TBG}}$ for $\hat{f} = \hat{c}$, and $\hat{H}_{\hat{f}} = \hat{H}_{D}$ for $\hat{f} = \hat{b}$. It follows that the single-particle band energies obey 
\begin{equation}
	\epsilon_{n,\eta}^{\hat{f}} \left( \vec{k} \right) = - \epsilon_{-n,\eta}^{\hat{f}} \left( - C_{2x} \vec{k} \right)
\end{equation} 
The action of the spatial many-body charge conjugation operator $\mathcal{P}$ on the projected single-particle contributions $H_{\mathrm{TBG}}$ and $H_{D}$ defined in \cref{eqn:proj_single_part_TBG,eqn:proj_single_part_Dirac} is then given by 
\begin{align}
	\mathcal{P} H_{\hat{f}} \mathcal{P}^{-1} &= \sum_{\vec{k}} \sum_{\substack{\abs{n}=1 \\ \eta,s}} \epsilon_{n,\eta}^{\hat{f}} \left( \vec{k} \right) \sum_{\substack{n_1',\eta_1' \\ n_2',\eta_2'}} \left[ B^{\mathcal{U}}_{\hat{f}} \left( \vec{k} \right) \right]_{n_1' \eta_1', n \eta} \des{f}{-C_{2x} \vec{k},n_1',\eta_1',s} \left[ B^{\mathcal{U}}_{\hat{f}} \left( \vec{k} \right) \right]^{*}_{n'_2 \eta'_2, n \eta} \cre{f}{-C_{2x} \vec{k},n_2',\eta_2',s} \nonumber \\
	&= -\sum_{\vec{k}} \sum_{\substack{\abs{n}=1 \\ \eta,s}} \epsilon_{-n,\eta}^{\hat{f}} \left(-C_{2x} \vec{k} \right) \des{f}{-C_{2x} \vec{k},-n,\eta,s} \cre{f}{-C_{2x} \vec{k},-n,\eta,s} \nonumber\\
	&= H_{\hat{f}} -\sum_{\vec{k}} \sum_{\substack{\abs{n}=1 \\ \eta,s}} \epsilon_{n,\eta}^{\hat{f}} \left( \vec{k} \right) = H_{\hat{f}}, \label{si:eqn:enMBC}
\end{align}
thus proving that $H_{\mathrm{TBG}}$ and $H_{D}$ are invariant under $\mathcal{P}$.
\subsubsection{Spatial many-body charge conjugation symmetry of $H^{\left( \hat{b}\hat{c} \right)}_{U}$ and $H^{\left( \hat{b} \right)}_{U}$}

The displacement field overlap matrix governing the projected displacement field Hamiltonian $H_{U}^{\left( \hat{b} \hat{c} \right)}$ matrix obeys
	\begin{align}
		&N^{\eta}_{mn} \left( \delta \vec{k} \right) = \frac{U}{2} \sum_{\alpha} u^{\hat{b}*}_{\eta \vec{q}_1 \alpha; m \eta} \left( \vec{k}_{\eta} \right) 
		u^{\hat{c}}_{\eta \vec{q}_1 \alpha; n \eta} \left( \vec{k}_{\eta} \right) \nonumber \\
		=&- \frac{U}{2} \sum_{\alpha,\eta_1,\eta_2} \left[ u^{\dagger \hat{b}}_{m \eta_1} \left( \vec{k}_{\eta} \right) \left[D^{\dagger \hat{b}} \left(\mathcal{U} \right) \right]_{\eta \eta_1} \right]_{\eta \vec{q}_1 \alpha} 
		\left[ \left[D^{\hat{c}} \left(\mathcal{U} \right) \right]_{\eta \eta_2} u^{\hat{c}}_{n \eta_2} \left( \vec{k}_{\eta} \right) \right]_{\eta \vec{q}_1 \alpha} \nonumber \\
		=& -\frac{U}{2} \sum_{\substack{\alpha \\ m',n'}} \left[B^{\dagger \mathcal{U}}_{\hat{b}} \left( \vec{k}_{\eta} \right) \right]_{m \eta, m' \eta} u^{\hat{b}}_{\eta \vec{q}_1 \alpha; m' \eta} \left( -C_{2x} \vec{k}_{\eta} \right) 
		 u^{*\hat{c}}_{\eta \vec{q}_1 \alpha; n' \eta} \left( -C_{2x} \vec{k}_{\eta} \right) \left[B^{\mathcal{U}}_{\hat{c}} \left( \vec{k}_{\eta} \right) \right]_{n' \eta,n \eta} \nonumber \\
		=& -\sum_{m',n'} \left[B^{\dagger \mathcal{U}}_{\hat{b}} \left( \vec{k}_{\eta} \right) \right]_{m \eta, m' \eta} N^{*\eta}_{m'n'} \left( -C_{2x} \delta \vec{k} \right) \left[B^{\mathcal{U}}_{\hat{c}} \left( \vec{k}_{\eta} \right) \right]_{n' \eta,n \eta}, \label{si:eqn:singl_part_N_gauge_U1} 
	\end{align}
where $\vec{k}_{\eta} = \delta \vec{k} + \eta \vec{q}_1$. Rearranging \cref{si:eqn:singl_part_N_gauge_U1} we find that  
\begin{equation}
	\label{si:eqn:singl_part_N_gauge_U2}
	\sum_{m,n} \left[B^{\mathcal{U}}_{\hat{b}} \left( \vec{k}_{\eta} \right) \right]_{m' \eta, m \eta} N^{\eta}_{mn} \left( \delta \vec{k} \right) \left[B^{\dagger \mathcal{U}}_{\hat{c}} \left( \vec{k}_{\eta} \right) \right]_{n \eta,n' \eta} = -N^{*\eta}_{m'n'} \left( -C_{2x} \delta \vec{k} \right).
\end{equation}
\Cref{si:eqn:singl_part_N_gauge_U2}, together with the reality condition \cref{si:eqn:singl_part_N_gauge_c2t_mat} implies that the projected displacement field Hamiltonian stays invariant under the spatial many-body charge conjugation transformation, \ie 
\begin{align}
	\mathcal{P} H^{\left( \hat{b}\hat{c} \right)}_U \mathcal{P}^{-1} &=  \sum_{\substack{\eta,s \\ \abs{n},\abs{m} = 1 \\ \abs{\delta \vec{k}_{\eta}} \leq \Lambda}} \sum_{\abs{n'},\abs{m'} = 1} \left\lbrace \left[B^{\mathcal{U}}_{\hat{b}} \left( \vec{k}_{\eta} \right) \right]_{m' \eta, m \eta} N^{\eta}_{mn} \left( \delta \vec{k} \right) \left[B^{\dagger \mathcal{U}}_{\hat{c}} \left( \vec{k}_{\eta} \right) \right]_{n \eta,n' \eta} \des{b}{-C_{2x} \vec{k}_{\eta},m',\eta,s} \cre{c}{ -C_{2x} \vec{k}_{\eta} ,n',\eta,s} + \mathrm{h.c.} \right \rbrace \nonumber \\
	&= \sum_{\substack{\eta,s \\ \abs{n},\abs{m} = 1 \\ \abs{\delta \vec{k}} \leq \Lambda}} \left[ N^{*\eta}_{mn} \left( -C_{2x} \delta \vec{k} \right) \cre{c}{ -C_{2x} \vec{k}_{\eta} ,n,\eta,s} \des{b}{-C_{2x} \vec{k}_{\eta},m,\eta,s} + \mathrm{h.c.} \right] = H^{\left( \hat{b}\hat{c} \right)}_U.\label{eqn:si:projU1_MBC}
\end{align}

Similarly to \cref{si:eqn:singl_part_N_gauge_U1}, the displacement field perturbation matrix obeys 
\begin{align}
		&\mathcal{B}^{\eta}_{n m} \left( \vec{k} \right) = \frac{U^2}{4} \sum_{i=1}^3 \sum_{\alpha,\beta} \frac{u^{*\hat{c}}_{\eta\vec{q}_i \alpha; n \eta} \left( \vec{k} \right) \left[ \left( \vec{k} - \eta \vec{q}_i \right) \pauliVecEta \right]_{\alpha \beta} u^{\hat{c}}_{\eta\vec{q}_i \beta; m \eta} \left( \vec{k} \right) }{\abs{\vec{k} - \eta \vec{q}_i}^2} \nonumber \\
		=&- \frac{U^2}{4} \sum_{i=1}^3 \sum_{\substack{\alpha,\beta \\ \eta_1,\eta_2}} \frac{ \left[ u^{\dagger \hat{c}}_{n \eta_1} \left( \vec{k} \right) \left[D^{ \dagger \hat{c}} \left( \mathcal{U} \right) \right]_{\eta \eta_1} \right]_{\eta \vec{q}'_i \alpha} 
		\left[ \left( \vec{k}' - \eta \vec{q}'_i \right) \pauliVecEta \right]^{*}_{\alpha \beta}
		\left[ \left[D^{\hat{c}}\left( \mathcal{U} \right) \right]_{\eta \eta_2} u^{\hat{c}}_{m \eta_2} \left( \vec{k} \right) \right]_{\eta \vec{q}'_i \beta} }{\abs{\vec{k}' - \eta \vec{q}_i'}^2} \nonumber \\
		=&- \frac{U^2}{4} \sum_{i=1}^3 \sum_{\substack{\alpha,\beta \\ m',n' }} \left[B_{\hat{c}}^{ \dagger \mathcal{U} } \left( \vec{k} \right) \right]_{m \eta, m' \eta} 
		\frac{
		u^{\hat{c}}_{\eta \vec{q}_i \alpha; m' \eta} \left( \vec{k}' \right) 
		\left[ \left( \vec{k}' - \eta \vec{q}_i \right) \pauliVecEta \right]^{*}_{\alpha \beta} 
		u^{*\hat{c}}_{\eta \vec{q}_i \beta; n' \eta} \left( \vec{k}' \right)}
		{\abs{\vec{k}' - \eta \vec{q}_i}^2}
		\left[B_{\hat{c}}^{ \mathcal{U} } \left( \vec{k} \right) \right]_{n' \eta,n \eta},
	\label{si:eqn:singl_part_B_gauge_U1} 
\end{align}
where, for the sake of brevity, we have introduced the notation $\vec{k}' \equiv - C_{2x} \vec{k} $. Rearranging \cref{si:eqn:singl_part_B_gauge_U1} we find that  
\begin{equation}
	\label{si:eqn:singl_part_B_gauge_U2}
	\sum_{m,n} \left[B^{\mathcal{U}}_{\hat{c}} \left( \vec{k} \right) \right]_{m' \eta, m \eta} \mathcal{B}^{\eta}_{mn} \left( \vec{k} \right) \left[B^{\dagger \mathcal{U}}_{\hat{c}} \left( \vec{k} \right) \right]_{n \eta,n' \eta} = -\mathcal{B}^{*\eta}_{m'n'} \left( -C_{2x} \vec{k} \right).
\end{equation}
By tracing over the band and valley indices in \cref{si:eqn:singl_part_B_gauge_U2}, we find that 
\begin{equation}
    \sum_{\substack{\abs{n}=1 \\ \eta}} \mathcal{B}^{\eta}_{nn} \left( \vec{k} \right) = -  \sum_{\substack{\abs{n}=1 \\ \eta}} \mathcal{B}^{*\eta}_{nn} \left( -C_{2x} \vec{k} \right).
\end{equation}
Together with the reality condition in \cref{si:eqn:singl_part_B_gauge_c2t_mat}, \cref{si:eqn:singl_part_B_gauge_U2} the second-order projected displacement field Hamiltonian is invariant under the many-body chage conjugation transformation, up to a constant
\begin{align}
	\mathcal{P} H^{\left(\hat{c} \right)}_{U} \mathcal{P}^{-1}  &= \sum_{\substack{\eta,s \\ \abs{n},\abs{m} = 1}} \sum_{\substack{\vec{k} \in \mathcal{C}_{\eta} \\ \abs{n'},\abs{m'} = 1}}  \left[B^{\mathcal{U}}_{\hat{c}} \left( \vec{k} \right) \right]_{m' \eta, m \eta} \mathcal{B}^{\eta}_{mn} \left( \vec{k} \right) \left[B^{\dagger \mathcal{U}}_{\hat{c}} \left( \vec{k} \right) \right]_{n \eta,n' \eta} \des{c}{-C_{2x} \vec{k},m',\eta,s} \cre{c}{-C_{2x}\vec{k},n',\eta,s} \nonumber \\
	&= -\sum_{\substack{\eta,s \\ \abs{n},\abs{m} = 1}} \sum_{\vec{k} \in \mathcal{C}_{\eta}}  \mathcal{B}^{*\eta}_{mn} \left( -C_{2x} \vec{k} \right) \des{c}{-C_{2x} \vec{k},m,\eta,s} \cre{c}{-C_{2x}\vec{k},n,\eta,s} \nonumber 
	\nonumber \\
	&= \sum_{\substack{\eta,s \\ \abs{n},\abs{m} = 1}} \sum_{\vec{k} \in \mathcal{C}_{\eta}}  \mathcal{B}^{\eta}_{nm} \left( \vec{k} \right) \cre{c}{\vec{k},n,\eta,s} \des{c}{ \vec{k},m,\eta,s} - \sum_{\substack{\eta,s \\ \abs{n} = 1}} \sum_{\vec{k} \in \mathcal{C}_{\eta}}  \mathcal{B}^{\eta}_{nn} \left( \vec{k} \right) = H^{\left(\hat{c} \right)}_{U}.
	\label{eqn:si:projU2_MBC}
\end{align}

\subsubsection{Spatial many-body charge conjugation symmetry of $H_I$}
To keep the discussion general, we consider the form factor matrix defined in \cref{si:eqn:str_fact_def} corresponding to a certain energy band creation operator $\cre{f}{} = \cre{b}{},\cre{c}{}$. As a consequence of the antiunitary single-particle transformation $\mathcal{U}$, the form factors obey
\begin{align}
		&M_{mn}^{\hat{f},\eta}\left( \vec{k},\vec{q}+\vec{G}\right) 
= \sum_{\alpha} \sum_{\vec{Q} \in \mathcal{Q}_{\hat{f},\eta}} u^{*\hat{f}}_{\vec{Q}-\vec{G} \alpha; m \eta} \left( \vec{k} + \vec{q} \right) u^{\hat{f}}_{\vec{Q} \alpha; n \eta} \left( \vec{k} \right) \nonumber \\
=& \sum_{\mathbf{Q}\in\mathcal{Q}_{\hat{f},\eta}} \sum_{\alpha,\eta_1,\eta_2} 
\left[u^{\dagger\hat{f}}_{m\eta_1} \left( \vec{k} + \vec{q} \right) \left[D_{\hat{f}}^\dagger \left( \mathcal{U} \right) \right]_{\eta \eta_1} \right]_{\vec{Q}-\vec{G} \alpha} 
\left[\left[D_{\hat{f}} \left( \mathcal{U} \right) \right]_{\eta \eta_2} u^{\hat{f}}_{n\eta_2}\left( \vec{k} \right)\right]_{\vec{Q} \alpha} \nonumber\\
=& \sum_{\substack{\alpha,\eta_1,\eta_2 \\ m',n'}} \left[B^{\dagger \mathcal{U}}_{\hat{f}} \left(\vec{k} + \vec{q} \right) \right]_{m \eta,m' \eta_1} \sum_{\vec{Q} \in \mathcal{Q}_{\eta_1}} u^{\hat{f}}_{\vec{Q} - \vec{G}' \alpha; m' \eta_1} \left( \vec{k}' + \vec{q}' \right) u^{*\hat{f}}_{ \vec{Q} \alpha; n' \eta_2} \left( \vec{k}' \right) \left[B^{\mathcal{U}}_{\hat{f}} \left( \vec{k} \right) \right]_{n' \eta_2,n \eta}, \label{si:eqn:f_form_gauge_U}
	\end{align}
where for simplicity we have defined $\vec{k}' \equiv -C_{2x} \vec{k}$, $\vec{q}' \equiv -C_{2x} \vec{q}$, and $\vec{G}' \equiv -C_{2x} \vec{G}$. Written in matrix form \cref{si:eqn:f_form_gauge_U} reads
\begin{equation}
	M^{\hat{f}}\left( \vec{k},\vec{q}+\vec{G}\right) = B^{\dagger \mathcal{U}}_{\hat{f}} \left( \vec{k} + \vec{q}\right) M^{* \hat{f}}\left( \vec{k}',\vec{q}'+\vec{G}' \right) B^{\mathcal{U}}_{\hat{f}} \left( \vec{k} \right),
\end{equation}
which after rearranging and using the Hermiticity condition in \cref{si:eqn:formFactHerm} leads to 
\begin{equation}
	\label{si:eqn:f_form_matrix_gauge_U}
	B^{\mathcal{U}}_{\hat{f}} \left( \vec{k} + \vec{q} \right) M^{\hat{f}}\left( \vec{k},\vec{q}+\vec{G}\right) B^{\dagger \mathcal{U}}_{\hat{f}} \left( \vec{k} \right) =  M^{* \hat{f}} \left( \vec{k}',\vec{q}' + \vec{G}' \right)  =  M^{T \hat{f}} \left( \vec{k}'+\vec{q}', - \vec{q}' - \vec{G}' \right).
\end{equation}
Together with the definitions from \cref{si:eqn:action_MB_PH}, \cref{si:eqn:f_form_matrix_gauge_U} and the reality of the form factors derived in Appendix \ref{app:interaction:gauge} implies that the action of the spatial many-body charge conjugation operation $\mathcal{P}$ on the $O^{\hat{f}}_{\vec{q},\vec{G}}$ operators is given by 
\begin{align}
	\mathcal{P} O^{\hat{f}}_{\vec{q},\vec{G}} \mathcal{P}^{-1} =& \sqrt{V \left( \vec{q} + \vec{G} \right)} \sum_{\substack{\eta, s \\ m,n}} \sum_{\vec{k}} M^{\hat{f},\eta}_{mn} \left( \vec{k},\vec{q}+\vec{G} \right) \left( \mathcal{P} \cre{f}{\vec{k}+\vec{q},m,\eta,s} \des{f}{\vec{k},n,\eta,s} \mathcal{P}^{-1} - \frac{1}{2} \delta_{\vec{q},\vec{0}} \delta_{m,n} \right) \nonumber \\
	=& \sqrt{V \left( \vec{q} + \vec{G} \right)} \sum_{\substack{\eta, s \\ m,n}} \sum_{\vec{k}} \left[ B^{\mathcal{U}}_{\hat{f}} \left( \vec{k} + \vec{q} \right) M^{\hat{f}}\left( \vec{k},\vec{q}+\vec{G}\right) B^{\dagger \mathcal{U}}_{\hat{f}} \left( \vec{k} \right) \right]_{m\eta,n\eta} \left( \des{f}{\vec{k}'+\vec{q}',m,\eta,s} \cre{f}{\vec{k}',n,\eta,s} - \frac{1}{2} \delta_{\vec{q},\vec{0}} \delta_{m,n} \right) \nonumber \\
	=&- \sqrt{V \left( \vec{q}' + \vec{G}' \right)} \sum_{\substack{\eta, s \\ m,n}} \sum_{\vec{k}} M^{\hat{f},\eta}_{nm}\left( \vec{k}'+\vec{q}', - \vec{q}' - \vec{G}' \right) \left( \cre{f}{\vec{k}',n,\eta,s} \des{f}{\vec{k}'+\vec{q}',m,\eta,s} - \frac{1}{2} \delta_{\vec{q}',\vec{0}} \delta_{m,n} \right) \nonumber \\
	=&- \sqrt{V \left(- \vec{q}' - \vec{G}' \right)} \sum_{\substack{\eta, s \\ m,n}} \sum_{\vec{k}} M^{\hat{f},\eta}_{nm}\left( \vec{k},-\vec{q}'- \vec{G}' \right) \left( \cre{f}{\vec{k}-\vec{q}',n,\eta,s} \des{f}{\vec{k},m,\eta,s} - \frac{1}{2} \delta_{\vec{q}',\vec{0}} \delta_{m,n} \right) \nonumber \\
	=&- O^{\hat{f}}_{-C_{2x}\vec{q},-C_{2x}\vec{G}}. \label{si:eqn:action_MB_PH_OqG}
\end{align}
In \cref{si:eqn:action_MB_PH_OqG}, the momentum $\vec{k}$ runs over the entire MBZ for $\hat{f}=\hat{c}$, and is restricted by the condition $\vec{k},\vec{k}+\vec{q} \in \mathcal{A}^{\eta}$ when $\hat{f} = \hat{b}$. In deriving \cref{si:eqn:action_MB_PH_OqG}, we have also used the invariance of the interaction potential under rotations in the plane of the graphene layers. Taken together with the definition of the projected interaction Hamiltonain in \cref{si:eqn:proj_int_hamiltonian}, \cref{si:eqn:action_MB_PH_OqG} implies that $H_I$ is symmetric under the spatial many-body charge conjugation symmetry
\begin{equation}
	\label{eqn:si:projHI_MBC}
	\left[\mathcal{P},H_I \right] = 0
\end{equation}

Finally, combining \cref{si:eqn:enMBC,eqn:si:projU1_MBC,eqn:si:projU2_MBC,eqn:si:projHI_MBC} implies that the projected fully-interacting TSTG Hamiltonian $H = H_0 + H_I $ is indeed invariant under the spatial many-body charge conjugation symmetry $\mathcal{P}$
\begin{equation}
	 \mathcal{P} H \mathcal{P}^{-1} = H.
\end{equation}

\subsection{Brief review of the $\UN{4}$ group}\label{app:fullSym:U4review} 

As it is featured extensively in this paper, this sections presents a brief review of the the $\UN{4}$ group and corresponding Lie algebra. The $\UN{N}$ group is defined by all the $N\times N$ unitary matrices $\mathcal{V}$ satisfying $\mathcal{V}^\dagger \mathcal{V}= \mathbb{1}_{N}$, where $\mathbb{1}_N$ is the identity matrix. The matrices $\mathcal{V}$ are generated by all the linearly independent $N\times N$ Hermitian matrices, thus the total number of generators is $N^2$. In particular, for the $\UN{4}$ group, the $16$ generators can be represented by the tensor product of two sets of $2\times2$ identity and Pauli matrices $\tau^a$ and $s^a$ as
\begin{equation}
	s_0^{ab}=\tau^a s^b ,
\end{equation}
where $a,b=0,x,y,z$. We denote their commutation relations as
\begin{equation}
\label{seq:U4-structure-const}
	\left[ s_0^{ab}, s_0^{cd} \right] = \sum_{e,f} f^{ab,cd}_{ef} s_0^{ef},
\end{equation}
with $f^{ab,cd}_{ef}$ of the $\UN{4}$ group's Lie algebra.

\subsection{Continuous symmetries of the mirror-antisymmetric sector in the $U=0$ case}\label{app:fullSym:cont_sym_b_operators}

In \cref{si:eqn:u2_generators_b} we have written down the generators of the $\left[ \UN{2} \times \UN{2} \right]_{\hat{b}}$ symmetry of the single-particle Hamiltonian $\hat{H}_D$. Here we show that in the absence of displacement field, this symmetry is not only inherited by the mirror-antisymmetric sector of the projected many-body TSTG Hamiltonian, but is also promoted to an enlarged continuous group. To keep the notation general, we introduce the operators
\begin{equation}
	\label{si:eqn:cont_b_generators}
	S_{\hat{b}}^{ a b}=\sum_{\substack{\abs{\delta \vec{k}} \leq \Lambda \\m,\eta,s\\n,\eta',s'}} \left(s_{\hat{b}}^{ab} \right)_{m \eta s, n \eta' s'} \cre{b}{\delta \vec{k} + \eta \vec{q}_1,m,\eta,s} \des{b}{\delta \vec{k} + \eta' \vec{q}_1,n,\eta',s'} ,
\end{equation}
representing the generators of the continuous symmetry group of the mirror-antisymmetric sector (which, for the moment, we denote by $\mathcal{G}$). For a certain pair of indices $a$ and $b$, the Hermitian matrices $s^{ab}_{\hat{b}}$ defined in the band, valley, and spin subspaces form the representation of the Lie algebra of the group $\mathcal{G}$. The definition in \cref{si:eqn:cont_b_generators} is the projected form of \cref{si:eqn:u2_generators_b} that has been further generalized to include arbitrary band, valley, and spin rotations. It is worth mentioning that the generators $S_{\hat{b}}^{ a b}$ always preserve the \emph{relative} momentum $\delta \vec{k}$, but change the \emph{actual} momentum $\vec{k}$ when the matrix $s_{\hat{b}}^{ab}$ is not diagonal in valley space. 

We first investigate the symmetries of the $O^{\hat{b}}_{\vec{q},\vec{G}}$ operators defined in \cref{si:eqn:oqg_definition} which govern the Coulomb interaction of the Dirac cone Fermions. The commutator of the generators $S_{\hat{b}}^{ab}$ with the $O^{\hat{b}}_{\vec{q},\vec{G}}$ operators is given by 

\begin{equation}
	\label{si:eqn:commut_oqg_bgenerators}
	\begin{split}
		\left[S_{\hat{b}}^{ a b}, O^{\hat{b}}_{\vec{q},\vec{G}}  \right] &= \sum_{\substack{\abs{\delta{\vec{k}}} \leq \Lambda \\ m,n,\eta,s \\n',\eta',s'}} \sqrt{V\left( \vec{q} + \vec{G} \right)} \left[ \left( s_{\hat{b}}^{ a b} \right)_{n \eta s, m \eta' s'} M_{m n'}^{\hat{b},\eta'} \left(\delta \vec{k} + \eta' \vec{q}_1,\vec{q}+\vec{G}\right) \right. \\
		&\left. - M_{n m}^{\hat{b},\eta} \left(\delta \vec{k} + \eta \vec{q}_1,\vec{q}+\vec{G}\right) \left( s_{\hat{b}}^{ a b} \right)_{m \eta s, n' \eta' s'} \right] \left(\cre{b}{\delta \vec{k}+\eta \vec{q}_1 + \vec{q},n,\eta,s} \des{b}{\delta \vec{k} + \eta' \vec{q}_1,n',\eta',s'} \right).
	\end{split}
\end{equation}

Similarly, the commutator between the generators $S_{\hat{b}}^{ab}$ and the Hamiltonian $H_{D}$ reads
\begin{equation}
	\label{si:eqn:commut_energy_bgenerators}
		\left[S_{\hat{b}}^{ a b}, H_{D}  \right] = \sum_{\substack{\abs{\delta{\vec{k}}} \leq \Lambda \\ n,\eta,s \\n',\eta',s'}} \left[s_{\hat{b}}^{ a b}, \epsilon^{\hat{b}} \left( \delta \vec{k} \right) \right]_{n \eta s, n' \eta' s'} \left(\cre{b}{\delta \vec{k} + \eta \vec{q}_1,n,\eta,s} \des{b}{\delta \vec{k} + \eta' \vec{q}_1, n',\eta',s'} \right),
\end{equation}
where $\epsilon^{\hat{b}} \left( \delta \vec{k} \right)$ is the Dirac energy band matrix introduced in Appendix \ref{app:gauge_single_part:energy}.
Aided by the parameterization of \cref{si:eqn:gen_form_b_M}, we find that the maximal set of generators that commute with the $O^{\hat{b}}_{\vec{q},\vec{G}}$ operators is given by 
\begin{equation}
	\label{si:eqn:cont_b_generators_big}
	S_{\hat{b} \pm}^{ a b}=\sum_{\substack{\abs{\delta \vec{k}} \leq \Lambda \\m,\eta,s\\n,\eta',s'}} \left(s_{\hat{b} \pm}^{ab} \right)_{m \eta s, n \eta' s'} \cre{b}{\delta \vec{k} + \eta \vec{q}_1,m,\eta,s} \des{b}{\delta \vec{k} + \eta' \vec{q}_1,n,\eta',s'},
\end{equation}
where $a,b=0,x,y,z$ and the representation matrices are defined by
\begin{equation}
	s_{\hat{b} \pm}^{a b} = \frac{1}{2} \left( \zeta^0 \pm \zeta^y \right) \tau^a s^b, \qquad \left(a,b = 0,x,y,z \right).
\end{equation}

In this form, the generators obey
\begin{align}
	\left[ S^{ab}_{\hat{b} \pm}, S^{cd}_{\hat{b} \pm} \right] &= \sum_{e,f} f^{ab,cd}_{ef} S^{ef}_{\hat{b} \pm}, \\
	\left[ S^{ab}_{\hat{b} +}, S^{cd}_{\hat{b} -} \right] &= 0
\end{align}
where $f^{ab,cd}_{ef}$ represent the structure factors of the $\UN{4}$ group. The symmetry group of the $O^{\hat{b}}_{\vec{q},\vec{G}}$ operators is thus seen to be $\left[ \UN{4} \times \UN{4} \right]_{\hat{b}}$, where the 16 generators $S^{ab}_{\hat{b} +}$ generate one $\UN{4}$ group, while the 16 generators $S^{ab}_{\hat{b} -}$ generate the other one. 

The large Fermi velocity of the single-particle Dirac cone Hamiltonian $\hat{H}$ implies that there is no \emph{flat} limit for the mirror-antisymmetric sector. Otherwise stated, neglecting the single-particle projected contribution $H_{D}$ is not a physically valid approximation. Even though the $O^{\hat{b}}_{\vec{q},\vec{G}}$ operators governing the interaction of the Dirac fermions are invariant under $\left[ \UN{4} \times \UN{4} \right]_{\hat{b}}$, the introduction of the kinetic term reduces the symmetry to $\left[ \UN{4} \right]_{\hat{b}}$. Using the parameterized form of the single-particle band-energy from \cref{si:eqn:param_e_b}, we see that only a subset of the 32 generators from  \cref{si:eqn:cont_b_generators_big} commute with the $H_{D}$. More precisely, we find that $\left[S^{ab}_{\hat{b} +} + S^{ab}_{\hat{b} -}, H_D  \right] = 0$, while $\left[S^{ab}_{\hat{b} +} - S^{ab}_{\hat{b} -}, H_D  \right] \neq 0$, for any $a,b=0,x,y,z$. We can therefore conclude that the mirror-antisymmetric sector of TSTG in the absence of displacement field enjoys an enhanced $\left[ \UN{4} \right]_{\hat{b}}$ symmetry for which the representation matrices are given by
\begin{equation}
	s^{ab}_{\hat{b}} = s^{ab}_{\hat{b}+}+ s^{ab}_{\hat{b}-} = \zeta^{0} \tau^{a} s^{b}, \qquad \left( a,b=0,x,y,z \right),
\end{equation}
and correspond to full $\UN{4}$ valley-spin rotations in the mirror-antisymmetric sector. The generators $S^{ab}_{\hat{b}}$ of this $\left[ \UN{4} \right]_{\hat{b}}$ symmetry obey the algebra defined in \cref{seq:U4-structure-const}
\begin{equation}
	\left[ S^{ab}_{\hat{b}}, S^{cd}_{\hat{b}} \right] = \sum_{e,f} f^{ab,cd}_{ef} S^{ef}_{\hat{b}}.
\end{equation}

\subsection{Continuous symmetries of the mirror-symmetric sector in the $U=0$ case}
\label{app:fullSym:cont_sym_c_operators}

In the absence of displacement field, the continuous symmetries of the mirror-symmetric sector of the projected many-body TSTG Hamiltonian are determined by the single-particle projected Hamiltonian $H_{\mathrm{TBG}}$, as well as by the $O^{\hat{c}}_{\vec{q},\vec{G}}$ operators governing the Coulomb interaction of the TBG fermions. The symmetries of $H_{\mathrm{TBG}}$ and $O^{\hat{c}}_{\vec{q},\vec{G}}$ have been derived and extensively discussed in Refs.~\cite{SEO19,KAN19,BER20a,BUL20}, in the context of ordinary TBG. 

Here, we will summarize the continuous symmetries of the mirror-symmetric sector of TSTG in the absence of displacement field and only briefly justify them from the parameterized forms of $H_{\mathrm{TBG}}$ and $O^{\hat{c}}_{\vec{q},\vec{G}}$, which were summarized respectively in Appendices \ref{app:gauge_single_part:energy} and \ref{app:interaction:gauge:c}. We refer the reader to Ref.~\cite{BER20a} for the detailed proofs. 

In analogy with the generators $S_{\hat{b}}^{ a b}$, we define the operators
\begin{equation}
	\label{si:eqn:cont_c_generators}
	S_{\hat{c}}^{ a b}=\sum_{\substack{\vec{k} \in \mathrm{MBZ} \\m,\eta,s\\n,\eta',s'}} \left(s_{\hat{c}}^{ab} \right)_{m \eta s, n \eta' s'} \cre{c}{\vec{k},m,\eta,s} \des{c}{\vec{k},n,\eta',s'},
\end{equation}
representing the generators of the various continuous symmetry groups pertaining to the mirror-symmetric sector of the TSTG many-body projected Hamiltonian. The commutators of the $O^{\hat{c}}_{\vec{q},\vec{G}}$ operators with the generators in \cref{si:eqn:cont_c_generators} are given by
\begin{equation}
	\label{si:eqn:commut_oqg_cgenerators}
	\left[S_{\hat{c}}^{ a b}, O^{\hat{c}}_{\vec{q},\vec{G}}  \right] = \sum_{\substack{\vec{k} \in \mathrm{MBZ} \\ n,\eta,s \\n',\eta',s'}} \sqrt{V\left( \vec{q} + \vec{G} \right)} \left[s_{\hat{c}}^{ a b}, M^{\hat{c}} \left( \vec{k},\vec{q}+\vec{G}\right) \right]_{n \eta s, n' \eta' s'} \left(\cre{c}{\vec{k} + \vec{q},n,\eta,s} \des{c}{\vec{k}, n',\eta',s'} \right).
\end{equation}
Similarly, the commutator between the generators defined in \cref{si:eqn:cont_c_generators} and the single-particle projected TBG Hamiltonian reads
\begin{equation}
	\label{si:eqn:commut_energy_cgenerators}
	\left[S_{\hat{c}}^{ a b}, H_{\mathrm{TBG}}  \right] = \sum_{\substack{\vec{k} \in \mathrm{MBZ} \\ n,\eta,s \\n',\eta',s'}} \left[s_{\hat{c}}^{ a b}, \epsilon^{\hat{c}} \left( \vec{k} \right) \right]_{n \eta s, n' \eta' s'} \left(\cre{c}{\vec{k},n,\eta,s} \des{c}{\vec{k}, n',\eta',s'} \right),
\end{equation}
where $\epsilon^{\hat{c}} \left( \vec{k} \right)$ is the TBG energy band matrix introduced in Appendix \ref{app:gauge_single_part:energy}.

We will now investigate the implications of \cref{si:eqn:commut_oqg_cgenerators,si:eqn:commut_energy_cgenerators} for the continuous symmetry group of the TBG fermions. It is worth noting that unlike the symmetry generators corresponding to the mirror-antisymmetric sector defined in \cref{si:eqn:cont_b_generators}, the generators related to the mirror-symmetric operators introduced in \cref{si:eqn:cont_c_generators} leave the momentum invariant. 

\subsubsection{$\left[ \UN{4} \times \UN{4} \right]_{\hat{c}}$ symmetry in the (first) chiral-flat limit}\label{app:fullSym:cont_sym_c_operators:cf}

In the (first) chiral-flat limit we approximate the TBG bands as being perfectly flat, completely neglecting the projected single-particle contribution $H_{\mathrm{TBG}}$ and therefore disregarding \cref{si:eqn:commut_energy_cgenerators}. Assuming that the (first) chiral condition holds ($w_0=0$), the $O^{\hat{c}}_{\vec{q},\vec{G}}$ operators can be parameterized according to \cref{si:eqn:gen_form_c_M_chiral}, and so \cref{si:eqn:commut_oqg_cgenerators} determines the maximal set of commuting generators to be
\begin{equation}
	\label{si:eqn:cont_c_generators_big_cf}
	S_{\hat{c} \pm}^{ a b}=\sum_{\substack{\vec{k} \in \mathrm{MBZ} \\m,\eta,s\\n,\eta',s'}} \left(s_{\hat{c} \pm}^{ab} \right)_{m \eta s, n \eta' s'} \cre{c}{\vec{k},m,\eta,s} \des{c}{\vec{k},n,\eta',s'},
\end{equation}
where $a,b=0,x,y,z$ and the representation matrices are defined by
\begin{equation}
	s_{\hat{c} \pm}^{a b} = \frac{1}{2} \left( \zeta^0 \pm \zeta^y \right) \tau^a s^b, \qquad \left(a,b = 0,x,y,z \right).
\end{equation}
The generators in \cref{si:eqn:cont_c_generators_big_cf} obey the commutation relations
\begin{equation}
	\left[ S^{ab}_{\hat{c} \pm}, S^{cd}_{\hat{c} \pm} \right] = \sum_{e,f} f^{ab,cd}_{ef} S^{ef}_{\hat{c} \pm},
\end{equation}
where $f^{ab,cd}_{ef}$ represent the structure factors of the $\UN{4}$ group and $\left[ S^{ab}_{\hat{c} +}, S^{cd}_{\hat{c} -} \right]=0$. The symmetry group TBG fermions is thus seen to be $\left[ \UN{4} \times \UN{4} \right]_{\hat{c}}$~\cite{BUL20,BER20a}, where the 16 generators $S^{ab}_{\hat{c} +}$ generate one $\UN{4}$ group, while the 16 generators $S^{ab}_{\hat{c} -}$ generate the other one. Using the Chern band basis defined in Appendix \ref{app:gauge:c}, we can also write the generators of the $\left[ \UN{4} \times \UN{4} \right]_{\hat{c}}$ symmetry as
\begin{align}	
	S_{\hat{c} \pm}^{ a b} &= \sum_{\substack{\vec{k} \in \mathrm{MBZ} \\ \eta,s \\\eta',s'}} \left(\tau^a s^b \right)_{\eta s, \eta' s'} \cre{d}{\vec{k},\pm 1,\eta,s} \des{d}{\vec{k},\pm 1,\eta',s'}.	
\end{align}

\subsubsection{$\left[ \UN{4} \right]_{\hat{c}}$ symmetry in the nonchiral-flat limit}\label{app:fullSym:cont_sym_c_operators:ncf}

Compared with the chiral-flat limit from Appendix \ref{app:fullSym:cont_sym_c_operators:cf}, in the nonchiral-flat limit we also neglect the dispersion of the TBG active bands, but we do not assume the (first) chiral condition ($w_0=0$) to hold. As such, the $O^{\hat{c}}_{\vec{q},\vec{G}}$ operators can be parameterized according to \cref{si:eqn:gen_form_c_M_nonchiral}, resulting in the maximal set of symmetry generators being given by \cref{si:eqn:cont_c_generators} for $a,b=0,x,y,z$ and having the representation matrices
\begin{equation}
	s_{\hat{c}}^{0 b} = \zeta^0 \tau^0 s^b, \qquad
	s_{\hat{c}}^{x b} = \zeta^y \tau^x s^b, \qquad
	s_{\hat{c}}^{y b} = \zeta^y \tau^y s^b, \qquad
	s_{\hat{c}}^{z b} = \zeta^0 \tau^z s^b,
\end{equation}
where $b=0,x,y,z$. The generators in \cref{si:eqn:cont_c_generators} obey the commutation relation 
\begin{equation}
	\left[ S^{ab}_{\hat{c}}, S^{cd}_{\hat{c}} \right] = \sum_{e,f} f^{ab,cd}_{ef} S^{ef}_{\hat{c}},
\end{equation}
where $f^{ab,cd}_{ef}$ represent the structure factors of the $\UN{4}$ group. The symmetry group pertaining to the TBG fermions in the nonchiral-flat is thus seen to be $\left[ \UN{4} \right]_{\hat{c}}$~\cite{KAN19,BER20a}.

\subsubsection{$\left[ \UN{4} \right]_{\hat{c}}$ symmetry in the chiral-nonflat limit}\label{app:fullSym:cont_sym_c_operators:cnf}

In the (first) chiral-nonflat limit, we assume the chiral condition to hold, but, in contrast to Appendices \ref{app:fullSym:cont_sym_c_operators:ncf} and \ref{app:fullSym:cont_sym_c_operators:cnf}, we also account for the non-zero dispersion of the TBG active bands. The parameterizations of $H_{\mathrm{TBG}}$ and $O^{\hat{c}}_{\vec{q},\vec{G}}$ given respectively in  \cref{si:eqn:param_e_c_chiral,si:eqn:gen_form_c_M_chiral} restrict the maximal set of generators in \cref{si:eqn:cont_c_generators} through \cref{si:eqn:commut_oqg_cgenerators,si:eqn:commut_energy_cgenerators}. The representation matrices in the chiral-nonflat limit read
\begin{equation}
	s_{\hat{c}}^{a b} = \zeta^0 \tau^a s^b, \qquad \left(a,b = 0,x,y,z \right),
\end{equation}
implying that the generators from \cref{si:eqn:cont_c_generators} obey the Lie algebra of the $\UN{4}$ group
\begin{equation}
	\left[ S^{ab}_{\hat{c}}, S^{cd}_{\hat{c}} \right] = \sum_{e,f} f^{ab,cd}_{ef} S^{ef}_{\hat{c}},
\end{equation}
As in the nonchiral-flat limit, the symmetry group of the TBG fermions in the chiral-nonflat limit is given by $\left[ \UN{4} \right]_{\hat{c}}$~\cite{BER20a}, but with different generators.

\subsubsection{$\left[ \UN{2} \times \UN{2} \right]_{\hat{c}}$ symmetry in the nonchiral-nonflat case}\label{app:fullSym:cont_sym_c_operators:ncnf}

When neither the chiral condition holds, nor the dispersion of the TBG active bands is ignored, the parameterizations of $H_{\mathrm{TBG}}$ and $O^{\hat{c}}_{\vec{q},\vec{G}}$ given respectively in \cref{si:eqn:oqg_definition,si:eqn:gen_form_c_M_nonchiral} imply that the TBG fermions have only the trivial $\left[\UN{2} \times \UN{2} \right]_{\hat{c}}$ symmetry, associated with the spin-charge conservation per valley. The generators of this symmetry are given by \cref{si:eqn:cont_c_generators} for $a=0,z$ and $b=0,x,y,z$, with the corresponding representation matrices being given by
\begin{equation}
	s_{\hat{c}}^{0 b} = \zeta^0 \tau^0 s^b, \qquad
	s_{\hat{c}}^{z b} = \zeta^0 \tau^z s^b,
\end{equation}
for $b=0,x,y,z$.

\subsection{Symmetries of the projected many-body TSTG Hamiltonian with displacement field}
\label{app:fullSym:cont_sym_withU}
The perpendicularly applied displacement field couples the mirror-symmetry sector fermions at the single-particle level through the contribution $H^{\left( \hat{b} \hat{c} \right)}_{U}$. As a result of this, the TBG and Dirac fermion flavors can no longer be independently rotated in the spin, valley, or band subspaces. We will therefore define the operators
\begin{equation}
	\label{si:eqn:cont_generators}
	S^{ a b}=\sum_{\substack{m,\eta,s\\n,\eta',s'}} \left[ \sum_{\abs{\delta \vec{k}} \leq \Lambda} \left(s_{\hat{b}}^{ab} \right)_{m \eta s, n \eta' s'} \cre{b}{\delta \vec{k} + \eta \vec{q}_1,m,\eta,s} \des{b}{\delta \vec{k} + \eta' \vec{q}_1,n,\eta',s'} + \sum_{\vec{k} \in \mathrm{MBZ} } \left(s_{\hat{c}}^{ab} \right)_{m \eta s, n \eta' s'} \cre{c}{\vec{k},m,\eta,s} \des{c}{\vec{k},n,\eta',s'} \right],
\end{equation}
representing the generators of the continuous symmetry group of the TSTG Hamiltonian. In appearance, \cref{si:eqn:cont_generators} represents just the sum of \cref{si:eqn:cont_b_generators,si:eqn:cont_c_generators}. Note however, that we have not made \emph{any} assumptions regarding the Hermitian matrices $s^{ab}_{\hat{b}}$ and $s^{ab}_{\hat{c}}$, other that the fact that they provide isomorphic representations for the Lie alegebra of the continuous symmetry group of the many-body projected TSTG Hamiltonian. 

The advantage of the notation in \cref{si:eqn:cont_generators} is that the commutation of the generators $S^{a b}$ with the various terms of the many-body projected TSTG Hamiltonian can be readily computed from \cref{si:eqn:commut_oqg_bgenerators,si:eqn:commut_oqg_cgenerators,si:eqn:commut_energy_bgenerators,si:eqn:commut_energy_cgenerators}

\begin{align}
	\left[S^{ a b}, O^{\hat{b}}_{\vec{q},\vec{G}}  \right] &= \sum_{\substack{\abs{\delta{\vec{k}}} \leq \Lambda \\ m,n,\eta,s \\n',\eta',s'}} \sqrt{V\left( \vec{q} + \vec{G} \right)} \left[ \left( s_{\hat{b}}^{ a b} \right)_{n \eta s, m \eta' s'} M_{m n'}^{\hat{b},\eta'} \left(\delta \vec{k} + \eta' \vec{q}_1,\vec{q}+\vec{G}\right) \right. \nonumber \\
	\label{si:eqn:commut_obqg_generators}	
		&\left. - M_{n m}^{\hat{b},\eta} \left(\delta \vec{k} + \eta \vec{q}_1,\vec{q}+\vec{G}\right) \left( s_{\hat{b}}^{ a b} \right)_{m \eta s, n' \eta' s'} \right] \left(\cre{b}{\delta \vec{k}+\eta \vec{q}_1 + \vec{q},n,\eta,s} \des{b}{\delta \vec{k} + \eta' \vec{q}_1,n',\eta',s'} \right), \\
	\label{si:eqn:commut_ocqg_generators}
	\left[S^{ a b}, O^{\hat{c}}_{\vec{q},\vec{G}}  \right] &= \sum_{\substack{\vec{k} \in \mathrm{MBZ} \\ n,\eta,s \\n',\eta',s'}} \sqrt{V\left( \vec{q} + \vec{G} \right)} \left[s_{\hat{c}}^{ a b}, M^{\hat{c}} \left( \vec{k},\vec{q}+\vec{G}\right) \right]_{n \eta s, n' \eta' s'} \left(\cre{c}{\vec{k} + \vec{q},n,\eta,s} \des{c}{\vec{k}, n',\eta',s'} \right), \\	
	\label{si:eqn:commut_enD_generators}
	\left[S^{ a b}, H_{D}  \right] &= \sum_{\substack{\abs{\delta{\vec{k}}} \leq \Lambda \\ n,\eta,s \\n',\eta',s'}} \left[s_{\hat{b}}^{ a b}, \epsilon^{\hat{b}} \left( \delta \vec{k} \right) \right]_{n \eta s, n' \eta' s'} \left(\cre{b}{\delta \vec{k} + \eta \vec{q}_1,n,\eta,s} \des{b}{\delta \vec{k} + \eta' \vec{q}_1, n',\eta',s'} \right), \\
	\label{si:eqn:commut_enTBG_generators}
	\left[S^{ a b}, H_{\mathrm{TBG}}  \right] &= \sum_{\substack{\vec{k} \in \mathrm{MBZ} \\ n,\eta,s \\n',\eta',s'}} \left[s_{\hat{c}}^{ a b}, \epsilon^{\hat{c}} \left( \vec{k} \right) \right]_{n \eta s, n' \eta' s'} \left(\cre{c}{\vec{k},n,\eta,s} \des{c}{\vec{k}, n',\eta',s'} \right).
\end{align}
To \cref{si:eqn:commut_obqg_generators,si:eqn:commut_ocqg_generators,si:eqn:commut_enD_generators,si:eqn:commut_enTBG_generators}, we add the commutators of $S^{ a b}$ with the projected displacement field contributions $H^{\left(\hat{b} \hat{c} \right)}_{U}$ and $H^{\left( \hat{c} \right)}_{U}$
\begin{align}
	\left[S^{ a b}, H^{\left(\hat{b} \hat{c} \right)}_{U}  \right] =& \sum_{\substack{\abs{\delta{\vec{k}}} \leq \Lambda \\ n,\eta,s \\n',\eta',s'}} \left\lbrace \left[ s^{ab}_{\hat{b}} N \left( \delta \vec{k} \right) \right]_{n \eta s, n' \eta' s'} \cre{b}{\delta \vec{k} + \eta \vec{q}_1,n,\eta,s} \des{c}{\delta \vec{k} + \eta' \vec{q}_1,n',\eta',s'} \right. \nonumber \\ 
	-& \left[ N \left( \delta \vec{k} \right) s^{ab}_{\hat{c}} \right]_{n \eta s, n' \eta' s'} \cre{b}{\delta \vec{k} + \eta \vec{q}_1,n,\eta,s} \des{c}{\delta \vec{k} + \eta \vec{q}_1,n',\eta',s'} \nonumber \\
	-& \left[ s^{ab}_{\hat{b}} N \left( \delta \vec{k} \right) \right]^*_{n \eta s, n' \eta' s'} \cre{c}{\delta \vec{k} + \eta' \vec{q}_1,n',\eta',s'} \des{b}{\delta \vec{k} + \eta \vec{q}_1,n,\eta,s} \nonumber \\ 
	+& \left. \left[ N \left( \delta \vec{k} \right) s^{ab}_{\hat{c}} \right]^*_{n \eta s, n' \eta' s'} \cre{c}{\delta \vec{k} + \eta \vec{q}_1,n',\eta',s'} \des{b}{\delta \vec{k} + \eta \vec{q}_1,n,\eta,s} \right\rbrace \label{si:eqn:commut_u1_generators}, \\
	\left[S^{ a b}, H^{\left(\hat{c} \right)}_{U}  \right] =& \sum_{\substack{n,\eta,s \\n',\eta',s'}} \left\lbrace \sum_{\vec{k} \in \mathcal{C}_{\eta'}} \left[ s^{ab}_{\hat{c}} \mathcal{B} \left( \vec{k} \right) \right]_{n \eta s, n' \eta' s'} \cre{c}{\vec{k},n,\eta,s} \des{c}{\vec{k},n',\eta',s'} \right. \nonumber \\
	-& \left. \sum_{\vec{k} \in \mathcal{C}_{\eta}} \left[ \mathcal{B} \left( \vec{k} \right) s^{ab}_{\hat{c}} \right]_{n \eta s, n' \eta' s'} \cre{c}{\vec{k},n,\eta,s} \des{c}{\vec{k},n',\eta',s'} \right\rbrace \label{si:eqn:commut_u2_generators}.
\end{align}
The careful analysis of the valley indices in \cref{si:eqn:commut_u1_generators} reveals that the commutator $\left[S^{ a b}, H^{\left(\hat{b} \hat{c} \right)}_{U}  \right]$ can only vanish if $s^{ab}_{\hat{b}}$ is diagonal in valley space, as the generators from \cref{si:eqn:cont_generators} do not otherwise preserve crystal momentum. Additionally, the vanishing of the commutator in \cref{si:eqn:commut_u1_generators} in conjunction with the parameterizations in Appendix \ref{app:gauge_single_part:displacement} imply that $s^{ab}_{\hat{b}} = s^{ab}_{\hat{c}}$. Therefore, the generators of continuous symmetries in the presence of displacement field are restricted to the following form
\begin{equation}
	\label{si:eqn:cont_generators_restricted}
	S^{ a b}=\sum_{\substack{m,\eta,s\\n,\eta',s'}} \left[ \sum_{\abs{\delta \vec{k}} \leq \Lambda} \left(s^{ab} \right)_{m \eta s, n \eta' s'} \cre{b}{\delta \vec{k} + \eta \vec{q}_1,m,\eta,s} \des{b}{\delta \vec{k} + \eta' \vec{q}_1,n,\eta',s'} + \sum_{\vec{k} \in \mathrm{MBZ} } \left(s^{ab} \right)_{m \eta s, n \eta' s'} \cre{c}{\vec{k},m,\eta,s} \des{c}{\vec{k},n,\eta',s'} \right],
\end{equation} 
which was obtained from \cref{si:eqn:cont_generators} by setting $s_{\hat{b}}^{ab} = s_{\hat{c}}^{ab} = s^{ab}$. We additionally require that the representation matrices $s^{ab}$ are diagonal in valley subspace, but we make no restriction on their action in the band or spin subspaces. 

Irrespective of the physical limit of TSTG we consider, the generators from \cref{si:eqn:cont_generators_restricted} must at least obey the commutation relations in \cref{si:eqn:commut_obqg_generators,si:eqn:commut_ocqg_generators,si:eqn:commut_enD_generators}. These are enough to restrict the possible representation matrices to the set
\begin{equation}
	s^{0 b} = \zeta^0 \tau^0 s^b, \qquad
	s^{z b} = \zeta^0 \tau^z s^b,
\end{equation}
for $b=0,x,y,z$, since they must be diagonal in valley space. The corresponding generators obtained from \cref{si:eqn:cont_generators_restricted} will also obey the commutation relations in \cref{si:eqn:commut_enTBG_generators,si:eqn:commut_u1_generators,si:eqn:commut_u2_generators}  under any of the physical limits considered. We conclude that the introduction of displacement field breaks the symmetries of the many-body projected TSTG Hamiltonian to the trivial spin-valley $\UN{2} \times \UN{2}$ rotation symmetry. 
\end{document}